\documentclass[phd,12pt]{psuthesis}
 \pdfoutput=1 

\usepackage[T1]{fontenc}
\usepackage{lmodern}
\usepackage{textcomp}
\usepackage{microtype}

\usepackage{amsmath}
\usepackage{amssymb}
\usepackage{amsthm}
\usepackage{subfigure}
\usepackage{eqlist} 
\usepackage[nosepfour,warning,np,debug,autolanguage]{numprint}
\usepackage{acro}
\usepackage{booktabs}
\usepackage[final]{graphicx}
\usepackage[dvipsnames]{color}
\DeclareGraphicsExtensions{.pdf, .jpg}

\usepackage{url}
\usepackage{bbm} 
\usepackage{wrapfig}
\usepackage{amsmath}
\usepackage{makecell}
\usepackage{multirow} 

\usepackage{cite}

\usepackage{titlesec}
\usepackage{float}

\usepackage{hyperref}

\input{UserDefinedCommands}

\title{Attractor-based models for sequences and pattern generation in neural circuits}

\author{Juliana Londono Alvarez}
\dept{Mathematics}
\degreedate{December 2024}
\copyrightyear{2024}

\bachelorsdegreeinfo{for a baccalaureate degree \\ in Engineering Science \\ with honors in Engineering Science}

\documenttype{Dissertation}

\submittedto{The Graduate School}

\collegesubmittedto{ }

\numberofreaders{4}

\honorsadvisor{Honors P. Advisor}
{Associate Professor of Engineering Science and Mechanics}
\honorsadvisortwo{Honors P. Advisor, Jr.}
{Professor of Engineering Science and Mechanics}

\secondthesissupervisor{Second T. Supervisor}

\escdepthead{Department Q. Head}
\escdeptheadtitle{P. B. Breneman Chair and Professor 
of Engineering Science and Mechanics
}

\advisor[Dissertation Advisor][Chair of Committee]{Carina Curto}{Professor of Mathematics}

\readerone{Vladimir Itskov}{Associate Professor of Mathematics}

\readertwo{Timothy Reluga}{Associate Professor of Mathematics and Biology}

\readerthree{Dezhe Z Jin}{Associate Professor of Physics}

\readerfour[Head of the Department of Mathematics]{Paul A Milewski}{Professor of Mathematics}

\definecolor{gray75}{gray}{0.75}
\newcommand{\hsp}{\hspace{15pt}}
\titleformat{\chapter}[display]{\fontsize{30}{30}\selectfont\bfseries\sffamily}{Chapter \thechapter\hsp\textcolor{gray75}{\raisebox{3pt}{|}}}{0pt}{}{}

\titleformat{\section}[block]{\Large\bfseries\sffamily}{\thesection}{12pt}{}{}
\titleformat{\subsection}[block]{\large\bfseries\sffamily}{\thesubsection}{12pt}{}{}

\usepackage{listings}
\begin{document}
\pagestyle{fancy}
\fancyhead[L,C,R]{}
\fancyfoot[L,R]{}
\fancyfoot[C]{\thepage}
\renewcommand{\headrulewidth}{0pt}
\renewcommand{\footrulewidth}{0pt}
\frontmatter

%

\psutitlepage

\psucommitteepage

%
\chapter*{Abstract}
    \begin{singlespace}
Neural circuits in the brain perform a variety of essential functions, including input classification, pattern completion, and the generation of rhythms and oscillations that support processes such as breathing and locomotion \cite{Marder-CPG}. There is also substantial evidence that the brain encodes memories and processes information via \textit{sequences} of neural activity. In this dissertation, we are focused on the general problem of how neural circuits encode rhythmic activity, as in central pattern generators (CPGs), as well as the encoding of sequences. Traditionally, rhythmic activity and CPGs have been modeled using coupled oscillators. Here we take a different approach, and present models for several different neural functions using threshold-linear networks. Our approach aims to unify attractor-based models (e.g., Hopfield networks)  which encode static and dynamic patterns as attractors of the network. 

In the first half of this dissertation, we present several attractor-based models. These include: a network that can count the number of external inputs it receives; two models for locomotion, one encoding five different quadruped gaits and another encoding the orientation system of a swimming mollusk; and, finally, a model that connects the fixed point sequences with locomotion attractors to obtain a network that steps through a sequence of dynamic attractors. In the second half of the thesis, we present new theoretical results, some of which have already been published in \cite{Parmelee2022}. There, we established conditions on network architectures to produce sequential attractors. Here we also include several new theorems relating the fixed points of composite networks to those of their component subnetworks, as well as a new architecture for layering networks which produces ``fusion'' attractors by minimizing interference between the attractors of individual layers.

    \end{singlespace}
\newpage

\setcounter{page}{5}

\thesistableofcontents

\begin{singlespace}
\renewcommand{\listfigurename}{\sffamily\Huge List of Figures}
\setlength{\cftparskip}{\baselineskip}
\addcontentsline{toc}{chapter}{List of Figures}
\listoffigures
\end{singlespace}
\clearpage

\begin{singlespace}
\renewcommand{\listtablename}{\sffamily\Huge List of Tables}
\setlength{\cftparskip}{\baselineskip}
\addcontentsline{toc}{chapter}{List of Tables}
\listoftables
\end{singlespace}
\clearpage


%
\chapter{Acknowledgments}
\begin{singlespace}
    I want to extend my deepest gratitude to my advisor, Prof. Carina Curto, for her constant support, guidance, and encouragement. Her expertise and mentorship have been invaluable in shaping not only my current research program but also my identity as a researcher. Her belief in me and the strong role model she is has been a truly transformative experience in this journey.

I would also like to express my appreciation to all my collaborators for their contributions to our shared projects, especially to Prof. Katie Morrison for her incredible support and valuable advice, which greatly contributed to the research presented in this dissertation.

Thanks also to labmates, but especially to Nikki Sanderson, for their continuous academic, professional, and personal support, which has been instrumental in navigating the challenges of graduate school.

I extend my appreciation to my committee members for taking the time to provide valuable feedback, constructive criticism, and insightful suggestions to this work. Their expertise has been instrumental in refining my ideas and strengthening the quality of my current and future work.

I would also like to acknowledge the Department of Mathematics for providing a nurturing environment for research, learning, and teaching. The resources offered by the department have played a crucial role in strengthening my professional profile. Special thanks to Profs. Cheryl Hile and Kathryn Stewart for their exceptional mentorship during my time as a graduate instructor. I am the most fortunate to have had such dedicated and inspiring teaching mentors.

I am also indebted to all those who have supported me personally during this time. Their presence in my life has been indispensable in keeping a healthy research-life balance.

Finally, I am immensely grateful for the funding provided by the Department of Mathematics, the NIH grant R01 EB022862 and NSF grant DMS-1951165, without which this research would not have been possible. The material in Chapter \ref{ch:open-questions} is based upon work supported by the National Science Foundation under Grant No. DMS-1929284 while the author was in residence at the Institute for Computational and Experimental Research in Mathematics in Providence, RI, during the Math + Neuroscience program. The findings and conclusions in this dissertation do not necessarily reflect the view of the funding agencies. 
\end{singlespace}


\thesismainmatter

\allowdisplaybreaks{
%

\chapter{Introduction} \label{ch:introduction}

Attractor neural networks play an important role in computational neuroscience by providing a rich framework for modeling the dynamic behavior of neural systems and giving insights into how the brain might process information and perform computations \cite{Amit-ANNs, Khona2022}. Originally devised as models of associative memory, these networks were designed to store static patterns representing discrete memories as attractors. Their ability to simultaneously encode multiple static patterns, represented as fixed points, made them ideal for this purpose. Perhaps the most well-known example is that of Hopfield networks, a foundational example of attractor neural networks \cite{Hopfield1}. As illustrated in Figure \ref{fig:sequential-control-cartoons-intro}A, in the classical Hopfield paradigm, memories are stored in the network as several coexistent stable fixed points, each one accessible via distinct input pulses (represented as color-coded pulses in the figure). The state space is partitioned into basins of attraction, and each input will start the trajectory into one of these basins. Coexistence of attractors, even if it's just multiple stable fixed points, requires nonlinear dynamics.
\begin{figure}[!h]
	\centering
	\includegraphics[width=\textwidth]{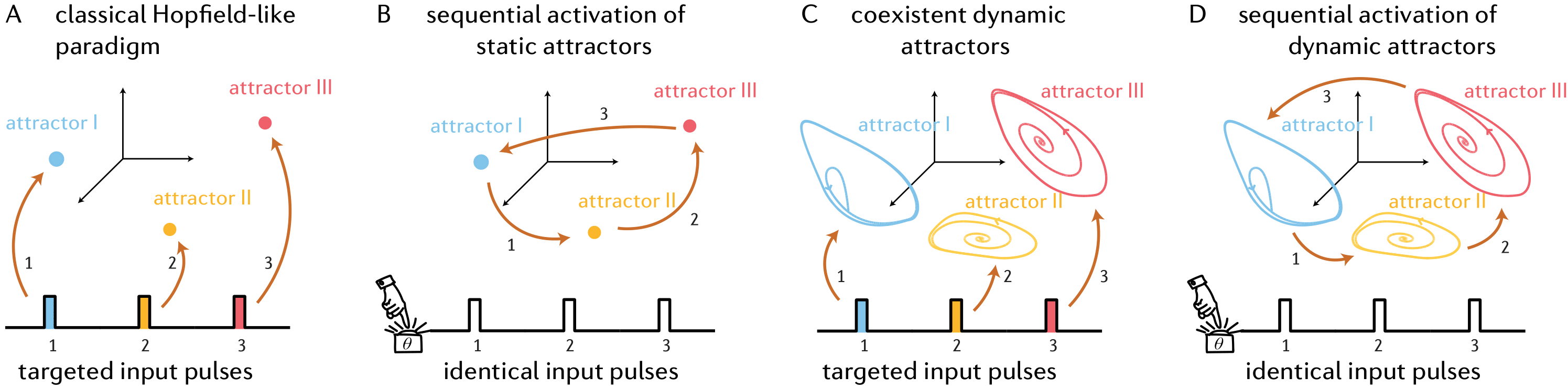}
	\caption[Summary of models as attractor networks]{\textbf{Summary of models as attractor networks}. 
		(A) Classical Hopfield-like paradigm. Multiple stable fixed points (stable attractors) are encoded in the same network, each one accessible via distinct (colored) input pulses.
		(B) Multiple stable fixed points are encoded in the same network, each one accessible via identical pulses. Since the pulses are all identical, the network is \emph{internally} encoding a sequence of fixed points. Each identical pulse causes one transition in the sequence, as indicated by the corresponding numbers in the pulses and arrows.  
		(C) Multiple \textit{dynamic} attractors encoded in the same network, each one accessible via distinct (colored) input pulses. 
		(D) Internally encoded sequence of dynamic attractors, each step of the sequence accessible via identical inputs.}
	\label{fig:sequential-control-cartoons-intro}
\end{figure}


While Hopfield networks are well-suited for encoding multiple static patterns, some patterns of neural activity are better stored as \emph{dynamic} attractors. Such is the case for the rhythms and oscillations produced by Central Pattern Generator circuits (CPGs). CPGs are neural circuits that generate rhythmic patterns that control movements like walking, swimming, chewing, and breathing \cite{Marder-CPG}. Unlike static patterns (such as images), these rhythmic processes require attractors whose neurons fire \textit{sequentially}, meaning that neurons take turns to fire. Furthermore, a single CPG network should potentially be able to encode multiple such patterns. Certainly, animals have several locomotive gaits, all of which activate the same limbs \cite{CPG-review,Inhibition-hippocampus}. How can all these different overlapping dynamic patterns be produced by the same network? Modeling this using attractors requires multi-stability of \textit{dynamic} attractors, which is known to be a difficult problem \cite{Pisarchik2014,Feudel2018}.

Traditional models of locomotion and other CPGs have tackled these challenges by using coupled oscillators \cite{Kopell1,Kopell2,Golubitsky-nature,Collins1993,Dutta2019,Ijspeert2008}. In these models, the parameters are typically adjusted depending on the desired pattern, effectively altering the dynamical system in use. For various reasons, this approach presents some challenges. For instance, while there is evidence of pacemaker neurons, not all neurons are intrinsic oscillators \cite{Penn2016}. Additionally, assuming that synaptic strengths change every time we transition between different locomotive gaits necessitates additional ingredients for the model, such as synaptic plasticity. Despite this, coupled oscillator models have remained popular for important reasons. One of the reasons they are so widely used is the availability of theoretical results coming from physics and mathematics \cite{Ashwin2016}. Indeed, many CPG models stemmed from the physics and mathematics communities \cite{Kopell1,Kopell2,Golubitsky-nature,Collins1993}. 

The availability of theoretical tools is indeed a very compelling argument for using coupled oscillators to model central pattern generators. Here we aim to take a different approach that also leverages recent theoretical advances, but within the world of attractor neural networks. The framework of attractor neural networks differs from traditional coupled oscillator models in two key ways. First, neurons do not intrinsically oscillate; instead, patterns of activity emerge as a result of connectivity. Second, these patterns are true attractors of the system, making them more robust and stable to noise and other perturbations. Additionally, providing models within this framework would unify our approach to CPGs with other classical models, like the aforementioned associative memory models, ring attractors, etc \cite{Khona2022,Amit-ANNs}. On the biological side, it has also been suggested that cortical circuits involved in associative memory encoding and retrieval have many features in common with CPGs \cite{Yuste-CPG}.

To accomplish this, we propose the use of Threshold-Linear Networks (TLNs). TLNs are recurrent neural networks with simple non oscillating units and piecewise linear activation \cite{Hartline1958}. This makes them focus on the role of connectivity in emergent behaviors. TLNs have a rich history as neural models \cite{AppendixE,Tsodyks1997,Seung-Nature,Lappalainen2023,Biswas2022}, and more importantly, they are supported by a wide array of theoretical results \cite{XieHahnSeung, Hahnloser2003-ma, net-encoding, book-chapter,stable-fp-paper}, many of them recent \cite{Curto2016,fp-paper,Parmelee2022,CTLN-preprint}.

One such key finding is that threshold-linear networks with symmetric connectivity matrices can only have stable fixed point attractors (static patterns) \cite{Hopfield1,HahnSeungSlotine}, which is why here we use non-symmetric TLNs, introduced in Chapter \ref{ch:background}. These are known to give rise to a rich variety of non-linear dynamics including multi-stability, limit cycles, chaos and quasi-periodicity. Therefore, we expect it to be possible to simultaneously encode multiple dynamic patterns of activity, whereas in classical Hopfield networks the stored patterns are all static. Within this unified framework, our aim is to provide attractor models for three broad neural functions, as summarized in Figure \ref{fig:sequential-control-cartoons-intro}, and detailed below:

First, we propose a simple neural integrator model that is both robust to noise and can count inputs, using fixed point attractors. Neural integration refers to the process in which information from various sources is combined to create an output. For instance, counting the number of left and right cues to make a decision is quite literally integration (summation) in the mathematical sense. Here, we propose to model a discrete counter as a sequence of static attractors, as shown in Figure \ref{fig:sequential-control-cartoons-intro}B.  While this concept is akin to the classical Hopfield model (Figure \ref{fig:sequential-control-cartoons-intro}A), our aim is to \emph{internally} encode the sequence of fixed points, meaning that the input pulses are all identical and contain no information about which fixed point comes next in the sequence (i.e. they are just like pushing a single input button).

Second, we aim to devise a (small) network that has attractors corresponding to 4-5 distinct, but overlapping, quadruped gaits. The goal is for the attractors to coexist in the same network so that they can be accessed by different initial conditions, and without changing parameters. This requires the presence of several coexistent \textit{dynamic} attractors, as pictured in Figure \ref{fig:sequential-control-cartoons-intro}C, arising as distinct limit cycles in state space. Recall that the simultaneous encoding of multiple dynamic attractors (non fixed point attractors) is a network is challenging, especially when the attractors have overlapping units.  While classic models circumvent this challenge by adjusting synaptic weights, we aim to obtain a single fixed network, which in turn means a simpler model with fewer control parameters. 

In addition to quadruped gaits, we will also model ``Clione's hunting system '' \cite{Panchin1995neuronalmechanisms}, which is a different CPG example, using the same framework. Previous models for it have also used intrinsically oscillating units and fine-tuned parameters \cite{Varona2002}. Here we intend to devise a more robust network for this using attractors to avoid the need for finely-tuned parameters.

Third, we combine the modeling approaches from panels B and C to devise a network that can step through a set of dynamic attractors \textit{sequentially}, as in Figure \ref{fig:sequential-control-cartoons-intro}D. Can different attractors be linked together so that they can be activated in sequence, where the sequence itself is stored within the network? This could be useful for modeling sequences of complex movements, such as a choreographed sequence of dance moves, for instance. 

\paragraph*{Sequences of sequential attractors.} Note that in Figure \ref{fig:sequential-control-cartoons-intro}C, we are dealing with dynamic attractors, that are sequential themselves, meaning attractors whose nodes activate in an ordered sequence \cite{Parmelee2022}. This definition does not completely exclude attractors in which there is some synchrony in the activations. So for example, we consider both attractors in Figure \ref{fig:sequential-attractors-examples} \emph{sequential attractors}, even though the attractor on the right has nodes 2 and 3 synchronized (we will later formalize the dynamic prescription of the nodes, for now note the sequential of activations of nodes).
\begin{figure}[!h]
	\centering
	\includegraphics[width=\textwidth]{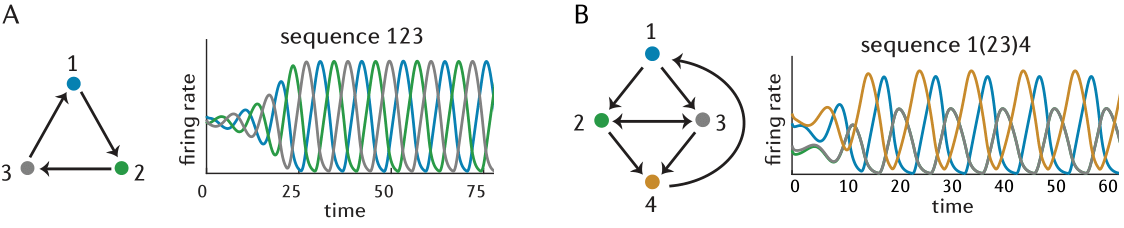}
	\caption[Examples of sequential attractors]{\textbf{Examples of sequential attractors}. 
		(A) Nodes 1,2,3 are activating in sequence, as seen by the curves to the left of the graph. 
		(B) Node 1 activates, then 2 and 3 simultaneously, then node 4.}
	\label{fig:sequential-attractors-examples}
\end{figure}

These are great to model rhythmic activations like those of CPGs. In contrast, Figure \ref{fig:sequential-control-cartoons-intro}D deals with \textit{sequences of dynamic attractors}, which means that different attractors, either static or dynamic, are activated in a specific ordered sequence (e.g., attractor A, then attractor B, then attractor C). With this distinction clear, our ultimate goal is to achieve an internally encoded sequence of sequential attractors.

Internally encoded sequences of sequential attractors are good models for complex sequences of movements, like choreographed dancing. These complex motor behaviors have also been modeled in the past using threshold-linear recurrent networks that choose and learn motor motifs, but whose choice mechanism requires plasticity, and where the sequence's order is externally encoded \cite{Logiaco2021}. It is not uncommon to think of broader cognitive sequential processes as a recombination of several pre-stored patterns, which offers an efficient alternative to re-encoding patterns with each occurrence. Studies in this vein include \cite{Mikhail2018}, where they utilize a combination of discrete metastable states, leveraging winnerless competition of oscillator neurons. Similar mechanisms have been explored using boolean and spiking networks \cite{Suresh2010,Cui2020}.

\paragraph*{Desired properties of models.}
As it turns out, the versatility of threshold linear networks will prove ideal for modeling both sequential attractors and sequences of them. To summarize the discussion above, we want our models to satisfy the following properties:
\begin{enumerate}
	\item Neurons are \textit{not} intrinsic oscillators.
	\item Stored static and dynamic patterns should emerge as \textit{attractors} of the network, rather than being fine-tuned trajectories. Static patterns should manifest as fixed points, while dynamic patterns should arise from non fixed point attractors, like limit cycles.
	\item A network's attractors should be accessible via different initial conditions, easily implemented via input pulses that target subsets of neurons.
	\item A sequence of attractors should be accessible via a series of identical inputs pulses, with the sequence itself stored within the network (possibly in a separate layer, as observed in some biological brains).
	\item The models should be mathematically tractable--that is, simple enough to be analyzed mathematically.
\end{enumerate}

These properties will distinguish our models from previous coupled oscillator models and position them within the framework of attractor neural networks. We begin by adhering to the last point above, by choosing a framework that fits into the attractor neural network paradigm and also provides mathematically tractable models. With tractability also come great simplifications. Although TLNs are inspired by networks of biological neurons, real neurons and their interactions are of course far more complex than TLNs paint them to be. TLNs remain useful however because they capture two fundamental pieces of biological networks: connectivity and threshold-activation. This is why here we also focus on another simplification of TLNs, known as Combinatorial Threshold-Linear Networks (CTLNs) \cite{CTLN-preprint, book-chapter,fp-paper}. CTLNs are a special family of TLNs, whose connectivity matrix is defined by a simple directed graph (giving rise to binary connections), as in Figure \ref{fig:sequential-attractors-examples}. Their added simplicity can be used to gain further theoretical results. 

\paragraph*{Summary of models.} The table below lists all the models included in the dissertation. Each row defines a single model/network, and each is an example of the attractor behaviors described by Figure \ref{fig:sequential-control-cartoons-intro}:
\begin{center}
	\begin{tabular}{c|c|c|c}
		Model & Network function & Type of attractor & Chapter \\ 
		\hline\hline
		Model 1a & counter & sequence of static attractors & \ref{ch:sequences-of-attractors}\\  
		\hline
		Model 1b & signed counter & sequence of static attractors & \ref{ch:sequences-of-attractors}\\  
		\hline
		Model 1c & dynamic attractor chain & \makecell{sequence of identical \\ dynamic attractors}  & \ref{ch:sequences-of-attractors}\\
		\hline
		Model 2a & quadruped gaits &  \makecell{coexisting distinct \\ dynamic attractors}  & \ref{ch:CPGs}\\
		\hline
		Model 3a & molluskan swimming & \makecell{coexisting identical \\ dynamic attractors}  & \ref{ch:CPGs}\\
		\hline
		Model 2b & \makecell{sequential control of \\ quadruped gaits} & \makecell{sequence of distinct \\ dynamic attractors}  & \ref{ch:sequential-control}\\
		\hline
		Model 3b & \makecell{sequential control of \\ molluskan swimming} & \makecell{sequence of identical \\ dynamic attractors} & \ref{ch:sequential-control}\\
	\end{tabular}
\end{center}

In Chapter \ref{ch:sequences-of-attractors}, we introduce three models for sequences of attractors. Models 1a and 1b are two counter networks that step through sequences of fixed points. Models 1a is shown in Figure \ref{fig:summary-of-models1}A, where we can see that identical pulses move the network into the next stable fixed point, where it stays until it receives another pulse. This network serves as robust discrete neural integrators of inputs, as we show in their chapter by doing a thorough robustness analysis. Additionally, we extend our work to dynamic attractors by presenting an additional network, Model 1c, capable of encoding a sequence of \emph{dynamic} attractors. These dynamic attractors are all qualitatively identical, and because of this, there are some symmetries in the basins of attractors, and thus all attractors easily accessible via distinct input pulses. However, to effectively model CPGs, we require \textit{different} types of patterns to coexist simultaneously. How can we achieve this?
\begin{figure}[!h]
	\centering
	\includegraphics[width=\textwidth]{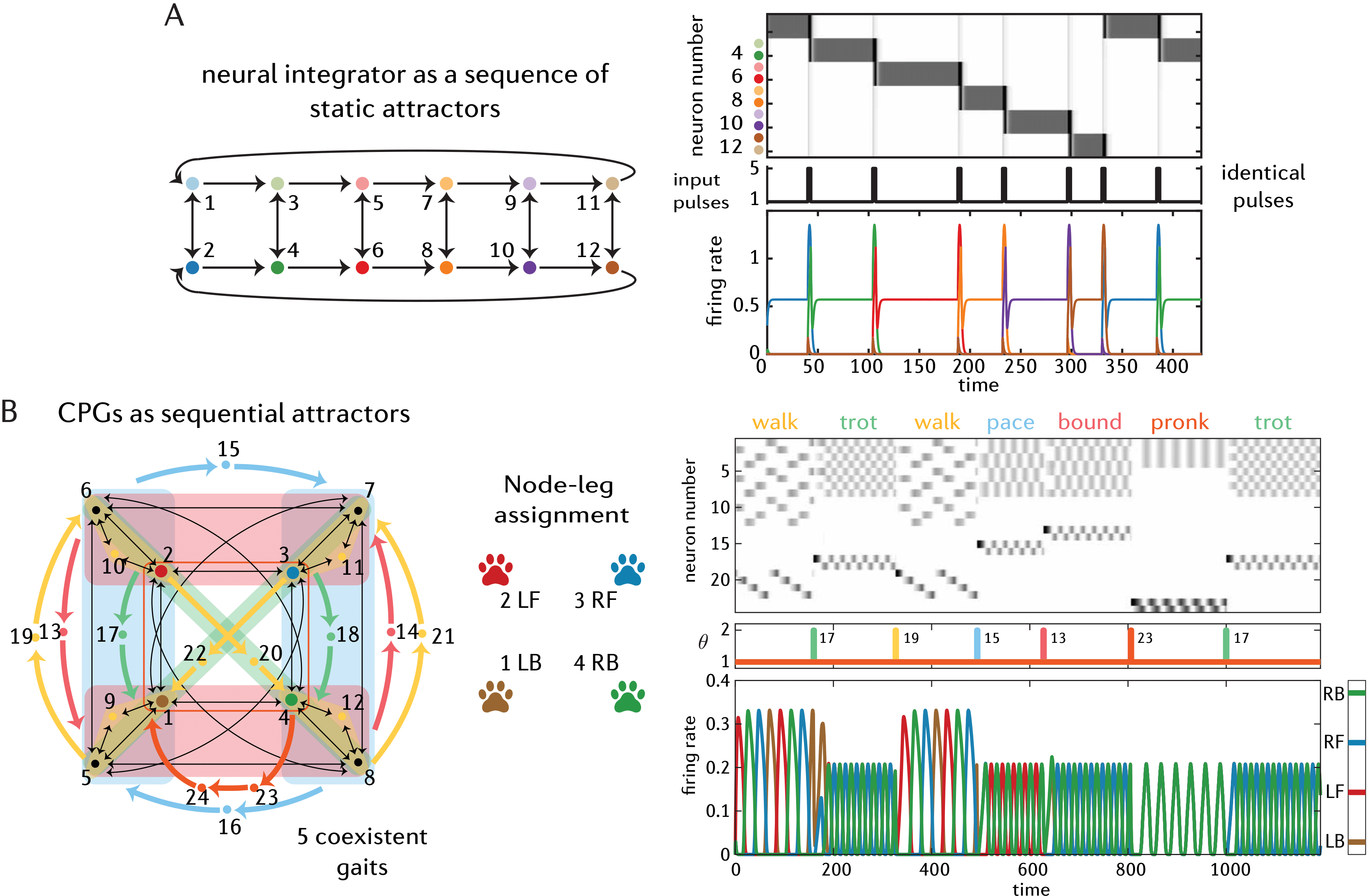}
	\caption[Summary of models I]{\textbf{Summary of models I.}
		(A) Model 1a: counter network from Chapter \ref{ch:sequences-of-attractors}. Pulses are all identical (in black).
		(B) Model 2a: quadruped gaits network from Chapter \ref{ch:CPGs}. Pulses are attractor-specific (colored according to stimulated node).
	}
	\label{fig:summary-of-models1}
\end{figure}
 
That is the content of Chapter \ref{ch:CPGs}, where we model two different CPGs, Model 2a and 3a, which require sequential activation of neurons. Model 2a, developed in in Section \ref{sec:quadruped-gaits}, is pictured in Figure \ref{fig:summary-of-models1}B. It consists of a network encoding five different quadruped gaits as coexistent limit cycles in a 24 unit network. There, we see that all gaits coexist and are accessible via gait-specific pulses. We do a thorough analysis of its dynamics via the set of fixed point supports, and the effect of parameters in modulating gait characteristics. Model 3a, in Section \ref{sec:molluskan-hunting}, consists of a network encoding swimming orientation of a marine mollusk (Clione). Since the attractors are all identical, we manage to prove symmetry of its basins of attraction in Theorem \ref{thm:octahedral-basins}.
\begin{figure}[!h]
	\centering
	\includegraphics[width=\textwidth]{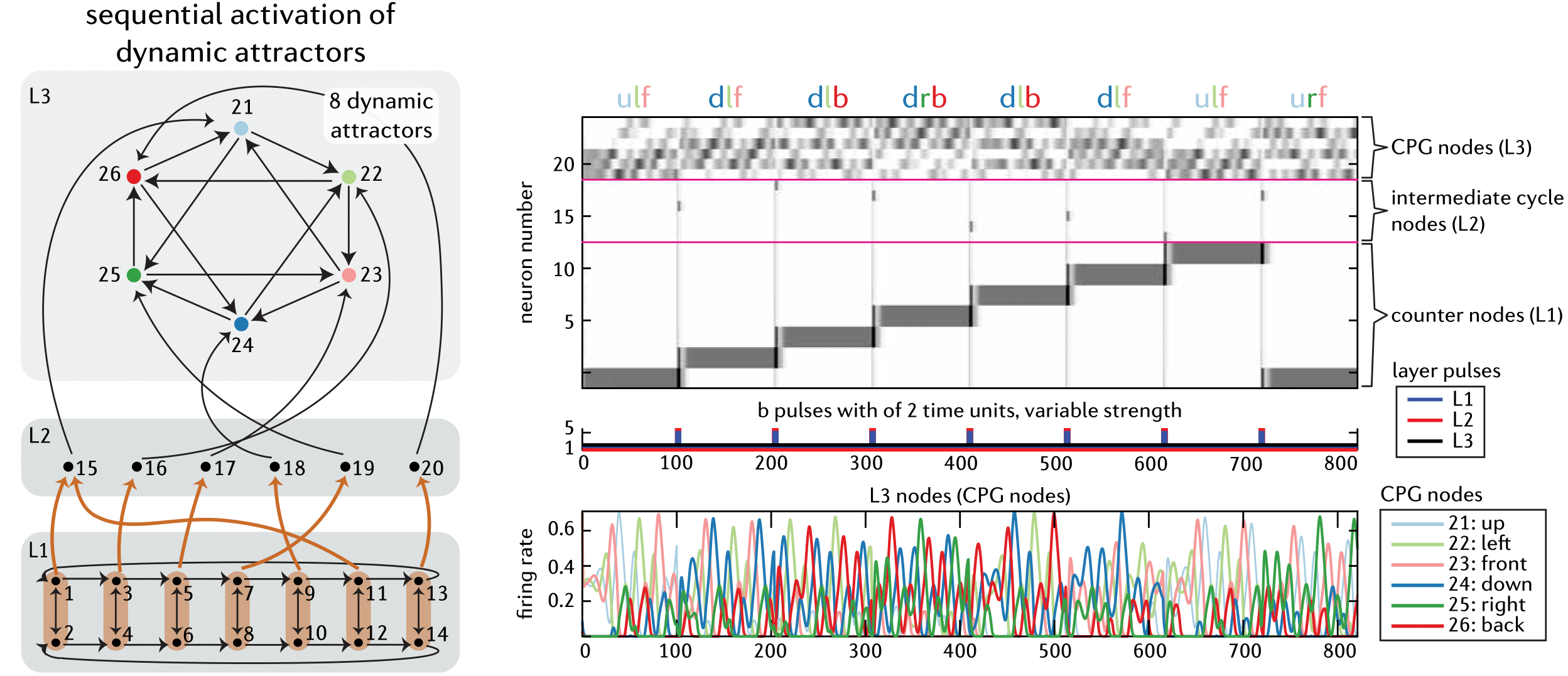}
	\caption[Summary of models II]{\textbf{Summary of models II.} 
		Model 3b: sequential control of molluskan swimming directions from Chapter \ref{ch:sequential-control}. Pulses are identical by layer. Attractors are labeled on top of the greyscale by which direction is active (up/down, left/right, front/back). These are also color-coded by node.
	}
	\label{fig:summary-of-models2}
\end{figure}

Finally, in Chapter \ref{ch:sequential-control}, we merge the concepts introduced in Chapters \ref{ch:sequences-of-attractors} and \ref{ch:CPGs} to achieve \emph{sequences of sequential attractors}. From Chapter \ref{ch:sequences-of-attractors}  we get the counter network that will encode the sequence transitions, and from Chapter \ref{ch:CPGs} we get the coexistent sequential attractors. From this integration we obtain Models 2b and 3b: a network for the sequential control of quadruped locomotion and for the sequential control directing swimming movements in Clione. The latter is the one pictured in Figure \ref{fig:summary-of-models2}, where we observe the attractors from the CPG network ``fuse'' with the attractors of the counter network, as both are simultaneously active, and look qualitatively like they did when isolated. Figure \ref{fig:summary-of-models2} shows the resulting network using Clione's model, but it can also be done, analogously, with the five-gait network. The fact that we could use the exact same construction with two different networks, led us to believe this is an even more general phenomenon, arising from some structural constraints on these networks, ad they were indeed built with similar principles.

Code to reproduce the plots in Figures \ref{fig:summary-of-models1} and \ref{fig:summary-of-models2}, and also all models listed in the table, can be found at \url{https://github.com/juliana-londono/phd-thesis-basic-plots}.

Note that in Figure \ref{fig:summary-of-models2}, we see  a ``blend'' of two different attractors: at the top of they greyscale we see the dynamic attractors coming from layer L3, and at the bottom we see fixed points coming from layer L1. This phenomenon was also observed in \cite{rule-of-thumb}, where it was called \textit{fusion attractors}. Fusion attractors offer a clean solution for managing sequences of static and dynamic attractors. Understanding the mechanisms behind this phenomenon motivates us to further explore the underlying structural constraints giving rise to it, from a theoretical standpoint. This is why we now transition from models to theory.

\paragraph*{New network theory.} We want to note that all the models we have developed thus far were built within the TLN framework, for which there are plenty of well-established theoretical results. This theoretical foundation made the process of building these models a lot easier. However, our models have now surpassed the available theory and so now they serve as sources of inspiration for the development of new theoretical results. This is why in the second part of the dissertation, we take a reverse approach: while theory initially guided our modeling efforts, now the models are leading the development of new theoretical results.

Chapter \ref{ch:new-theoretical-results} presents original theoretical contributions, including several results recently published in the paper I co-authored: "Sequential Attractors in Combinatorial Threshold-Linear Networks" \cite{Parmelee2022}. This chapter is divided in three parts. First, in Section \ref{sec:technical-background}, we establish some necessary technical results, some of which are earlier version of results that we end up generalizing in this chapter. Then, in Section \ref{sec:simply-embedded-CTLNs}, we derive new structural theorems for CTLNs supporting sequential attractors. All of the results within this section are my contribution to \cite{Parmelee2022}, which contains several other architectures that support sequential attractors. All theorems I proved are in bold. Most of these results relate the fixed point supports of a network to the fixed point supports of component subnetworks, as follows:
\begin{itemize}
	\item \textbf{Theorem \ref{thm:menu-CTLN}} for ``simply-embedded partitions'',  constrains the possible fixed point supports of a network to unions of fixed points chosen from a particular \emph{menu} of component subnetwork fixed point supports. This is generalizing results from \cite{fp-paper}. In the same section, \textbf{Theorem \ref{thm:removables}} and \textbf{Corollary \ref{cor:removables}} give conditions on when a node can be removed from a network without changing the set of fixed points supports. We include here a new result on removable nodes, that has not been published: \textbf{Theorem \ref{thm:simply-embedded-domination}}.
	\item \textbf{Theorem \ref{thm:linear-chain}} for ``simple linear chains'', showing that the set of fixed point supports of a simple linear chain network is closed under unions of ``surviving'' component fixed point supports.
	\item \textbf{Theorem \ref{thm:bidir-sa-partition}} for ``strongly simply-embedded partitions'', showing that the set of fixed point supports of a network can be fully determined from knowledge of the component fixed point supports together with knowledge of which of those component fixed points ``survive'' in the full network. 
\end{itemize}

Finally,  in Section \ref{sec:layered-ctlns}, we extend some of these theorems to TLNs and provide theoretical explanations for the fusion attractors observed in Chapter \ref{ch:sequential-control}, culminating with:
\begin{itemize}
	\item \textbf{Theorem \ref{thm:simply-embedded-si-factorization}}, which is an important technical result generalizing previous theorems on certain determinant factorizations that control the set of fixed point supports of a network. It relies on a new determinant factorization lemma, \textbf{Lemma \ref{lemma:det-linear-algebra}}, which I have also proven. \textbf{Theorem \ref{thm:simply-embedded-si-factorization}} is then used as a crucial ingredient in the proofs of: \textbf{Theorem \ref{thm:menu}}, generalizing Theorem \ref{thm:menu-CTLN} above. \textbf{Theorem \ref{thm:strongly-simply-layered}}, explaining how the fixed points of some special networks are formed from fixed point of smaller component networks. That theorem generalizes both Theorem \ref{thm:bidir-sa} (from two components to $N$ components) and Theorem \ref{thm:bidir-sa-partition} (from several CTLN components to several TLN components). And finally, we present a similar result but for ``nested'' component fixed point supports in \textbf{Theorem \ref{thm:simply-embedded-nested}}.   
\end{itemize} We also show that the networks in Chapter \ref{ch:sequential-control} satisfy these conditions, thus explaining the fusion attractors observed there.

In this dissertation's final chapter, Chapter \ref{ch:open-questions}, we present partial and further theoretical results derived from projects in sections \ref{sec:quadruped-gaits} and \ref{sec:layered-ctlns}. In Section \ref{sec:degeneracy}, Lemma \ref{lemma:vector-field-interpolation}, gives conditions under which the same attractor can arise from two different networks. This phenomenon is known as degeneracy.  All code from this section is available online at \url{https://github.com/juliana-londono/TLN-attractor-interpolation}. In Section \ref{sec:chirotope}, Lemma \ref{lemma:vanishingdetsv2}, gives a new way to think about certain Cramer's determinants, which are at the core of the dynamics of TLNs.

The rest of this dissertation is organized as follows: Chapter \ref{ch:background} introduces the framework, including firing rate models, attractor neural networks, TLNs, and CTLNs. Chapter \ref{ch:sequences-of-attractors} provides models for sequences of static and dynamic attractors, both internally and externally encoded. In Chapter \ref{ch:CPGs}, we provide two CPG models of locomotion, each consisting of several coexistent dynamic attractors, easily accessible via initial conditions or inputs. Chapter \ref{ch:new-theoretical-results} explores new architectures and theoretical results, focusing on sequential dynamics complex and networks made up of simpler subnetworks. Finally, Chapter \ref{ch:open-questions} discusses further theoretical results derived from the presented projects, suggesting avenues for further exploration. Appendix \ref{ch:appendixA} contains the matrices and parameters used to construct all the models, along with some technical calculations of the fixed points of the five-gait quadruped network. That's all. We hope you enjoy reading this dissertation. We encourage you to keep in mind Figure \ref{fig:sequential-control-cartoons-intro}, which is the road map guiding us through the chapters on models.

\chapter{Review of relevant background}\label{ch:background}

This section offers a broad overview of the mathematical and historical background necessary for understanding and contextualizing the subject at hand. More detailed technical background specific to each chapter is provided later, as needed. None of the results in this chapter are original work of my own. Thus, the proofs of these results are not included here and can be found in their original publications, as cited.

\section{Firing rate models and attractors}
In this dissertation, we deal with the dynamics of recurrent neural networks. A recurrent neural network consists of a directed graph along with a prescription of the nodes' dynamics. The nodes are thought of as neurons and edges represent synapses between them. One way to interpret the dynamics of the nodes is as firing rates, which indicate the average frequency at which the neuron generates action potentials (or ``fires'') \cite{Dayan2001}. The dynamics of a firing rate network model can be described by a system of $n$ coupled differential equations:
\begin{equation}\label{eq:ctrnn}
	\tau_i\frac{dx_i}{dt} = -x_i + \varphi\left(\sum_{j=1}^n W_{ij}x_j+b_i\right), \quad i=1,\dots,n,
\end{equation}
where $x_i=x_i(t)$ represents the firing rate of neuron $i$,  $W = [W_{ij}]$ is a matrix prescribing the interaction strengths between neurons, and $b_i = b_i(t)$ is some external input to neuron $i$ (which might vary in time), for $n$ recurrently connected neurons. In Equation \ref{eq:ctrnn}, $\tau_i$ is referred to as the time constant, representing the rate of decay when there is no input to neuron $i$, and $\varphi$ is an activation function (e.g. sigmoid, ReLU). 
 
Also, Equation \ref{eq:ctrnn} indicates that we are viewing recurrent neural networks here as dynamical systems, contrasting with the typical  machine learning perspective that treats neural networks as learning algorithms or black-box function approximators. 

While firing rate models are clearly a simplification of biological neural dynamics, they allow us to focus on factors such as the interactions between neurons, activation functions, and external inputs, and help bridge the gap between detailed spiking neuron models and large-scale network behavior. This provides a valuable framework for understanding the functional roles of these factors in neural computations.

Indeed, firing rate models of recurrent neural networks are a popular tool for studying nonlinear dynamics in neuroscience \cite{Dayan2001,Durstewitz2023,Ermentrout2010,AppendixE}, particularly within the context of attractor neural networks. Attractor neural networks have emerged as a framework for studying neural dynamics, covering various cognitive processes like memory recall, decision-making, and perception \cite{Khona2022,Amit-ANNs}. Understanding how these networks compute is useful for advancing our understanding of how biological networks compute. 

Under this framework, attractors of the system are thought of as representations of some cognitive process or pattern encoded in the system. Hopfield networks are a classical example \cite{Hopfield1}. In these networks, fixed points of the network are interpreted as encoded memories. In Hopfield networks, units can only be in one of two states, and the dynamics are discrete. Importantly, if the connections between neurons are symmetric, then the network is guaranteed to converge to a stable fixed point, which happens to be the minimum of an energy function, pictured as an energy landscape in Figure \ref{fig:attractor-neural-networks}A \cite{Hopfield1,Hopfield2}.
\begin{figure}[!h]
	\begin{center}
		\includegraphics[width=0.65\textwidth]{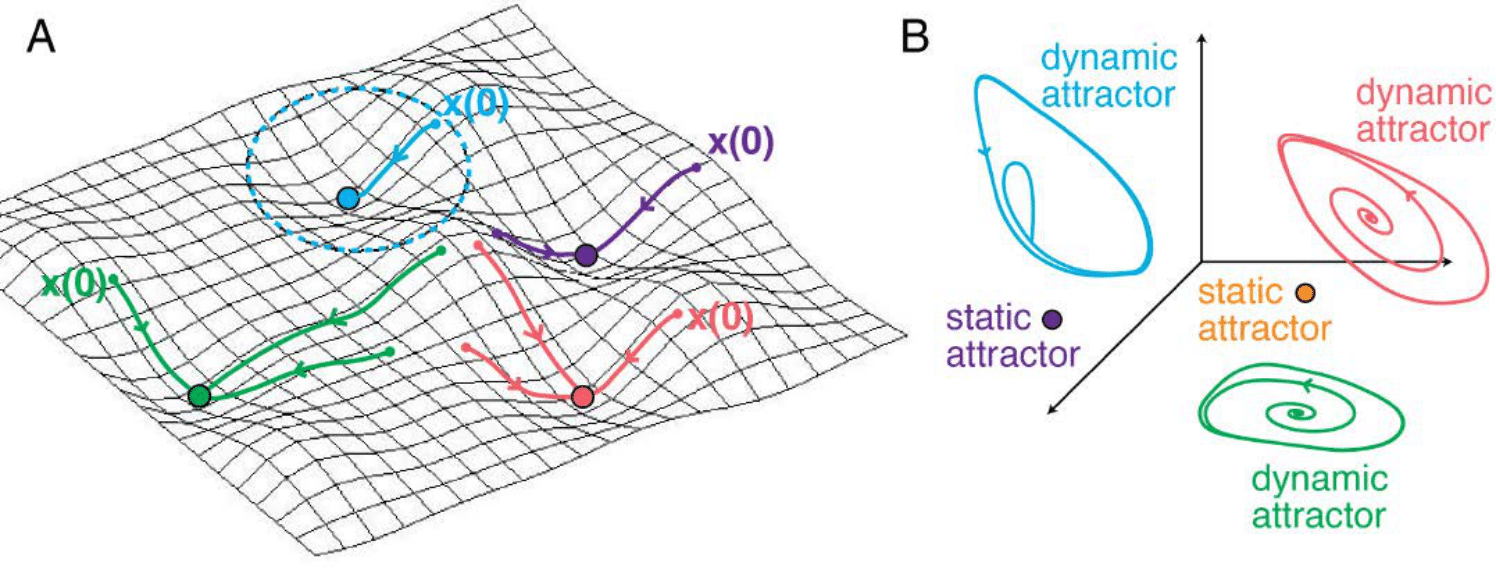}
	\end{center}
	\caption[Attractor neural networks]{(A) Minima of energy landscape for symmetric network. (B) Diversity of attractors of non-symmetric networks. Adapted from \cite{Notices}.}
	\label{fig:attractor-neural-networks}
\end{figure} 

Such theoretical results are not only available for Hopfield networks, but also more broadly for certain continuous-time recurrent neural networks, which are the main subject of this dissertation: threshold-linear networks (TLNs). In the case of TLNs, $\varphi(y) = \max(0,y) = [y]_+$ in Equations \ref{eq:ctrnn} (a.k.a. ReLU activation function, Fig. \ref{fig:relu}). TLNs were introduced around the 50s \cite{Hartline1958}, and have been widely used in computational and mathematical neuroscience \cite{AppendixE,Tsodyks1997,Seung-Nature,Lappalainen2023,Biswas2022} since.
 \begin{figure}[!h]
	\begin{center}
		\includegraphics[width=0.5\textwidth]{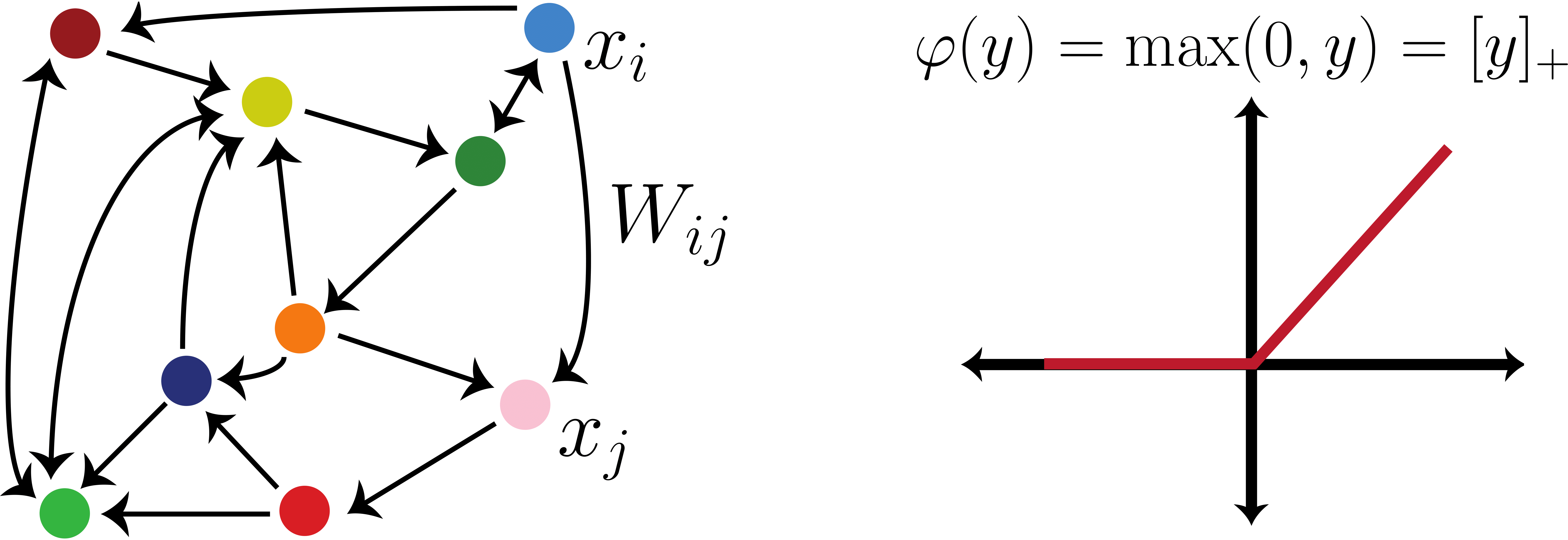}
	\end{center}
	\caption[A recurrent neural network]{A recurrent neural network, and the ReLU activation function.}
	\label{fig:relu}
\end{figure}

A crucial aspect of TLNs is their piecewise linear activation function, which greatly simplifies mathematical analysis. This mathematical tractability has given rise to numerous theoretical results. Notably, as with Hopfield networks, \emph{symmetric} TLNs were shown to converge to stable fixed points under some extra restrictions on $W$ (copositivity) \cite{HahnSeungSlotine}.  Furthermore, there exist constraints on which sets of neurons can be co-active at a stable steady state (forbidden and permitted sets in \cite{HahnSeungSlotine}). These results have been further explored in subsequent studies \cite{net-encoding,Curto2016}.

We are however interested in a broader set of attractors, not only stable fixed points. That is why we focus here in \emph{non-symmetric} inhibition-dominated TLNs, for which $W$ is non-symmetric and non-positive. This choice is motivated by the fact that inhibition fosters competition between neurons, and thus neurons tend to alternate in reaching peak activity levels, leading to sequential behaviors within limit cycles. 

Indeed, these networks, even though still piecewise linear, nonetheless permit rich non-linear dynamics like multi-stability, limit cycles, chaos and quasi-periodicity to arise and, moreover, coexist \cite{CTLN-preprint,rule-of-thumb,book-chapter}, as in Figure \ref{fig:attractor-neural-networks}B. This, along with extra simplifying assumptions (particularly on $W$), has given rise to a robust body of theoretical work characterizing the dynamics of TLNs \cite{XieHahnSeung, Hahnloser2003-ma, net-encoding, Curto2016, CTLN-preprint, book-chapter,fp-paper, stable-fp-paper, Parmelee2022}. 


In particular, a main focus of this dissertation is on a special type of non-symmetric TLNs, called combinatorial threshold-linear networks (CTLNs), introduced in \cite{CTLN-preprint}. CTLNs are TLNs where the connectivity matrix $W$ is prescribed by a simple directed graph, and often the input is assumed to be uniform across neurons. 

Below, we formally introduce TLNs and CTLNs, and present some results that we will use throughout the dissertation. The presentation of these results is adapted from \cite{fp-paper, Parmelee2022}, with slight modifications. Since CTLNs are a special family of TLNs, they inherit many properties from TLNs. Hence, we first provide an overview of the necessary background shared with TLNs, and then state results specific to CTLNs.

\section{Threshold-linear networks (TLNs)}

A threshold-linear network (TLN) is a continuous-time recurrent neural network (Eqns. \ref{eq:ctrnn}) where $\varphi= \max(0,y) = [y]_+$. In addition, we assume for the time being that the input is constant in time $b_i(t) = b_i$ and the timescales are constant and uniform across neurons (without loss of generality, we assume $\tau_i = 1$). More precisely:

\begin{definition}
A threshold linear network on $n$ neurons is a system of ordinary differential equations \begin{equation}\label{eq:TLNdynamics}
		\dfrac{dx_i}{dt} = -x_i + \left[\sum_{j=1}^n W_{ij}x_j+b_i \right]_+, \quad i = 1,\ldots,n,
\end{equation}
where $W_{ij},b_i \in \RR$ for all $i,j = 1,\dots,n$. This can also be written in vector form as $$\frac{dx}{dt} = -x + [Wx+b]_+,$$ where $W$ is an $n \times n$ real matrix and $b \in \RR^{n}$.
\end{definition}

Under these circumstances, a given TLN is completely determined by the choice of the connectivity matrix $W$ and its vector of inputs $b$, so we denote it by $(W,b)$. Unless otherwise specified, $n$ is the total number of neurons in the network. As mentioned above, here we only consider inhibition-dominated TLNs (a.k.a. competitive). This means we assume $W_{ij}\leq 0$ for all $i,j=1,\dots,n$. In addition, we pose some extra restrictions on degeneracies. The following definition requires some Cramer's determinants of some sub-matrices to not vanish, we denote by $((A_\sigma)_i;b_\sigma)$ the matrix the matrix $A_\sigma$ where the column corresponding to the index $i \in \sigma$ has been replaced by $b_\sigma$, where the subindex denotes restriction to the rows/columns given by $\sigma$. In the definition below, and in all that follows, $[n]$ denotes the set of indices $\{1,\dots,n\}$.
\begin{definition}[{\cite[Definitions 1 and 2]{fp-paper}}]\label{def:competitive}\label{def:nondegenerate}
We say that a TLN $(W,b)$ is {\it competitive} if $W_{ij}\leq0$, $W_{ii} = 0$ and $b_i\geq 0$ for all $i,j\in[n]$. We say that it is {\it non-degenerate} if 
	\begin{itemize}
		\item $b_i>0$ for at least one $i \in [n]$
		\item $\det(I-W_\sigma) \neq 0$ for each $\sigma \subseteq [n]$, and 
		\item for each $\sigma \subseteq [n]$ such that $b_i>0$ for all $i \in \sigma$,  the corresponding Cramer's determinant is nonzero: $\det((I-W_\sigma)_i;b_\sigma) \neq 0$. 
	\end{itemize}
\end{definition} Unless otherwise noted, all TLNs here are competitive and non-degenerate.

A good point of entry to explore the dynamics of a competitive TLN are its fixed points: a fixed point $x^*$ of a TLN is a solution that satisfies $dx_i/dt|_{x=x^*} = 0$ for each $i \in [n]$. Per equations \ref{eq:TLNdynamics}, this translates into $$x_i^* = \left[\sum_{j=1}^n W_{ij}x^*_j+b_i \right]_+ \text{ for all } i = 1,\ldots,n.$$ By the defintiion of the $[\cdot]_+$ function, when $y_i = \sum_{j=1}^n W_{ij}x^*_j+b_i \leq 0$,  $y_i$ will evaluate to 0. Thus, these $y_i$ define hyperplanes that divide state space into \textit{chambers}, and inside each of those chambers, Equations \ref{eq:TLNdynamics} define a linear system of ODEs. Under the non-degeneracy assumption, each of those can have exactly one fixed point, though its fixed point might not lie inside the correct chamber.

\begin{figure}[!h]
	\begin{center}
		\includegraphics[width=0.5\textwidth]{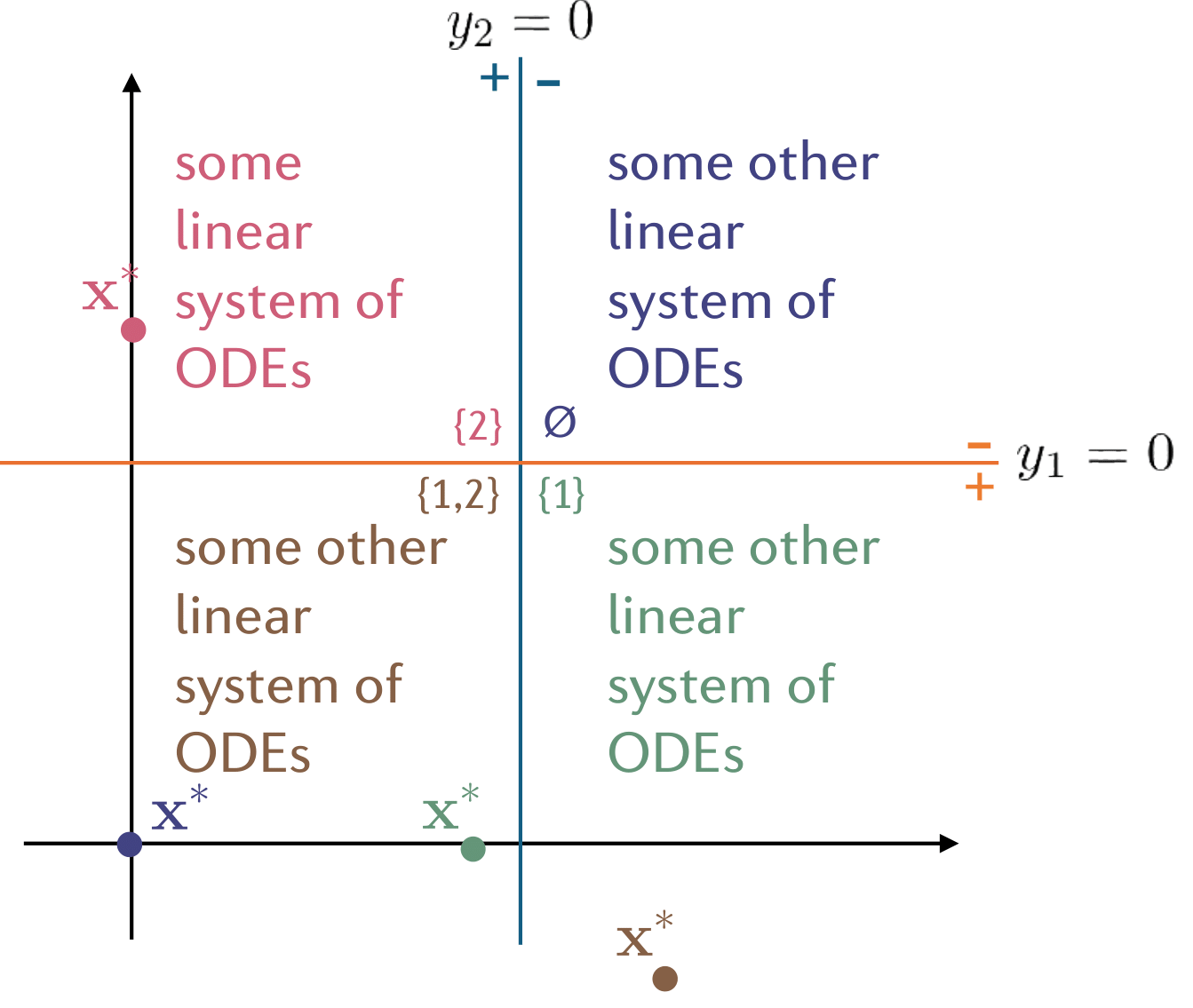}
	\end{center}	
	\caption[Hyperplane chambers cartoon]{\textbf{Hyperplane chambers cartoon.} 2-dimensional state space is divided in $2^2$ chambers by the hyperplanes $y_i = \sum_{j=1}^n W_{ij}x^*_j+b_i=0$, inside each the dynamics are truly linear. The fixed points of the linear systems are color-coded by chamber.}	
	\label{fig:chambers-cartoon}
\end{figure}
In Figure \ref{fig:chambers-cartoon}, for example, we have marked the four fixed points associated to the 4 linear systems with corresponding colors, but only the fixed point in chamber $\{2\}$ (the pink one) is in its correct chamber. This makes it the only true fixed point of the TLN pictured there. Since there is at most one fixed point per chamber, we can label all the fixed points of a network by their \emph{support} $\supp{x^*} = \{i \mid x^*_i>0\}$. We gather the supports of all fixed points into a set
\begin{equation}\label{eq:FPWb}
	\FP(W,b) \od \{\sigma \subseteq [n] ~|~  \sigma \text{ is a fixed point support of } (W,b) \}.
\end{equation}
We have experimentally observed that often $\FP(W,b) \neq [n]$, although no precise number for that ``often'' are provided. Thus, this set is often non trivial and so it already contains quite a bit of information about the dynamics of $(W,b)$, as we will see later. We refer to this set many times throughout this dissertation, so it is good to keep it in mind. Finally, we give an important result about belonging to $\FP(W,b)$.

\begin{corollary}[\cite{fp-paper}]\label{cor:inheritance}
	Let $(W,b)$  be a TLN on $n$ neurons, and let $\sigma\subseteq [n]$. The following are equivalent:
	\begin{enumerate}
		\item $\sigma \in \FP(W,b)$
		\item $\sigma \in \FP(W|_{\tau},b|_{\tau})$ for all $\sigma \subseteq \tau \subseteq [n]$
		\item $\sigma \in \FP(W|_{\sigma},b|_{\sigma})$ and $\sigma \in \FP(W|_{\sigma \cup k},b|_{\sigma \cup k})$ for all $k \notin \sigma$
		\item $\sigma \in \FP(W|_{\sigma \cup k},b|_{\sigma \cup k})$ for all $k \notin \sigma$
	\end{enumerate}
\end{corollary}

This concludes the necessary background on TLNs. Next, we formally introduce CTLNs and review some results particular to them.

\section{Combinatorial threshold-linear networks (CTLNs)}
A big part of this dissertation focuses on a special family of TLNs, called combinatorial threshold-linear networks (CTLNs). In this case,  in addition to the network being a competitive non-degenerate TLN, the connectivity matrix $W$ in Equation \ref{eq:TLNdynamics} is prescribed by a simple directed graph $G$, as shown in Figure \ref{fig:global-inhibition}. We are thinking of the graph as retaining only the excitatory neurons and the connections between them, where the inhibitory neurons and their background inhibition are not represented in the graph, but are included in the equations defining the dynamics. 
\begin{figure}[!h]	
	\begin{center}
		\includegraphics[width=0.4\textwidth]{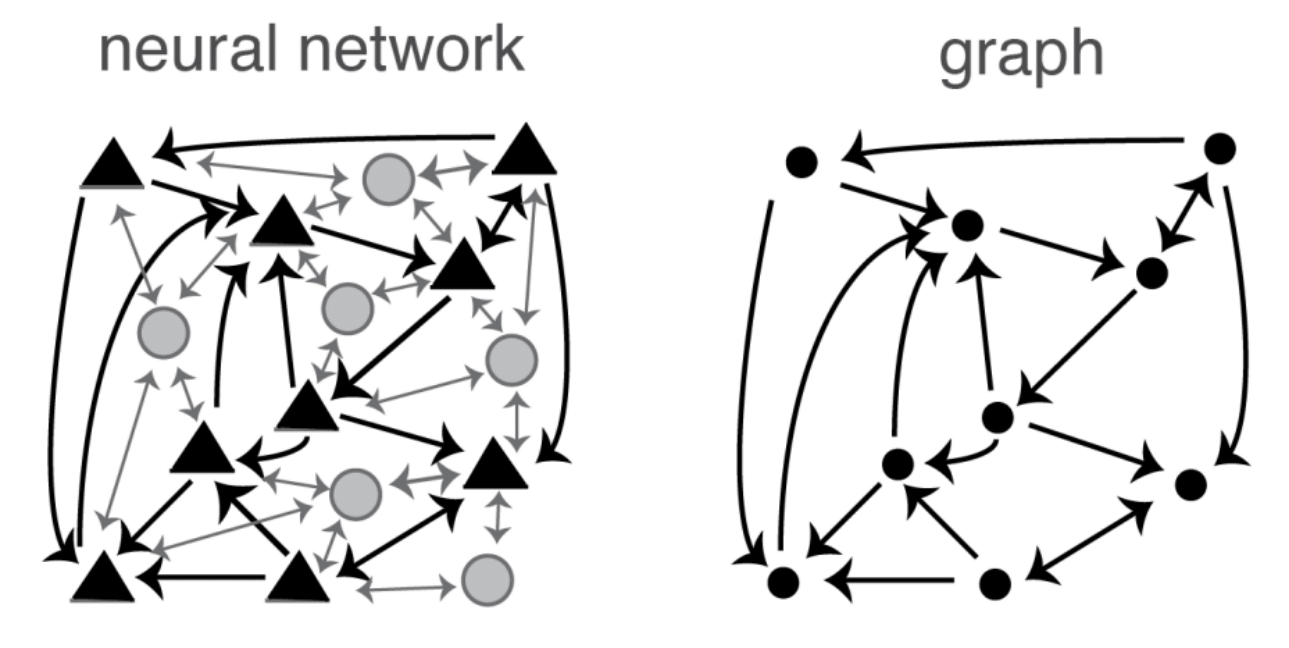}
	\end{center}	
	\caption[CTLNs]{\textbf{CTLNs}. Right: A neural network with excitatory neurons (black) and inhibitory neurons (gray), producing global inhibition. Left: Only the excitatory connections remain in our graph representation of the network. From \cite{Notices}.
	}	
	\label{fig:global-inhibition}
\end{figure} 

More precisely, if we denote $j \not\rightarrow i \text{ in } G$ by $j \rightarrow i \text{ not in } G$, we have the following definition
\begin{definition}	
	A \textit{combinatorial threshold-linear network} is a threshold-linear network $(W,b)$ whose connectivity matrix $W$ is prescribed by a simple directed graph $G$ as
	\begin{equation} \label{eq:binary-synapse}
		W_{ij} = \left\{\begin{array}{ll} \phantom{-}0 & \text{ if } i = j, \\ -1 + \varepsilon & \text{ if } j \rightarrow i \text{ in } G,\\ -1 -\delta & \text{ if } j \not\rightarrow i \text{ in } G. \end{array}\right. \quad \quad \quad \quad
	\end{equation} where $\varepsilon,\delta \in \RR$, and the input values $b_i = \theta$ are kept constant across neurons. When $\theta>0$, $\delta >0$, and $0 < \varepsilon < \frac{\delta}{\delta+1}$, we say that the parameters are in the \emph{legal range}.
\end{definition}

Although this puts us further from reality, this simplification allows us to retain fundamental properties while being able to derive relationships between structure and function more easily. We are assuming input to be constant across neurons to further isolate the role of connectivity in the network. 

When $\theta>0$, $\delta >0$, and $0 < \varepsilon < \frac{\delta}{\delta+1}$, we say that the parameters are in the \emph{legal range}. These are chosen both to have all inhibitory connections, but also to satisfy some conditions in some proofs. Parameters in this dissertation are always chosen from the legal range, unless otherwise noted. Before we show an example, it is important to clarify that our convention for the adjacency matrix of a graph $G$ is: \begin{equation*}
	A_{ij} =  \left\{\begin{array}{ll} 
	1 & \text{ if } j \rightarrow i \text{ in } G,\\ 
	0 & \text{ if } j \not\rightarrow i \text{ in } G.
\end{array}\right.
\end{equation*} to match the $W$ connectivity matrix with the $W_{ij}$ conventions.
 
\begin{figure}[!h]	
	\begin{center}
		\includegraphics[width=\textwidth]{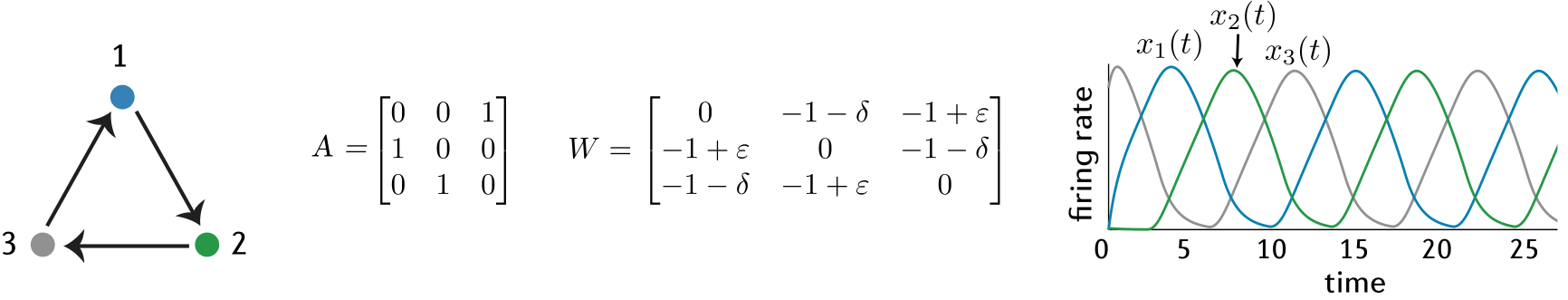}
	\end{center}	
	\caption[CTLN example]{\textbf{CTLN example}. From a 3-cycle graph we obtain its adjacency matrix $A$, from which we obtain the resulting connectivity matrix $W$ using Equation \ref{eq:binary-synapse}, from which we obtain the dynamics of the system using Equation \ref{eq:TLNdynamics}. The activations follow the arrows of the graph. Modified from \cite{rule-of-thumb}.
	}	
	\label{fig:ctln-small-example}
\end{figure} 
Figure \ref{fig:ctln-small-example} shows an example of a CTLN where the defining graph consists of a 3-cycle. From it, we get the transposed adjacency matrix, and from it we get the connectivity matrix $W$ as defined by Equation \ref{eq:binary-synapse} with parameters $\varepsilon = 0.25$ and $\delta = 0.5$. To get the rate curves in the Figure \ref{fig:ctln-small-example}, we simulate the system of equations \ref{eq:TLNdynamics} with $b = \theta\mathbbm{1}$. The network activity follows the arrows in the graph. Peak activity occurs sequentially in the cyclic order 123.

The dependence of the dynamics of a CTLN on its defining graph $G$ has given rise to extensive theory relating the dynamics and the combinatorial properties of $G$ through the set of its fixed point supports, which we now denote as $\FP(G, \varepsilon, \delta)$\footnote{The set $\FP(G)$ does not depend on $\theta$, see \cite{book-chapter}.} or $\FP(G)$, if clear from the context:
$$\FP(G) = \FP(G, \varepsilon, \delta)\od \{\sigma \subseteq [n] ~|~  \sigma \text{ is a fixed point support of } W(G,\varepsilon,\delta) \}.$$ 
This set contains information about all the fixed points of the network, giving insights into the dynamics of the network through a special subset, as it will be seen next. Indeed, we will see that the fixed points shape the dynamics of the network, whether or not they are stable. In the sections that follow, we show connections between this set and the dynamics of the network, and our main entry point are the the core motifs, introduced in the next section. There are special minimal fixed point supports which have been conjectured to correspond to attractors. Since these have proven to be insightful, we provide rules to find them in Subection \ref{sec:graph-rules}. We conclude the chapter by showing how to build in certain core motifs into the network, hoping to see the desired attractors, using constructions like cyclic and clique unions \cite{fp-paper}.

\subsection{Core motifs}\label{sec:core-motifs}

In prior work \cite{Parmelee2022}, it has been conjectured that the dynamic attractors of a network $G$ correspond to supports $\sigma \in \FP(G)$ that are \emph{core motifs}: 
\begin{definition}[\cite{Notices}]
Let $G$ be the graph of a CTLN on $n$ nodes. An induced subgraph $G|_\sigma$ is a \textit{core motif} of the network if $\FP(G|_{\sigma}) = \{\sigma\}$.
\end{definition}
We refer to both the support of the fixed point $\sigma$ and the induced subgraph $G|_{\sigma}$ as the core motif. Note that a necessary condition to be a core motif of the network is that $
\sigma$ is minimal in $\FP(G)$ by inclusion. Core motifs have been observed to be useful in predicting the dynamics of a network. For example, consider Figure \ref{fig:ctln-attractors}. The CTLN defined by the graph in panel A has two core motifs in it, $123$ and $4$. $1234$ is not a core motif, since it is not minimal in $\FP(G)$, by inclusion. Panel B shows the results of simulating the CTLN using $ \theta = 1, \varepsilon = 0.25$, and $\delta = 0.5$ under two different initial conditions. In the top, the initial condition is a small perturbation of the fixed point supported on $123$. The activity spirals out of the unstable fixed point and converges to a limit cycle where the high-firing neurons are the ones in the fixed point support. In the bottom, the initial condition is a small perturbation of the fixed point supported on $4$, which is a stable fixed point. Thus, the activity converges to a static attractor where the high-firing neuron is the one in the fixed point support. In panel C, we see that the 1234 fixed point is fundamentally different: small perturbations of the fixed point supported on $1234$ produce solutions that either converge to the limit cycle shown in panel B, or to the stable fixed point. This support therefore does not ``correspond'' to any attractor, but rather acts as a ``tipping point'' between two distinct attractors. 
\begin{figure}[!h]
	\begin{center}
		\includegraphics[width=\textwidth]{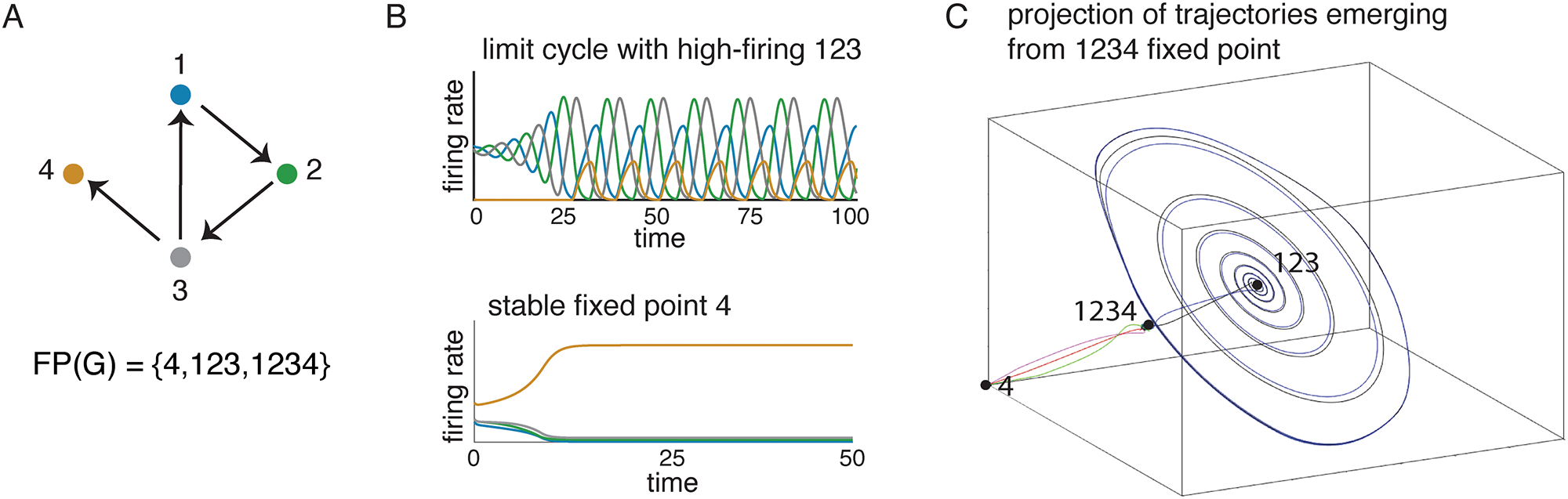}
	\end{center}
	\caption[CTLN attractors]{\textbf{CTLN attractors}.
		(A) Graph of a CTLN and its $\FP(G)$ set. 
		(B) Solutions to the CTLN with the graph in panel A using two different initial conditions, which are perturbations of the fixed points supported in $123$ (top) and $4$ bottom.
		(C) Fixed points and example trajectories which are perturbations of the fixed point supported in $1234$, depicted in a three-dimensional projection of the four-dimensional state space. 
		From \cite{rule-of-thumb}.
	}
	\label{fig:ctln-attractors}
\end{figure}

Less informally, we will say that a support $\sigma$ \textit{corresponds} to an attractor if initial conditions that are small perturbations from the fixed point lead to solutions that converge to the attractor. Heuristically, the high-firing neurons in the attractor tend to match the support of the fixed point. Core motifs often correspond to attractors \cite{Parmelee2022} and, consequently, understanding the core motifs of a network becomes useful when predicting the dynamics of a given CTLN. We make use of this heuristic often in the dissertation and so we give the set of core motifs of a given network $G$ its own notation:
$$\FP_\text{core}(G) = \FP_\text{core}(G, \varepsilon, \delta)\od \{\sigma\in\FP(G) ~|~  \sigma \text{ is a core motif of } W(G,\varepsilon,\delta) \}.$$

An example where the core-attractor correspondence is perfect is for cycle graphs, that is, a graph (or an induced subgraph) where each node has exactly one incoming and one outgoing edge, and they are all connected in a single directed cycle. First, it is a fact that all cycle graphs are core motifs:
\begin{theorem}
	If $G|_{\sigma}$ is a cycle, then $G|_{\sigma}$ is a core-motif.
\end{theorem}
Indeed, it was recently proven that a graph that is an $3$-cycle has a corresponding attractor \cite{Horacio-paper}.

\subsection{Graph rules}\label{sec:graph-rules}

The connection between attractors and the fixed points has motivated an extensive research program where \emph{graph rules} were developed \cite{rule-of-thumb,fp-paper,Parmelee2022,extended-notices}. These refer to theoretical results directly connecting the structure of $G$, to the set $\FP(G)$. Several of those results are independent of the choice of $\varepsilon, \delta,$ and thus useful for engineering \emph{robust} networks with prescribed attractors. We use of some these in the modeling section, and so we reproduce them  below. 

The first graph rule, central to our work here, concerns a special type of graph. We say that $G|_\sigma$ has \textit{uniform in-degree $d$} if every node $i\in\sigma$ has $d$ incoming edges from within $G|_\sigma$. In that case, we have:

\begin{theorem}[uniform in-degree, \cite{fp-paper}]\label{thm:uniform-in-deg}
	Let $G$ be a graph on $n$ nodes and $\sigma \subseteq [n]$. Suppose $G|_\sigma$ has uniform in-degree $d$.  For each $k \notin \sigma$, let $d_k \od |\{i \in \sigma \mid i \to k \text{ in } G\}|$ be the number of edges $k$ receives from $\sigma$. Then 
	$$\sigma \in \FP(G) \;\; \Leftrightarrow \;\; d_k \leq d \;\text{ for all }\; k \not\in \sigma.$$
	Furthermore, if $|\sigma|>1$ and $d<|\sigma|/2$, then the fixed point is unstable. If $d=|\sigma|-1$, then the fixed point is stable.
\end{theorem}

From this theorem, we easily get the following Rules for families of uniform in-degree graphs that we use in our models:
\begin{figure}[!h]
	\begin{center}
		\includegraphics[width=0.75\textwidth]{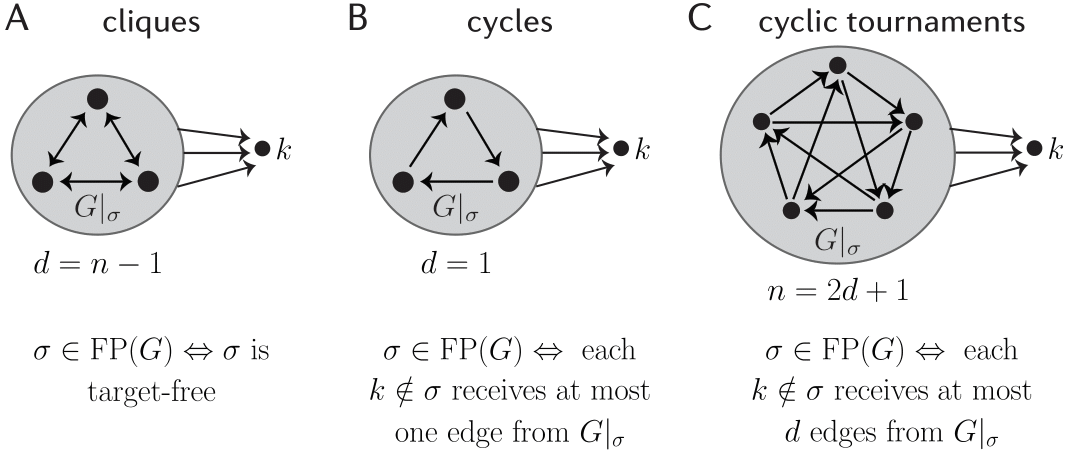}
	\end{center}
	\caption[Three families of uniform in-degree graphs]{\textbf{Three families of uniform in-degree graphs}. Three example families of uniform in-degree graphs, and corresponding Rules.
	}
	\label{fig:graph-rules-cartoon}
\end{figure}

Cliques are all-to-all connected graphs and therefore uniform in-degree with $d=n-1$, where $n$ is the total number of nodes in the clique. 
\begin{rules}[Cliques, Fig. \ref{fig:graph-rules-cartoon}A]\label{rule:cliques}
If $G|_\sigma$ is a clique, then $\sigma$ supports a \emph{stable} fixed point if and only if for all $k \not\in \sigma$, $k$ receives at most $n-1$ edges from $G|_\sigma$. When no external node receives $n$ edges from $G|_\sigma$, we say that $G|_\sigma$ is \emph{target-free}.
\end{rules}

Cycles are graphs whose vertices are connected in a closed chain and therefore uniform in-degree with $d=1$.
\begin{rules}[Cycles, Fig. \ref{fig:graph-rules-cartoon}B]\label{rule:cycles}
If $G|_\sigma$ is a cycle, then $\sigma$ supports an \emph{unstable} fixed point if and only if for all $k \not\in \sigma$, $k$ receives at most one edge from $G|_\sigma$.
\end{rules}

A tournament is an orientation of a (undirected) complete graph. A tournament $G|_\sigma$ is called cyclic if the set of all automorphisms of $G|_\sigma$ contains $(1, 2, \ldots, n) \in S_n$. All cyclic tournaments must have an odd number of nodes $n=2d+1$, and have uniform in-degree $d$ \cite{tesis-joaquin}. 
\begin{rules}[Cyclic tournaments, Fig. \ref{fig:graph-rules-cartoon}C]\label{rule:cyclic-tournaments}
If $G|_\sigma$ is a cyclic tournament, then $\sigma$ supports an \emph{unstable} fixed point if and only if for all $k \not\in \sigma$, $k$ receives at most $d = \frac{n-1}{2}$ edges from $G|_\sigma$, where $n$ is the number of nodes in $G$.
\end{rules}

All of these graphs are core motifs \cite{fp-paper,tesis-joaquin} and they indeed have a corresponding attractor, as expected. In the case of cliques, the fixed point is stable. In the case of cycles and cyclic tournaments, their unique fixed point is unstable and the dynamics reveal a corresponding limit cycle attractor. 

In addition to providing tools for engineering certain patterns  into a network, graph rules can also be used to prove that a given support $\sigma$ belongs to $\FP(G)$ in more general cases. Indeed, in Chapter \ref{ch:new-theoretical-results}, we use such rules to prove new theoretical results. A key one derives from the concept of graphical domination, introduced in \cite{fp-paper} and summarized in Figure \ref{fig:domination-cartoon}:

\begin{definition}
	We say that $k$ \textit{graphically dominates $j$ with respect to $\sigma$} if the following three conditions hold:
	\begin{enumerate}
		\item For each $i \in \sigma \setminus\{k,j\}$, if $i \rightarrow j$ then $i \rightarrow k$.
		\item If $j \in \sigma$, then $j \rightarrow k$.
		\item If $k \in \sigma$, then $k \not \rightarrow j$.
	\end{enumerate}
\end{definition}

If there is graphical domination within a graph, a lot can be said about its fixed points:
\begin{figure}[!h]
	\begin{center}
		\includegraphics[width=\textwidth]{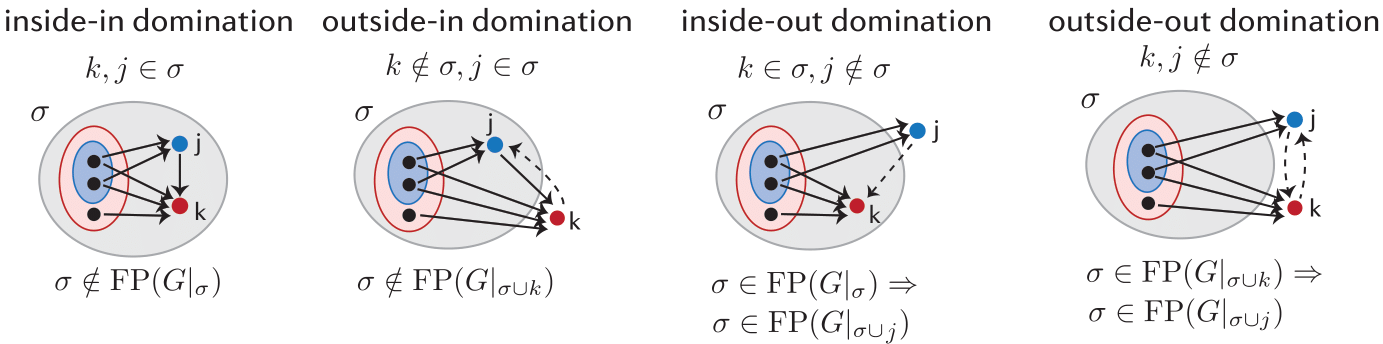}
	\end{center}
	\caption[Graphical domination]{\textbf{Graphical domination}. Four types of graphical domination from Rule \ref{rule:graph-domination}. Modified from \cite{Notices}. Dashed arrows represent optional edges. 
	}
	\label{fig:domination-cartoon}
\end{figure}

\begin{rules}[graphical domination, \cite{Notices}]\label{rule:graph-domination} Let $G$ be a graph on $n$ nodes, and $\sigma \subseteq [n]$. Suppose $k$ graphically dominates $j$ with respect to $\sigma$.  Then the following statements hold: 
	\begin{enumerate}
		\item[1.] (inside-in) If $j,k \in \sigma$, then $\sigma \notin \FP(G|_\sigma)$, and thus $\sigma \notin \FP(G)$. 
		
		\item[2.] (outside-in) If $j \in \sigma$ and $k \notin \sigma$, then $\sigma \notin \FP(G|_{\sigma \cup \{k\}})$, and thus $\sigma \notin \FP(G)$. 
		
		\item[3.] (inside-out) If $k \in \sigma$ and $j \notin \sigma$, then $\sigma \in \FP(G|_\sigma)$ implies $\sigma \in \FP(G|_{\sigma \cup \{j\}})$.
		
		\item[4.] (outside-out) If $j,k \notin \sigma$, then $\sigma \in \FP(G|_{\sigma \cup \{k\}})$ implies $\sigma \in \FP(G|_{\sigma\cup \{j\}})$.
		
	\end{enumerate}
\end{rules}

A good example of the use of Rule \ref{rule:graph-domination} in proving more general results on the $\FP(G)$, is the rule of \textit{sources}. A source is a node that has in-degree 0. The we have the following result:

\begin{rules}[Sources, \cite{fp-paper}]
Let $G$ be a graph on $n$ nodes and $\sigma \subseteq[n]$. If there exists an $\ell \in[n]$ such that $\sigma$ contains a proper source in $\left.G\right|_{\sigma \cup\{\ell\}}$, then $\sigma \notin \mathrm{FP}(G)$.
\end{rules}

\begin{proof}
Suppose there exists an $\ell \in[n]$ such that $k \in \sigma$ is a proper source in $\left.G\right|_{\sigma \cup\{\ell\}}$ with $k \rightarrow \ell$. Then $\ell$ graphically dominates $k$ since $k$ has no inputs in $\left.G\right|_{\sigma \cup\{\ell\}}$ and $k \rightarrow \ell$. Hence, $\sigma \notin \operatorname{FP}\left(\left.G\right|_{\sigma \cup\{\ell\}}\right)$ by Theorem \ref{rule:graph-domination}, and so $\sigma \notin \operatorname{FP}(G)$ by Corollary \ref{cor:inheritance}. 
\end{proof}

We conclude this section with a cool application of Rule \ref{rule:graph-domination}, which we use in Subection \ref{sec:updated-SIADs-results} to present an alternative proof of one of the results in \cite{Parmelee2022}. The proof of the result below can be found in \cite{new-cores-paper}.

\begin{theorem}\label{thm:new-domination}
	Let $G$ be a graph on $n$ nodes, and suppose there is $j,k \in [n]$ such that $k$ graphically dominates $j$ with respect to $[n]$. Then $\FP(G) = \FP(G|_{[n]\setminus\{j\}})$.
\end{theorem}

\subsection{Cyclic unions and sequential attractors}\label{sec:cyclic-unions}

These graph rules allow us to construct graphs with prescribed fixed point supports $\sigma$, and facilitate the analysis of $\FP(G)$ in terms of the structure of $G$. This in turns allows for the derivation of more general structure theorems, supported on simpler building blocks. One of these structures, which does the heavy lifting of pattern generation in Chapter \ref{ch:CPGs}, is the \emph{cyclic union}, pictured in Figure \ref{fig:cyclic-unions-small}A:
\begin{figure}[!h]
	\begin{center}
		\includegraphics[width=0.9\textwidth]{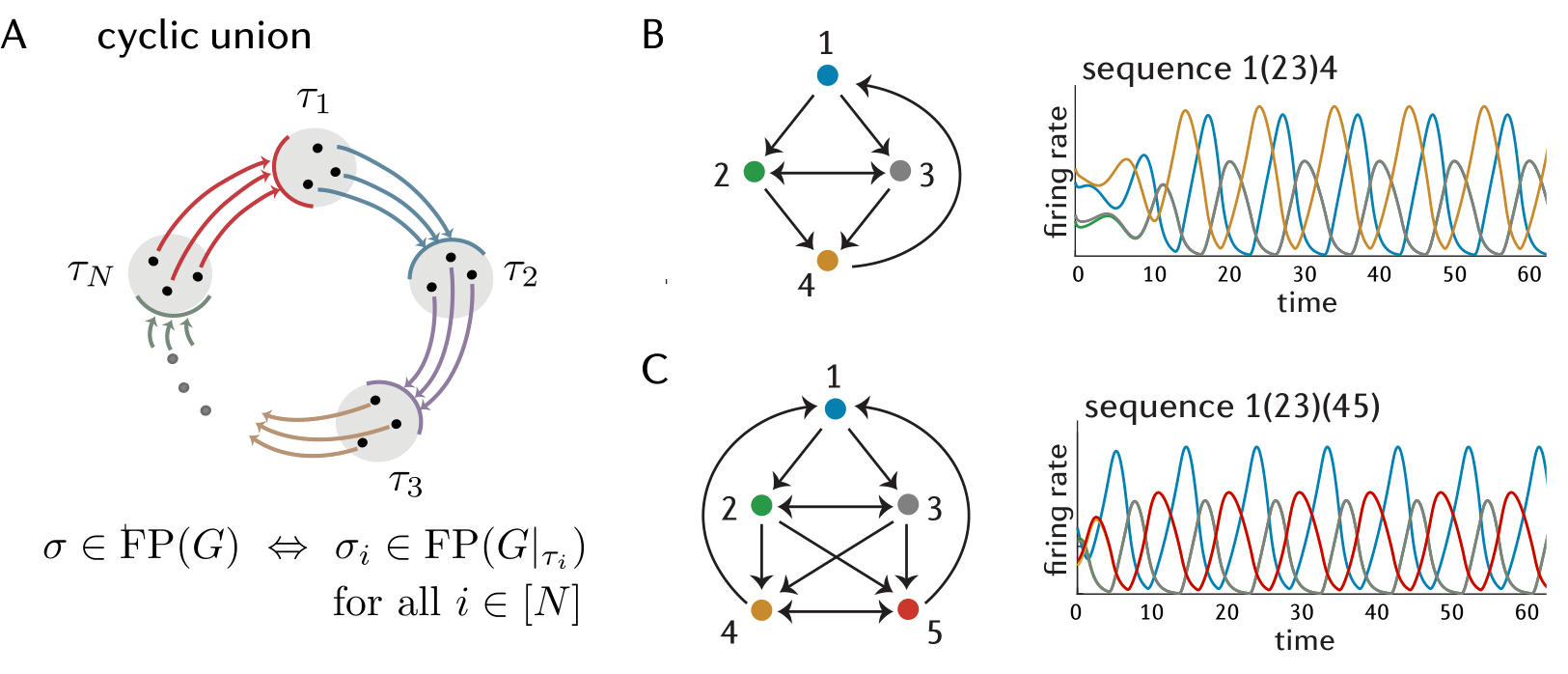}
	\end{center}
	\caption[Cyclic unions]{\textbf{Cyclic unions}.
		(A) A cyclic union of $N$ components, and Theorem \ref{thm:cyclic-unions}.
		(B - C) Two examples of a cyclic union, and its sequential activation of the nodes. Nodes whose activations are synchronized appear in parenthesis.
	}
	\label{fig:cyclic-unions-small}
\end{figure}

\begin{definition}\label{def:cyclic-union}
	Given a set of component subgraphs $G|_{\tau_1}, \ldots, G|_{\tau_N},$ on subsets of nodes $\tau_1, \ldots, \tau_N$, the \emph{cyclic union} of is constructed by connecting these subgraphs in a cyclic fashion so that there are edges forward from every node in $\tau_i$ to every node in $\tau_{i+1}$ (cyclically identifying $\tau_N$ with $\tau_0$), and there are no other edges between components
\end{definition}

Cyclic unions are great for pattern generation because, as it can be seen in Figure \ref{fig:cyclic-unions-small}B-C, they give rise to sequential attractors whose order of activation coincides with the overall structure and direction of the cyclic union. Indeed, this is a general fact, that derives from the way in which the $\FP(G)$ set is made up for cyclic unions:
\begin{theorem}[cyclic unions, theorem 13 in \cite{fp-paper}]\label{thm:cyclic-unions}
	Let $G$ be a cyclic union of component subgraphs $G|_{\tau_1}$, $\ldots$, $G|_{\tau_N}.$ For any $\sigma \subseteq [n]$, we have
	$$\sigma \in \FP(G) \quad \Leftrightarrow \quad \sigma \cap \tau_i \in \FP(G|_{\tau_i})~~\text{ for all } i \in [N].$$ 
	Moreover, 
	$$\sigma \in \FP_{\text{core}}(G) \quad \Leftrightarrow \quad \sigma \cap \tau_i \in \FP_{\text{core}}(G|_{\tau_i})~~\text{ for all } i \in [N].$$ 
\end{theorem}

This means that the global fixed point supports can be completely understood in terms of the component fixed point supports. Theorem \ref{thm:cyclic-unions} guarantees that every fixed point of $G$ hits every component. In simulations we have observed that this ensures that the every component is active in corresponding attractors. Moreover, neurons are activated in the order of the components in cyclic order. 
\begin{figure}[!h]
	\begin{center}
		\includegraphics[width=\textwidth]{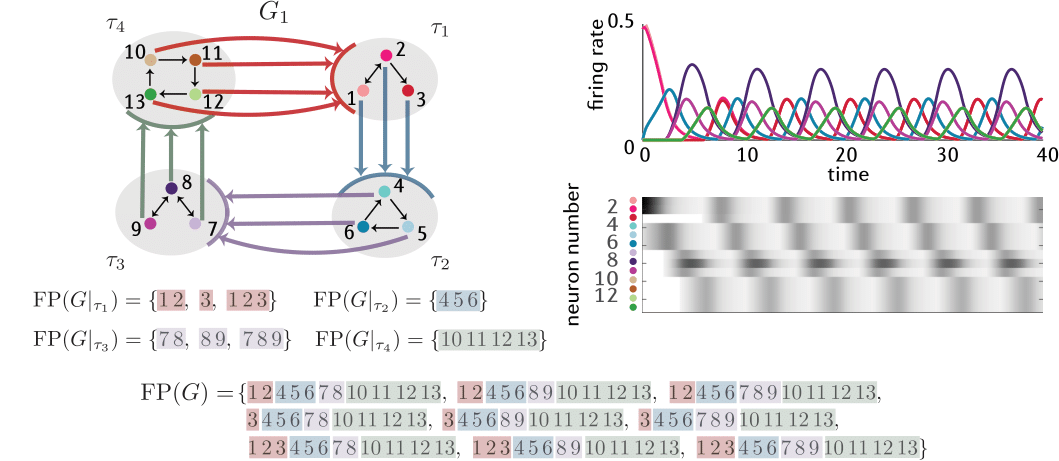}
	\end{center}
	\caption[$\FP(G)$ of cyclic union]{\textbf{$\FP(G)$ of cyclic union}.
		Example of a cyclic union, how its $\FP(G)$ set is made up of from component $\FP(G|_{\tau_i})$ pieces (color-coded), and how the sequential activations of the nodes follow the cyclic union structure. Modified from \cite{Parmelee2022}.
	}
	\label{fig:cyclic-union-example}
\end{figure}

Indeed, Figure \ref{fig:cyclic-union-example} shows an example of this. There, $G$ is a cyclic union of four component subgraphs $G|_{\tau_{i}}$, $i = 1,2,3,4$. Thick colored edges from a node to a component indicate that the node send edges out to all the nodes in the receiving component. $\FP(G|_{\tau_{i}})$ can be easily computed using graph rules, and it is shown below the graph in color-coded components. To simplify notation for $\FP(G)$, we denote a subset $\{i_1, \ldots, i_k\}$ by $i_1\cdots i_k$.  For example, $12$ denotes the set $\{1,2\}$. $\FP(G)$ follows the same color convention and consists of unions of component fixed point supports, exactly one per component. A solution for the corresponding CTLN is pictured. It shows that the attractor visits every component, in the cyclic union order.

There exists generalization of cyclic unions that also give rise to sequential attractors, see \cite{Parmelee2022}.

Theorem \ref{thm:cyclic-unions} is thus helpful when engineering networks that must follow a sequential activation, because the order of activation of neurons will match the direction of the cyclic union. This is exactly what we look for in central pattern generators models, which are the main topic of Chapter \ref{ch:CPGs}.

\subsection{Fusion attractors}\label{sec:fusion-attractors}

In the previous subsection, we saw by example that cyclic unions naturally gave rise to sequential attractors. Cyclic unions are one of the three block-structures first introduced in \cite{fp-paper}, that relate the parts to the whole. The two other structures are the clique union and the disjoint union of component networks.
\begin{figure}[!h]
	\begin{center}
		\includegraphics[width=0.48\textwidth]{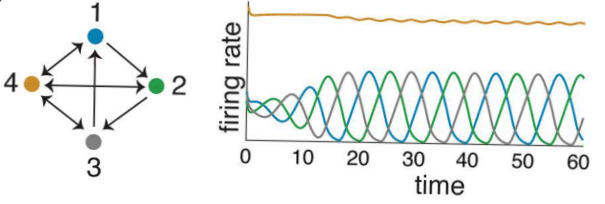}
	\end{center}	
	\caption[Fusion 3-cycle]{The fusion 3-cycle from \cite{rule-of-thumb}.}	
	\label{fig:fusion3-cycle}
\end{figure} 

We review the clique union here because it is special in that it gives rise to ``fusion attractors'', introduced in \cite{rule-of-thumb}. Clique unions are built by partitioning the vertices of the graph into components $\tau_1,\dots,\tau_N$, and putting all edges between all pairs of components. For example, the fusion 3-cycle from \cite{rule-of-thumb}, in Figure \ref{fig:fusion3-cycle}, is the clique union of $\tau_1=\{123\}$ and $\tau_2=\{4\}$. What we observe in the dynamics is the ``fusion'' of what seems to be two different attractors: the stable fixed point supported in 4, and the limit cycle supported in 123. That's why it is called fusion attractor. However clique unions are a bit too restrictive. How can we relax the assumptions on the structure of the network, and still get fusion attractors? This is partly the topic of Chapter \ref{ch:new-theoretical-results}. Later, we will generalize this concept for layered TLNs to get attractors for sequential control of quadruped gaits and other CPGs.



\chapter{Sequences of attractors}\label{ch:sequences-of-attractors} 
Recall  that in Hopfield networks \cite{Hopfield1}, memories are encoded as fixed points. Many memories can be encoded simultaneously, but accessing these fixed points usually requires attractor-specific inputs (equivalently, attractor-specific initial conditions), as illustrated in Figure \ref{fig:sequential-control-cartoons-B-highlight}A. Thus, we obtain a sequence of fixed points with whose transitions are controlled by external inputs, this means that, in a way, the order of the sequence is encoded in the external pulses, i.e. \emph{externally} encoded. Would it be possible to have the sequence's order \emph{internally} encoded, as in Figure \ref{fig:sequential-control-cartoons-B-highlight}B, where the transitions happen in response to identical external pulses who do not know what comes next? This could be useful to model highly stereotyped sequences, like songbird songs \cite{Long2010} or as a counting mechanism, like in discrete neural integrators \cite{goldman_neural_2010}.

\begin{figure}[!h]
	\begin{center}
		\includegraphics[width=\textwidth]{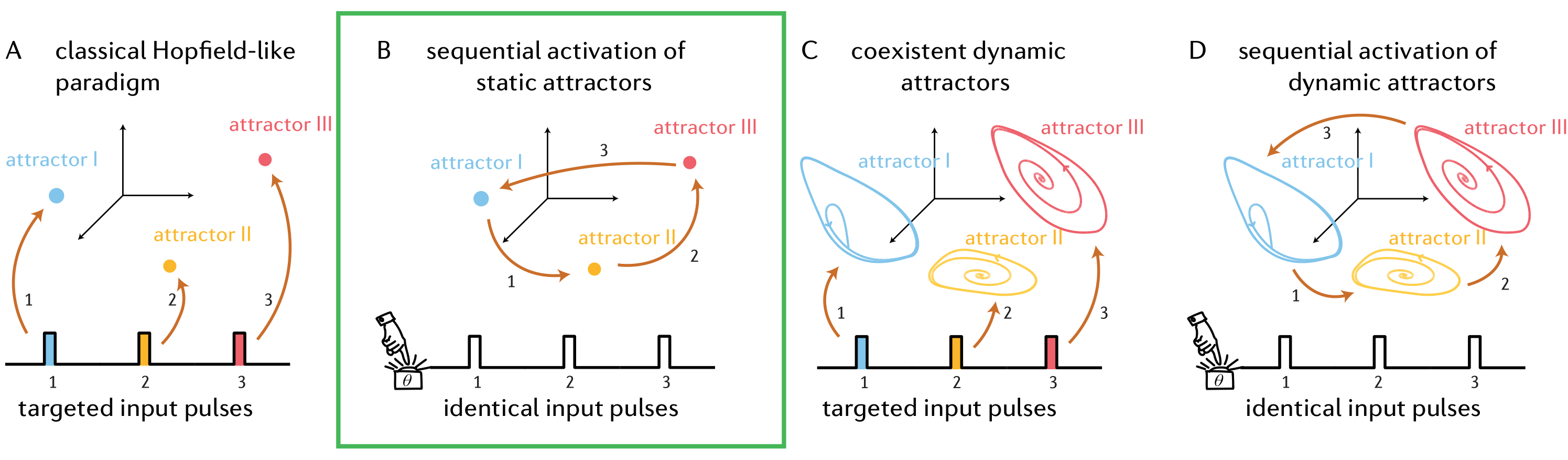}
	\end{center}
	\caption[Sequences of attractors]{\textbf{Sequences of attractors}. Reproduction of Figure \ref{fig:sequential-control-cartoons-intro}. The focus of this chapter is Panel B: Internally encoded sequence of stable fixed points, each step of the sequence accessible via identical inputs. We also briefly touch on the situations of Panel C and D.
	}
	\label{fig:sequential-control-cartoons-B-highlight}
\end{figure}

The first half of this chapter focuses on internally encoded sequences of static attractors, as depicted in Figure \ref{fig:sequential-control-cartoons-B-highlight}B. The second half will attempt to model the situation in Figures \ref{fig:sequential-control-cartoons-B-highlight}C,D, where several dynamic attractors coexist and are easily accessed via attractor-specific inputs. Our model will have a catch though, and it is that all attractors are identical to each other. Later, in Chapter \ref{ch:CPGs}, we will focus on modeling the situation of Figure \ref{fig:sequential-control-cartoons-B-highlight}C again, but this time all dynamic attractors are different. Finally, in Chapter \ref{ch:sequential-control}, we join the models from this chapter and Chapter \ref{ch:CPGs} to present a model/mechanism for the situation of Figure \ref{fig:sequential-control-cartoons-B-highlight}D, where we have an internally encoded sequence of diverse dynamic attractors.

This is, again, a good place to recall that we are using sequence/sequential in two different contexts. On one hand, we have \emph{sequential attractors} (Figs. \ref{fig:sequential-control-cartoons-B-highlight}C,D), which refer to attractors whose nodes fire in sequence e.g. limit cycles. On the other hand, we have \emph{sequences of attractors}, where the elements of the sequence are not the nodes, but the attractor themselves (Figs. \ref{fig:sequential-control-cartoons-B-highlight}B,D).

In this section, we leverage theoretical results from \cite{fp-paper} to engineer networks with prescribed $\FP(G)$ sets. In doing so, we have a strong indication that the network will support the desired attractor. Throughout the chapter, we hope the important role of the theoretical results in facilitating the engineering work will become evident. Indeed, we begin the work by recalling the rules from Chapter \ref{ch:background} that followed from Theorem \ref{thm:uniform-in-deg}, and that we use in the sections that follow to design our attractors:

\noindent\textbf{Rule \ref{rule:cliques}.} If $G|_\sigma$ is a clique, then $\sigma$ supports a \emph{stable} fixed point if and only if for all $k \not\in \sigma$, $k$ receives at most $n-1$ edges from $G|_\sigma$. When no external node receives $n$ edges from $G|_\sigma$, we say that $G|_\sigma$ is \emph{target-free}.

\noindent\textbf{Rule \ref{rule:cycles}.} If $G|_\sigma$ is a cycle, then $\sigma$ supports an \emph{unstable} fixed point if and only if for all $k \not\in \sigma$, $k$ receives at most one edge from $G|_\sigma$.

\noindent\textbf{Rule \ref{rule:cyclic-tournaments}.} If $G|_\sigma$ is a cyclic tournament, then $\sigma$ supports an \emph{unstable} fixed point if if and only if for all $k \not\in \sigma$, $k$ receives at most $d$ edges from $G|_\sigma$.
\section{Sequences of fixed point attractors}\label{sec:fixed-point-counters}

In this section we model a discrete counter as a sequence of static attractors, each accessible via identical inputs. Neural integration is an important correlate of processes such as oculomotor and head orientation control, short term memory keeping, decision making, and estimation of time intervals \cite{goldman_neural_2010,BUTTNER20061, SEUNG2000259,Mazurek2003,Cain2012}. Understanding how neurons integrate information has been a longstanding problem in neuroscience. How exactly do brain circuits generate a persistent output when presented with transient synaptic inputs? 

Many of models have been proposed to model neural integrators, but classic models are known to be very fine-tuned, requiring exact values for the parameters in order to achieve perfect integration. By contrast, some robuster models tend to be rather insensitive to weaker inputs, requiring strong inputs to switch between adjacent states \cite{goldman_neural_2010,SEUNG2000259}. The models we propose here represent a very simple alternative to classic discrete neural integrators. Our models are both robust and respond well to a wide variety of input strengths. Importantly, since they are encoding sequences of stable fixed points, they could be also useful to model highly stereotyped sequences, like songbird songs \cite{Long2010}.

\paragraph*{CTLN counter.}
Our initial model is a simple and robust CTLN that can keep a count of the number of input pulses it has received via well separated \emph{discrete} states, providing a straightforward readout mechanism of the encoded count. We henceforth refer to this network, informally, as a ``counter network''. Counting the number of inputs can be achieved with an ordered sequence of stable fixed points, each representing one position in the count. Each transition between fixed points indicates an increase in the count, and so in order to be integrating the external inputs, we want them to cause the fixed point transitions.

First, to encode a set of stable fixed points in a network we can appeal to Rule \ref{rule:cliques}, reproduced again in Figure \ref{fig:counter-construction}A, which says that a clique will yield a stable fixed point if it is target free. Thus, our network must have as many cliques as desired stable fixed points, and we need to embed them in the network in such a way that they will \textit{survive} as fixed points, that is $\sigma \in \FP(G|_{\sigma})$ but also $\sigma \in \FP(G)$. 
\begin{figure}[!h]
	\begin{center}
		\includegraphics[width=0.9\textwidth]{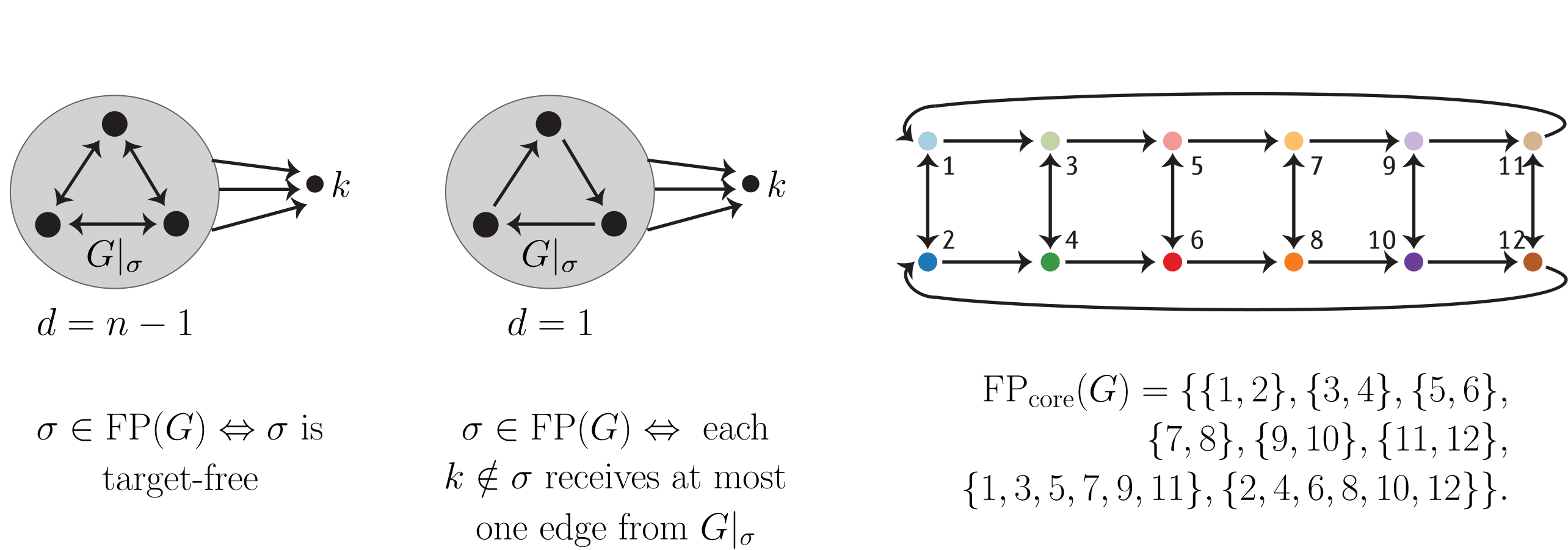}
	\end{center}
	\caption[Construction of counter]{\textbf{Construction of counter.} 
		(A) Rule \ref{rule:cliques} from Chapter \ref{ch:background}. A clique will yield a stable fixed point if and only if it is target free.
		(B) Rule \ref{rule:cycles} from Chapter \ref{ch:background}. A cycle will be a fixed point support if and only if it every other node in the network receives at most one edge from the cycle.
		(C) Resulting construction of the CTLN counter network.
	}
	\label{fig:counter-construction}
\end{figure}

We also need some mechanism to be able to transition between these \emph{stable} fixed points. Emphasis on stable, as this indicates we will need a strong enough perturbation to leave that steady state. We can potentially achieve this by adding some edges in the graph that will permit the activity to flow from clique to clique in response to input pulses. Graph-wise, these are all the ingredients we need to construct our network.

We get to work and assemble the network by chaining together several target-free 2-cliques, as shown in Figure \ref{fig:counter-construction}C. Each of these cliques is embedded in a target-free way. More specifically, Rule \ref{rule:cliques} says that the $2$-cliques will survive as fixed point supports as long as they do not have more than $d= 2-1 =1$ outgoing edge to any other node in the graph, and this is in fact the case for all of them. This would not be the case if, for example, we were to add the edge $1 \rightarrow 4$ in the network of Figure \ref{fig:counter-construction}C, because then node $4$ would be a target of $\{1,2\}$. 

Note that the top and bottom cycles ($\{1,3,5,7,9,11\},\{2,4,6,8,10,12\}$) also each support a fixed point of the network by Rule \ref{rule:cycles} (Fig. \ref{fig:counter-construction}B), since no node receives more than one edge from these cycles either. This cycles will provide the perturbation we are looking for to be able to exit the fixed points we just built-in. We made the cycles wrap around to ensure activity will not stale in the last clique of the chain, and it will continue cycling. Adding six 2-cliques was an arbitrary decision, we could have had more or less. 

In this way, we have built a network whose $\FP(G)$, by design, should contain all the cliques ($\{1,2\},\{3,4\},\{5,6\},\{7,8\},\{9,10\},\{11,12\}$) and the bottom and top cycles as core fixed point supports. Computationally, we confirm these inclusions and find, more precisely that $|\FP(G)| = 141$ (yes, $141$ fixed points, this is not that crazy because there are up to $2^{12}$ linear systems that could potentially lead to a fixed point of the network) and that
\begin{multline}\label{eq:TLN-counter-FPcore}
	\FP_\text{core}(G) = \{\{1,2\},\{3,4\},\{5,6\},\{7,8\},\{9,10\},\{11,12\},\\
	\{1,3,5,7,9,11\},\{2,4,6,8,10,12\}\}.
\end{multline}
Based on the core motif and attractor heuristic correspondence, this is indicative that we will have a stable fixed point attractor per clique, and two unstable fixed points, most likely giving rise to dynamic attractors (but also maybe not, as the heuristic is a heuristic and not a theorem or perfect correspondence, only the simulation will tell).

We have designed this network to have 6 stable fixed points, which we computationally checked that indeed it has. Since the $\FP(G)$ set does not depend on $\theta$, so far no mention of it has been necessary, and those stable fixed points will be there regardless of the value of $\theta$, as long as it is in the legal range. However, if we aim to transition between \emph{stable} fixed points, we will need exercise the perturbation somehow. And that's where hysteresis comes. The perturbation will come in the form of external pulses, or $\theta$-pulses, defined below.

External pulses are communicated the network by transiently changing the value of the external input $b_i$ in Equations \ref{eq:TLNdynamics}, which we recall to be given by
\begin{equation*}
	\dfrac{dx_i}{dt} = -x_i + \left[\sum_{j=1}^n W_{ij}x_j+b_i \right]_+, \quad i = 1,\ldots,n.
\end{equation*}

\begin{figure}[!h]
	\begin{center}
		\includegraphics[width=0.85\textwidth]{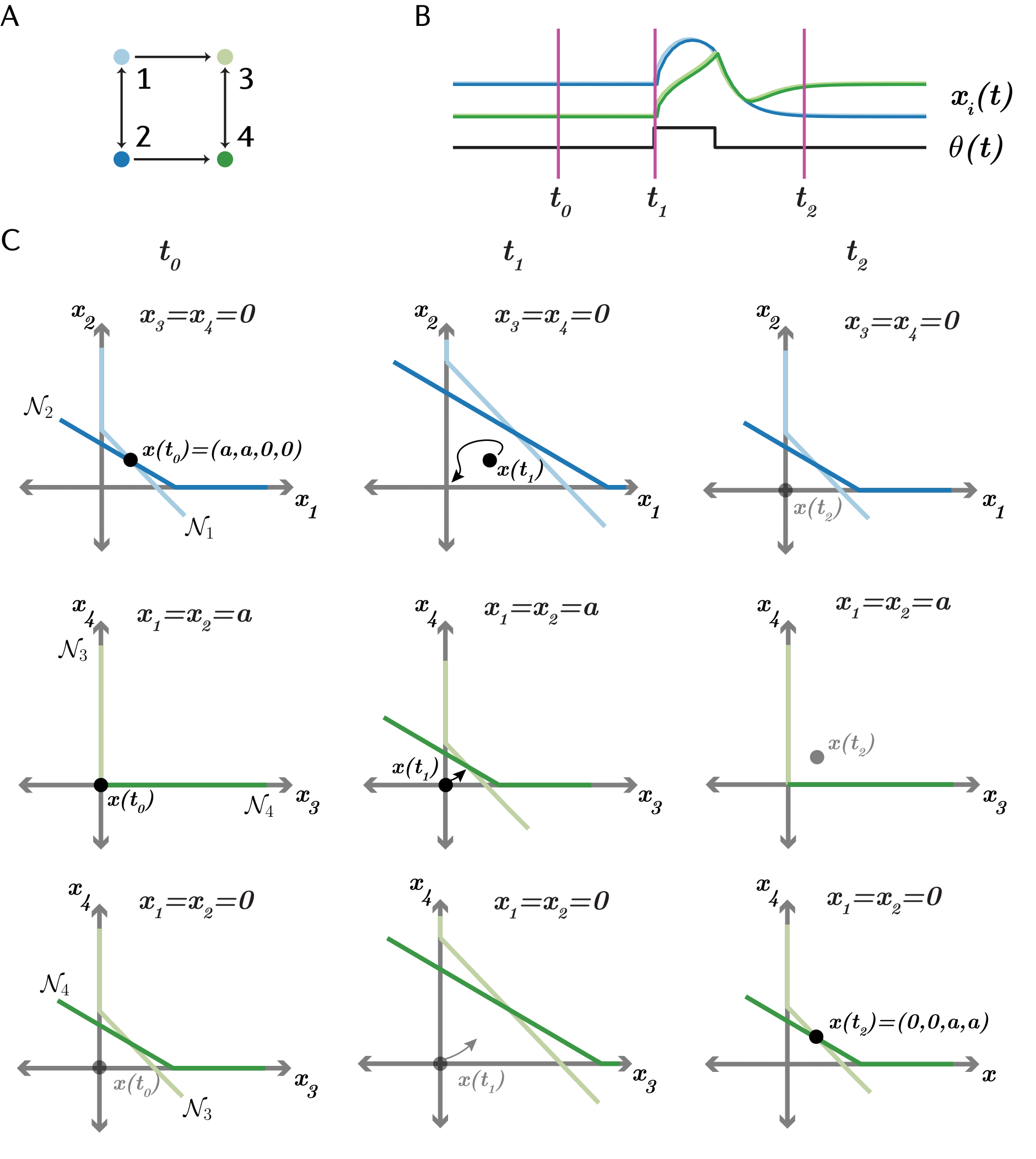}
	\end{center}
	\caption[Hysteresis]{\textbf{Hysteresis.} 
		(A) 2-clique counter network.
		(B) Close-up of the pre- and post-pulse dynamics of counter network.
		(C) Three different 2-dimesional slices of the 4-dimensional state, at three time points: before the pulse, $t_0$, when the network is settled in the fixed point where the only high-firing neurons are the blue ones. Right at the beginning of the pulse, $t_1$, when the hyperplanes move, and the trajectory has to adapt to this change. And after the pulse has ceased, at $t_2$, where the network stabilized in a different fixed point, the one where the green nodes are the high-firing. Nullclines are colored piecewise lines. The $\theta$-pulse causes a translation of the nullclines, pushing the state of the network into the basin of attraction of a different stable fixed point. Black points indicate current state of the network, grey points indicate the current state of the network when the state does not lie in that plane, but it is projected to that plane. 
	}
	\label{fig:hysteresis}
\end{figure}

Because we are using CTLNs, $W$ is prescribed by the $G$ graph of Figure \ref{fig:hysteresis}A, using the rule 
\begin{equation*}
	W_{ij} = \left\{\begin{array}{ll} 
		\phantom{-}0 & \text{ if } i = j, \\
		 -1 + \varepsilon & \text{ if } j \rightarrow i \text{ in } G,\\
		  -1 -\delta & \text{ if } j \not\rightarrow i \text{ in } G. 
		  \end{array}\right. 
\end{equation*}

Now, in the CTLN setup the external input $b_i(t)$ would typically constant, but because we want to use pulses, the external input can no longer be constant. Because now we will have a temporary identical perturbation for all the neurons, $b_i(t) = \theta(t)$ above must be defined as a step function, pictured right at the very top of Figure \ref{fig:hysteresis}B, that also does not depend on $i$, because we assume the pulses to be identical across neurons.

\begin{equation*}
b_i(t)= \theta(t) = \left\{\begin{array}{ll} 
	\theta_0 & \text{ for } t \text{ outside of pulses} \\
	\theta_1 & \text{ for pulse times } t. \\
\end{array}\right. 
\end{equation*} 

We refer to $\theta_0$ as the baseline and to $\theta_1$ as the pulse. It is important to note that this varying external pulse will cause our network to no longer be a CTLN (which requires external input to be constant and uniform across neurons), but rather a TLN, or a piecewise CTLN, if you wish. But because we will only need two different $\theta$ values, and we designed the network with CTLN principles, and because it really is a piecewise CTLN, we continue to call this network a CTLN. 

Now, how does hysteresis arise from the $\theta_1$ pulse? Figure \ref{fig:hysteresis} explores how this transient $\theta$ change is affecting the dynamics of the network, at three time-stamps. First, note that the nullclines of the system are piecewise linear functions given by
\begin{equation*}
	\mathcal{N}_i : x_i = \left\{\begin{array}{ll} 
	\sum_{j=1}^n W_{ij}x_j+\theta_i \quad &\text{ if } \sum_{j=1}^n W_{ij}x_j+\theta_i > 0\\
	0 &\text{ if } \sum_{j=1}^n W_{ij}x_j+\theta_i \leq 0 \\
	\end{array}\right. 
\end{equation*}  
That is, because TLNs are piecewise linear, the nullclines are piecewise linear as well. These are pictured in Figure \ref{fig:hysteresis}C for neurons 1 to 4 of the network from panel A of the same figure, using corresponding colors.  

By changing the value of $\theta$, we are actually causing the nullclines of the system to  shift in state space. More specifically, looking at Figure \ref{fig:hysteresis}C, we see that at $t_0$ we are at the baseline $\theta_0$, and the system is at rest in the stable fixed point supported on $\{1,2\}$ (cf. panel B). When the pulse arrives, at time $t_1$, the nullclines experience a translation in state space, causing the trajectory to move towards the new fixed point (always at the intersection of the nullclines). When the pulse ceases, at $t_2$, it is too late for the trajectory to go back, it has fallen in the basin of attraction of a different fixed point, the one supported on $\{3,4\}$!

It is important to note that newer, more robust models of neural integration, also make use of hysteresis, but it typically arises from the use of bistable units \cite{goldman_robust_2003,Koulakov2002,Nikitchenko2008}. In our case, hysteresis emerged as a property of connectivity, since CTLN units are not bistable by themselves (as they just die off), only when connected to each other, and so the multistability in our model is a property of the network, not of each unit.

What we explained above is exactly what we observe in simulations. We simulated the network using the parameters $\varepsilon = 0.25$, $\delta = 0.5$, a baseline $\theta_0 = 1$, and an input pulse $\theta_1 = 5$. However, note that since Theorem \ref{thm:uniform-in-deg} does not depend on the values of these parameters, any (C)TLN with dynamics prescribed by Equation \ref{eq:TLNdynamics}, whose connectivity matrix $W$ is build  from the graph of Figure \ref{fig:fixed-point-counters}A according to the binary synapse prescription of Equation \ref{eq:binary-synapse}, will have the supports of  Eqn \ref{eq:TLN-counter-FPcore}. This is true for any values of $\varepsilon$ and $\delta$, provided they are within the legal range. 
\begin{figure}[!h]
	\begin{center}
		\includegraphics[width=\textwidth]{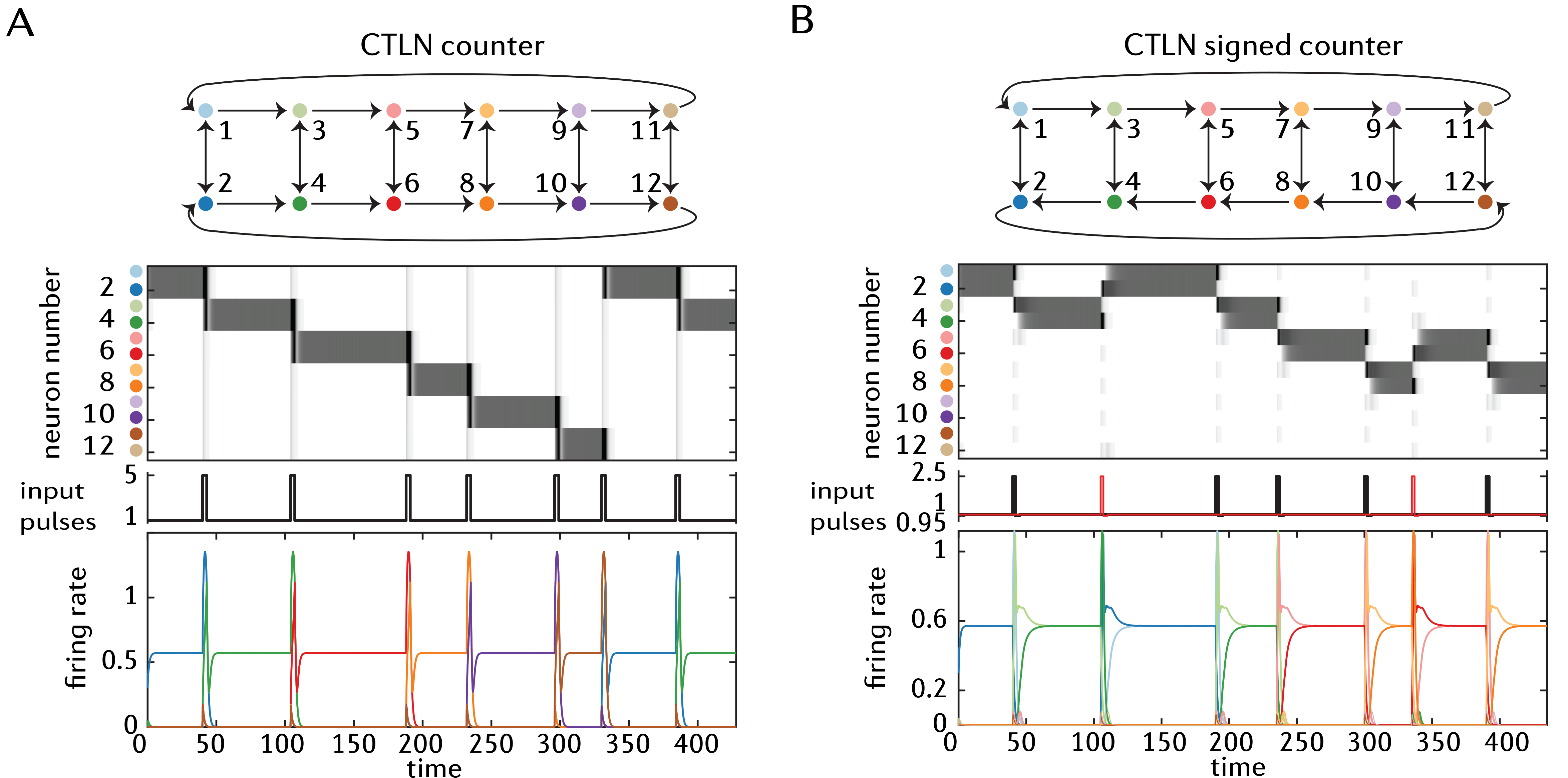}
	\end{center}
	\caption[CTLN counters]{\textbf{CTLN counters.} (A) Counter. Pulses shown in the middle plot are sent to all neurons in the network. Activity slides to the next clique (stable fixed point) after reception of the pulse. Pulse duration is 3 time units, with no refractory period.
		(B) Signed counter. Black pulses are sent to odd-numbered neurons and red pulses are sent to even-numbered neurons. Activity slides to the right for black pulses, and to the left for red pulses. Pulse duration is 2 time units, with 3 time units of refractory period where $\theta = 0.95$.
	}
	\label{fig:fixed-point-counters}
\end{figure}

The $\FP_\text{core}(G)$ predictions about the dynamics are partly confirmed by simulations. We do observe the stable fixed points arise, but not the limit cycle. In any case, the rate curves of Figure \ref{fig:fixed-point-counters}A confirm the counting mechanism works: activity remains in the stable fixed point that the network was initialized to, unless a uniform $\theta$-pulse is sent to all neurons in the network. When the network receives the pulse, activity slides down to the next stable state in the chain. Activity is maintained in this state indefinitely until future pulses are provided to the system. Indeed, the network is effectively counting the number of input pulses it has received via the position of the attractor in the linear chain of attractor states. This network is a very simple alternative to the neural integrator models often used to maintain a count in working memory of some number of input cues. The matrix and parameters needed to reproduce this Figure are available in Appendix \ref{ch:appendixA}, Equation \ref{eq:counter-matrix}. 

Good! We have successfully encoded a sequence of fixed point attractors, all accessible via identical pulses. But can we adapt this construction to be able to decrease the count when presented with a negative input? 

\paragraph*{CTLN signed counter.}
The architecture that made the transitions possible before, the cycles, can be so slightly modified to allow for the count to be decreased: making the bottom cycle travel in the opposite direction, as seen in Figure \ref{fig:fixed-point-counters}B, allows us to perturb  the network with pulses so that the fixed points can also transition backwards in the chain, decreasing the count. This means that the pulses are now \emph{signed}, and thus we informally refer to this modified network as ``signed counter network''. The core motif analysis in this case is the same, as the cliques remain target-free and the two cycles again survive by Rule \ref{rule:cycles}. Indeed, computationally, we found $|\FP(G)| = 369$ and
\begin{multline}\label{eq:TLN-signed-counter-FPcore}
	\FP_\text{core}(G) = \{\{1,2\},\{3,4\},\{5,6\},\{7,8\},\{9,10\},\{11,12\},\\
	\{1,3,5,7,9,11\},\{2,4,6,8,10,12\} \}.
\end{multline}

Again the predicted behavior of the network is confirmed by the rate curves of Figure \ref{fig:fixed-point-counters}B. The dynamics of the signed counter are analogous to those of the previous counter, except now pulses are signed, and the count can be decreased. More specifically, neurons are divided into two opposite populations (top and bottom cycles or, equivalently, odd and even nodes), and pulses are sent to either one of them. Note that pulses are now followed by a brief refractory period where $\theta = 0.95$ to allow the system to reset. The color of the pulse in Figure \ref{fig:fixed-point-counters}B indicates which group of neurons received the pulse. Black pulses are sent to the top cycle (odd-numbered nodes), and red pulses are sent to the bottom cycle (even-numbered nodes). When the network receives a black pulse, the attractor slides to the right. If it receives a red pulse, the attractor slides to the left. That is, the activity travels in the direction of the cycle that received the pulse. This network can not only keep a count, but also store net displacement, position on a line, or relative number of left and right cues. The matrix and parameters needed to reproduce this Figure are available in Appendix \ref{ch:appendixA}, Equation \ref{eq:signed-counter-matrix}. 

Both counter models are simple in that they only require twice as many neurons as stable states, and pulses input to the system are given to \emph{all} neurons on the network. Although both counter networks pictured in Figure \ref{fig:fixed-point-counters} are able to keep a count modulo 6 (number of chained cliques), this can be generalized to an arbitrary (but finite) number of cliques, allowing to keep a memory of an arbitrary large count. This means that the number of possible states approaches infinity as the number of  links in the chain approaches infinity as well, and so the count is effectively continuous. In addition, since the networks are able to keep a modular count of the number of pulses, it could be useful to estimate modular counts like time intervals or angles. Both counters thus constitute a simple alternative to traditional discrete neural integrators, but are they really that much robust?

\paragraph*{Robustness of CTLN counters.} In general, CTLNs are expected to perform robustly, since a lot of their dynamic information is contained in $\FP_\text{core}(G)$ and we know this set is preserved across the legal range. 

To asses the performance of both counters across parameter space, we ran the simulations of Figure \ref{fig:fixed-point-counters} using various pairs of $(\varepsilon,\delta)$ parameters, and various combinations of pulses strengths and durations. We found three different possible outcomes, exemplified in Figure \ref{fig:good-parameter-grid-TLN-counters}A. We classified these according to whether the function is preserved (keeps the correct count), corrupted (counts in multiples of two) or lost (does not keep a consistent count) as $\varepsilon, \delta, \theta$ vary. 
\begin{figure}[!h]
	\begin{center}
		\includegraphics[width=\textwidth]{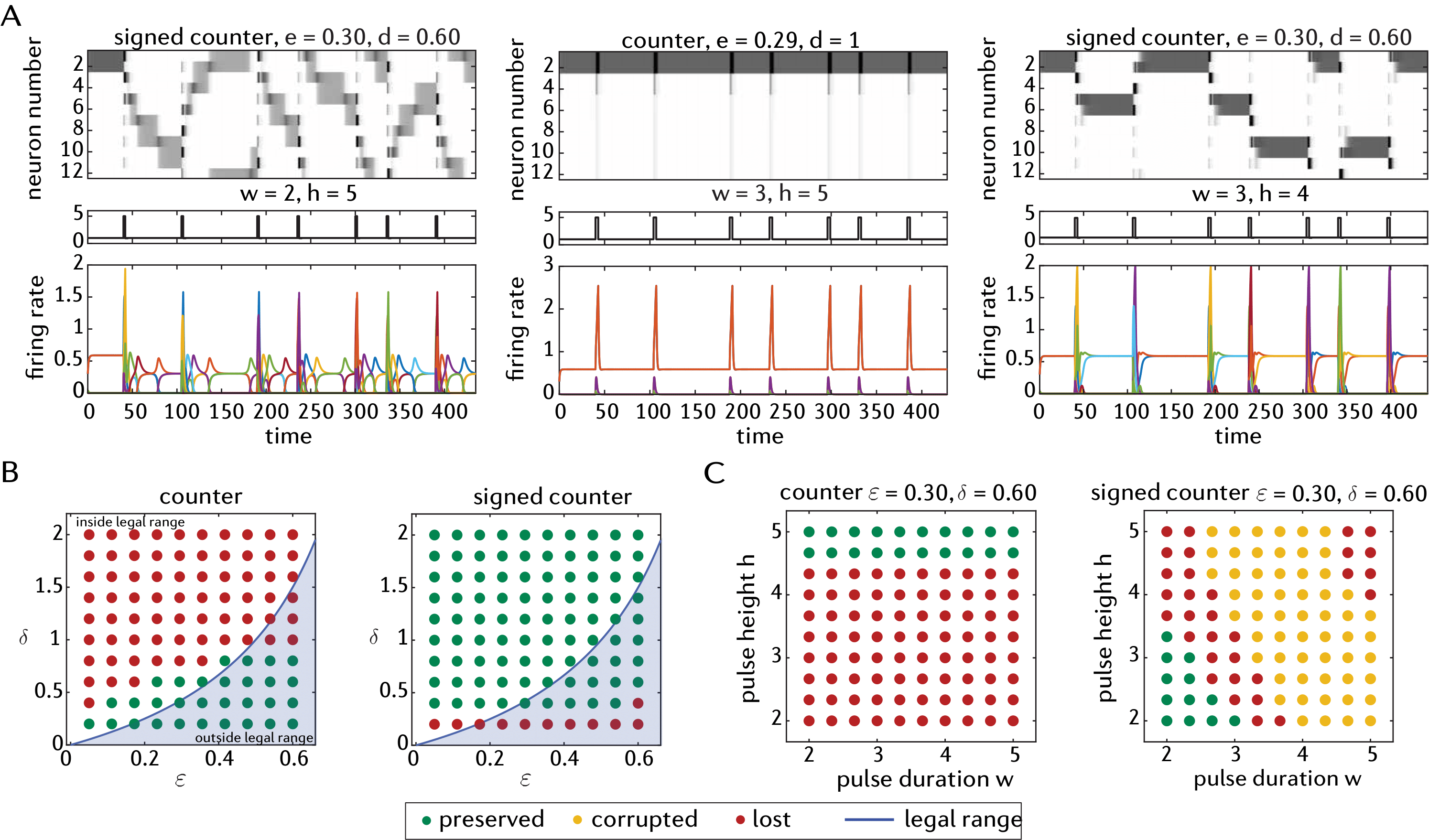}
	\end{center}
	\caption[Good parameter grid for CTLN counters]{\textbf{Good parameter grid for CTLN counters.} 
		(A) Examples of what can go wrong in a single simulation. In the first plot the signed counters enter a ``roulette'' behavior (function is lost), in the second plot the pulses do not move the counter from the current fixed point (function is lost), and in the third plot the counter consistently slides two positions (function is corrupted).
		(B) Various counter behaviors for fixed values of baseline $\theta = 1$, with pulses $\theta = 5$ for counter, $\theta = 2.5$ and $\theta = 0.95$ for signed counter. $\varepsilon, \delta$ vary according to the axes. Shaded in blue are parameter pairs outside the legal range $\varepsilon < \frac{\delta}{\delta+1}$.
		(C) Various counter behaviors for a fixed values of  $\varepsilon = 0.30, \delta = 0.60$ and baseline $\theta = 1$, with pulse height and duration varying.}
	\label{fig:good-parameter-grid-TLN-counters}
\end{figure}

The results for all pairs $(\varepsilon,\delta)$ of parameters and various combinations of pulses strengths and durations are recorded in the bottom row of Figure \ref{fig:good-parameter-grid-TLN-counters}, where green indicates preserved, yellow indicates corrupted and red indicates lost performance. One dot corresponds to one simulation of $7$ pulses, with a running time of $T = 428$ time units for the counter and $T = 435$ time units for the signed counter. We found that there is a good range of parameters where the counters behave as expected, successfully keeping the count of the number of input pulses received. 

This robustness across parameter space was given to us by the theoretical results. But will the performance of the counters survive to added noise in the input and connections? To verify robustness to noise, we computed the proportion of failed transitions among $m = 100$ pulses in both fixed point counters. This was done by introducing varying percentages of noise into the $\theta$ input, varied every $dt = 0.1$ of a time unit, and into the connectivity matrix $W$, as specified in the axes of Figure \ref{fig:noisy-TLN-counters}. A failed transition is defined as any behavior that does not advance the counter a single clique forward/backwards, depending on the sign of the pulse. Figure \ref{fig:noisy-TLN-counters} contains examples of all possible transition failures, as well as examples successful transitions in noisy conditions, as detailed below. All plots simulations used to quantify the failures were done with $\varepsilon = 0.25, \delta = 0.5$.
\begin{figure}[!h]
	\begin{center}
		\includegraphics[width=\textwidth]{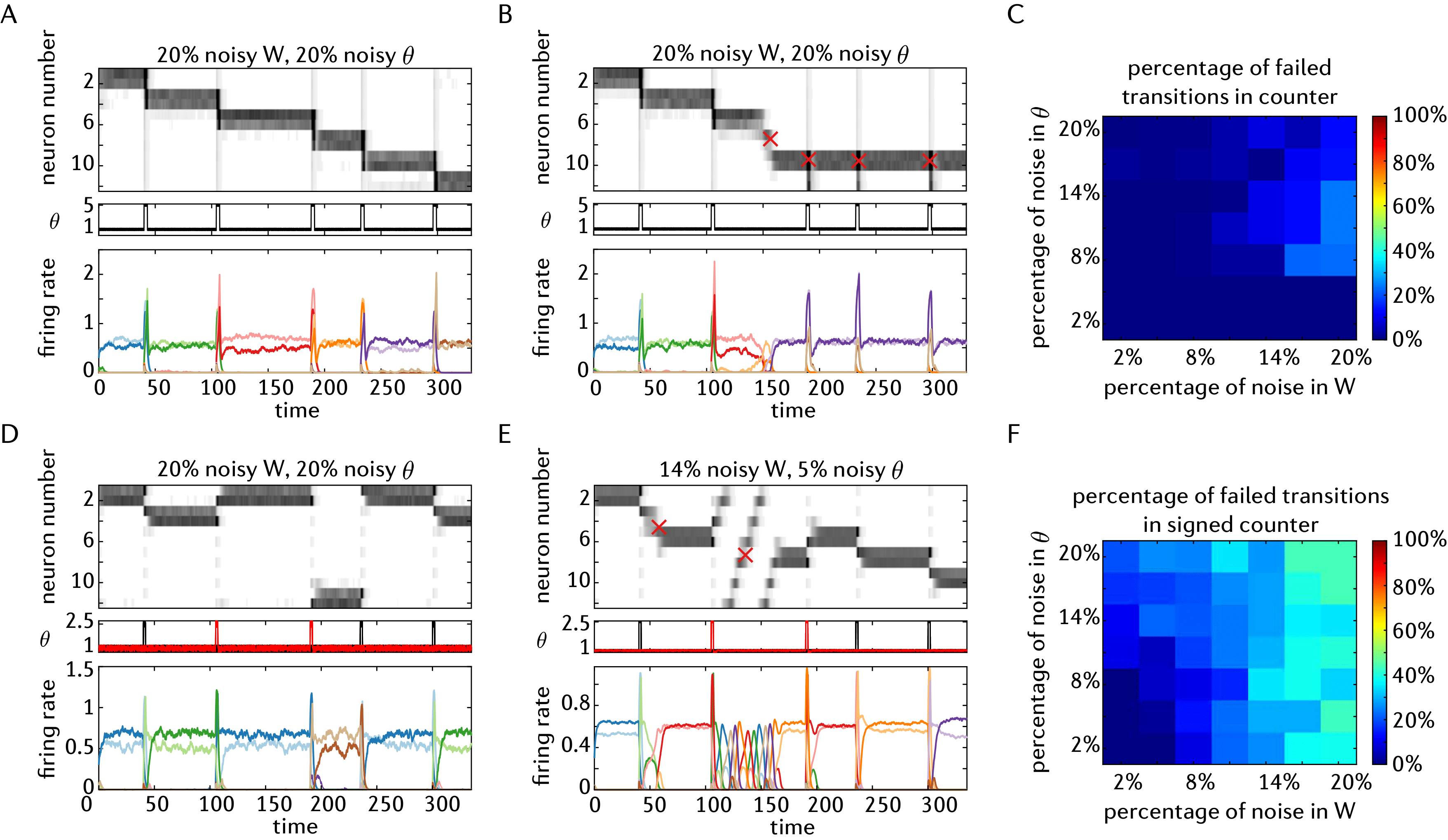}
	\end{center}
	\caption[Noisy CTLN counters]{\textbf{ Noisy CTLN counters.} 
		(A) Example of a 20\% noisy counter that performs well. 
		(B) Example of a noisy counter with four failures, marked with red crosses: first it slides too many cliques down, and then it gets stuck in the same clique indefinitely.
		(C) Percentage of failed transitions for the counter in 100 trials, for several percentages of noise in $W$ and $\theta$.
		(D) Example of a 20\% noisy signed counter that performs well. 
		(E) Example of a noisy signed counter with two failures, marked with red crosses: first it slides too many cliques down, and then it rolls around the counter until it settles in some arbitrary clique.
		(F) Percentage of failed transitions for the signed counter in 100 trials, for several percentages of noise in $W$ and $\theta$.}
	\label{fig:noisy-TLN-counters}
\end{figure}

Panel A of Figure \ref{fig:noisy-TLN-counters} is an example of a very noisy counter that still preforms well. Each pulse advances the counter a single clique forward, as expected. Panel B of Figure \ref{fig:noisy-TLN-counters} shows a noisy counter with two types of failures: first the counter advances too many steps, and then it gets stuck in one of the cliques (altogether these would count as four failures in out analysis, even though there is probably just two defective cliques). Panel D of Figure \ref{fig:noisy-TLN-counters} is an example of a very noisy signed counter that performs perfectly well. Panel E of Figure \ref{fig:noisy-TLN-counters} is an example of things that can go wrong in  the signed counter. The first failure advances the counter one more than expected, the second failure makes the counter go into this kind of roulette behavior until it stops in a given clique. We have also observed the signed counter getting stuck at some point, as in Figure  \ref{fig:noisy-TLN-counters}B.

The percentages of noise in Figure \ref{fig:noisy-TLN-counters} were calculated as follows: The noise in the external input $\theta$ is i.i.d random noise from the interval $(-1,1)$, and it was introduced to vary every $0.1$ fraction of a unit time. The noise in the connectivity matrix $W$ is obtained by perturbing the adjacency matrix $A$ with i.i.d random noise from the interval $(0,1)$. More precisely, if denote by $\widetilde{A},\widetilde{W}, \widetilde{\theta}$ the noisy versions of $A,W,\theta$; where $A$ is the transpose of the adjacency matrix of the graph $G$ defining the network, then the noisy versions are:
\begin{align*} 
	\widetilde{A} &= A + pR \\
	\widetilde{W} &= (-1+\varepsilon)\widetilde{A} +(-1+\delta)(\mathbf{1}-I-\widetilde{A})\\ 
	\tilde{\theta} &= \theta + qS
\end{align*}
where $p,q$ are the percentages of noise introduced, and $S, R$ are random matrices the same size as $W, \theta$ with entries between $(0,1)$ and $(-1,1)$, respectively, and $\mathbf{1}$ is a matrix of $1$'s the same size as $A$. This is equivalent to perturbing $W$ by an amount proportional to length of the interval $[-1-\delta,-1+\varepsilon]$, that is:
\begin{align*} 
	\widetilde{W} &= (-1+\varepsilon)\widetilde{A} +(-1+\delta)(\mathbf{1}-I-\widetilde{A})\\ 
	&= W + p[(-1+\varepsilon)-(-1-\delta)]R
\end{align*}
We choose to perturb the adjacency matrix instead for computational easy. For each percentage of noise pair $(p,s)$, we ran 20 simulations, each one consisting of 5 (signed) pulses in the CTLN (signed) counter, as exemplified in Figure \ref{fig:noisy-TLN-counters}. The results of counting over all these pulses are summarized in panels C and F of Figure \ref{fig:noisy-TLN-counters}. Each pixel represents the fraction of failed transitions for each pair $(p,s)$. We found that discrete counters perform perfectly well up to 5\% noise, proving that perfect synapses are not necessary. When the noise limits are pushed beyond, counters lose stability and start to slide too many attractors, or to get stuck in one of the cycles. The (unsigned) counter, not surprisingly, proved to be a lot more robust, being barely affected by the amount of noise in $W$. Although not as robust, the signed counter accurately kept the count 50\% of times under 20\% noise. Not too bad.

Having successfully encoded sequences of stable fixed points, it is natural to now ask: can we use the same ideas to construct a network that encodes a sequence of dynamic attractors instead? Maybe by chaining together other kinds of core motifs, instead of cliques? 

\section{Sequences of dynamic attractors of the same type}\label{sec:dynamic-counter}

Here we use one of the core motifs giving rise to a dynamic attractor from Chapter \ref{ch:background} to construct a network that steps through a sequence of dynamic attractors. In the previous section, we chained together repeated core motifs that we knew gave rise to stable fixed points to obtained a network that steps through a sequence of stable fixed points. We use the same idea as before, but a cyclic tournament instead of a 2-clique. 

\begin{figure}[!h]
	\begin{center}
		\includegraphics[width=0.9\textwidth]{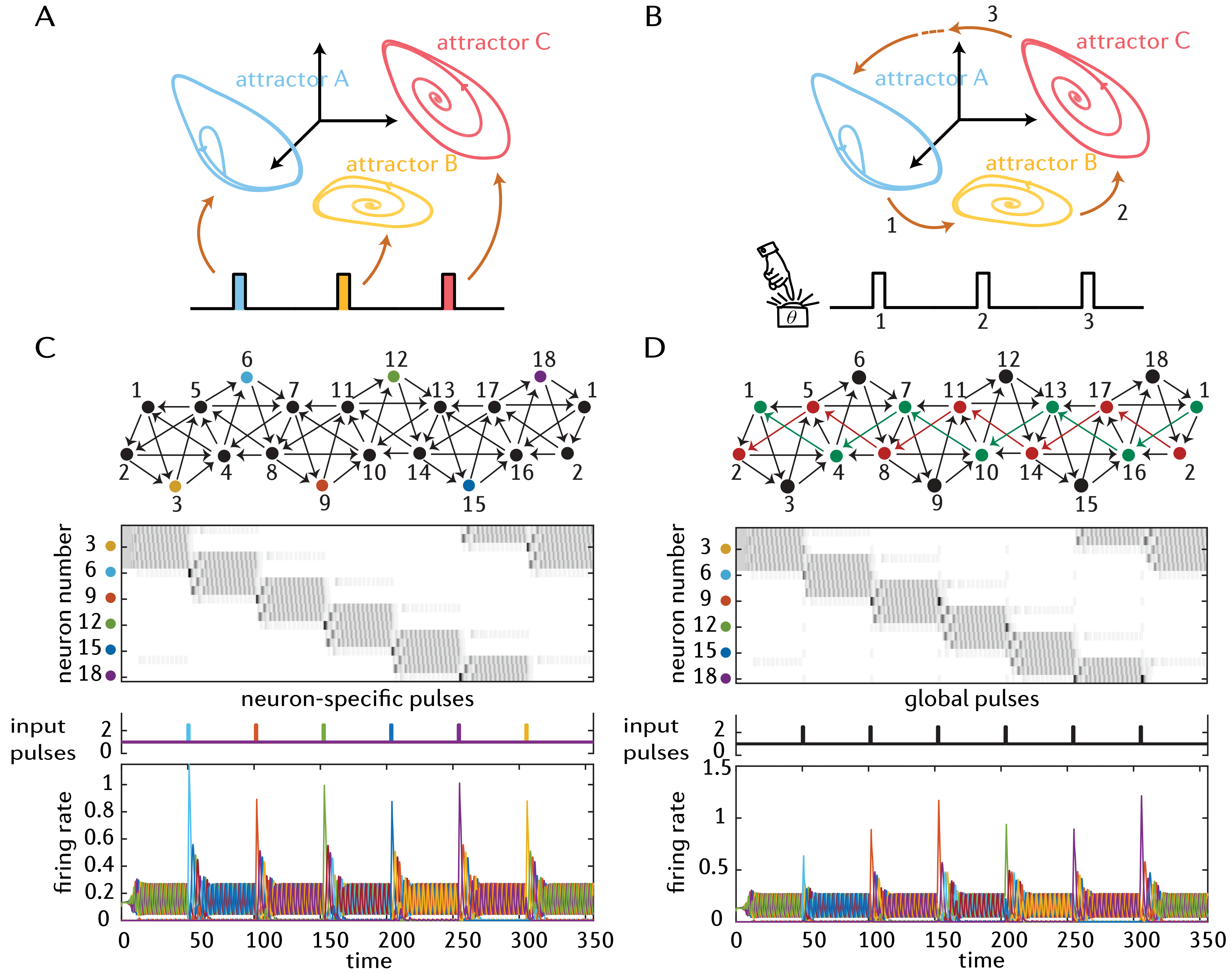}
	\end{center}
	\caption[Dynamic attractor chains]{\textbf{Dynamic attractor chains.}
		(A) Coexistent dynamic attractors. Each attractor is accessed via attractor-specific pulses. The sequential information is externally encoded.
		(B) Coexistent dynamic attractors. Each attractor is accessed via identical pulses sent to the network. The sequential information is internally encoded.
		(C) Patchwork of six 5-stars. Pulses are sent, one at the time, to neurons $6,9,12,15,18,3$ (colored) in that order. Each pulse makes the attractor slide to the next limit cycle down the chain. Pulse duration is 1 time unit.
		(D) Same network as in panel A receives simultaneous pulses to neurons $3,6,9,12,15,18$ (bold nodes). Pulse duration is 1 time unit. Two cycles go through the network in the direction opposite to attractor sliding. Despite their apparent symmetry, the green cycle is a core motif of the network, but the red cycle is not (it dies by Rule \ref{rule:cycles}).
		}
	\label{fig:dynamic-attractor-counters}
\end{figure}

To construct the network, we chain together overlapping 5-stars, which are cyclic tournaments. These were introduced in Chapter \ref{ch:background} and are the motifs composing the networks shown in Figures \ref{fig:dynamic-attractor-counters}C, D (a single 5-star is, for instance, the subgraph induced by 1 to 5). Rule \ref{rule:cyclic-tournaments} says that each 5-star will yield an unstable fixed point support, since each 5-star is connected in such a way that it survives to the larger network. Thus, we should get:
\begin{multline*}
	\{ \{1, 2, 3, 4, 5\}, \{4, 5, 6, 7, 8\}, \{7, 8, 9, 10, 11\},\\
	\{10, 11, 12, 13, 14\},\{13, 14, 15, 16, 17\}, \{1, 2, 16, 17, 18\}\}\subseteq \FP_\text{core}(G)
\end{multline*}
Indeed, computational work shows that $\FP_\text{core}(G)$ contains precisely the six supports above, each for one 5-star. We also found $|\FP(G)| = 361$ (which again, is not surprising since there are up to $2^{18}$ linear systems) and that
\begin{multline}\label{eq:dynamic-chain-FPcore}
	\FP_\text{core}(G) = \{ \{1, 2, 3, 4, 5\}, \{1, 2, 16, 17, 18\}, \{4, 5, 6, 7, 8\}, \{7, 8, 9, 10, 11\},\\
	\{10, 11, 12, 13, 14\},\{13, 14, 15, 16, 17\}, \{1, 4, 7, 10, 13, 16\}\}
\end{multline}
Notice that we got an extra core motif, the last one in Equation \ref{eq:dynamic-chain-FPcore}. This is a cycle that is also uniform in-degree, and that survives by Rule \ref{rule:cycles} (colored in green in Figure \ref{fig:dynamic-attractor-counters}D). In contrast, the colored red cycle of Figure \ref{fig:dynamic-attractor-counters}D, at first glance symmetric to the green cycle, will not support a fixed point, by Rule \ref{rule:cycles} again (because, for instance, node 7 has two inputs from it). The attractor corresponding to the core motif $\{1, 4, 7, 10, 13, 16\}$ has not yet been found computationally.

We therefore expect to see at least six limit cycles, each one corresponding to a 5-star. Simulations confirm these predictions. In this case, both selective stimulation of specific nodes (Fig. \ref{fig:dynamic-attractor-counters}C) and identical stimulation of all multiple-of-three nodes (Fig. \ref{fig:dynamic-attractor-counters}D) can move the attractor to the next limit cycle in the chain. Indeed, in Figure \ref{fig:dynamic-attractor-counters}C pulses are sent to specific nodes (color coded) which activate the attractor associated with the core motif that the node belongs to. For example, stimulating node $9$ will activate the limit cycle associated with the core motif $\{7, 8, 9, 10, 11\}$, to which $9$ belongs.  Note that the order in which pulses are sent matches the order in which the attractors are chained. Simulations show that is it not always possible to jump between non-adjacent 5-stars. The matrix and parameters needed to reproduce this Figure are available in Appendix \ref{ch:appendixA}, Equation \ref{eq:dynamic-attractor-chain-matrix}. 

By contrast, in Figure \ref{fig:dynamic-attractor-counters}D, pulses are identically sent to the nodes $\{3,6,9,12,15,18\}$ (yes, all the multiples of 3), each producing a transition to the next limit cycle down the chain, analogous to the mechanism observed in the previous two counters, where pulses contain no information about which attractor comes next, but the state of the network counts pulses.This difference is more clearly conceptualized in the cartoon above each type of stimulation (Fig. \ref{fig:dynamic-attractor-counters}A vs Fig. \ref{fig:dynamic-attractor-counters}B).

\paragraph*{Robustness of dynamic attractor chain.}

We performed similar robustness simulations for both kinds of stimulation types in the dynamic attractor chain as we did for the fixed point counters. That is, we ran simulations for several parameter pairs $(\varepsilon,\delta)$, and several percentages of noise $(p,q)$ in $W$ and $\theta$. The results of these noise simulations are summarized in Figure \ref{fig:dynamic-attractor-counters-robustness}. 
\begin{figure}[!h]
	\begin{center}
		\includegraphics[width=\textwidth]{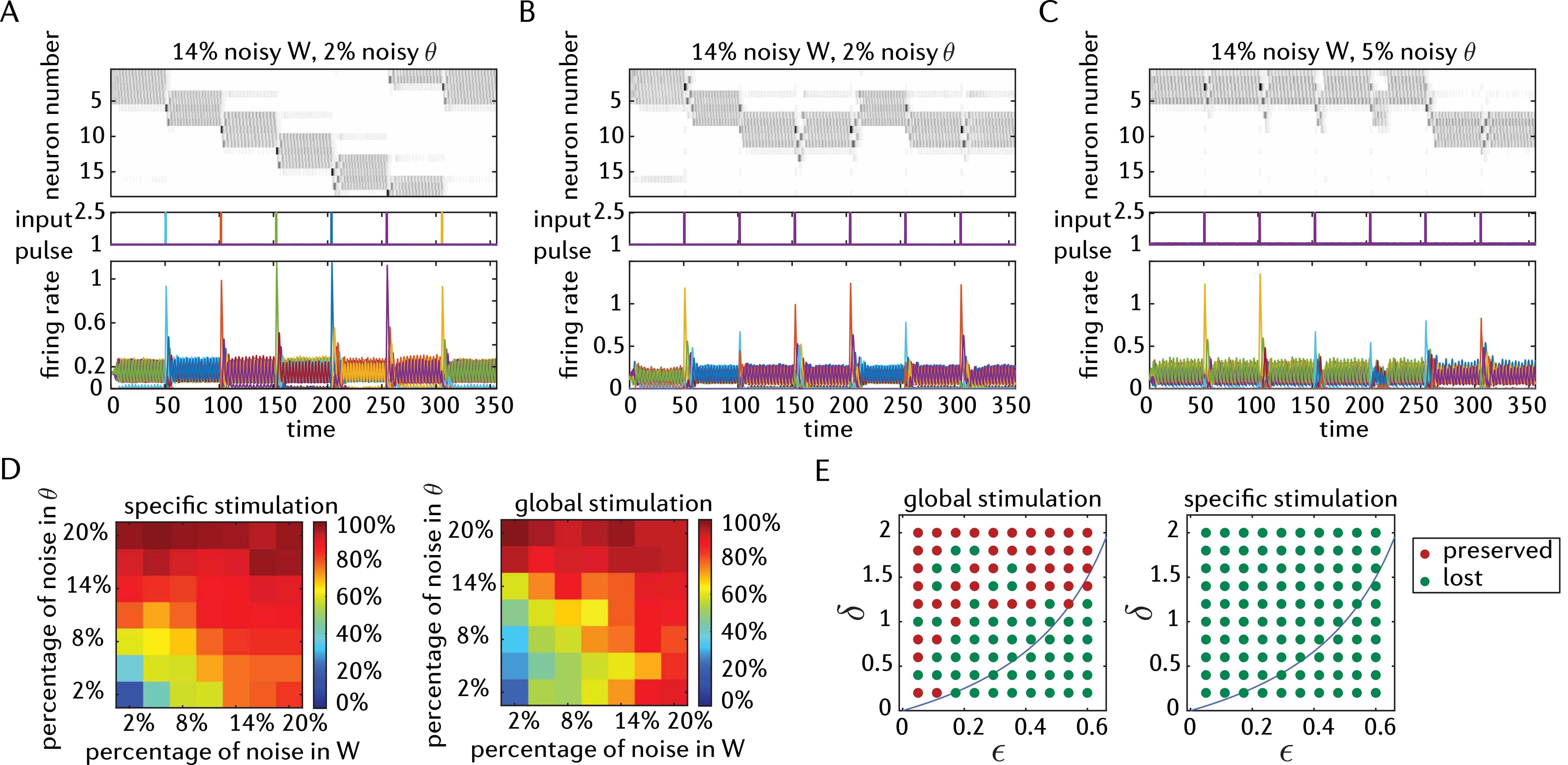}
	\end{center}
	\caption[Noisy dynamic attractors chain]{\textbf{Noisy dynamic attractors chain.} 
	(A) Example of a noisy dynamic attractor chain network that performs well for specific stimulation of all multiple-of-three nodes. 
	(B) Example of a noisy dynamic attractor chain network with 3 failures, global stimulation.
	(C) Example of a noisy dynamic attractor chain network with 4 failures, global stimulation.
	(D) Percentage of failed transitions for specific and global stimulation in the dynamic attractor chain network in 120 trials for several percentages of noise in $W$ and $\theta$.
	(E) Different behaviors of the dynamic attractor chain for various pairs of $\varepsilon,\delta$ parameter values. 
	}
	\label{fig:dynamic-attractor-counters-robustness}
\end{figure}

First, we analyze the performance under varying levels of noise. All the noise simulations were done with $\varepsilon = 0.51, \delta = 1.76$. The robustness to noise of the dynamic attractor chain, in both types of stimulation, was calculated as in Section \ref{sec:fixed-point-counters}. Recall that this was done by computing the proportion of failed transitions among $m$ pulses in both fixed point counters. This was done by introducing varying percentages of noise into the $\theta$ input, varied every $dt = 0.1$ of a time unit, and into the connectivity matrix $W$, as specified in the axes of Figure \ref{fig:dynamic-attractor-counters-robustness}D. This time however, $m=120$ to accommodate the 6-pulse simulation, for 20 simulations per percentage of noise pair $(p,s)$. As expected, our dynamic attractor chain is not nearly as robust as our CTLN counters of static attractors. Interestingly, global stimulation performed slightly better than specific stimulation in the presence of noise, but even this was barely robust at 5\% noise. Nevertheless, although not very robust to noise, the dynamic attractor chain still performs well under a wide range of $\varepsilon,\delta,\theta$ parameters, as seen in Figure \ref{fig:dynamic-attractor-counters-robustness}E, where the red dot indicates that the attractor was qualitatively identical to the ones originally observed in Figure \ref{fig:dynamic-attractor-counters}C,D.

By chaining together identical graphs, we have successfully constructed a network that steps through a set of dynamic attractors that were all repetitions of motifs of the same type, namely the 5-stars. Can we patch together a more diverse set of dynamic attractors in a single network, all accessible via initial conditions and/or pulses? In the next section, we present a toy model that requires coexistence of diverse dynamic attractors, where each attractor is accessible via targeted pulses. Later, we will put these networks together with a counter to step through a set of disparate dynamic attractors.



\chapter{Central pattern generators}\label{ch:CPGs}


In the last chapter, we leveraged network symmetry to provide an example of a network that can support several coexistent dynamic attractors, all easily accessible via attractor-specific inputs, and also accessible in sequence. A natural neuroscience application of coexistent dynamic attractors, and the one we seek to model in this section, is pattern generation. Central Pattern Generators (CPGs) are networks of neurons that can generate rhythmic output in the absence of rhythmic driving input \cite{Marder-CPG}.  However, unlike the dynamic attractor chain of the previous section, CPGs may need to support many different types of dynamic patterns simultaneously. So, in addition to coexistence of dynamic attractors, we now also want diversity of said attractors, as schematized in Figure \ref{fig:sequential-control-cartoons-C-highlight}C.
\begin{figure}[!h]
	\begin{center}
		\includegraphics[width=\textwidth]{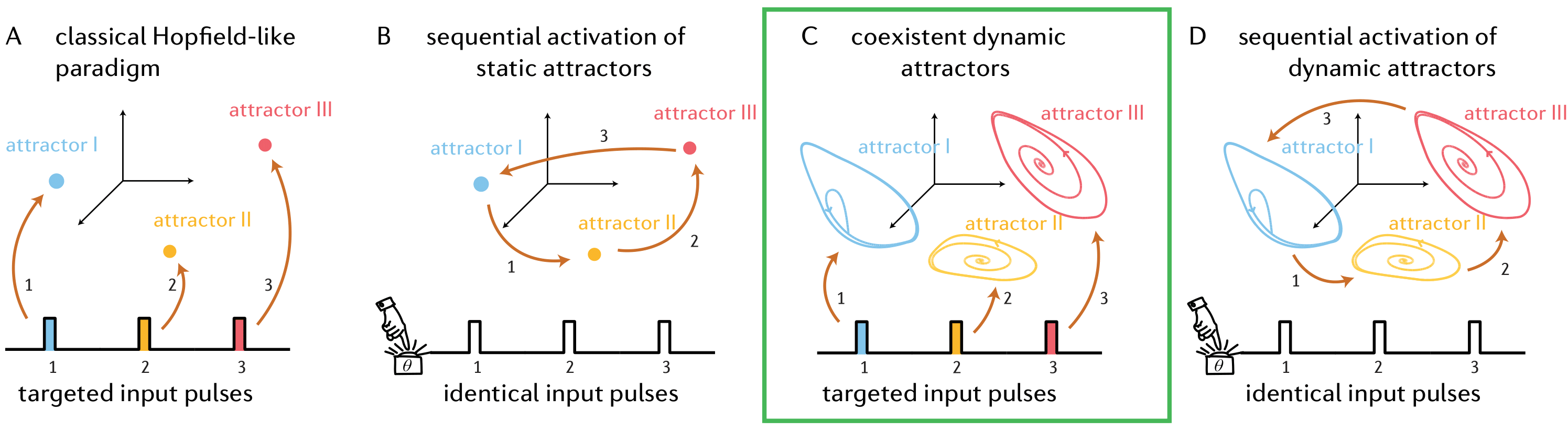}
	\end{center}
	\caption[Coexistent dynamic attractors]{\textbf{Coexistent dynamic attractors}. Reproduction of Figure \ref{fig:sequential-control-cartoons-intro}. The focus of this chapter is Panel C: Multiple dynamic attractors encoded in the same network, each one accessible via targeted (colored) inputs.
	}
	\label{fig:sequential-control-cartoons-C-highlight}
\end{figure}

In this chapter we focus on CPGs that have been of interest in neuroscience for a long time, and recently in robotics: animal locomotion \cite{Marder-CPG,Dzeladini2018,Dutta2019,Varona2002,OliveiraSantos2023}. Quadruped locomotion is probably the most popular example of a CPG in neuroscience \cite{Golubitsky1998,Ijspeert2008,Dutta2019}, and beyond \cite{Gambaryan1974,Muybridge1891}. Modeling the different modes of locomotion, called gaits, has been challenging because of the overlap in units controlling legs between different gaits. Most models overcome this difficulty by requiring changes in network parameters, such as synaptic weights \cite{Golubitsky1998,Dutta2019}, in order to transition between different gaits. In contrast, here we present a CTLN capable of reproducing \emph{five} coexistent quadrupedal gaits (bound, pace, trot, walk and pronk). Gaits coexist as distinct limit cycle attractors in the network, without any change of parameters needed to access different gaits. Instead, different gaits are obtained from different initial conditions, or by $\theta$-stimulation of a specific neuron involved in the gait. Also, where most models for locomotion use oscillating units, the oscillatory behavior we obtain is a result of connectivity alone, since none of our units are intrinsically oscillating.

Although several types of architectures have been proposed to model distinct CPGs (recurrent neural network based, half-center oscillator based and abstract oscillator based CPGs)  \cite{Dzeladini2018}, coexistence of different attractors in the same circuit  (as would be needed to model different modes of locomotion, controlled by the same set of neurons) has been challenging to many models. In this chapter we leverage cyclic unions (Thm. \ref{thm:cyclic-unions}, Ch. \ref{ch:background}), to present two different locomotion CPGs where patterns arise as attractors, and are thus very robust to noise and perturbations: a network that supports five different quadrupedal gaits and molluskan hunting mechanism. 

\section{Cyclic unions as pattern generators}
\begin{figure}[!h]
	\begin{center}
		\includegraphics[width=0.9\textwidth]{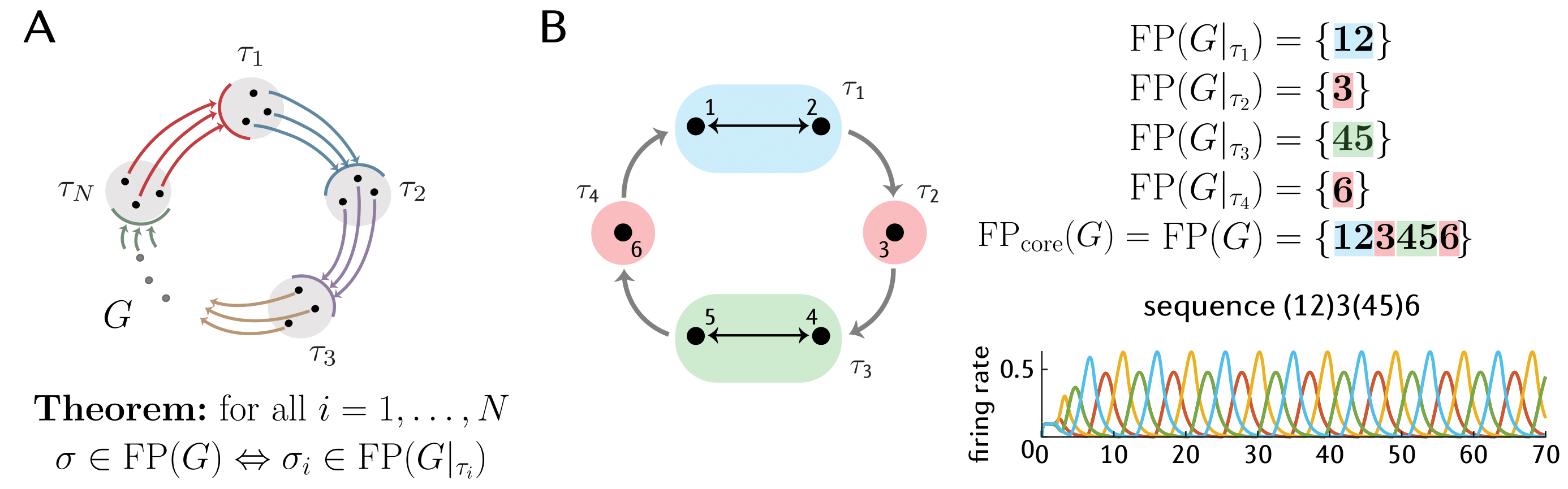}
	\end{center}
	\caption[Cyclic unions as pattern generators]{\textbf{Cyclic unions as pattern generators}. (A) Cyclic unions, and Theorem \ref{thm:cyclic-unions}.
		(B) Example of a cyclic union giving rise to sequential activation of the nodes. }
	\label{fig:cyclic-unions-as-pattern-generators}
\end{figure} 
Recall from Chapter \ref{ch:background} that for cyclic unions (reproduced in Fig. \ref{fig:cyclic-unions-as-pattern-generators}A), the neural activity flows through the components in cyclic order. Thus, cyclic unions are particularly well suited to model networks that must follow a sequential activation of the nodes, like CPGs, because attractors themselves will follow the direction of the cyclic union. This is true for other, less constricted architectures. These are the subject of \cite{Parmelee2022}, some of whose results are work of my own and thus presented in Chapter \ref{ch:new-theoretical-results}.

Recall also from Theorem \ref{thm:cyclic-unions} in Chapter \ref{ch:background} that the core motifs of the cyclic unions are made up of core motifs coming from each component and thus cyclic union of core motifs is core. For instance, in Figure \ref{fig:cyclic-unions-as-pattern-generators}B, we have an example of a cyclic union of four core motifs: two 2-cliques ($\{1,2\}$, $\{4,5\}$) and two single nodes ($\{3\}$, $\{6\}$). The $\FP(G)$ set of the whole graph is made up of pieces taken from each $\FP(G_{\tau_i})$, colored-coded by component. Because all components were core motifs, the whole graph is now core motif, as seen by its $\FP(G)$ in Figure \ref{fig:cyclic-unions-as-pattern-generators}B. The activity of the network, pictured below its $\FP(G)$ set, follows the direction of the cyclic union. Note that neurons 1 and 2 are synchronized, neurons 5 and 4 are synchronized, and then there is a successively cycling through these pairs, half a period apart. 
\begin{figure}[!h]
	\begin{center}
		\includegraphics[width=0.9\textwidth]{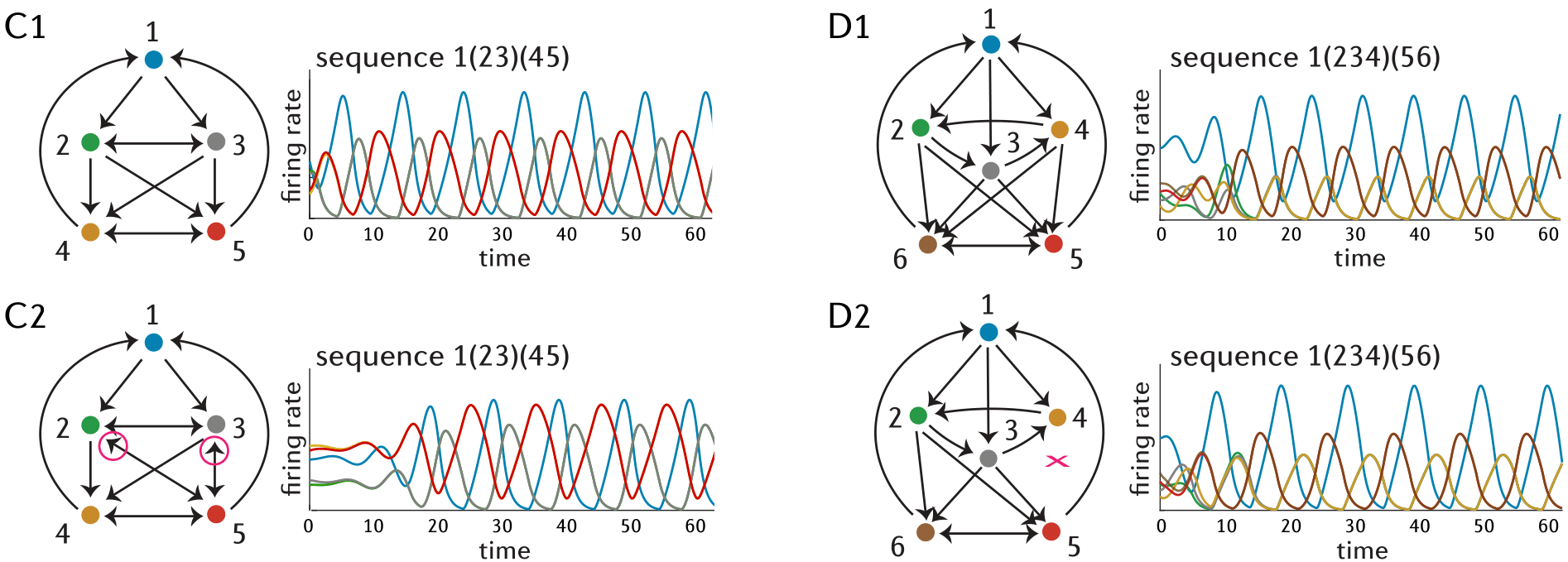}
	\end{center}
	\caption[Cyclic unions and generalizations]{\textbf{Cyclic unions and generalizations}. (C1,D1) Cyclic unions with firing rate plots showing solutions to the corresponding CTLN.  (C2,D2) These graphs are all variations on the cyclic unions above them, with some edges added or dropped (in magenta). Solutions of the bottom CLTNs qualitatively match the solutions of top CTLNs. Modified from \cite{Parmelee2022}.}
	\label{fig:cyclic-union-generalizations}
\end{figure}

As mentioned earlier, not only cyclic unions yield sequential attractors, as seen in Figure \ref{fig:cyclic-union-generalizations}: both the graphs at the top of the figure are examples of cyclic unions with three components, and the activity for these networks traverses the components in cyclic order. Compare these with the the bottom graphs of the figure, which do not have a perfect cyclic union structure (each graph has some added back edges or dropped forward edges highlighted in magenta), but have very similar dynamics to the ones above them. Despite the deviations from the cyclic union architecture, these graphs still produce sequential dynamics. 

\section{Quadruped gaits}\label{sec:quadruped-gaits}
In this section we present a CTLN capable of reproducing \emph{five} coexistent quadrupedal gaits: bound, pace, trot, walk and pronk. Figure \ref{fig:gaits-phase-relations} shows how the gaits are characterized by their \emph{relative phases}. For example, the bound gait is characterized by having both front legs synchronized, both back legs synchronized, and then successively cycling through these pairs, half a period apart. This is why back legs have a relative phase of 0, and front legs both have a relative phase of $0.5$. The rest of the gaits in Figure \ref{fig:gaits-phase-relations} are read similarly. Yes, pronk is a little all-legs jump. Gazelles do it.
\begin{figure}[!h]
	\begin{center}
		\includegraphics[width=\textwidth]{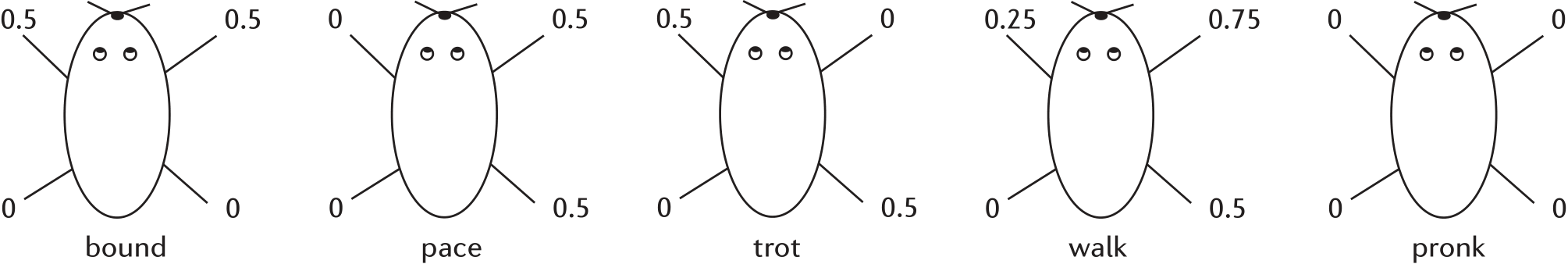}
	\end{center}
	\caption[Gait phase relations]{\textbf{Gait phase relations}. Diagram showing the relative phases of the limbs for each gait considered here. Modified from \cite{Alexander1984,Buono2001}.}
	\label{fig:gaits-phase-relations}
\end{figure}

We begin by modeling each one of these on its own, using cyclic unions, as foreshadowed by the example of Figure \ref{fig:cyclic-unions-as-pattern-generators}B.

\subsection{Construction of gaits}

\paragraph*{Single gaits} In the example of Figure \ref{fig:cyclic-unions-as-pattern-generators}B, we actually constructed the bound gait: two pairs of nodes are synchronized, half a period apart. There, we did it using only six nodes. In what follows, we aim to glue several gaits together, so it is not convenient to have such a small number of nodes per gait (very few nodes with a lot of edges will form cliques, which  yield stable fixed points, which we do not want for gaits), since adding new gaits will results in more and more connections between leg nodes and this would make everything collapse into a giant clique, resulting in a stable fixed point. This is why we now propose a slightly different construction for bound, and other gaits, by adding a few extra auxiliary neurons to overcome this issue. 

\begin{figure}
	\begin{center}
		\includegraphics[width=0.8\textwidth]{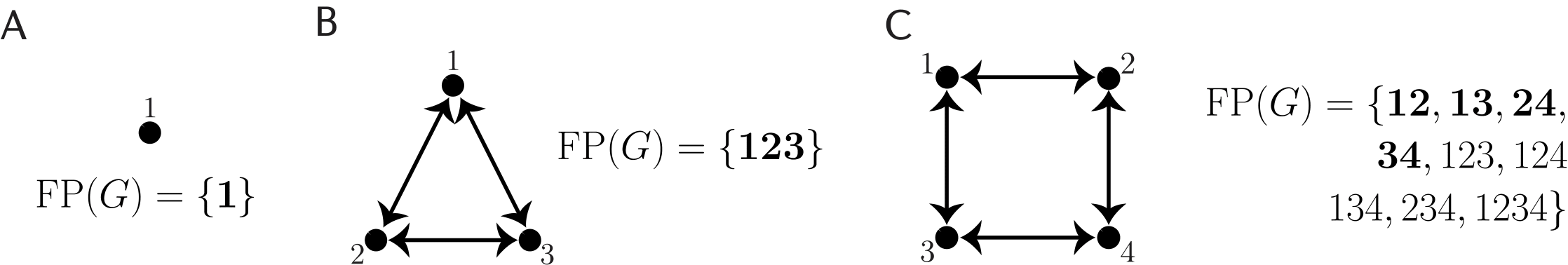}
	\end{center}
	\caption[Building blocks]{\textbf{Building blocks.} 
	(A) An isolated node.
	(B) A 3-clique.
	(C) A ``square'' made out of 2-cliques.
	Only A and B are core motifs.
	}
	\label{fig:building-blocks}
\end{figure} 

To re-model the bound gait, we replace the old 2-clique components by a square made up of four 2-cliques, like that of Figure \ref{fig:building-blocks}C. Why? The $\FP_{\text{core}}(G)$ of such a square is given by all the individual 2-cliques composing its sides, as seen in Figure \ref{fig:building-blocks}C. And since cyclic unions will grab a piece from every component, we are sure that a given pair of legs will still be part of some core motif of the network. More precisely: since the bound gait is characterized by having both front legs synchronized, both back legs synchronized, and then successively cycling through these pairs, half a period apart, as in Figure \ref{fig:design-of-isolated-quadruped-gaits}A, it is natural to group front legs in a single component and back legs in a single component, and to connect them through two auxiliary 1-node components. 
\begin{figure}[!h]
	\begin{center}
		\includegraphics[width=\textwidth]{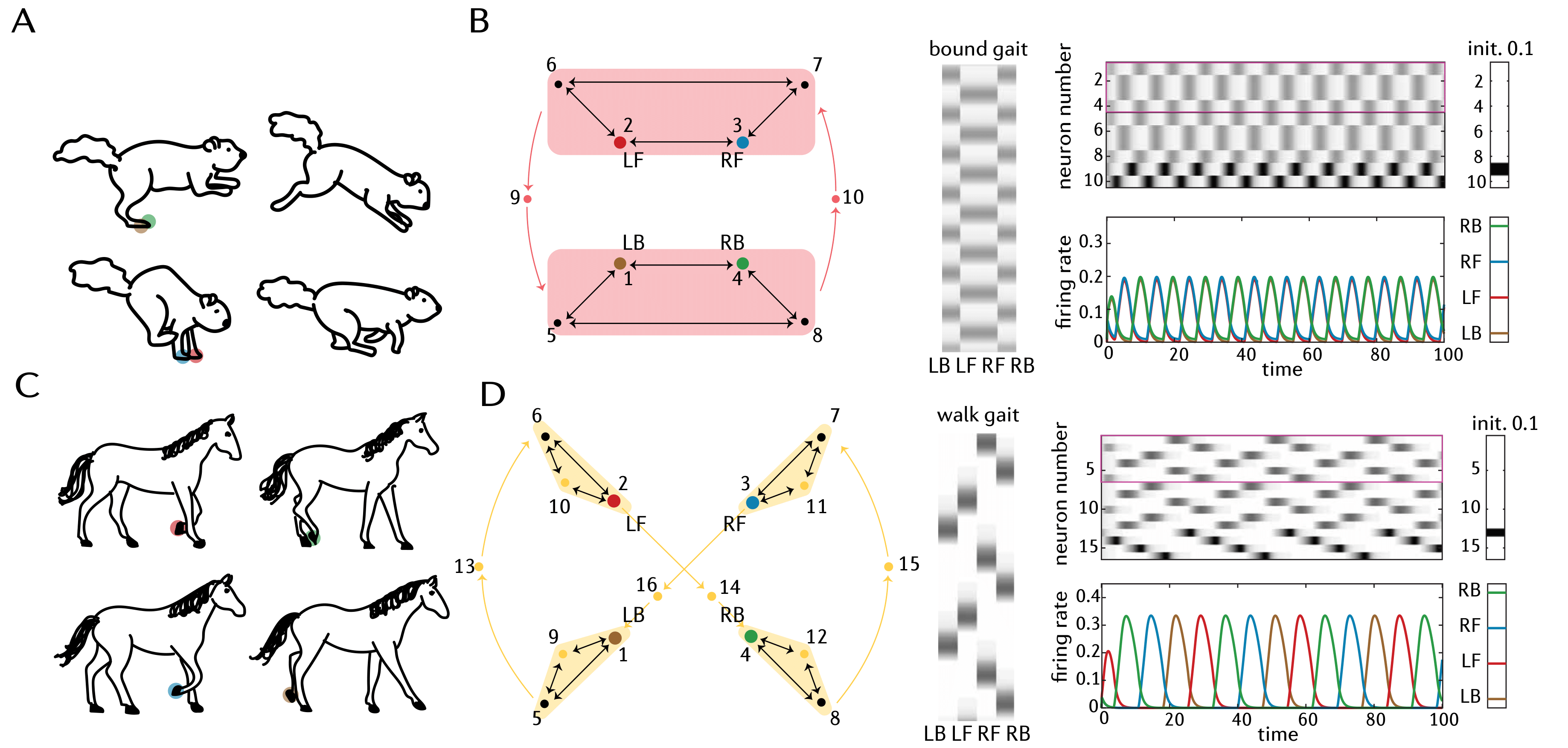}
	\end{center}
	\caption[Design of two different quadrupedal gaits]{\textbf{Design of two different quadrupedal gaits.} Colored circles represent node-leg assignment. Shading groups nodes in the same component. Thick colored arrows represent edges from/to all nodes in a single component. Nodes 1 through 4 control the legs (LB, LF, RF, RB). The first four rows of the greyscale, squared in pink and reproduced vertically, show the activity of the legs.
		(A) Souslik's bound. Pictured modified from \cite{Gambaryan1974}.  
		(B) Bound network and its corresponding CTLN solution. Nodes 5 to 10 are auxiliary nodes.The network was initialized at $x_9(0)=0.1$ and all other neurons at 0. 
		(C) Horse's walk. Pictured modified from \cite{Gambaryan1974}.
		(D) Walk network and its corresponding CTLN solution.  Nodes 5 to 16 are auxiliary nodes. The network was initialized at $x_{14}(0)=0.1$ and all other neurons at 0. 
	}
	\label{fig:design-of-isolated-quadruped-gaits}
\end{figure}

To achieve this, we split the nodes into four components, as shown Figure \ref{fig:design-of-isolated-quadruped-gaits}B. The resulting components are $\tau_1 = \{2,3,6,7\}$, $\tau_2  = \{10\}$, $\tau_3  = \{1,4,5,8\}$, and  $\tau_4  = \{9\} $. Theorem \ref{thm:cyclic-unions} now says that the $\FP_{\text{core}}(G)$ set of the bound gait of Figure \ref{fig:design-of-isolated-quadruped-gaits}B is obtained by selecting a support $\sigma_i \in \FP_{\text{core}}(G|_{\tau_i})$ from each component $\tau_i$ and forming their union: $\sigma_1 \cup \{9\} \cup \sigma_2 \cup \{10\}$, where $\sigma_1 \in \{\{2,3\},\{2,6\},\{6,7\},\{3,7\}\}$ and $\sigma_2 \in \{\{1,4\},\{1,5\},\{5,8\},\{4,8\}\}$. This will include, among others, the support $\{1,2,3,4,13,14\}$ corresponding to the bound gait attractor seen in the rate curves of Figure \ref{fig:design-of-isolated-quadruped-gaits}B. It is because we made sure to built-in the support $\{1,2,3,4,13,14\}$ that we can easily see and access the attractor reproducing the bound gait.

A slightly different construction consists of the gait walk, where legs move cyclically with a quarter period difference, as seen in Figure \ref{fig:design-of-isolated-quadruped-gaits}C. In the same manner, we use Theorem \ref{thm:cyclic-unions} to group each leg in a component along with its auxiliary neurons and add four auxiliary neurons to ensure that the phase between legs is a quarter period apart (Figure \ref{fig:design-of-isolated-quadruped-gaits}D). The building block for the leg components is now the 3-clique of Figure \ref{fig:building-blocks}B. Since the building block is a core motif, the $\FP_{\text{core}}(G)$ of the walk gait consists of a single element, formed by taking a core motif from each component in Figure \ref{fig:design-of-isolated-quadruped-gaits}D, resulting in $\FP(G) = \FP_{\text{core}}(G) = \{1,2,3,4,5,6,7,8,9,10,11,12,13,14,15,16\}$. This core motif promises to give rise to the attractor reproducing the walk gait. Indeed, the resulting dynamics of this network are depicted in Figure \ref{fig:design-of-isolated-quadruped-gaits}D, and the network once again behaves as expected, reproducing the walk gait. The matrix and parameters needed to reproduce these simulations, and other individual gaits, are available in Appendix \ref{ch:appendixA}, Equations \ref{eq:bound-matrix} to \ref{eq:pronk-matrix}. 

Other individual gaits can be similarly constructed, as shown in Figure \ref{fig:five-gait-network-and-transitions}A. In pace, left legs are synchronized, right legs are synchronized, half a period apart, and so we group neurons into appropriate components, in the same way we did bound. Similarly with trot, where diagonal legs are synchronized, half a period apart. The core motifs of pace and trot are similarly derived, using the ``square'' building block of Figure \ref{fig:building-blocks}C, along with Theorem \ref{thm:cyclic-unions}. The last gait our model will include is pronk, where all legs are synchronized, requiring now a single component to group all legs together along with a pair of auxiliary nodes that control the frequency of the pronking movement. 
\begin{figure}[!h]
	\begin{center}
		\includegraphics[width=\textwidth]{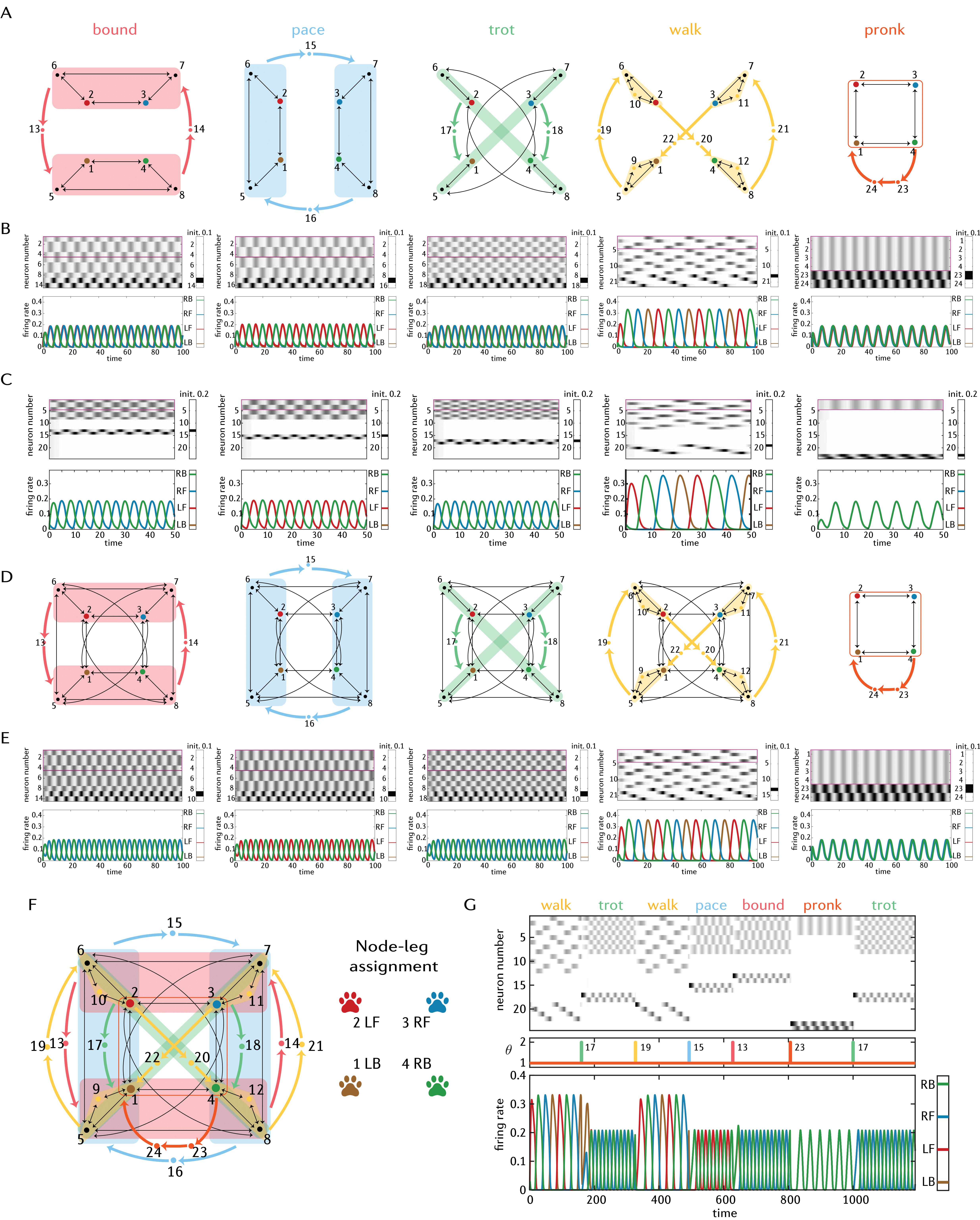}
	\end{center}
	\vspace{-5pt}
	\caption[Five-gait network and transitions]{
		(A) Graphs for individually constructed gaits. 
		(B) Solution for isolated gaits.
		(C) Solution for embedded gaits.
		(D) Induced subgraphs of embedded gaits.
		(E) Solution for subgraphs of embedded gaits.
		(F) Five-gait network.
		(G) Gait transitions via $\theta$-pulses.
	}
		\vspace{-5pt}
	\label{fig:five-gait-network-and-transitions}
\end{figure}

Note that all gaits have been constructed as a cyclic union of the three basic building blocks of Figure \ref{fig:building-blocks}: a ``square'' made out of 2-cliques, a 3-clique and an isolated node. From these pieces, and by Theorem \ref{thm:cyclic-unions}, the $\FP(G)$ set for each gait $G$ can be readily obtained as we did with bound: by selecting a support $\sigma_i \in \FP(G_i)$ from each component $G_i$ and forming their union $\sigma = \cup_{i=1}^{4} \sigma_i \in \FP(G)$. For a detailed analysis of the $\FP(G)$ sets see Tables \ref{tab:quadrupedgaitsFP} and  \ref{tab:isolated-gaits-fp} in Appendix \ref{ch:appendixA}. 

Since each gait was constructed to have $\FP_{\text{core}}(G)$ include the leg nodes, along with the appropriate auxiliary nodes, we expect the graphs of Figure \ref{fig:five-gait-network-and-transitions}A to give rise to a CLTN that reproduces these gaits. Indeed, in Figure \ref{fig:five-gait-network-and-transitions}B we have simulated the individually constructed gait networks. As expected, we see an attractor reproducing the respective gait. Initial conditions are plotted to the right of each greyscale, and were chosen to have all neurons off, except one colored auxiliary neuron on.

Matrices and parameters needed to reproduce the simulations of Figure \ref{fig:five-gait-network-and-transitions} are available in Appendix \ref{ch:appendixA}.

Now, can we make these attractors we are seeing here coexist into a single network, with leg nodes overlapping? Said differently, can we glue together (in topology sense) the networks in Figure \ref{fig:five-gait-network-and-transitions}A by leg nodes, and other shared nodes? As we will see below, the answer is yes.

\subsection{Coexistent gaits} 

To construct the five-gait network, we glued together the networks in Figure \ref{fig:five-gait-network-and-transitions}A by identifying common neurons ($1$ through $6$) and induced edges between them. Figure \ref{fig:five-gait-network-and-transitions}F shows the resulting glued network. More precisely, this gluing identifies all nodes from $1$ to $6$ coming from each isolated gait into a single one in the glued network. This operation does not exactly correspond to any graph operation, and it is more like a gluing in a topological way, where the gluing instructions are given by the vertex numbers and the edges between them.

As mentioned before, the inclusion of several auxiliary nodes was necessary to merge the five networks into a single one. Indeed, nodes 1 to 4 are shared by all gaits and represent the leg assignment (1. LB - left back, 2. LF - left front, 3. RF - right front, 4. RB - right back), but nodes 5 to 8 are shared by bound, pace and trot, while not representing any leg assignment. The rest of the nodes are gait-specific, meaning that they are uniquely associated to a single gait and therefore only active when such gait is active, and off otherwise. These nodes play the very important roles of keeping the phase between foot steps in each gait and activating a specific gait by stimulating one of them. These nodes are: nodes 9 to 12 which are there to augment the walk cliques, and so they are specific to walk; and nodes 13 to 23 which are specific to a unique gait, as color coded in Figure \ref{fig:five-gait-network-and-transitions}A.


Although the complex structure of the merged network makes it hard to analytically derive the $\FP(G)$, we found computationally that $|\FP(G)| = 875$ (again, coming from up to $2^{24}$ linear systems), and 
\begin{multline}\label{eq:gait-cores}
\FP_{\text{core}}(G) = \{\{1, 2, 23, 24\},\{1, 4, 23, 24\},\{2, 3, 23, 24\},\{3, 4, 23, 24\}, \\ \{1, 6, 7, 8, 13, 16, 18\}, \{2, 5, 7, 8, 14, 16, 17\}, \{3, 5, 6, 8, 14, 15, 18\}, 
\{4, 5, 6, 7, 13, 15, 17\}\}.
\end{multline}
Although sadly none of the core motifs in Equation \ref{eq:gait-cores} there correspond to the gait attractors, in Figure \ref{fig:five-gait-network-and-transitions}C it can be seen that all gaits are actually preserved when embedded in the big network with all gaits, as the dynamics of each gait when embedded in the glued network are indeed remarkably the same as when isolated (cf. panel B). The greyscale and rate curves for each gait embedded within the glued network are plotted in the same column as its corresponding graph in Figure \ref{fig:five-gait-network-and-transitions}A. Also, it is important to note that even though the isolated gait networks (Fig. \ref{fig:five-gait-network-and-transitions}A) are \emph{not} the same as the induced gait networks (Fig. \ref{fig:five-gait-network-and-transitions}D), they support the same (qualitatively) attractor, as seen in the rate curves of Figure \ref{fig:five-gait-network-and-transitions}E.

To see that each gait is accessible via different initial conditions when embedded in the five-gait network, notice in Figure \ref{fig:five-gait-network-and-transitions}B that when all neurons, but neuron 13, are initialized to zero, the system evolves into the bound gait. This is because neuron 13 corresponds \emph{uniquely} to bound. Similarly, when the network is initialized in neuron 15, it goes into pace. When it is initialized in neuron 17, it goes into trot. And so on. Recall that walk has two different sets of gait-specific neurons (9 - 12 and 19 - 22), and both kinds will send the network into the walk gait. Hence, our five-gait network successfully encodes these coexistent dynamic attractors, each accessible via initial conditions, that have several overlapping active nodes (nodes 1 to 8), with no interference between attractors. Moreover, the attractors are not overly sensitive to the value of initial conditions, as long as the highest firing neuron in the initial condition is the one associated with the desired gait.

In particular, since all gaits coexist as distinct limit cycle attractors in the network, each accessible via different initial conditions, it is also possible to smoothly transition between gaits by means of a gait-specific external pulse $\theta$. To effectively change gaits, it suffices to stimulate a the appropriate gait-specific auxiliary neuron with a pulse. The network quickly settles into the dynamic attractor corresponding to the gait of the auxiliary neuron stimulated, as seen in Figure \ref{fig:five-gait-network-and-transitions}G. There, pulses of $2$ time units are sequentially sent to neurons 17,19,15,13,23 and 17, producing the expected transitions between gaits. The network smoothly transitions to the corresponding gait. The order in which neurons are stimulated does not matter, any sequence of pulses sent to gait-specific auxiliary neurons will produce the respective transitions, regardless of which gait comes before or after. 

However, not all gaits are created equal, and this becomes evident in some transitions. From the construction of the graph we suspect that bound, pace and trot are on equal footing, since they were created analogously, only differing by which pairs of legs are synchronized. Consequently, we expect some symmetry in their basins of attraction, so that transitioning to and from these gaits is easier, and this has been indeed observed in simulations. By contrast, we have observed that walk and pronk (both at the boundary of how many legs can be synchronized with 1 leg and 4 legs respectively) seem to have a bigger basins of attraction, since the network requires a higher $\theta$ stimuli to transition \emph{out} of these two gaits. All in all, this network provides a straightforward mechanism to switch between the desired attractors, fulfilling our initial goal of controlling coexistent dynamic attractors in a single network.

\subsection{Parameter analyses}\label{sec:parameter-analyses}

\paragraph*{Good parameters.} A common drawback in many models is that sometimes they require very fine-tuned values of their parameters to perform as expected. Also since dynamic attractors are known to be less robust than fixed point attractors, it makes sense to assess the robustness of the network in Figure \ref{fig:five-gait-network-and-transitions}F. Individual gaits were designed as a cyclic union (Thm. \ref{thm:cyclic-unions}) of building blocks (Fig. \ref{fig:building-blocks}) whose $\FP(G)$ is parameter independent \cite{fp-paper}, meaning that its $\FP(G)$ set is preserved across the \emph{legal range} of parameters $\varepsilon < \frac{\delta}{\delta+1}$. Since cyclic unions of parameter independent components are also parameter independent, all isolated gaits are parameter independent as well. Indeed, it is a simple consequence of Theorem \ref{thm:cyclic-unions}:
\begin{cor}
	Let $G$ be a cyclic union of component subgraphs $G|_{\tau_1},\dots,G|_{\tau_N}$. Then $\FP(G)$ is parameter independent if and only if $\FP(G|_{\tau_i})$ is parameter independent for all $i\in [N]$.
\end{cor}
\begin{proof}
	Suppose $\sigma \in \FP(G,\varepsilon,\delta,\theta)$ for all $\varepsilon,\delta,\theta$ in the legal range. By Theorem \ref{thm:cyclic-unions}, $\sigma \cap \tau_i \in \FP(G|_{\tau_i})$ for all $\varepsilon,\delta,\theta$ in the legal range and all $i \in [N]$, meaning that $\FP(G|_{\tau_i})$ is also parameter independent. The other implication is analogous.  
\end{proof}

This implies that the $\FP(G)$ set does not vary with parameter changes within the legal range, and so we can expect the gait attractors to be present in each individual gait network for a wide range of parameter values, so each gait, \emph{isolated}, turns out to be very robust! The same unfortunately cannot be said about the glued five-gait network, since it is not truly a cyclic union, but rather a gluing of cyclic unions, for which we do not have theoretical results yet. So, how sensitive is our network to the choice of $\varepsilon,\delta,\theta$ parameters? Are the five attractors still present, and easily accessible, if we vary our parameter values? 

To answer this question, we tried to reproduce the simulations of Figure \ref{fig:five-gait-network-and-transitions}C using different $\varepsilon,\delta,\theta$ parameters. First, we observed how the attractor corresponding to each gait behaved under several values of $\varepsilon, \delta$ in the five-gait network. We found that there were three different possibilities: either the attractor is \emph{preserved}, meaning that it is identical in qualitative behavior to the attractors in Figure \ref{fig:five-gait-network-and-transitions}C (up to a change in amplitude or period); \emph{corrupted}, indicating that the auxiliary neurons corresponding uniquely to that gait are still active and firing cyclically, but the leg nodes have lost its symmetry; or \emph{lost}, when the auxiliary neurons corresponding uniquely to that gait are not firing cyclically (this usually translated into a gait becoming a different attractor, which could either be another gait, a stable fixed point or a chaotic-looking attractor).  

An example of each of these outcomes is exemplified in Figure \ref{fig:good-parameter-grid-gaits}A, where only the last 200 time units of the simulation are shown, for clarity.  In the left example, the network was initialized to the bound gait, and the gait is perfectly preserved. In the middle plot, the network was initialized to the pronk gait, and even though the auxiliary neurons corresponding to pronk are still active, the leg neurons lost their symmetry towards the end. In the right plot, the network was initialized to the trot gait, but near the end of the simulation the network settles into a corrupted form of pronk, as indicated by the auxiliary neurons of pronk being active. The results of all those simulations are summarized in Figure \ref{fig:good-parameter-grid-gaits}B. Green denotes preserved attractor, yellow denotes corrupted attractor and red denotes lost attractor. The intersection of the green dotted regions is the range in which all gaits are preserved: $0.05<\varepsilon<0.23$, $0.2<\delta<0.6$, for at least the first 500 units of time. These are the ``good parameters''.
\begin{figure}[!h]
	\begin{center}
		\includegraphics[width=\textwidth]{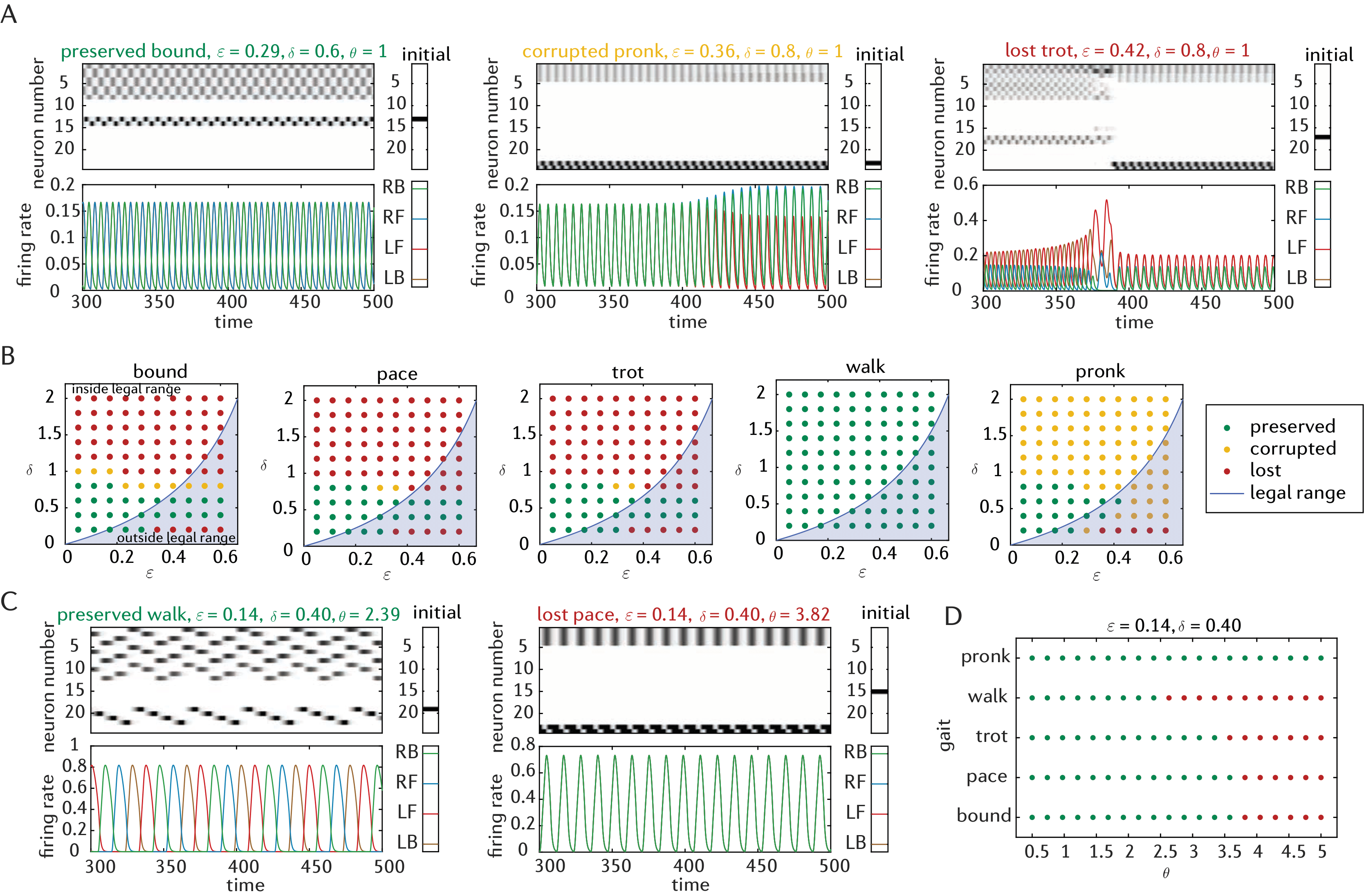}
	\end{center}
	\caption[Gaits survival under several values of $\varepsilon,\delta,\theta$]{\textbf{Gaits survival under several values of $\varepsilon,\delta,\theta$.} All plots were generated with a running time of $500$ time units.
		(A) Three examples of what can happen to an attractor under three different parameter values. 
		(B) Gaits are classified according to whether the attractors are either preserved (green dots), corrupted (yellow dots) or  lost to a different attractor (red dots) when $\varepsilon, \delta$ vary.  All the dots were generated using $\theta = 1$ fixed. 
		(C) Two examples of what can happen to an attractor under two different $\theta$ parameter values.
		(D)  Gaits are classified according to whether the attractors are either preserved (green dots), corrupted (yellow dots) or  lost to a different attractor (red dots) when $\theta$ varies.  All the dots were generated using $\varepsilon = 0.14$ and $\delta = 0.40$.}
	\label{fig:good-parameter-grid-gaits}
\end{figure}

We then did the same thing but varying the $\theta$ parameter only. Figure \ref{fig:good-parameter-grid-gaits}C shows examples of what can happen in this case. In the first example, the network was initialized to the walk gait, and the gait is perfectly preserved. In the right example, on the other hand, the network was initialized to the pace gait, but immediately went into pronk. Figure \ref{fig:good-parameter-grid-gaits}D shows the same summary of results, but carried out for constant values $\varepsilon = 0.14$, $\delta = 0.40$, and varying $\theta$. These values were arbitrarily chosen from the $0.05<\varepsilon<0.23$, $0.2<\delta<0.6$ good parameters from above. 

In conclusion, we have computationally found that all the attractors corresponding to the five gaits in our network are preserved for all $0.05<\varepsilon<0.23$, $0.2<\delta<0.6$ and $0.5 < \theta < 2.5$, for at least 500 units of time. Having a range of parameters available raises the question: what part do they play in modulating the non-qualitative characteristics of our gaits such as period and amplitude? These are important questions because period and amplitude in firing rates affect muscle tension \cite{purves2001,Williams2010}. 

\paragraph*{Parameter modulation.}
\begin{figure}[!h]
	\begin{center}
		\includegraphics[width=0.85\textwidth]{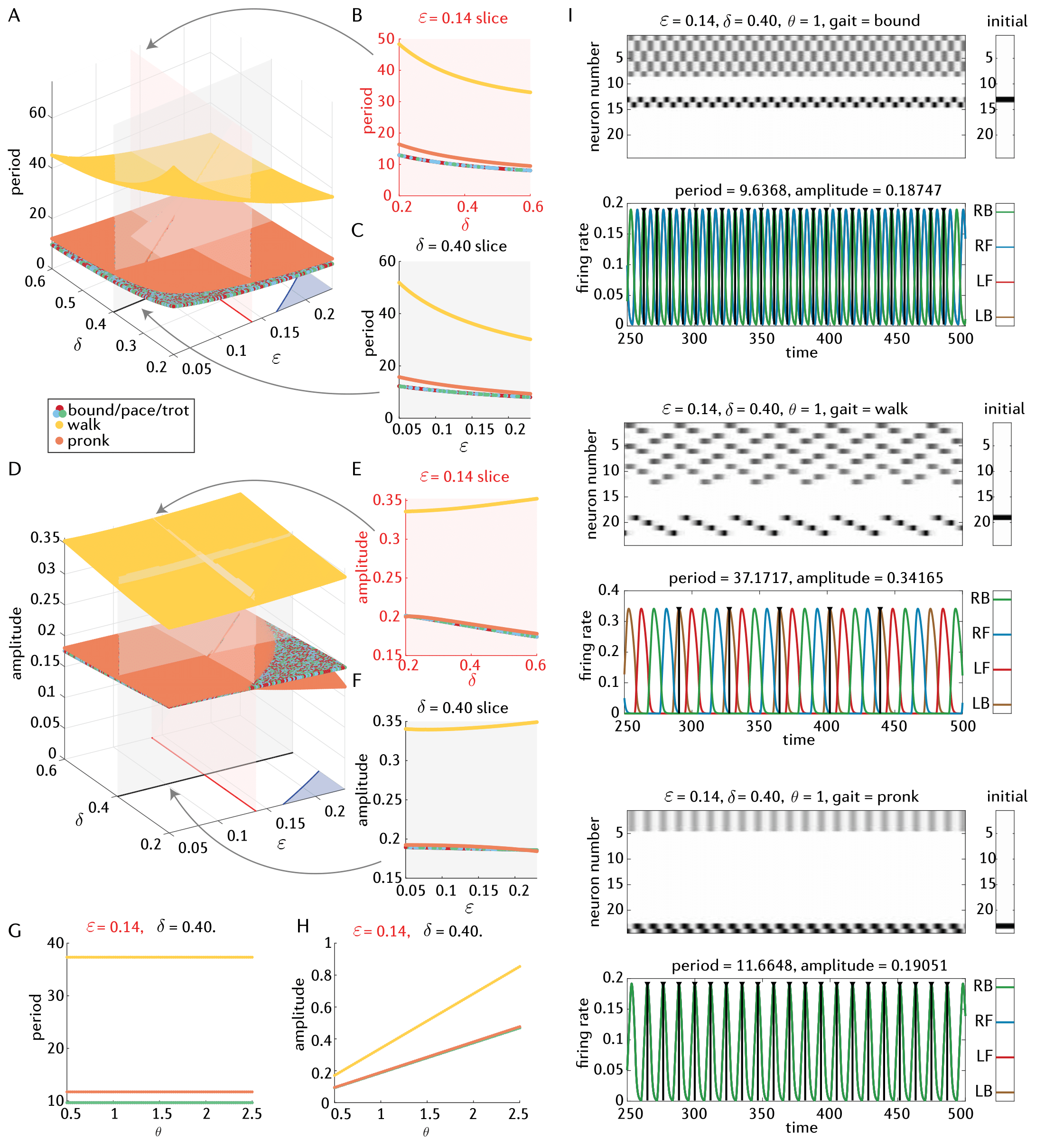}
	\end{center}
	\caption[Amplitude and period of the gaits in the five-gait network]{\textbf{Amplitude and period of the gaits in the five-gait network}. Values of period and amplitude for a $(0.05,0.23)\times(0.2,0.6)$ grid of $(\varepsilon,\delta)$ values and all gaits.
		(A) Period values. The data points for bound, pace and trot are superimposed.
		(B) Cross section of plot in panel A corresponding to $\varepsilon = 0.14$. 
		(C) Cross section of plot in panel A corresponding to $\delta = 0.40$.
		(D) Amplitude values. The data points for bound, pace and trot are superimposed.
		(E) Cross section of plot in panel D corresponding to $\varepsilon = 0.14$. 
		(F) Cross section of plot in panel D corresponding to $\delta = 0.40$.
		(G) Values of period for the cross sections in panels B and C ($\varepsilon = 0.14, \delta = 0.40$) and varying values of $\theta$.
		(H) Values of amplitude for the cross sections in panels E and F ($\varepsilon = 0.14, \delta = 0.40$) and varying values of $\theta$.
		(I) Three examples of how period and amplitude were computed.}
	\label{fig:amplitude-period-gaits}
\end{figure}

In this section, we compute the period and amplitude, of each gait within the five-gait network, using the good parameters from above. To do so, we ran a single simulation for $T=500$ time units and then we used the MATLAB function findpeaks to find the location and value of the peaks of a single rate curve. Examples of this can be seen in Figure \ref{fig:amplitude-period-gaits}I. The black bars in the rate curves show the amplitude, and the black triangles mark the time of the peaks, both calculated using only the first leg ($x_1(t)$ - LB). We found these peaks only during the last half of the simulation to avoid taking into account the time before settling into the attractor. To find the periods we subtracted all the consecutive peak locations. We had to get rid of the first and last period and amplitude data point because they became corrupted by chopping the curve. Then we averaged the periods and peak values to obtain a single period and a single amplitude per simulation. This is the value shown in Figures \ref{fig:amplitude-period-gaits}A-H.

There, each point corresponds to a single simulation. The color of the point corresponds to a given gait, as specified in the label of Figure \ref{fig:amplitude-period-gaits}A. Note that, since bound, pace and trot have virtually the same structure, the data point for these are overlapped in Figures \ref{fig:amplitude-period-gaits}A-H. We also show crossections for Figures \ref{fig:amplitude-period-gaits}A,D. The blue shaded corner of the 3-dimensional plots corresponds to values outside the legal range. In  Figures \ref{fig:amplitude-period-gaits}A-F we are varying only $\varepsilon$ and $\delta$, with a fixed value of $\theta=1$. In  Figures \ref{fig:amplitude-period-gaits}G-H, we use the fixed values of $\varepsilon,\delta$ in the slices above to vary the value of $\theta$.

Notice that for fixed values of $\varepsilon,\delta, \theta$, period decreases with $\varepsilon$ and $\delta$; whereas amplitude increases for walk, and decreases for all other gaits. Period is seen to be invariant to $\theta$ changes, and amplitude increases with increasing values of $\theta$. It is known that an increased firing rate produces an increase in muscle tension by summation of more frequent successive motor contractions \cite{purves2001,Williams2010}. This implies that in our model, period would correspond to how fast/slow muscles contract and relax (bigger period means muscles contract and relax faster). Period therefore is controlling how fast the movement is executed, that is, the speed of the gait. Amplitude of firing rate corresponds to tension in muscle, because the more action potentials, the more muscle units are activated. 

Great. But while we have showed that there is a fairly wide range of parameters under which our network behaves as expected, and where we can module the properties of the firing rates, there is still the assumption that all synaptic connections are perfect and identical: either $-1+\varepsilon$ or $-1-\delta$. Is our network robust to noise in the synapses? What about the inputs? Are transitions still possible in the light of noisy connections? 

\subsection{Robustness to noise} 
To explore these questions, we added uniform noise to $W$ and $\theta$ and computationally tested whether the attractor was lost. A gait is said to be lost when the attractor corresponding to said gait degenerates into a different attractor, which is clear from which auxiliary neurons are active. We did not measure if the attractor was corrupt or not, because any amount of noise will break the symmetry of the legs. 
\begin{figure}[!h]
	\begin{center}
		\includegraphics[width=\textwidth]{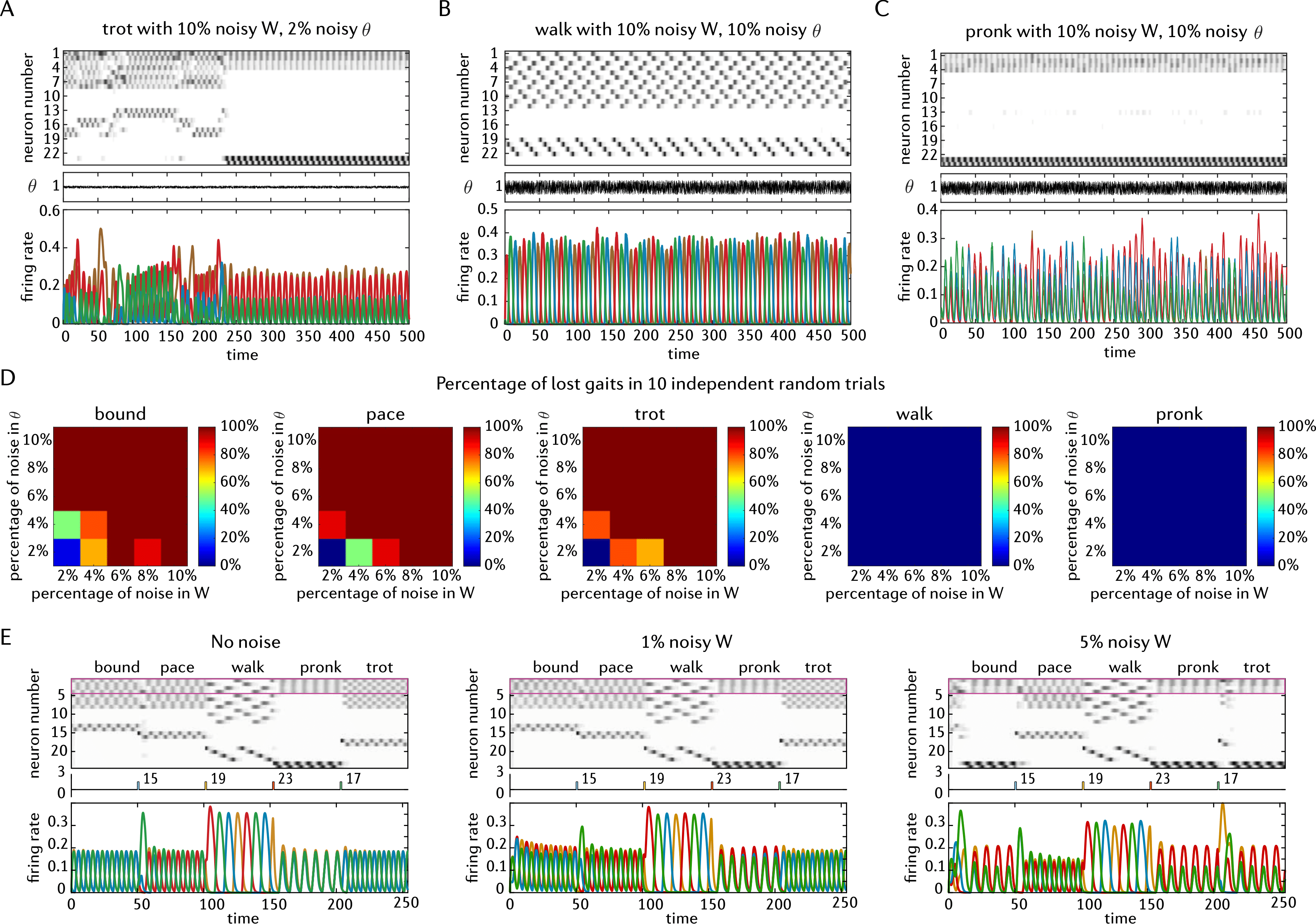}
	\end{center}
	\caption[Percentage of lost gaits in noisy five-gait networks]{\textbf{Percentage of lost gaits in noisy five-gait networks.} 
		(A) Example of lost trot.  
		(B) Example of a very noisy network, where walk is not lost, but it is corrupt. 
		(C) Example of a very noisy network, where pronk is not lost, but it is corrupt. 
		(D) Percentage of lost gaits in 10 trials, for several percentages of noise in $W$ and $\theta$.
		(E) Examples of noisy transitions.
	}
	\label{fig:noisy-gaits-and-transitions}
\end{figure} 
This was done in the same way as we did in Chapter \ref{ch:sequences-of-attractors}. We ran a single noisy simulation and observed if the attractor was lost or not. Examples of the behaviors we observed can be found in Figure \ref{fig:noisy-gaits-and-transitions}A-C. In A, the networks was initialized to trot, but it chaotically jumps between different attractors, until it settles to pronk. In B, the network was initialized to walk, and it remains there, as good as it can. This means, the legs lost their symmetry, but the high-firing neurons are the same as in walk, in the same order. This classifies as not lost in our analysis. In C, we observe the same situation but with pronk. Here, the asymmetry of the legs is more notable. These are wobbly gaits. 

Figure \ref{fig:noisy-gaits-and-transitions}D summarizes our findings. We simulated each gait 10 times, and then calculated the percentage of lost gaits in those 10 trials, for varying percentages of noise in $W$ and $\theta$. The color of the pixel in Figure \ref{fig:noisy-gaits-and-transitions}D represents the percentage of lost gaits out of these 10 trials. A trial corresponds to a single noisy pair $(W,\theta)$, which was used to test all gaits once. All simulations used to quantify the failures were done with $\varepsilon = 0.25, \delta = 0.5$ and $\theta$-noise with $dt = 0.1$.  

In around 90\% of trials, gaits were not lost at 2\% noisy $W$ and $\theta$, indicating that that the perfect binary synapses of CTLNs are not necessary for this network to produce the desired gaits, but bigger levels of noise will break the network's performance anyway. Interestingly, at bigger levels of noise, the gaits were lost most commonly to pronk, again. Pronk also happens to be one of the two most robust gaits (see pronk in Figure \ref{fig:noisy-gaits-and-transitions}D). This supports our suspicion that pronk's basin of attraction is bigger. It is also evident again the way in which the construction of the first three gaits differs from the last two gaits. This raises the question of what would happen to the robustness of the network if there was no pronk or walk, so that the network is perfectly symmetric.

Finally, we also tested the ability of the network to transition under 1\% and 5\% noise in $W$, as shown in Figure \ref{fig:noisy-gaits-and-transitions}E. Transitions are robust at 1\% added noise, but past 5\% gaits are commonly lost to pronk and thus transitions are not successful.

\section{Molluskan hunting}\label{sec:molluskan-hunting}

A dynamically different example of a CPG is Clione's hunting mechanism. Clione is a marine mollusk without a visual system, so when it chemically senses its prey, it must explore its surroundings by swimming in its vicinity until it finds it. The direction of swimming is controlled by Clione’s tail, which receives input from Clione’s gravitational sensory organs, the statocyst. The statocyst is covered by mechanoreceptors that receive inputs from the statolith, a stone-like structure moving inside the statocyst under the effect of gravity \cite{Panchin1995neuronalmechanisms}. 

The model we propose here is inspired by the model in \cite{Varona2002}. There, the authors propose a six-neuron model, with Lotka-Volterra-type units and inhibitory non-symmetric connections. This fosters competition and lead to a state of winnerless competition, where no single neuron dominates the activity. This in turns result in complex spatiotemporal patterns that mimic random changes in direction in the gravitational field. These patterns drive the unpredictable and random-like hunting behavior observed in Clione.

In that model, each node represents a receptor neuron in the gravity sensory organ, the statocyst. These receptor neurons are part of the sensory network that processes information related to the body's position relative to the gravitational field and are influenced by the central hunting neuron during hunting behavior in Clione. Here we propose a very similar model, also consisting of six-units, connected via inhibitory non-symmetric connections. However, our units are not intrinsic oscillators, and so the mechanism of pattern generation in our case arises purely from connectivity. 

\begin{figure}[h!]
	\begin{center}
		\includegraphics[width=\textwidth]{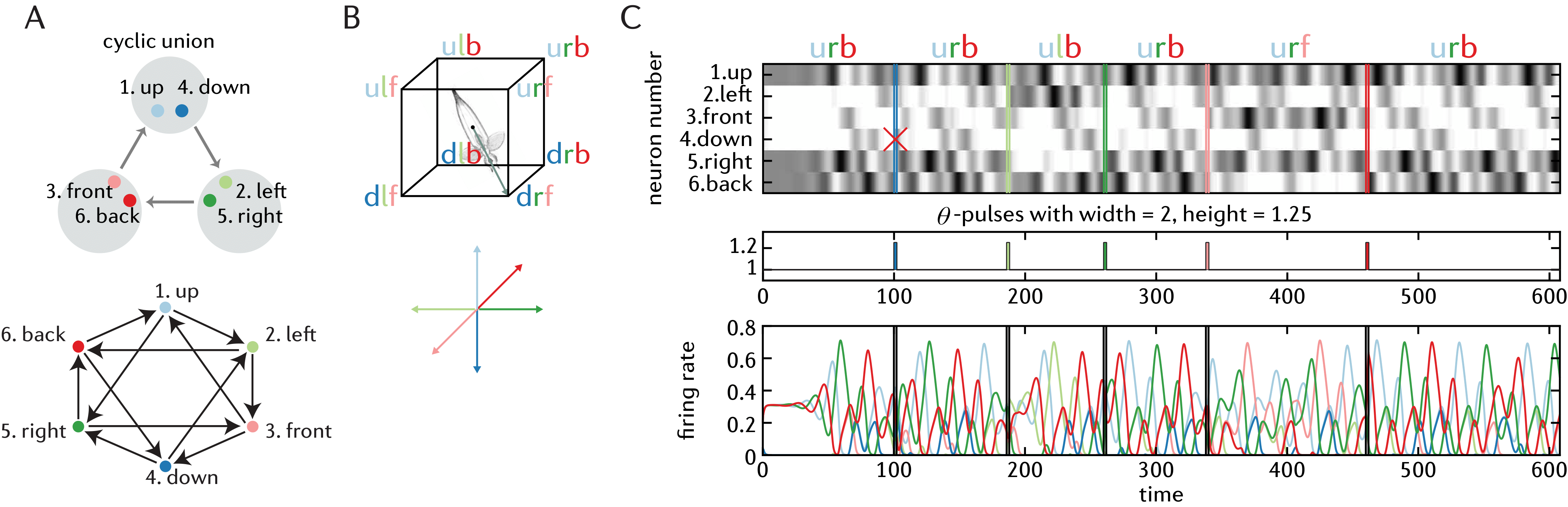}
	\end{center}
	\caption[Clione's hunting mechanism]{\textbf{Clione's hunting mechanism}.
		(A) Clione's network as a cyclic union and as an octahedron.
		(B) All possible swimming directions in 3-dimensional space, each one corresponding to an attractor as labeled.
		(C) CTLN solution for Clione's network. Attractor-specific pulses are sent to the network. Often producing a change in the attractor, labeled at the top.
		}
	\label{fig:clione-transitions}
\end{figure}
Recall that cyclic unions are particularly well-suited to model CPGs, so here we construct our network as a cyclic union of three components: each component is an independent set of opposing directions, as shown in Figure \ref{fig:clione-transitions}A. This will create competition between opposing directions, and  force the choice among them (thus avoiding invalid combinations where two opposing directions are simultaneously high-firing). This should result in the 8 possible combinations of directions in 3-dimensional space (up/down, right/left, front/back), as shown in Figure \ref{fig:clione-transitions}B. 


Each neuron represents swimming in a given direction: up, down, right, left, front, and back. Any valid combination of these three directions will uniquely define a direction in 3-dim space in which Clione will find itself swimming. All possible swimming directions will then correspond to the eight octants of 3-dimensional space, as illustrated in Figure \ref{fig:clione-transitions}B. Antagonist directions are colored in the same color with different shades. A more intuitive view of the network is given by the octahedral shape in Figure \ref{fig:clione-transitions}A, hence we also sometimes refer to this network as ``octahedral network''.

Since our network is a cyclic union of independent sets, the computation of core motifs becomes straightforward. By Theorem \ref{thm:cyclic-unions}, the core motifs are obtained by choosing a core motif from each component. The core motifs of an independent set are each singleton node \cite[Rule 3]{fp-paper}. This results in the following $2^3=8$ core motifs:
\begin{equation}\label{eq:octahedron-FPcore}
	\FP_\text{core}(G) = \{\{1,2,3\},\{1,2,6\},\{1,3,5\},\{1,5,6\},\{2,3,4\},\{2,4,6\},\{3,4,5\},\{4,5,6\}\},
\end{equation} which correspond to the 8 possible swimming directions, as expected. Computationally, we also found that $|\FP(G)| = 27$, as expected from Theorem \ref{thm:cyclic-unions}, as well.

Simulations show that each core motif in Equation \ref{eq:octahedron-FPcore} corresponds to an attractor, and that each attractor is accessible via initial conditions, as shown in the simulations of Figure \ref{fig:clione-transitions}. In addition, as in the five-gait network, it is possible for the octahedral network to transition between attractors by sending a $\theta$-pulse to the direction we want to switch to. For example, in Figure \ref{fig:clione-transitions}C, the network was was initialized in the attractor corresponding to up-right-back, and even though a first pulse fails to transition the network, the second pulse, sent to neuron 2 (left) changes the attractor in this single direction, that is from up-\textit{right}-back to up-\textit{left}-back. We observe that subsequent pulses do produce the desired effect of transitioning a single direction. The matrix and parameters needed to reproduce the simulations of Figure \ref{fig:clione-transitions} are available in Appendix \ref{ch:appendixA}, Equation \ref{eq:mollusk-matrix}.

But why is it that the first pulse failed to make the network transition attractors? Figure \ref{fig:clione-basins} offers some insights. Note that in Figure \ref{fig:clione-basins}B, neuron 2 (left) received a pulse, thereby sending the network into the down-left-back attractor, as wanted and expected. However in Figure \ref{fig:clione-transitions}C, a pulse was sent to switch from up to down, and nothing really happened. 
\begin{figure}[h!]
	\begin{center}
		\includegraphics[width=\textwidth]{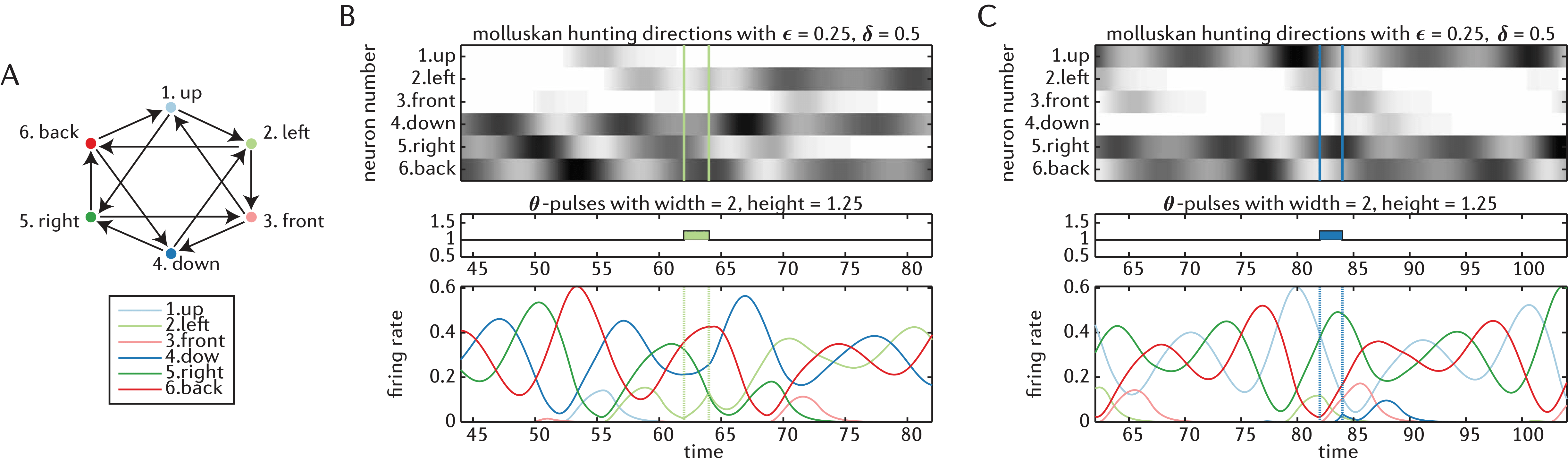}
	\end{center}
	\caption[Clione's basins of attraction]{\textbf{Clione's basins of attraction}.
		(A) Reproduction of Clione's network.
		(B) Effective pulse transition.
		(C) Ineffective pulse transition.
	}
	\label{fig:clione-basins}
\end{figure}

Looking closely, we observe that in D, the dark green rate curve surpasses the light green rate curve, and the transition is successful. In E, by contrast, the dark blue rate curve was very close to surpassing the light blue rate curve, however, at the end of the pulse the dark rate curve was still below the light blue curve, so the transition fails. It can be proven that this is a general phenomenon: the network will transition attractors into an opposing direction only if the neuron that received the pulse ``wins'' at end of the pulse. 

\subsection{Analysis of basins of attraction}

Indeed, the next theorem shows that the network will change from direction $i$ to direction $j$ at time $t$ only if $x_i(t)<x_j(t)$, for $i,j$ opposite directions. To formalize the opposite directions notation we introduce a permutation $\pi = (14)(25)(36)$ that prescribes which nodes are opposite. We then have:
\begin{theorem}\label{thm:octahedral-basins}
	Let $\mathcal{I}$ denote the core-motifs of the octahedral network ($\FP_\text{core}(G)$ above) and suppose there exists an attractor $\mathcal{A}_{\sigma}$ of the octahedral network, corresponding to one of the core motifs $\sigma \in \mathcal{I}$, and that there are no attractors that do not correspond to a core motif. Then there are exactly 8 attractors, one for each core motif, and their respective basins of attraction $\mathcal{B}_{\sigma}$ are contained in the sets
	\begin{equation}\label{eq:octahedral-basins}
		C_{ijk} = \{(x_1,x_2,x_3,x_4,x_5,x_6)  \in \RR^6 \, | \, x_i \leq x_{\pi(i)} ,\, x_j \leq x_{\pi(j)},\, x_k \leq x_{\pi(k)}\},
	\end{equation}  
	where $\pi = (14)(25)(36)$.
\end{theorem}
\begin{proof}
	Note that the octahedral network is symmetric under the following permutations in $S_6$:
	\begin{align*}
		\pi_1 &= (14) \\
		\pi_2 &= (25) \\
		\pi_3 &= (36) 
	\end{align*} so that $\pi = \pi_1 \pi_2 \pi_3$. Therefore the group of symmetries of the network is $\langle \pi_1, \pi_2, \pi_3 \rangle  \cong \ZZ_2 \times \ZZ_2 \times \ZZ_2$.  As a consequence of these symmetries, the existence of an attractor $\mathcal{A}_{\sigma}$ corresponding to a core motif $\sigma \in \mathcal{I}$ implies the existence of all other attractors corresponding to the rest of the core motifs in $\mathcal{I}$. Since there are no other attractors by assumption, the state space must be partitioned into exactly eight symmetric basins of attraction $\mathcal{B}_{\sigma}$. 
	
	On the other hand, since $\ell$ and $\pi(\ell)$ receive input from the same neurons (for $\ell = 1,\dots,6$), it is impossible to cross the hyperplanes $x_\ell = x_{\pi(\ell)}$, as we show next. Suppose that $x_\ell(t_0) = x_{\pi(\ell)}(t_0)$ for some $t_0$. The dynamic equations for neurons $\ell$ and $\pi(\ell)$ are
	\begin{align*}
		\frac{dx_\ell}{dt} &= -x_\ell + \left[\sum_{j \neq \ell,\pi(\ell)} W_{ij}x_j + (-1-\delta)x_{\pi(\ell)} + \theta \right]_+ \\
		\frac{dx_{\pi(\ell)}}{dt} &= -x_{\pi(\ell)} + \left[\sum_{j \neq \ell,\pi(\ell)} W_{ij}x_j + (-1-\delta)x_{\ell} + \theta \right]_+.
	\end{align*} Both summation terms are the same, since $\ell$ and $\pi(\ell)$ receive input from the same neurons. Since $x_\ell(t_0) = x_{\pi(\ell)}(t_0)$ by assumption, we have that $\frac{dx_\ell}{dt} = \frac{dx_{\pi(\ell)}}{dt}$, and therefore $x_\ell(t) = x_{\pi(\ell)}(t)$ for all $t\geq t_0$. This proves that it is impossible to cross the hyperplanes  $x_\ell = x_{\pi(\ell)}$. This implies that the sets $$C_{ijk} = \{(x_1,x_2,x_3,x_4,x_5,x_6)  \in \RR^6 \, | \, x_i \leq x_{\pi(i)} ,\, x_j \leq x_{\pi(j)},\, x_k \leq x_{\pi(k)}\}$$ are forward-invariant and also partition the state space into eight sets. Since the basins of attraction $\mathcal{B}_{\sigma}$ are also forward-invariant and partition the state space into eight sets, we get the desired result.
\end{proof}

This theorem says that, under some reasonable assumptions, the basins of attraction partition state space symmetrically, a direct consequence of the symmetry of the graph. A notable consequence of the theorem is that the transitions depend on the intensity and duration of the pulse and in turn, of the timing of the pulse arrival. This would partially explain the observed randomness in direction switching of Clione. If Clione receives an external signal (from a prey encounter, for instance), then this signal must arrive at the right time and have the right duration and strength for Clione to change directions. This introduces a great amount of randomness into the system, because several conditions must align for Clione to effectively change directions.

\chapter{Sequential control of dynamic attractors}\label{ch:sequential-control}

Let us recall the path that brought us here, as re-told by Figure \ref{fig:sequential-control-cartoons-D-highlight}. Networks that support many stable fixed point attractors, each accessible via attractor-specific inputs, as in Figure \ref{fig:sequential-control-cartoons-D-highlight}A, have been around for a while. In Chapter \ref{ch:sequences-of-attractors}, Section \ref{sec:fixed-point-counters}, we built two networks that internally encode a sequences of fixed point attractors, each accessible via identical inputs, as in Fig. \ref{fig:sequential-control-cartoons-D-highlight}B. In the same chapter, in Section \ref{sec:dynamic-counter}, we also presented a network supporting several dynamic attractors, each accessible via both targeted and identical inputs, as in Fig. \ref{fig:sequential-control-cartoons-D-highlight}B,C. Those attractors were all of the same type, and so in Chapter \ref{ch:CPGs} we gave an example model of a neural function that required coexistence and accessibility of several different dynamic attractors, also as in Fig. \ref{fig:sequential-control-cartoons-D-highlight}C. 
\begin{figure}[!h]
	\begin{center}
		\includegraphics[width=\textwidth]{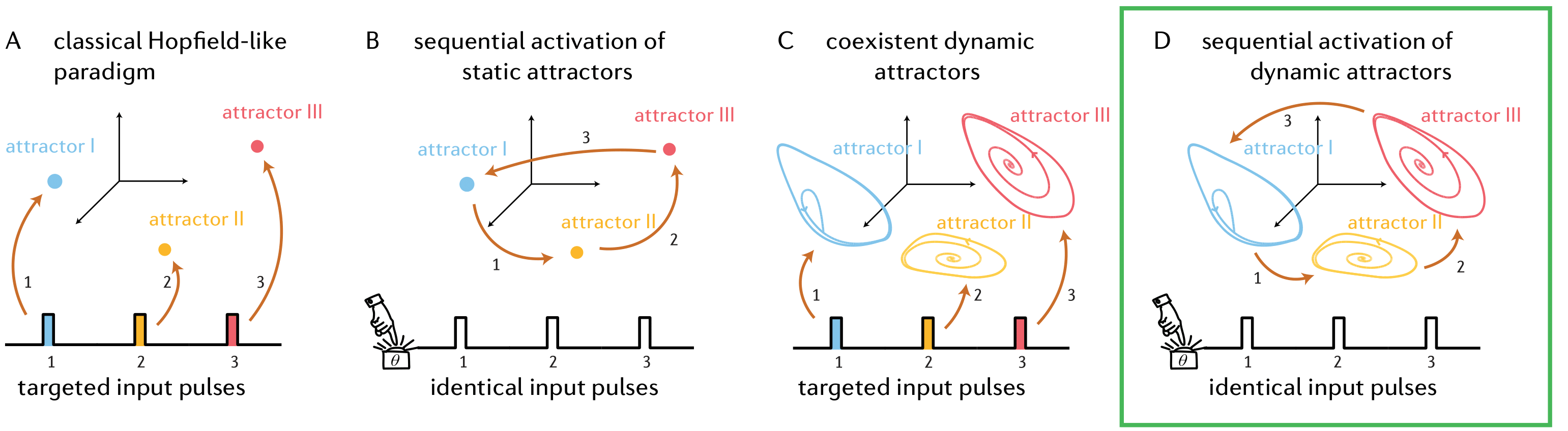}
	\end{center}
	\caption[Coexistent dynamic attractors]{\textbf{Coexistent dynamic attractors}.  Reproduction of Figure \ref{fig:sequential-control-cartoons-intro}. The focus of this chapter is Panel D: Internally encoded sequence of dynamic attractors, each step of the sequence accessible via identical inputs.}
	\label{fig:sequential-control-cartoons-D-highlight}
\end{figure}

Now, in this chapter, motivated by the hierarchical complex motor control of biological brains \cite{Sakai2003}, we aim to combine Figs. \ref{fig:sequential-control-cartoons-D-highlight}B and \ref{fig:sequential-control-cartoons-D-highlight}C to obtain Fig. \ref{fig:sequential-control-cartoons-D-highlight}D, but whose attractors can be \emph{flexibly} recombined. A key difference between Fig. \ref{fig:sequential-control-cartoons-D-highlight}D as done in Chapter \ref{ch:sequences-of-attractors}, and the network we build in this chapter, is that in the dynamic attractor chain network of Fig. \ref{fig:dynamic-attractor-counters}D, the order of the sequence was built into the network, meaning we could reorder attractors arbitrarily, and each new element of the sequence required storing the same pattern in the network again in the form of an extra motif.

By contrast, the construction we propose here encodes the order of the sequence in a network separate from that supporting the dynamic attractors that will be the sequence elements. This very akin to how birds and humans sequence complex movements \cite{Long2010,Yokoi2019}. This approach has many advantages, among which we have the capacity to recombine attractors in any order and that to add an attractor we do not need to make a whole new copy of it and embedded into the network.
\begin{figure}[h!]
	\begin{center}
		\includegraphics[width=\textwidth]{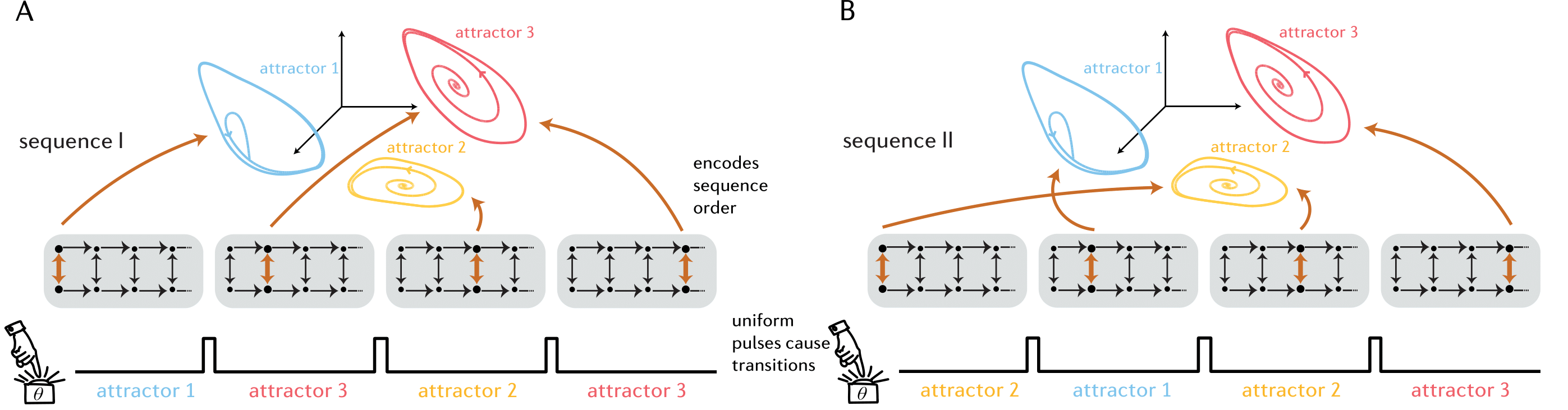}
	\end{center}
	\caption[Flexible sequencing of dynamic attractors]{\textbf{Flexible sequencing of dynamic attractors}.
		(A) Schematic of sequential control when the attractors are encoded independently from sequential order information. 
		(B) A different sequence encoded in the same structure. Note that the only difference between sequences I and II are the orange arrows.
	}
	\label{fig:dynamic-attractors-wired-counter}
\end{figure}

The general construction of such a network, illustrated in Figure \ref{fig:dynamic-attractors-wired-counter}, goes as follows: the CTLN fixed point counter network from Chapter \ref{ch:sequences-of-attractors} will encode the order of the sequence. A separate network will encode the dynamic attractors that are to be the elements of the sequence. In Figure \ref{fig:dynamic-attractors-wired-counter} we see this as several coexistent (dynamic) attractors, each one accessed via global $\theta$-pulses sent to the CTLN counter network (shaded in gray). We represent these global pulses by a hand pushing a button in Figure \ref{fig:dynamic-attractors-wired-counter}. Each pulse activates the next clique down the chain,  which will activate a different attractor in the network above, as prescribed by the orange arrows. With this construction, the order of the sequence is encoded in the orange connections, separately from the attractors themselves. Notice that the only difference between panels A and B are the orange arrows, and each wiring will result in two different sequences, whose elements are the same, but accessed in different order.

To give a proof of concept of this idea, we illustrate the construction using both networks from Chapter \ref{ch:CPGs} as our networks encoding the dynamic attractors that are to be the sequence elements.

\section{Sequences of quadruped gaits}\label{sec:sequential-control-gaits}
In this section, we apply the construction outlined in the previous paragraph by using our five-gait network from Section \ref{sec:quadruped-gaits} as the network supporting the attractors will be the sequence elements. The resulting network, consisting of less than 50 units, is shown in Figure \ref{fig:sequential_control_gaits}A. The network is composed of three layers: L1, the unsigned CTLN counter from Section \ref{sec:fixed-point-counters}, with as many cliques as sequence steps. L2, a intermediate layer with one node per attractor in the sequence. And L3, the five-gait network of Section \ref{sec:quadruped-gaits}, encoding the set of dynamic attractors that are the elements of the sequence. 
\begin{figure}[h!]
	\begin{center}
		\includegraphics[width=\textwidth]{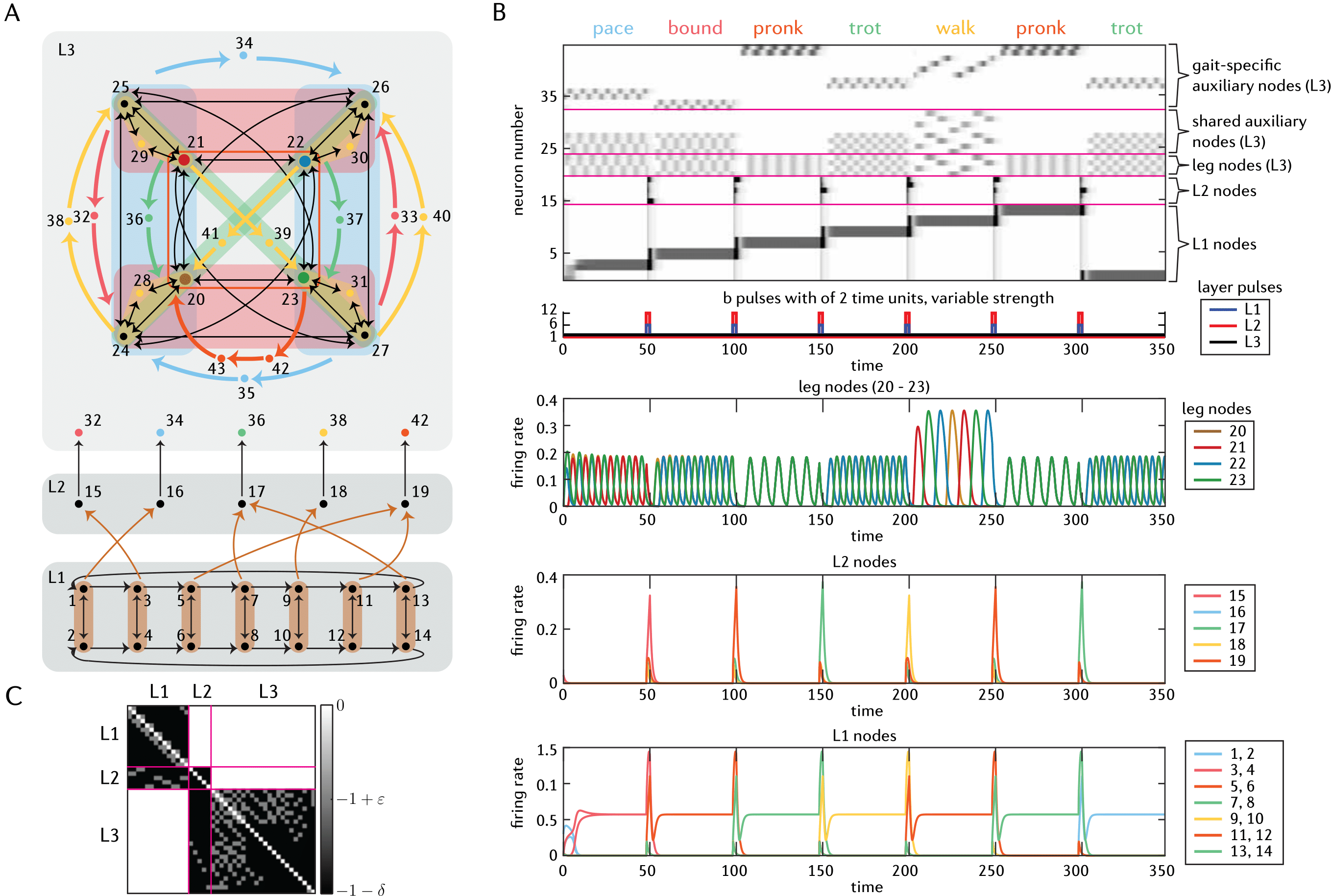}
	\end{center}
	\caption[Sequential control of quadruped gaits]{\textbf{Sequential control of quadruped gaits}.
		(A) Layered network for sequential control. 
		(B) Greyscale, pulses and rate curves for layered network of panel A.
		(C) $W$ matrix for the network in panel A.
	}
	\label{fig:sequential_control_gaits}
\end{figure}

Layers are connected in such a way that each node in L2 is connected to a node in L3 activating one of the sequence elements; and the connections from L1 to L2 determine the order in which each attractor in L3 will be accessed. The shaded squares in L1 indicate that all the shaded nodes send edges, which are drawn as a single thick arrow stemming from the shaded region. Each of these layers, and the connections between them, play an important role, as described below:

L1 in Figure \ref{fig:sequential_control_gaits}A, the counter network layer, advances the sequence upon reception of each input pulse. Recall that when the neurons in the counter receive a uniform input pulse, the activity slides to the next stable fixed point in the chain, thus advancing the sequence one step by activation of the next clique. This shift in activity is effectively communicated as a pulse to the intermediate layer L2.

L2 in Figure \ref{fig:sequential_control_gaits}A, because each of its nodes is connected to a single clique from L1, is sort of summarizing the input from L1 to communicate it to L3. This ``summary'' of activity will prevent L1, and its high firing rates, to interfere with L3, and its attractors. Neurons in this layer must also receive a pulse simultaneously with L1 to successfully advance the sequence, but not as strong.

L3's job is to support the dynamic attractors. Recall that each of these can be accessed via $\theta$-stimulation of a specific neuron uniquely associated with each gait. This fact is precisely what makes this construction possible. Otherwise it would not be possible for L2 to select a specific attractor in L3 by communicating the ``pulses'' coming from L1 to the specific neuron associated with the desired attractor. In the five-gait network, these are the auxiliary nodes corresponding to different gaits, which are redrawn at the bottom of L1 for clarity in Figure \ref{fig:sequential_control_gaits}A (nodes 32,34,36,38,42).

Finally, connections between L1 and L2 is where the order of the sequence is encoded: the order in which the nodes in L1 connect to the nodes in L2 will dictate the order in which the attractors are accessed. In Figure \ref{fig:sequential_control_gaits}A, for instance, the cliques connect to L2, in such a way that nodes in L2 activate the auxiliary nodes of L3 corresponding to pace (34), bound (32), pronk (42), trot (36), walk (38), pronk (42), and trot (36), and so the sequence will be executed in that exact order. 

The formal layering scheme is prescribed by the matrix in Figure \ref{fig:sequential_control_gaits}C, and formalized in the next section. This matrix, along with Equations \ref{eq:TLNdynamics} and pulses in Figure \ref{fig:sequential_control_gaits}B, is what we used to simulate the dynamics. The dynamics behave as we hoped when we first put together the layering scheme in our imaginations, as seen in Figure \ref{fig:sequential_control_gaits}C, and as described below:

The first plot shows the greyscale of all neurons. Below the greyscale, the pulses that L1 and L2 receive are pictured, colored by layered. L1, pictured in blue, has a baseline of $b=1$ and a pulse $b=6$. L2's pulse, colored in red has a baseline $b=0$ and a pulse $b=12$, double that of L1. The idea behind having a 0 baseline, is that neurons in L2, the only ones connecting to L3, will die out and won't interfere with the dynamics of L3 for long. Finally, L3 has $b=1$ constant, so this layer is not receiving external pulses at all. Instead, external pulses are sent to all neurons in layers L2 and L3, as opposed to specific neurons controlling each gait (as in Ch. \ref{ch:CPGs}). This means that the sequence is fully encoded within the network, and the pulse themselves carry no information about what the next step is. The ``pulses'' that L3 is receiving from L2 are attractor-specific, even though the pulses sent to the network are not. 

All of these predictions about the behavior can be clearly seen in the rate curves of Figure \ref{fig:sequential_control_gaits}B. Since the baseline $b$ is zero for neurons in L2, the rate curves of L2 nodes show peaks only at pulse times, but otherwise neurons die off. We also observe the dynamics of L1, remarkably the same as in \ref{sec:fixed-point-counters}; each pulse moves L1 one step to the right, activating the next sequence element, L2 communicates this transition to L3. The sequence of gaits can be read off from the rate curves of the leg nodes, as labeled above the greyscale, and it is exactly the same sequence encoded in the connection from L1 to L2. Transitions seem as effective as sending specific pulses to the five-gait network on its own, as in Section \ref{sec:quadruped-gaits}. Note that the color of the rate curves of L1 and L2 match those of the auxiliary neurons on Figure \ref{fig:sequential_control_gaits}A. The matrix needed to reproduce the simulation of Figure \ref{fig:sequential_control_gaits}B is available in Appendix \ref{ch:appendixA}, Equation \ref{eq:sequential-control-matrices}.

As L2 effectively acts as transient pulses to L3, it seems like all it takes to make this construction possible is to have the attractors in L3 be easily accessible via targeted pulses. If that is the case, can we generalize this architecture to other CPGs? As we will see below, the answers is yes, at least for the CPG all the CPGs we presented in Chapter \ref{ch:CPGs}.

\section{Sequences of molluskan hunting}\label{sec:sequential-control-mollusk}

Here, we offer a proof of concept by using the molluskan hunting network of Section \ref{sec:molluskan-hunting}. Since all attractors of the molluskan hunting network are accessible via changes in initial conditions, or targeted (long or strong) $\theta$-pulses, we are able to use an analogous construction to internally encode a sequence of its attractors.
\begin{figure}[h!]
	\begin{center}
		\includegraphics[width=\textwidth]{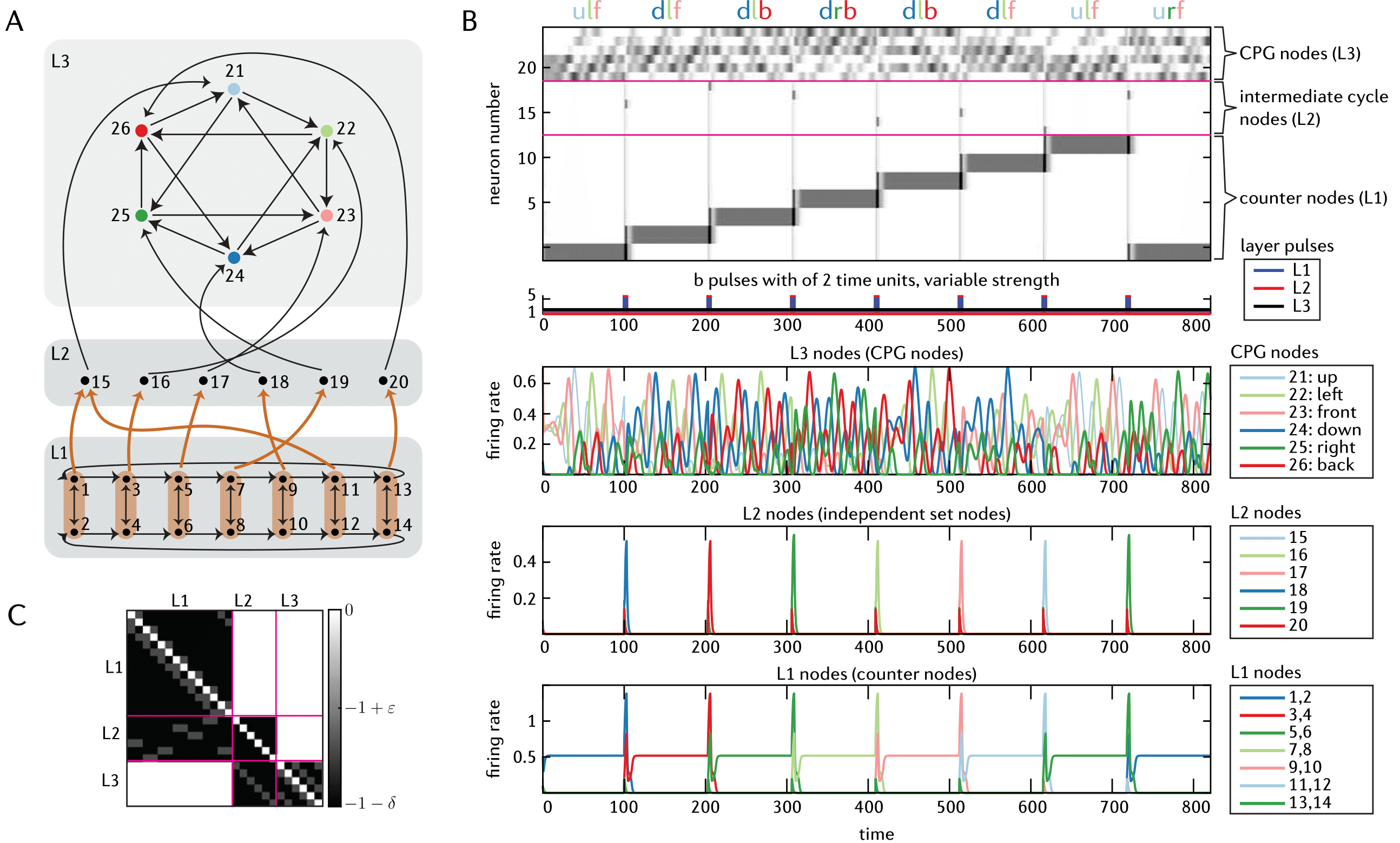}
	\end{center}
	\caption[Sequential control of swimming directions]{\textbf{Sequential control of swimming directions}.
		(A) Layered network for sequential control.
		(B) Greyscale, pulses and rate curves for layered network of panel A.
		(C) $W$ matrix for the network in panel A.
	}
	\label{fig:clione-wired-counter}
\end{figure}

Indeed, in Figure \ref{fig:clione-wired-counter}A we layered our octahedral network with the same L1 and L2 layers from last section to obtain a network that can encode sequences of attractors in Clione's swimming direction. We have prescribed the order of the attractors (orange arrows from L1 to L2) in such a way that a transition is always required, so that it is clearly seen that transitions are indeed taking place. 

Again, network performs as in last section, as seen in Figure \ref{fig:clione-wired-counter}B. Attractors are labeled by the initials of up/down, left/right and front/back, as before, and it can be seen above the greyscale that they are accessed in the order prescribed by the orange arrows. Only the stimulated direction is making the switch. Also, in the rate curves it can be seen that there are no failed transitions, and this partly thanks to Theorem \ref{thm:octahedral-basins} as well; because now we can make sure to send pulses that are long/strong enough to make the transition happen. It is important here that other nodes in L2 are not overly active, because again by Theorem \ref{thm:octahedral-basins}, it could mean that the wrong node is being stimulated. The matrix needed to reproduce the simulation of \ref{fig:clione-wired-counter}B is available in Appendix \ref{ch:appendixA}, Equation \ref{eq:sequential-control-matrices}.

This second example shows that this construction is truly versatile, as long as attractors in L3 are easily accessible. This construction also comes with great advantages:

\paragraph*{Sequential information is internally encoded.} Since our counter network from Section \ref{sec:fixed-point-counters} sequentially steps through stable fixed points, we leveraged this sequence of fixed point activations to encode a sequence of pulses as wirings between layers that are to receive uniform pulses, and the CPG layer.

To transition CPG attractors, we simply send a uniform, non-specific pulse to all neurons in the counter layer (and cycle layer) so as to advance the counter, and therefore send a targeted pulse to the next gait down the chain. 

This means that pulses are attractor-specific \emph{locally} in the CPG layer, but not globally, as the counter and cycle layers are the ones encoding the sequential information, and they only require uniform, non-specific pulses. The sequence is therefore truly encoded within the network, and the pulse themselves carry no information about what the next step is. This fact contrasts with other models for sequential control where the information about the order of the sequence is not encoded in the network itself, and so it requires motif-specific inputs to activate each element in the sequence \cite{Latorre2019,Logiaco2021}. Having the sequence encoded within the network would be advantageous for highly stereotyped patterns, like songbird, choreographed dance or other sequential tasks that reuse motifs. 

\paragraph*{Sequential information is independently encoded from motor commands.} As mentioned before, the order of the sequence is encoded in the connections from L2 to L3, whereas the dynamic attractors are completely encoded within L3. This means that sequential information is independently encoded from motor commands and so to access an attractor multiple times in the sequence, it suffices to create a new connection from L2 to L3. In Figure \ref{fig:sequential_control_gaits}C, for example, both pronk and trot are activated twice in the sequence, but there was no need to add a whole new set of neurons that are re-encoding pronk or trot. That is, we are re-using our attractors in an efficient and flexible manner. This is remarkable property of this construction, as dynamic attractors are usually very sensitive to any interference. 

This also implies that to add a new element to the sequence, it is only necessary to add three extra neurons and a few connections between them (an extra 2-clique, one cycle node, and connect this to the appropriate neuron in the five-gait network). This greatly improves efficiency, since the repeated pattern does not have to be stored again, separately into the network.  Also, new elements in the sequence do not do not interfere with other elements, controlled by the same network. This is because since the cycle neurons are active one at the time, the new inhibition stemming from the new sequence element will not affect the CPG dynamics whatsoever until the new neuron is active. When the new neuron is active, it will uniformly inhibit everyone in the CPG, except the aux neuron it is connected to. At pulse reception, when the new neuron is active, the jump in neural activity will simply act as pulse to transition the CPG into the appropriate gait. This model can thus generate all kinds of transitions, and the sequence can be as long as desired. 

This mechanism is not unlike to how our brain represents complex sequences of movement \cite{Lashley1951,Wong2019}. Also, the structure of our model resembles very much they way in which songbird is encoded, where propagation of syllables is mediated by a synaptic chain of neurons \cite{Long2010}.  Also, and although not anatomically separated, similar mechanisms of complex motor sequence generation have also been found in humans' neocortex \cite{Yokoi2019} and basal ganglia \cite{Graybiel1998}. 

\paragraph*{Timing and sequential information are independent.} Notice also that timing of the sequence is controlled by these external pulses and therefore a sequence's execution can be sped up or slowed, or rhythm altered altogether, without interfering with attractors or sequence order. Moreover, there is no need to change synaptic time constants, physiological variables, to alter the timing of the sequence. This disassociation also implies efficiency. This is an important characteristic of sequential control in premotor cortex areas and serves, among others, the purpose of independently controlling the speed of movements from the order of execution \cite{Kornysheva2014}. Notice also that timing of the sequence is controlled by these external pulses and therefore a sequence's execution can be sped up or slowed, or rhythm altered altogether, without interfering with attractors or sequence order. 

Can this construction be generalized to other CPG networks? We conjecture that the five-gait network could potentially be replaced by any other network that has coexistent attractors, each accessible via changes in initial conditions or specific stimulation of neurons. The mechanism would be the same: the order of the sequence is determined by the connections from the cycle layer (L2) to the multi-attractor network, and any sequential ordering is possible, with as many sequence steps as desired.

From the matrices in Figures \ref{fig:sequential_control_gaits}B, it can be seen that these networks are not CTLNs, but rather a TLN whose ``layers'' are CLTNs. Can we find theoretical grounds to support this seemingly generalizable construction? We should be able to leverage all those zeros!


\chapter{New theoretical results}\label{ch:new-theoretical-results}

Why are the networks of Chapter \ref{ch:sequential-control} so neat? In this chapter, we provide theoretical explanations on why we see the ``fusion attractors'' of Chapter \ref{ch:sequential-control} arise so seamlessly. First, we introduce some technical background that will help us prove new theoretical results, along with some extra CTLN background theorems. Then, we present new CTLN results generalizing certain structures from \cite{fp-paper} (these ``certain structures'' are reviewed in the technical background section). The new CTLN results are published in \cite{Parmelee2022}, and only my contributions to that paper are included here, with only small changes to the presentation. Finally, we generalize some of those results to TLNs. Since the new TLN results are generalizing the CTLN results, we have omitted the proofs for the following CTLN results: Lemma \ref{lemma:simply-added}, Theorem \ref{thm:menu-CTLN}, Lemma \ref{lemma:full-factorization-CTLN} and Theorem \ref{thm:bidir-sa-partition}. 

\section{Technical background}\label{sec:technical-background}

A characterization of $\FP(W,b)$ that we often use in the proofs that follow, developed in \cite{fp-paper}, relies on Cramer's determinants. For any $\sigma \subseteq [n],$ define $s_i^\sigma$ to be the relevant Cramer's determinant:
\begin{equation}\label{eq:s_i}
s_i^\sigma \od \det((I-W_{\sigma\cup\{i\}})_i;b_{\sigma\cup\{i\}}), \;\; \text{for each} \;\; i \in [n],
\end{equation} 
where $\det(A_i; b)$ denotes the determinant obtained by replacing the $i^{\text{th}}$ column of $A$ with the vector $b$ and $((A_\sigma)_i;b_\sigma)$ denotes the matrix obtained from the restricted matrix $A_\sigma$ by replacing the column corresponding to the index $i \in \sigma$ with the restricted vector $b_\sigma$. Note that $i$ might ot might not be in $\sigma$, both are possible.
%

In \cite[Lemma 2]{fp-paper}, a formula for $s_k^\sigma$ was proven that directly connects it to the relevant quantity in the ``off"-neuron condition:

\begin{equation}\label{eq:s_k^sigma}
	s_k^\sigma = \sum_{i \in \sigma} W_{ki}s_i^\sigma + \theta \det(I-W_\sigma) \text{ for any } k \in [n].
\end{equation}

Combining this with Cramer's rule, it was shown that $\FP(G)$ can be fully characterized in terms of the \emph{signs} of the $s_i^\sigma$.  It turns out these signs are also connected to the \emph{index} of a fixed point. For each fixed point of a CTLN $W= W(G, \varepsilon, \delta)$, labeled by its support $\sigma \in \FP(G)$, we define the {\em index} as
$$\idx(\sigma) \od \sgn \det(I-W_\sigma).$$
Since we assume our CTLNs are nondegenerate, $\det(I-W_\sigma) \neq 0$ and thus $\idx(\sigma) \in \{\pm 1\}$.  

For each fixed point of a TLN labeled by its support $\sigma \in \FP(W,b)$, we define the {\em index} as $\idx(\sigma) \od \sgn \det(I-W_\sigma).$ In \cite{fp-paper} it was shown that $\FP(W,b)$ can be fully characterized in terms of the \emph{signs} of the $s_i^\sigma$, which are also connected to the index of the fixed point. This is the tool we use constantly in this chapter to prove that some support $\sigma$ belongs to $\FP(W,b)$, so it is worth keeping in mind:

\begin{theorem}[sign conditions,\cite{fp-paper}]\label{thm:sgn-conditions}
Let $(W,b)$ be a TLN on $n$ neurons. For any nonempty $\sigma \subseteq [n]$,
$$\sigma \in \FP(W_\sigma,b_\sigma)  \;\; \Leftrightarrow \;\; \sgn s_i^\sigma = \sgn s_j^\sigma
\text{ for all } i,j \in \sigma.$$
In that case we say that $\sigma$ is permitted (and forbidden otherwise), and 
$\sgn s_i^\sigma = \sgn\det(I-W_\sigma) = \idx(\sigma)$ for all $i \in \sigma$. Furthermore,
$$\sigma \in \FP(W,b)  \;\; \Leftrightarrow \;\; \sgn s_i^\sigma = \sgn s_j^\sigma = -\sgn s_k^\sigma
\text{ for all } i,j \in \sigma,\; k \not\in \sigma.$$
\end{theorem} 

Also recall from Chapter \ref{ch:background}:

\begingroup
\def\thetheorem{\ref{cor:inheritance}}
\begin{corollary}
	Let $(W,b)$  be a TLN on $n$ neurons, and let $\sigma\subseteq [n]$. The following are equivalent:
	\begin{enumerate}
		\item $\sigma \in \FP(W,b)$
		\item $\sigma \in \FP(W|_{\tau},b|_{\tau})$ for all $\sigma \subseteq \tau \subseteq [n]$
		\item $\sigma \in \FP(W|_{\sigma},b|_{\sigma})$ and $\sigma \in \FP(W|_{\sigma \cup k},b|_{\sigma \cup k})$ for all $k \notin \sigma$
		\item $\sigma \in \FP(W|_{\sigma \cup k},b|_{\sigma \cup k})$ for all $k \notin \sigma$
	\end{enumerate}
\end{corollary}
\addtocounter{theorem}{-1}
\endgroup

Note from the above corollary that a given $\sigma \subseteq [n]$ might be permitted, but it can be the case that $\sigma \notin \FP(W,b)$. This is because since $\sigma$ might not \emph{survive} the addition of extra neurons to the network. In that case, we say that $\sigma$ \emph{dies} in the larger network. To make this distinction formal, we define, for any $\sigma \subseteq [n]$, the sets of surviving ($S_\sigma$) and dying fixed points supports ($D_\sigma$): 
$$S_\sigma \od \FP(W|_\sigma,b|_\sigma) \cap \FP(W,b), \quad \text{and} \quad D_\sigma \od \FP(W|_\sigma,b|_\sigma)\setminus S_\sigma.$$

It is possible to characterize exactly the fixed points supports of some networks in terms of its dying and surviving fixed points. These networks are the transversal topic of this chapter. To begin exploring the special structures of those, we go back to when the simply-added structure was introduced in \cite{fp-paper}:

\begin{definition}[simply-added split]\label{def:simply-added-split}
Let $G$ be a graph on $n$ nodes.  For any nonempty $\omega,~ \tau\subseteq [n]$ such that $\omega \cap \tau= \emptyset$, we say $\omega$ is \emph{simply-added} onto $\tau$ if for each $j \in \omega$, either $j$ is a \emph{projector} onto $\tau$, i.e., $j \to k$ for all $k \in \tau$, or $j$ is a \emph{non-projector} onto $\tau$, so $j \not\to k$ for all $k \in \tau$. In this case, we say that $\tau$ is \emph{simply-embedded} in $G$, and we say that $(\omega, \tau)$ is a \emph{simply-added split} of the subgraph $G|_\sigma$, for $\sigma = \omega\cup\tau $.
\end{definition}

This structure is beneficial because when the graph has a simply-added split, the $s_i^\sigma$'s easily factor, which makes it straightforward to compute their signs and apply Theorem \ref{thm:sgn-conditions}.

\begin{theorem}[\cite{fp-paper}]\label{thm:simply-added}
Let $G$ be a graph on $n$ nodes, and let $\omega,~ \tau\subseteq [n]$ be such that $\omega$ is simply-added to $\tau$. For $\sigma \subseteq \omega \cup \tau$, define $\sigma_\omega \od \sigma \cap \omega$ and $\sigma_\tau \od \sigma \cap \tau$.  Then 
$$s_i^\sigma = \frac{1}{\theta}s_i^{\sigma_\omega} s^{\sigma_\tau}_i = \alpha s^{\sigma_\tau}_i \quad\text{for each $i \in \tau$,}$$ 
where $\alpha = \frac{1}{\theta}s_i^{\sigma_\omega}$ has the same value for every $i \in \tau$.
\end{theorem}

Notice that being simply-added is not a bidirectional property, and Theorem \ref{thm:simply-added} only provides factorization for each $i \in \tau$. A factorization that holds for every node requires the simply-added property be bidirectional, i.e.:

\begin{definition}[bidirectional simply-added split]
Let $G$ be a graph on $n$ nodes.  For any nonempty $\omega,~ \tau\subseteq [n]$ such that $[n] = \omega \cup \tau$ and $\omega \cap \tau= \emptyset$, we say that $G$ has a \emph{bidirectional simply-added split} $(\omega, \tau)$ if $\omega$ is simply-added onto $\tau$ and $\tau$ is simply-added onto $\omega$.  In other words, for all $j \in \omega$, either $j \to k$ for all $k \in \tau$ or $j \not\to k$ for all $k \in \tau$, \emph{and} for all $k \in \tau$, either $k \to j$ for all $j \in \omega$ or $k \not\to j$ for all $j \in \omega$.  
\end{definition}

In this case, the $s^\sigma_i$ factor for every $i \in [n]$\footnote{This foreshadows the importance of this factorization and will come back to us in Chapter \ref{ch:open-questions}, in unexpected ways.}.  With this structure, we can see exactly how the sets of fixed points supports is formed:

\begin{theorem}[\cite{fp-paper}]\label{thm:bidir-sa}
Let $G$ be a graph with bidirectional simply-added split $[n] = \omega \cup \tau$. For any nonempty $\sigma \subseteq [n]$, let $\sigma = \sigma_\omega \cup \sigma_\tau$ where
$\sigma_\omega \od \sigma \cap \omega$ and $\sigma_\tau \od \sigma \cap \tau$. Then
$\sigma \in \FP(G)$ if and only if one of the following holds:
\begin{itemize}
	\item[(i)]$ \sigma_\tau \in S_\tau \cup \{\emptyset\}\;\; \text{and}\;\; \sigma_\omega \in S_\omega \cup \{\emptyset\}, \;\; \text{or}$
	\item[(ii)] $\sigma_\tau \in D_\tau \;\; \text{and}\;\; \sigma_\omega \in D_\omega.$
\end{itemize}
In other words, $\sigma \in \FP(G)$ if and only if $\sigma$ is either a union of surviving fixed points $\sigma_i$, at most one from $\omega$ and at most one from $\tau$, or it is a union of dying fixed points, exactly one from $\omega$ and one from $\tau$.
\end{theorem}

In the sections that follow, we generalize this theorem to more than two components, and to TLNs.

Finally, some of the results we use only hold for TLNs with \emph{uniform} external input $\theta$. That is, $(W,b)$ is such that for every $i \in [n]$, $b_i = \theta$. In this case, we abuse notation and denote the TLN $(W,\theta\mathbbm{1})$ by just $(W,\theta)$. The next theorem is one of those results and it relates the \emph{magnitudes} of the $s_i^\sigma$'s to $\FP(W,\theta)$. This is a TLN version of Rule \ref{rule:graph-domination}. For this, \cite{fp-paper} defines the \emph{domination quantity} for $j \in [n]$:
$$w_j^{\sigma} = \sum_{i \in \sigma}\widetilde{W}_{ji}|s_i^\sigma|,$$
where $\widetilde{W} = -I+W$. We say that \emph{$k$ dominates $j$ with respect to $\sigma$}, if $w_k^\sigma > w_j^\sigma$.  The theorem then states that $\sigma \in \FP(W,\theta)$ precisely when these domination quantities are perfectly balanced within $\sigma$, so that $\sigma$ is \emph{domination-free}, and when every external node $k \notin \sigma$ is ``inside-out'' dominated by nodes inside $\sigma$:

\begin{theorem}[general domination, Theorem 15 in \cite{fp-paper}] \label{thm:gen-domination}
Let $(W,\theta)$ be a TLN with uniform input, and let $\sigma \subseteq [n]$. Then
$$\sigma \in \FP(W_\sigma,\theta) \quad \Leftrightarrow \quad w_i^\sigma = w_j^\sigma \text{ for all } {i,j\in\sigma}.$$ 
That is, $\sigma$ is permitted if and only if $\sigma$ is domination-free.
If $\sigma \in \FP(W_\sigma,\theta)$, then $\sigma \in \FP(W,\theta)$ if and only if for each $k \notin \sigma$, there exists $j \in \sigma$ such that $w_j^\sigma > w_k^\sigma$, i.e.\ such that $j$ inside-out dominates $k$.
\end{theorem}

That is all the extra technical background that we will reference in the proofs below (in addition to that of Ch. \ref{ch:background}).

\section{Simply-embedded CLTN structures}\label{sec:simply-embedded-CTLNs}

The results in this section are my contribution to \cite{Parmelee2022}. Again, Lemma \ref{lemma:simply-added}, Theorem \ref{thm:menu-CTLN}, Lemma \ref{lemma:full-factorization-CTLN} and Theorem \ref{thm:bidir-sa-partition} are presented with no proof, because the proofs now follow from more general theorems in Section \ref{sec:layered-ctlns}. Old proofs can be found in the original publication. The exposition and figures below come from that paper, with small modifications. 

The aim of this section is to generalize Theorem \ref{thm:cyclic-unions} by further constraining the simply-added architecture. The simply-added structure of cyclic unions was restrictive enough to allow us to completely characterize the fixed point supports of the graph in term of its components. However, it can be relaxed a little further to obtain similar results.

\subsection{Simply-embedded partitions}

We begin by generalizing simply-added splits (Def. \ref{def:simply-added-split}) and introduce the more general notion of a simply-embedded \emph{partition}. Recall that given a graph $G$ and a partition of its nodes into two components, $\{\omega | \tau\}$, we say that $\tau$ is \emph{simply-embedded} in $G$ if $\omega$ is \emph{simply-added} onto $\tau$. We can generalize this idea to more components by making every $\tau_i$ simply-embedded in $G$:

\begin{definition}[simply-embedded partition]\label{def:sa-partition}
Given a graph $G$, a partition of its nodes $\{\tau_1|\cdots|\tau_N\}$ is called a \emph{simply-embedded partition} if every $\tau_i$ is simply-embedded in $G$.  In other words, for each $\tau_i$ and each $k \notin \tau_i$, either $k \to j$ for all $j \in \tau_i$ or $k \not\to j$ for all $j \in \tau_i$.  
\end{definition}

Notice that the definition is trivially satisfied when there are no $k \notin \tau_i$ or there is only a single $j\in \tau_i$ for every $i$.  Thus, every graph has two trivial simply-embedded partitions: one where all the nodes are in one component and one where every node is in its own component.  Neither of these partitions is useful for giving information about the structure of $G$.  But when a graph has a nontrivial simply-embedded partition, this structure is sufficient to dramatically constrain the possible fixed point supports of $G$ to unions of fixed points chosen from a \emph{menu} of component fixed point supports, $\FP(G|_{\tau_i})$. To prove this fact, we need a lemma connecting the $s_j^\sigma$ values to the $s_j^{\sigma_i}$ values from the component subgraphs.

\begin{lemma}\label{lemma:simply-added}
Let $G$ have a simply-embedded partition $\{\tau_1|\cdots|\tau_N\}$, and consider $\sigma \subseteq [n]$. Let $\sigma_i \od \sigma \cap \tau_i$.  Then for any $\sigma_i \neq \emptyset$,
$$\sgn s_j^\sigma = \sgn s^\sigma_k \quad \Leftrightarrow \quad \sgn s_j^{\sigma_i} = \sgn s_k^{\sigma_i}, \quad\text{for all $j,k \in \tau_i$}.$$
\end{lemma}

Combining this lemma with the sign conditions characterization of fixed point supports (Thm. \ref{thm:sgn-conditions}), we get the desired result:

\begin{theorem}[$\FP(G)$ menu for simply-embedded partitions]\label{thm:menu-CTLN}
Let $G$ have a simply-embedded partition $\{\tau_1|\cdots|\tau_N\}$.  For any $\sigma \subseteq [n]$, let $\sigma_i \od \sigma \cap \tau_i$.  Then 
$$\sigma \in \FP(G) \quad \Rightarrow \quad \sigma_i \in \FP(G|_{\tau_i})\cup \{\emptyset\}~~\text{ for all } i \in [N]. $$ 
In other words, every fixed point support of $G$ is a union of component fixed point supports $\sigma_i$, at most one per component.
\end{theorem}


\begin{figure}[!h]
\begin{center}
	\includegraphics[width=\textwidth]{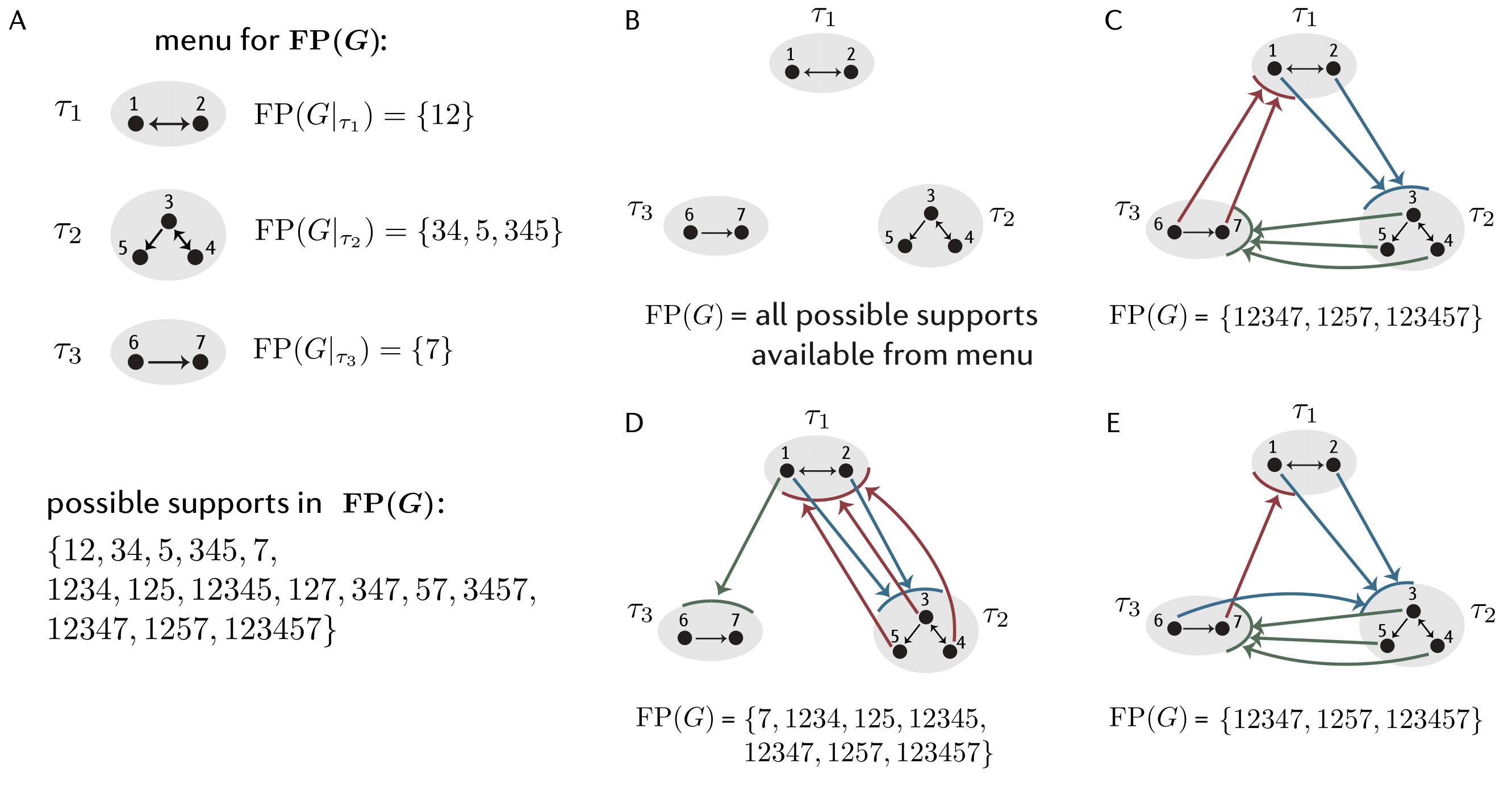}
\end{center}
\caption[Graphs with a simply-embedded partition]{\textbf{Graphs with a simply-embedded partition from Example~\ref{ex:sa-partitions}.} (A) (Top) A collection of component subgraphs with their $\FP(G|_{\tau_i})$.  (Bottom) The set of possible fixed point supports for any graph that has these subgraphs in the simply-embedded partition. (B-E) Example graphs with a simply-embedded partition with the component subgraphs from A, together with their $\FP(G)$.  In (C-E), thick colored edges from a node to a component indicate that the node projects edges out to all the nodes in the receiving component. Figure reproduced from \cite{Parmelee2022}.
}
\label{fig:sa-partition-examples}
\end{figure}

\begin{example}[Example 3.1 in \cite{Parmelee2022}]\label{ex:sa-partitions}
Consider the component subgraphs shown in Figure~\ref{fig:sa-partition-examples}A together with their $\FP(G|_{\tau_i})$.  By Theorem~\ref{thm:menu-CTLN}, any graph $G$ with a simply-embedded partition of these component subgraphs has a restricted menu for $\FP(G)$ consisting of the component fixed point supports (the set of all possible supports derived from this menu is shown on the bottom of panel A).  Note that an arbitrary graph on 7 nodes could have up to $2^7 -1 = 127$ possible fixed point supports, but the simply-embedded partition structure narrows the options to only 15 candidate fixed points.  Figure~\ref{fig:sa-partition-examples}B-E show four possible graphs with simply-embedded partitions of these component subgraphs, together with $\FP(G)$ for each of the graphs.  

Observe that the graph in Figure~\ref{fig:sa-partition-examples}B is a \emph{disjoint union} of its component subgraphs.  For this graph, $\FP(G)$ consists of all possible unions of at most one fixed point support per component subgraph (see \cite[Thm. 11]{fp-paper}).  Thus, every choice from the menu provided by Theorem~\ref{thm:menu-CTLN} does in fact yield a fixed point for $G$.  

In contrast, the graph in Figure~\ref{fig:sa-partition-examples}C is a \emph{cyclic union} of the component subgraphs.  For this graph, $\FP(G)$ only has sets that contain a fixed point support from every component, i.e., $\sigma_i \neq \emptyset$ for all $i \in [N]$ (by Thm. \ref{thm:cyclic-unions}).  Thus, any subset from the menu of Theorem~\ref{thm:menu-CTLN} that does not intersect every $\tau_i$ does not produce a fixed point for $G$.

Meanwhile, the graph in Figure~\ref{fig:sa-partition-examples}D is a simply-embedded partition with heterogeneity in the outgoing edges from a component (notice different nodes in $\tau_1$ treat $\tau_3$ differently).  $\FP(G)$ has a mixture of types of supports: there are some $\sigma \in \FP(G)$ that do not intersect every component, and others that do. 

Finally, the graph in Figure~\ref{fig:sa-partition-examples}E is another simply-embedded partition with heterogeneity (notice different nodes in $\tau_3$ treat $\tau_1$ and $\tau_2$ differently). However, for this graph, there is a uniform rule for the fixed point supports: every fixed point consists of exactly one fixed point support per component subgraph (identical to $\FP(G)$ for the graph in panel C).
\end{example}

Even though Theorem~\ref{thm:menu-CTLN} is not sufficient to fully determine $\FP(G)$, it significantly limits the options for fixed point supports (menu).  In particular, one direct consequence of Theorem~\ref{thm:menu-CTLN} is that if there is some node $j \in \tau_i$ in $G$ that does not participate in any fixed points of $G|_{\tau_i}$, then $j$ cannot participate in any fixed point of the full graph $G$. Thus the supports of all the fixed points of $G$ are confined to $[n]\setminus \{j\}$.  It turns out that if the removal of node $j$ does not change the fixed points of the component subgraph, i.e.\ if $\FP(G|_{\tau_i}) = \FP(G|_{\tau_i\setminus\{j\}})$, then we can actually remove $j$ from the full graph $G$ without changing $\FP(G)$.  Thus we have the following theorem.

\begin{theorem}[removable nodes]\label{thm:removables}
Let $G$ have a simply-embedded partition $\{\tau_1|\cdots|\tau_N\}$.  Suppose there exists a node $j \in \tau_i$ such that $\FP(G|_{\tau_i}) = \FP(G|_{\tau_i\setminus\{j\}})$.  Then $\FP(G) = \FP(G|_{[n]\setminus\{j\}}) $. 
\end{theorem}

\begin{proof}
To see that $\FP(G)\subseteq \FP(G|_{[n]\setminus\{j\}})$, notice that for all $\sigma \in \FP(G)$, we have $\sigma \subseteq [n] \setminus \{j\}$ by Theorem~\ref{thm:menu-CTLN}.  Then by Corollary \ref{cor:inheritance}(2), we must have $\sigma \in \FP(G|_{[n]\setminus\{j\}})$, and so $\FP(G)\subseteq \FP(G|_{[n]\setminus\{j\}})$.  

For the reverse containment, we will show that every fixed point in $\FP(G|_{[n]\setminus\{j\}})$ survives the addition of node $j$ by appealing to Theorem~\ref{thm:sgn-conditions} (sign conditions).   There are two cases to consider: $\sigma_i = \emptyset$ and $\sigma_i \neq \emptyset$, where $j \in \tau_i$ and $\sigma_i \od \sigma \cap \tau_i$.

\noindent\underline{Case 1:} $\sigma_i = \emptyset$.  Since $j$ is not contained in the support of any fixed point of $G|_{\tau_i}$, there must be at least one other node $k$ in $\tau_i$, since $\FP(G|_{\tau_i})$ cannot be empty.  Since $G$ is a simply-embedded partition, we have that $[n] \setminus \tau_i$ is simply-embedded onto $\tau_i$ meaning that every node in $\tau_i$ receives identical inputs from the rest of the graph. Recall from Equation \ref{eq:s_k^sigma}, that $s_j^\sigma = \sum_{\ell \in \sigma} W_{j\ell} s_\ell + \theta\det(I-W_\sigma)$.  Then since $\sigma \subseteq [n] \setminus \tau_i$, we have that $j$ and $k$ receive identical inputs from $\sigma$, so $W_{j\ell} = W_{k \ell}$ for all $\ell \in \sigma$, and thus $s_j^\sigma = s_k^\sigma$.  Since $\sigma \in \FP(G|_{[n]\setminus\{j\}})$, we have $\sgn s^\sigma_k = -\sgn s^\sigma_\ell$ for all $\ell \in \sigma$ by Theorem~\ref{thm:sgn-conditions} (sign conditions).  Thus, we also have $\sgn s^\sigma_j = -\sgn s^\sigma_\ell$ and $\sigma$ survives the addition of node $j$, so $\sigma \in \FP(G)$.    

\noindent\underline{Case 2:} $\sigma_i \neq \emptyset$.  First observe that $G|_{[n]\setminus\{j\}}$ has the same simply-embedded partition structure as $G$, but with $\tau_i \setminus\{j\}$ rather than $\tau_i$.  Thus $\sigma \in \FP(G|_{[n]\setminus\{j\}})$ implies that $\sigma_i \in \FP(G|_{\tau_i \setminus \{j\}})$ by Theorem~\ref{thm:menu-CTLN} (menu).  By hypothesis, $\FP(G|_{\tau_i \setminus \{j\}}) = \FP(G|_{\tau_i})$, and so $\sigma_i \in \FP(G|_{\tau_i})$.  Then by Theorem~\ref{thm:sgn-conditions} (sign conditions), since $j \not\in \sigma_i$, we have $\sgn s_j^{\sigma_i}= - \sgn s_\ell^{\sigma_i}$ for all $\ell \in \sigma_i$.  And by Lemma \ref{lemma:simply-added}, this ensures $\sgn s_j^{\sigma}= - \sgn s_\ell^{\sigma}$ for all $\ell \in \sigma_i$.  Since $\sigma \in \FP(G|_{[n]\setminus\{j\}})$, we have that $\sgn s_\ell^\sigma$ is identical for all $\ell \in \sigma$, not just $\ell \in \sigma_i$, and so $\sgn s_j^{\sigma}= - \sgn s_\ell^{\sigma}$ for all $\ell \in \sigma$.  Thus by Theorem~\ref{thm:sgn-conditions} (sign conditions), $\sigma$ survives the addition of node $j$, so $\sigma \in \FP(G)$.    
\end{proof}

Theorem~\ref{thm:removables} shows that if a node $j$ is locally removable without altering fixed points of its component, then node $j$ is also globally removable without altering the fixed points of the full graph $G$.  This result gives a new tool for determining that two graphs have the same collection of fixed points. 

\begin{corollary}\label{cor:removables}
Let $G$ have a simply-embedded partition $\{\tau_1|\cdots|\tau_N\}$ and suppose there exists $j\in\tau_i$ such that $\FP(G|_{\tau_i}) = \FP(G|_{\tau_i\setminus\{j\}})$. Let $G'$ be any graph that can be obtained from $G$ by deleting or adding outgoing edges from $j$ to any other component without altering the simply-embedded structure of $G$.  Then $\FP(G') = \FP(G)$.
\end{corollary}
\begin{proof}
Observe that by deleting all the outgoing edges from $j$ to a component $\tau_k$, node $j$ has simply changed from a projector onto $\tau_k$ to a non-projector.  Alternatively, by adding all the outgoing edges to $\tau_k$, node $j$ switches from being a non-projector onto $\tau_k$ to being a projector.  In either case, $j$ is still simply-added onto $\tau_k$, and so $G'$ has the same simply-embedded partition $\{\tau_1|\cdots|\tau_N\}$ as $G$ had.  Additionally, since no edges within $\tau_i$ have been altered, we have that 
$\FP(G'|_{\tau_i}) = \FP(G|_{\tau_i})= \FP(G|_{\tau_i\setminus\{j\}}) = \FP(G'|_{\tau_i\setminus\{j\}}).$
Thus both $G$ and $G'$ satisfy the hypotheses of Theorem~\ref{thm:removables}.  Moreover, $G|_{[n]\setminus\{j\}} = G'|_{[n]\setminus\{j\}}$ since the only differences between $G$ and $G'$ were in edges involving node $j$, which has been removed.  Thus, by Theorem~\ref{thm:removables},  $\FP(G) = \FP(G|_{[n]\setminus\{j\}}) = \FP(G').$
\end{proof}

The result below is new, but fits into the narrative of \cite{Parmelee2022}, which is the paper where the above results are published. That is why we include it here.

\begin{theorem}\label{thm:simply-embedded-domination}
	If $\tau$ is simply-embedded in $G$, then if $k$ dominates $j$ in $G|_{\tau}$, $\FP(G) = \FP(G|_{[n]\setminus\{j\}})$
\end{theorem}

\begin{proof}
	If  $k$ dominates $j$ in $G|_{\tau}$, because  $\tau$ is simply-embedded, both $k$ and $j$ receive the same inputs from the rest of the graph. Thus,  $k$ dominates $j$ with respect to $[n]$ and by Theorem \ref{thm:new-domination} $\FP(G) = \FP(G|_{\tau\setminus\{j\}})$. 
\end{proof}

\subsection{Simple linear chains}\label{subsec:simple-linear-chains}

In this section we add a chain-like architecture to a simply-embedded partition, and term this new architecture \emph{simple linear chains}:

\begin{definition}[simple linear chain] \label{def:linear-chain}
Let $G$ be a graph with node partition $\{\tau_1|\cdots|\tau_N\}$.  We say that $G$ is a \emph{simple linear chain} if the following two conditions hold:
\begin{enumerate}
	\item the only edges between components go from nodes in $\tau_i$ to $\tau_{i+1}$, and 
	\item for every $j \in \tau_i$, either $j \to k$ for every $k \in \tau_{i+1}$ or $j \not\to k$ for every $k \in \tau_{i+1}$.
\end{enumerate}
\end{definition}

A key structural advantage of linear chains is that if $\sigma_i \in \FP(G|_{\tau_i \cup \tau_{i+1}})$, then it turns out that $\sigma_i \in \FP(G)$; in other words, survival of the addition of the next component is sufficient to guarantee survival in the full network.  This occurs because $\sigma_i$ has no outgoing edges to any nodes outside of $\tau_i \cup \tau_{i+1}$. Lemma~\ref{lemma:no-out-edges} shows that whenever a permitted motif has no outgoing edges to a node $k$, then it is guaranteed to survive the addition of node $k$.

\begin{lemma} \label{lemma:no-out-edges}
Let $G$ be a graph on $n$ nodes,  let $\sigma \subseteq [n]$ be nonempty, and $k \in [n]\setminus \sigma$.  If $i \not\to k$ for all $i \in \sigma$, then 
$$\sigma \in \FP(G|_{\sigma \cup \{k\}}) \quad \Leftrightarrow \quad \sigma \in \FP(G|_{\sigma}).$$
In other words, if $\sigma$ has no outgoing edges to node $k$ then $\sigma$ is guaranteed to survive the addition of node $k$ whenever $\sigma$ is permitted.  
\end{lemma}

\begin{proof}
For any $j \in \sigma$, we have that $j$ inside-out dominates $k$. Thus by Rule~\ref{rule:graph-domination}c,  $\sigma \in \FP(G|_{\sigma \cup \{k\}})$ if and only if $\sigma \in \FP(G|_{\sigma})$.
\end{proof}

It turns out that the simply-embedded partition structure of the simple linear chain with the added restriction that $\tau_i$ does not send edges to any $\tau_k$ other than $\tau_{i+1}$ gives significant structure to the values of $s_i^\sigma$ and thus to the domination quantities $w_j^\sigma$.  This structure is the key to the proof of the following theorem.

\begin{theorem}[simple linear chains]\label{thm:linear-chain}
Let $G$ be a simple linear chain with components $\tau_1, \ldots, \tau_N$.  
\begin{enumerate}
	\item[(i)] If $\sigma \in \FP(G)$, then $\sigma_i \in \FP(G|_{\tau_i}) \cup \{\emptyset\}$ for all $i \in [N]$, where $\sigma_i = \sigma \cap \tau_i$. 
	\item[(ii)] Consider a collection $\{\sigma_i\}_{i \in [N]}$ of $\sigma_i \in \FP(G|_{\tau_i}) \cup \{\emptyset\}$.  If additionally $\sigma_i \in \FP(G|_{\tau_i\cup\tau_{i+1}}) \cup \{\emptyset\}$ for all $i \in [N]$, then \\
	$$\bigcup_{i \in [N]} \sigma_i \in \FP(G).$$
\end{enumerate}
In other words, $\FP(G)$ is closed under unions of component fixed point supports that survive in $G|_{\tau_i\cup\tau_{i+1}}$.
\end{theorem}
\begin{proof}
(i) follows directly from Theorem~\ref{thm:menu-CTLN} by noting that the simple linear chain structure endows $G$ with a simply-embedded partition: 
for every $\tau_i$, the nodes in $\tau_{i-1}$ are each either a projector or non-projector onto $\tau_i$, while all nodes outside of $\tau_{i-1}$ are all non-projectors onto $\tau_i$.

To prove (ii), consider $\{\sigma_i\}_{i \in [N]}$ where $\sigma_i\in\FP(G|_{\tau_i \cup \tau_{i+1}}) \cup \{\emptyset\}$ for all $i \in [N]$.   Notice that by Lemma \ref{lemma:no-out-edges}, the fact that $\sigma_i\in\FP(G|_{\tau_i \cup \tau_{i+1}})$ implies that $\sigma_i \in \FP(G)$ since $\sigma_i$ has no outgoing edges to any external node $k$ outside of $\tau_i \cup \tau_{i+1}$.   Thus, we may assume $\sigma_i \in \FP(G) \cup\{\emptyset\}$ for all $i \in [N]$.  We will prove that this guarantees that $\displaystyle \cup_{i \in [N]} \sigma_i \in \FP(G)$ by induction on the number $N$ of components of the simple linear chain. 

For $N = 1$, the result is trivially true. For $N = 2$, observe that the simple linear chain on $\{\tau_1~|~\tau_2\}$ actually has the structure of a bidirectional simply-embedded split $(\tau_1, \tau_2)$, and thus Theorem~\ref{thm:bidir-sa} gives the complete structure of $\FP(G)$ in terms of the surviving fixed points of the component subgraphs $S_{\tau_i}$ and the dying fixed points $D_{\tau_i}$.  The sets of interest here, $\sigma_i \subseteq \tau_i$ with $\sigma_i \in \FP(G)$, are precisely the elements of $S_{\tau_i}$.  Theorem~\ref{thm:bidir-sa}(1) then guarantees that $\sigma_1 \cup \sigma_2 \in \FP(G)$ whenever $\sigma_i \in \FP(G)$, and so the result holds when $N=2$.  

Now, suppose the result holds for any simple linear chain with $N-1$ components. For ease of notation, denote $\sigma_{1\cdots N-1}\od\sigma_1\cup\cdots \cup\sigma_{N-1}$ and let $\sigma \od \cup_{i \in [N]}\sigma_i$.  We will show the result holds for any simple linear chain $G$ with $N$ components.  

Observe that if $\sigma_N = \emptyset$, we have $\sigma = \sigma_{1 \cdots N-1} \in \FP(G|_{\tau_{1 \cdots N-1}})$ by the inductive hypothesis, and we need only show that this implies that $\sigma_{1\cdots N-1} \in \FP(G)$.  On the other hand, if $\sigma_N \neq \emptyset$, then $\sigma = \sigma_{1\cdots N-1} \cup \sigma_N$, where $\sigma_N \in \FP(G)$ by Lemma~\ref{lemma:no-out-edges}, since $\sigma_N \in \FP(G|_{\tau_N})$ and $\sigma_N$ has no outgoing edges to any external nodes outside of $\tau_N$.  Notice that the simple linear chain structure of $G$ ensures that $(\tau_{1 \cdots N-1}, \tau_N)$ is a bidirectional simply-embedded split.  Thus by Theorem~\ref{thm:bidir-sa}, since $\sigma_N$ is a surviving fixed point support, $\sigma_{1\cdots N-1} \cup \sigma_N \in \FP(G)$ if and only if $\sigma_{1\cdots N-1} \in \FP(G)$.  Therefore for any $\{\sigma_i\}_{i \in [N]}$, it suffices to show that $\sigma_{1\cdots N-1} \in \FP(G)$, and the result will follow.  

Notice that by the inductive hypothesis, $\sigma_{1 \cdots N-1} \in \FP(G|_{\tau_{1 \cdots N-1}})$, and thus to show $\sigma_{1\cdots N-1} \in \FP(G)$, we need only show that $\sigma_{1\cdots N-1}$ survives the addition of the nodes in $\tau_N$.  There are two cases to consider here based on whether $\sigma_{1\cdots N-1}$ intersects $\tau_{N-1}$ or not.  Observe that if $\sigma_{1\cdots N-1} \cap \tau_{N-1} = \emptyset$, then $\sigma_{1\cdots N-1}$ has no outgoing edges to $\tau_N$ since only nodes in $\tau_{N-1}$ can send edges forward to $\tau_N$ by the linear chain structure.  In this case, we have $i \not\to k$ for all $i \in \sigma_{1\cdots N-1}$ and all $k \in \tau_N$, and so Lemma~\ref{lemma:no-out-edges} guarantees that $\sigma_{1\cdots N-1} \in \FP(G)$ since we already had $\sigma_{1 \cdots N-1} \in \FP(G|_{\tau_{1 \cdots N-1}})$.  

For the other case where $\sigma_{1\cdots N-1} \cap \tau_{N-1} \neq \emptyset$, we will prove $\sigma_{1\cdots N-1} \in \FP(G)$ by appealing to Theorem~\ref{thm:gen-domination} (general domination) and demonstrating that each $k \in \tau_N$ is \emph{inside-out dominated} by some node $j \in \sigma_{1\cdots N-1}$.  First notice that $\sigma_{1\cdots N-1} = \sigma_{1\cdots N-2} \cup \sigma_{N-1}$ and by the simple linear chain structure of $G$, we have that $\tau_{1\cdots N-2}$ is simply-embedded onto $\tau_{N-1}$.  Thus by Theorem~\ref{thm:simply-added},
\begin{equation}\label{eqn:bidir-sa-split}
	s_i^{\sigma_{1\cdots N-1}} = \frac{1}{\theta} s_i^{\sigma_{1\cdots N-2}} s_i^{\sigma_{N-1}} = \alpha s_i^{\sigma_{N-1}} \textrm{ for all } i \in \sigma_{N-1},
\end{equation}
where $\alpha = \frac{1}{\theta}s_i^{\sigma_{1\cdots N-2}}$ has the same value for every $i \in \sigma_{N-1}$.  Using this, we can now compute the domination quantities $w_j^{\sigma_{1\cdots N-1}}$ and $w_k^{\sigma_{1\cdots N-1}}$ for $j \in \sigma_{N-1}$ and $k \in \tau_N$.  For $j \in \sigma_{N-1}$, we have:
\begin{align*} 
	w_j^{\sigma_{1\cdots N-1}} &\od \sum_{i \in \sigma_{1\cdots N-1}} \widetilde{W}_{ji} |s_i^{\sigma_{1\cdots N-1}}|\\ 
	&=  \sum_{i \in \sigma_{1\cdots N-2}} \widetilde{W}_{ji} |s_i^{\sigma_{1\cdots N-1}}| + \sum_{i \in \sigma_{N-1}} \widetilde{W}_{ji} |s_i^{\sigma_{1\cdots N-1}}| \\
	&=  \sum_{i \in \sigma_{1\cdots N-2}} \widetilde{W}_{ji} |s_i^{\sigma_{1\cdots N-1}}| + \sum_{i \in \sigma_{N-1}} \widetilde{W}_{ji} |\alpha s_i^{\sigma_{N-1}}| \quad\text{by \eqref{eqn:bidir-sa-split}}\\
	&=  \sum_{i \in \sigma_{1\cdots N-2}} \widetilde{W}_{ji} |s_i^{\sigma_{1\cdots N-1}}| + |\alpha|\sum_{i \in \sigma_{N-1}} \widetilde{W}_{ji} |s_i^{\sigma_{N-1}}| \\
	&=  \sum_{i \in \sigma_{1\cdots N-2}} \widetilde{W}_{ji} |s_i^{\sigma_{1\cdots N-1}}| + |\alpha|w_j^{\sigma_{N-1}}
\end{align*}
On the other hand, for $k \in \tau_N$ we have the following formula for $w_k^{\sigma_{1\cdots N-1}}$, where we use the fact that $\widetilde{W}_{ki} = -1-\delta$ for all $i \in \sigma_{1\cdots N-2}$ since there are no edges from nodes in $\tau_{1 \cdots N-2}$ to $\tau_N$:
\begin{align*} 
	w_k^{\sigma_{1\cdots N-1}} &\od  \sum_{i \in \sigma_{1\cdots N-1}} \widetilde{W}_{ki} |s_i^{\sigma_{1\cdots N-1}}|\\ 
	&=  \sum_{i \in \sigma_{1\cdots N-2}} \widetilde{W}_{ki} |s_i^{\sigma_{1\cdots N-1}}| + \sum_{i \in \sigma_{N-1}} \widetilde{W}_{ki} |s_i^{\sigma_{1\cdots N-1}}| \\
	&=  \sum_{i \in \sigma_{1\cdots N-2}} (-1-\delta)|s_i^{\sigma_{1\cdots N-1}}| + \sum_{i \in \sigma_{N-1}} \widetilde{W}_{ki} |\alpha s_i^{\sigma_{N-1}}| \\
	&=  \sum_{i \in \sigma_{1\cdots N-2}} (-1-\delta) |s_i^{\sigma_{1\cdots N-1}}| + |\alpha|\sum_{i \in \sigma_{N-1}} \widetilde{W}_{ki} |s_i^{\sigma_{N-1}}|\\
	&=  \sum_{i \in \sigma_{1\cdots N-2}} (-1-\delta) |s_i^{\sigma_{1\cdots N-1}}| + |\alpha|w_k^{\sigma_{N-1}}.
\end{align*}  
Moreover, since $\sigma_{N-1} \in \FP(G)$, we have that $j \in \sigma_{N-1}$ must inside-out dominate the external node $k$, so $ w_j^{\sigma_{N-1}} >  w_k^{\sigma_{N-1}}$.  Combining this with the fact that $\widetilde{W}_{ji} \geq -1-\delta$, we see that
\begin{align*} 
	w_k^{\sigma_{1\cdots N-1}} &\leq  \sum_{i \in \sigma_{1\cdots N-2}} \widetilde{W}_{ji} |s_i^{\sigma_{1\cdots N-1}}| + |\alpha|w_k^{\sigma_{N-1}} \\
	&< \sum_{i \in \sigma_{1\cdots N-2}} \widetilde{W}_{ji} |s_i^{\sigma_{1\cdots N-1}}| + |\alpha|w_j^{\sigma_{N-1}} = w_j^{\sigma_{1\cdots N-1}}
\end{align*}  
Thus $w_j^{\sigma_{1\cdots N-1}} > w_k^{\sigma_{1\cdots N-1}}$ and so $j$ inside-out dominates $k$ for all $k \in \tau_N$.  Thus by Theorem~\ref{thm:gen-domination}, $\sigma_{1\cdots N-1} \in \FP(G)$, and so $\cup_{i \in [N]} \sigma_i  = \sigma_{1\cdots N-1} \cup \sigma_N \in \FP(G)$ as desired.  
\end{proof}

\begin{figure}[!h]
\begin{center}
	\includegraphics[width=5.5in]{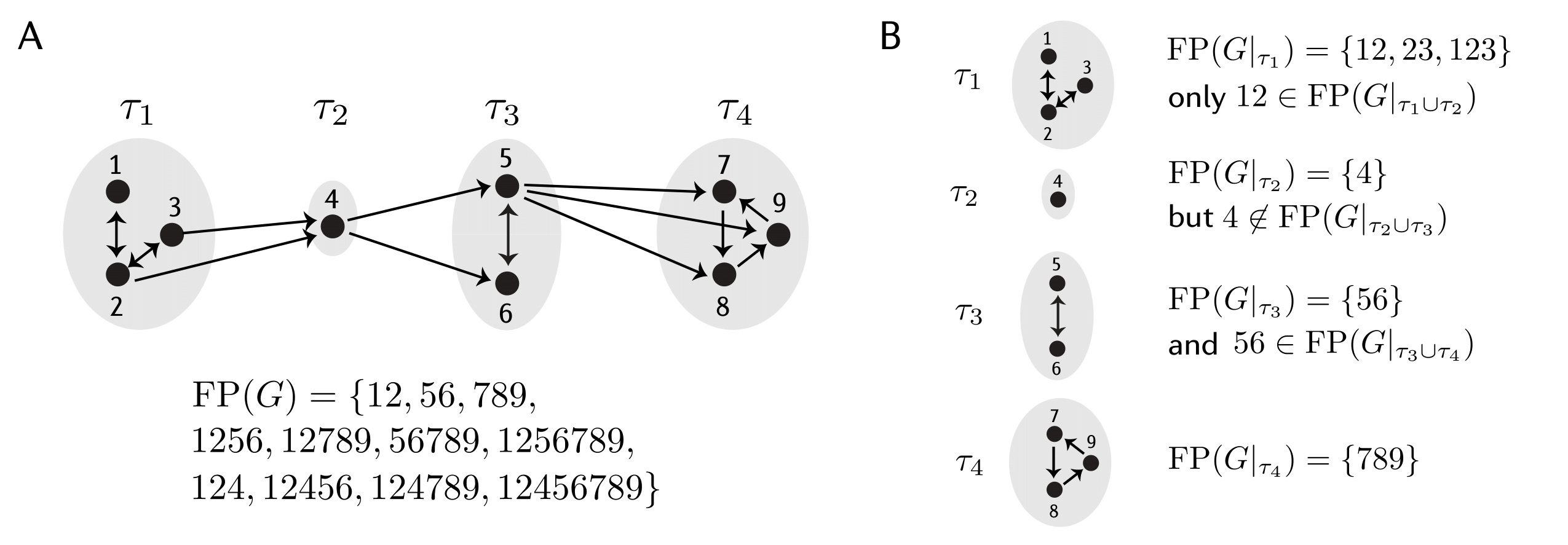}
\end{center}
\caption[Simple linear chain]{ \textbf{Simple linear chain.} (A) An example simple linear chain together with its $\FP(G)$.  The first row of $\FP(G)$ gives the surviving fixed points from each component subgraph; the second row shows that all unions of these component fixed points are also in $\FP(G)$ (Theorem~\ref{thm:linear-chain}(ii)); the third row shows the additional fixed point supports in $\FP(G)$ that arise from the broader menu (Theorem~\ref{thm:linear-chain}(i)).  (B) $\FP(G|_{\tau_i})$ for each component subgraph from A, and the list of which of these supports survive the addition of the next component in the chain. Figure reproduced from \cite{Parmelee2022}.}
\label{fig:linear-chain-thm-example}
\end{figure}

Figure~\ref{fig:linear-chain-thm-example} illustrates Theorem~\ref{thm:linear-chain} with an example simple linear chain.  By Theorem~\ref{thm:linear-chain}(i), every fixed point support in $\FP(G)$ restricts to a fixed point in $\FP(G|_{\tau_i})$.  Next consider a collection of $\sigma_i$ such that $\sigma_i \in \FP(G|_{\tau_i\cup\tau_{i+1}}) \cup \emptyset$ for all $i \in [N]$.  First observe that each $\sigma_i \in \FP(G|_{\tau_i \cup \tau_{i+1}})$ actually survives to the full network, and so $\sigma_i \in \FP(G)$.  This is guaranteed because $\sigma_i$ has no outgoing edges to nodes outside of $\tau_i \cup \tau_{i+1}$ (Rule~\ref{rule:graph-domination}C).  Moreover, by Theorem~\ref{thm:linear-chain}(ii), we see that every union of surviving component fixed points yields a fixed point of the full network, but additional fixed point supports are also possible.

\subsection{Strongly simply-embedded partitions}\label{subsec:strongly-simply-embedded}

Recall that the difference between a simply-added split and a bidirectional simply-added split is that, for bidirectional splits, not only $\tau$ is simply-embedded in $G$, but $[n]\setminus\tau$ is also simply-embedded in $G$. To achieve an analogous bidirectionality in the case of simply-embedded \emph{partitions} we must now additionally require that every $[n]\setminus\tau_i$ is simply-embedded in $G$ as well. This means that not only $\tau_i$ is treated the same by the rest of the graph, but it must also treat the rest of the graph the same. We term this new, more rigid, partition structure a \emph{strongly simply-embedded partition}.

\begin{definition}[strongly simply-embedded partition]\label{def:strongly-simply-embedded}
Let $G$ be a graph with a partition of its nodes $\{\tau_1|\cdots|\tau_N\}$.  The partition is called \emph{strongly simply-embedded} if for every node $j$ in $G$, either $j \to k$ for all $k \notin \tau_i$ or $j \not\to k$ for all $k \notin \tau_i$, where $\tau_i$ is the component containing $j$.
\end{definition}

The simplest examples of graphs with a strongly simply-embedded partition are \emph{disjoint unions} and \emph{clique unions}, which are building block constructions first studied in \cite{fp-paper}.  In a \emph{disjoint union} of component subgraphs $G|_{\tau_1}, \ldots, G|_{\tau_N}$, there are no edges between components. In this case, every node in $G$ is a non-projector onto the rest of the graph.  At the other extreme, a \emph{clique union} has bidirectional edges between every pair of nodes in different components.  In a clique union, every node is a projector onto the rest of the graph.  

The key fact that allowed to characterize the fixed points of disjoint and clique unions in \cite{fp-paper} was a complete complete factorization of the $s_j^\sigma$ values in terms of the $s_j^{\sigma_i}$ of the component fixed point supports. But recall that we can only get this complete factorization when the simply-added structure is bidirectional. Strongly simply-embedded partitions now satisfy that $(\tau_i, [n] \setminus \tau_i)$ is also a bidirectional simply-added split. This means we can prove a result analogous to Theorem \ref{thm:simply-added}, but for a partition, and for every $j\in[n]$. Moreover, the $s_j^{\sigma_i}$ values are fully determined by whether $\sigma_i$ is a surviving or a dying fixed point of $G|_{\tau_i}$.  Recall that we denote the sets of surviving and dying fixed points as:
$$S_{\tau_i} \od \FP(G|_{\tau_i}) \cap \FP(G) \quad \text{and} \quad D_{\tau_i} \od \FP(G|_{\tau_i}) \setminus S_{\tau_i}.$$


\begin{lemma}\label{lemma:full-factorization-CTLN}
Let $G$ be a graph on $n$ nodes with a strongly simply-embedded partition $\{\tau_1|\dots|\tau_N\}$. For any $\sigma \subseteq [n]$, denote $\sigma_i \od \sigma \cap \tau_i$, and $\sigma_{i_1\dotsi_k} \od \sigma_{i_1} \cup \dots \cup \sigma_{i_k}$ and let $I = \{i \in [N]~|~ \sigma_i \neq \emptyset\}$. Then for every $j\in [n]$, $$s_j^\sigma = \frac{1}{\theta^{|I|-1}}\prod_{i\in I} s_j^{\sigma_i},$$ 
where $s_j^{\sigma_i}$ has the same value for every $j \in [n]\setminus \tau_i$.\\
Moreover, for any $\sigma_i \in \FP(G|_{\tau_i})$ and $j \in \tau_i$:
$$\sgn s_j^{\sigma_i} = \begin{cases} 
	\phantom{-}\idx(\sigma_i) & \text{if } j \in \sigma_i \\
	-\idx(\sigma_i) & \text{if } j \in \tau_i\setminus \sigma_i
\end{cases}$$ 
while for any $k\notin \tau_i$, 
$$\sgn s_k^{\sigma_i} = \begin{cases} 
	-\idx(\sigma_i) & \text{if } \sigma_i \in S_{\tau_i}\\
	\phantom{-}\idx(\sigma_i) & \text{if } \sigma_i \in D_{\tau_i}
\end{cases}$$
\end{lemma}

Similar to simple linear chains, it turns out that strongly simply-embedded partitions also have the property that $\FP(G)$ is closed under unions of surviving fixed point supports of the component subgraphs.  With the added structure of the strongly simply-embedded partition, though, we can actually say something stronger -- $\FP(G)$ can be fully determined from knowledge of the component fixed point supports together with knowledge of which of those component fixed points survive in the full network.  This complete characterization of $\FP(G)$ is given in Theorem~\ref{thm:bidir-sa-partition} below.  

\begin{theorem}\label{thm:bidir-sa-partition}
Suppose $G$ has a strongly simply-embedded partition $\{\tau_1|\dots|\tau_N\}$, and let $\sigma_i \od \sigma \cap \tau_i$ for any $\sigma \subseteq [n]$.  
Then $\sigma \in \FP(G)$ if and only if $\sigma_i \in \FP(G|_{\tau_i}) \cup \{\emptyset\}$ for each $i \in [N]$, and either
\begin{enumerate}
	\item[(a)] every $\sigma_i$ is in $\FP(G) \cup \{ \emptyset\}$, or 
	\item[(b)] none of the $\sigma_i$ are in $\FP(G) \cup \{ \emptyset\}$.
\end{enumerate}
In other words, $\sigma \in \FP(G)$ if and only if $\sigma$ is either a union of surviving fixed points $\sigma_i$, at most one per component, or it is a union of dying fixed points, exactly one from every component.
\end{theorem}

This theorem generalizes Theorem~\ref{thm:bidir-sa}, characterizing every element of $\FP(G)$ in terms of the sets of surviving and dying component fixed points supports, $S_{\tau_i}$ and $D_{\tau_i}$.  Notice that in the statement of Theorem~\ref{thm:bidir-sa-partition}, all the fixed point supports of type (a) have the form $\bigcup_{i \in I} \sigma_i$ for $\sigma_i \in S_{\tau_i}$ and $I \subseteq [N]$, while those of type (b) have the form $\bigcup_{i=1}^N \sigma_i$ for $\sigma_i \in D_{\tau_i}$.

More generally, though, a strongly simply-embedded partition can have a mix of surviving and dying component fixed points, so that $\FP(G)$ has a mix of both type (a) and type (b) fixed point supports.  Figure~\ref{fig:bidir-sa-part-FP}A gives an example strongly simply-embedded partition, and panel B shows both the set of component fixed point supports, $\FP(G|_{\tau_i})$, and the subset of those that survive to yield fixed points of the full network.  Since there are dying fixed points in every component, we see that $\FP(G)$ has a mix of both type (a) and type (b) fixed point supports.  
\begin{figure}[!h]
\begin{center}
	\includegraphics[width=.8\textwidth]{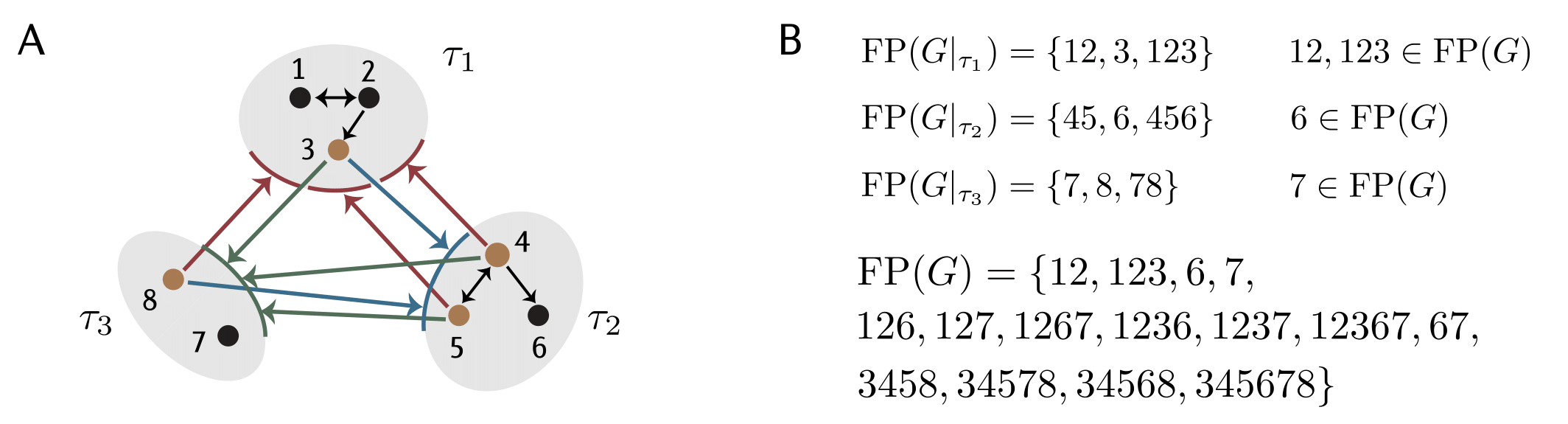}
\end{center}
\caption[Strongly simply-embedded partition]{ \textbf{Strongly simply-embedded partition with $\boldsymbol{\FP(G)}$.} (A) A graph with a strongly simply-embedded partition $\{\tau_1|\tau_2|\tau_3\}$.  Projector nodes are colored brown. (B) (Top) $\FP(G|_{\tau_i})$ for each component subgraph together with the supports from each component that survive within the full graph.  (Bottom) $\FP(G)$ for the strongly simply-embedded partition graph.  The first two lines of $\FP(G)$ consist of unions of surviving fixed points, at most one per component.  The third line gives the fixed points that are unions of dying fixed point supports, exactly one from every component. Figure reproduced from \cite{Parmelee2022}.}
\label{fig:bidir-sa-part-FP}
\end{figure}

A special case of a bidirectional simply-added split occurs whenever a graph contains a node that is projector/non-projector onto the rest of the graph.  Specifically, since any subset is always simply-added onto a single node $j$ trivially, we see that we have a bidirectional simply-added split $(\{j\}, [n]\setminus \{j\})$ whenever $j$ is either a projector or a non-projector onto the rest of the graph.  Recall that if $j$ is a non-projector onto $[n]\setminus \{j\}$, then $j$ has no outgoing edges in $G$, and so it is a \emph{sink}.  Moreover, we have seen that sinks are the only single nodes that can support fixed points since a singleton $\{j\}$ is trivially uniform in-degree 0, and thus only survives when it has no outgoing edges, by Theorem \ref{thm:uniform-in-deg}.  Combining this observation with the bidirectional simply-added split for a sink, we see there is certain internal structure that must be present in $\FP(G)$ whenever it contains any singleton sets.  

\begin{proposition}\label{prop:FP-with-singleton}
Let $G$ be a graph such that there is some singleton $\{j\} \in \FP(G)$.  Then for any $\sigma \in \FP(G)$ (with $\sigma \neq \{j\}$), 
\begin{enumerate}
	\item If $j \notin \sigma$, then $\sigma \cup \{j\} \in \FP(G)$; i.e., $\FP(G)$ is closed under unions with singletons.
	\item If $j \in \sigma$, then $\sigma \setminus \{j\} \in \FP(G)$; i.e., $\FP(G)$ is closed under set differences with singletons.
\end{enumerate}
\end{proposition}

\begin{proof}
First notice that since $\{j\} \in \FP(G)$, $j$ is a sink in $G$ by Theorem \ref{thm:uniform-in-deg} (since a singleton is trivially uniform in-degree 0, and thus survives exactly when it has no outgoing edges), and therefore $(\{j\}, [n]\setminus \{j\})$ is a bidirectional simply-added split. 

To prove (1), suppose $j\notin\sigma$. Since $(\{j\}, [n]\setminus \{j\})$ is a bidirectional simply-added split, Theorem~\ref{thm:bidir-sa} guarantees that $\sigma\cup \{j\} \in \FP(G)$ if and only if $\{j\}, \sigma$ both survive or both die.  By assumption, both sets are in $\FP(G)$, so both survive.  Thus,  $\sigma \cup \{j\} \in \FP(G)$.

To prove (2), suppose $j \in \sigma$. By Theorem \ref{thm:bidir-sa}, $\sigma \in \FP(G)$ if and only if $\{j\}, \sigma \setminus \{j\}$ both survive or both die. By assumption, $\{j\} \in \FP(G)$, and so $\sigma \setminus \{j\} \in \FP(G)$ as well.
\end{proof}

\begin{corollary}
Let $G$ be a graph such that $\FP(G)$ contains singleton sets  $\{j_1\}, \{j_2\}, \ldots, \{j_\ell\}$, and let $\S=\{j_1, \ldots, j_\ell\}$ be the set of singletons. Then for any $\sigma \in \FP(G)$ and any $\omega \subseteq \S$
$$\sigma \cup \omega \in \FP(G).$$
Moreover, let $\tau = [n] \setminus  \mathcal{S}$.  Then $\FP(G)$ has the direct product structure:
$$\FP(G) \cup \{\emptyset\}  \cong  \left(\{\sigma \in \FP(G|_{\tau})~|~ \sigma\in \FP(G) \} \cup \{\emptyset\} \right) \times \mathcal{P}( \mathcal{S}),$$
where $\mathcal{P}( \mathcal{S})$ denotes the power set of $\S$.  In other words, every fixed point support in $\FP(G)$ has the form $\sigma \cup \omega$ where $\sigma \in \FP(G|_{\tau}) \cup \{\emptyset\}$ and $\omega \subseteq  \S$.
\end{corollary}

\begin{proof}
The first statement follows by iterating Proposition~\ref{prop:FP-with-singleton}(1) $|\omega|$ times for each of the added singletons in $\omega$. To prove the second statement, we will show that every $\nu \in \FP(G)$ is the union of a surviving fixed point $\sigma \subseteq \tau$ (or the empty set) with a subset of $\S$ (including empty set); moreover, every such union yields a fixed point (other than $\emptyset \cup \emptyset$).  The direct product structure of $\FP(G)$ immediately follows from this decomposition of the fixed point supports.  By the first result, we see that every such union is contained in $\FP(G)$.  Thus, all that remains to show is that every element of $\FP(G)$ is such a union.  Let $\nu \in \FP(G)$ and let $\sigma = \nu \cap \tau$ and $\omega = \nu \cap \S$, so that $\nu = \sigma \cup \omega$.  If $\sigma$ or $\omega$ are empty, then we're done, so suppose both are nonempty.  Then we can iteratively apply Proposition~\ref{prop:FP-with-singleton}(2) $|\omega|$ times to see that $\sigma \in \FP(G)$.  Thus, every fixed point support arises as a union of some $\sigma \subseteq \tau$ with an arbitrary subset of $\S$, where $\sigma\in \FP(G)\cup \{\emptyset\}$ (and for every $\sigma \in \FP(G)$, we have $\sigma \in \FP(G|_\tau)$ as well by Corollary~\ref{cor:inheritance}(2)). 
\end{proof}

As an application of Theorem~\ref{thm:bidir-sa-partition}, we can immediately recover characterizations of the fixed points of disjoint unions and clique unions previously given in \cite[Theorems 11 and 12]{fp-paper}.  In a disjoint union, every component fixed point support survives to the full network since it has no outgoing edges (by Rule~\ref{rule:graph-domination}: inside-out domination).  Thus, for a disjoint union, $\FP(G)$ consists of all the fixed points of type (a) from Theorem~\ref{thm:bidir-sa-partition}: unions of (surviving) component fixed points $\sigma_i$, at most one per component.  In contrast, in a clique union, every component fixed point support dies in the full network since it has a target that outside-in dominates it (in fact, every node outside of $\tau_i$ is a target of any subset of $\tau_i$).  Thus, for a clique union, $\FP(G)$ consists of all the fixed points of type (b): unions of (dying) component fixed points $\sigma_i$, exactly one from every component.  Both the disjoint union and clique union characterizations of $\FP(G)$ \cite[Theorems 11 and 12]{fp-paper} are now immediate corollaries of Theorem~\ref{thm:bidir-sa-partition}.

\begin{corollary}\label{cor:disjoint-clique-unions}
Let $G$ be a graph with partition $\{\tau_1|\cdots|\tau_N\}$.  
\begin{itemize}
	\item[(a)] If $G$ is a disjoint union of $G|_{\tau_1}, \ldots, G|_{\tau_N}$, then $\sigma \in \FP(G)$ if and only if $\sigma_i \in \FP(G|_{\tau_i}) \cup \{\emptyset\}$ for all $i \in [N]$.
	\item[(b)] If $G$ is a clique union of $G|_{\tau_1}, \ldots, G|_{\tau_N}$, then $\sigma \in \FP(G)$ if and only if $\sigma_i \in \FP(G|_{\tau_i})$ for all $i \in [N]$.
\end{itemize}
\end{corollary}

\section{Layered CTLNs}\label{sec:layered-ctlns}

The results here are inspired by the sequential construction of Chapter \ref{ch:sequential-control}. Our goal is to provide theoretical explanations for why the networks of Chapter \ref{ch:sequential-control} work so well. Leaving the world of CTLNs, we allowed off-diagonal blocks to not be strictly constrained by a graph. Consequently, the networks in Chapter \ref{ch:sequential-control} are best described as \emph{layered} CTLNs, because the diagonal blocks are CTLNs but the off-diagonal blocs are not. Here we derive Theorem \ref{thm:bidir-sa-partition} for TLNs in general, not only layered CTLNs. The diagonal blocks, or layers, can also be TLNs. It is convenient to choose CTLNs as the layers, because we have plenty of results to build in the desired attractors.

Throughout the chapter, TLNs with a vector $b$ of heterogeneous (but constant in time) inputs are denoted as $(W,b)$, as usual. When the input is also constant across neurons, we abuse notation and denote it as $(W,\theta)$ instead of $(W,\mathbbm{1}\theta)$. We start by showing that any node $k$ with non-positive input $b_k \leq 0$ cannot be involved in any fixed points of the network. This will allow us to restrict, moving forward, to networks where all inputs are positive.

\begin{definition}
We say that $(W,b)$ is a competitive TLN on $n$ neurons if $W_{ij} \leq 0$ and $W_{ii} = 0$ for all $i,j \in [n]$.
\end{definition}

Attention! This definition is different from the one in \cite{fp-paper}, which requires $b_i \geq 0$ for all $i \in [n]$ too.

\begin{lemma}\label{lemma:restricted-fixed-point}
Let $(W,b)$ be a competitive non-degenerate TLN on $n$ neurons, and suppose that for some $k \in [n]$ we have $b_k\leq 0$. Then, $x^*$ is a fixed point of $(W, b)$ if and only if $x_k^*=0$ and $\widehat{x}^*$ is a fixed point of $(W_{[n]\setminus\{k\}},b|_{[n]\setminus\{k\}})$, where $\widehat{x}^*:=x^*|_{[n]\setminus\{k\}}$.
\end{lemma}

\begin{proof}
Denote $\widehat{W} = W_{[n]\setminus\{k\}}$, $\widehat{b} = b|_{[n]\setminus\{k\}}$ and let $\widehat{x}^*:=x^*|_{[n]\setminus\{k\}}$.

($\Rightarrow$) Suppose $x^*$ is a fixed point of $(W, b)$. We have that for all $i \in [n]$, $\left.\frac{d x_i}{d t}\right|_{x=x^*}=0$, and thus
$$x_i^*=\left[\sum_{j\in[n]} W_{i j} x_j^*+b_i\right]_{+}.$$
Since $b_k \leq 0$ and $W_{ij} \leq 0$, we obtain $x_k^*=\left[\sum_{j\in[n]} W_{kj} x_j + b_k\right]_+=0$, as desired. 

Now let's see that $\widehat{x}^*$ is a fixed point of $(\widehat{W}, \widehat{b})$. Indeed, for all $i \neq k$ we have that 
\begin{align*}
	\left.\frac{d x_i}{d t}\right|_{x=\widehat{x}^*} 
	& =-\widehat{x}_i^*+\left[\sum_{j \in [n]\setminus\{k\}} \widehat{W}_{i j} \widehat{x}_j^*+\widehat{b}_i\right]_{+} 
	=-x_i^*+\left[\sum_{j \in [n]\setminus\{k\}} W_{i j} x_j^*+b_i\right]_{+} \\
	& =-x_i^*+\left[\sum_{j\in[n]} W_{i j} x_j^*+b_i\right]_+ 
	=\left.\frac{d x_i}{d t}\right|_{x=x^*} = 0
\end{align*} where the second equality follows from the definition of $\widehat{W}, \widehat{b},\widehat{x}$ and the third one because $x_k^*=0$.

($\Leftarrow$) Suppose $\widehat{x}^*$ is a fixed point of $(\widehat{W}, \widehat{b})$ and $x_k^*=0$. Then we have that for all $i \neq k$,  $\left.\frac{d x_i}{d t}\right|_{x=\widehat{x}^*}=0$. Let's see that $x^*$ is a fixed point of $(W,b)$. Indeed,
\begin{align*}
	\left.\frac{d x_i}{d t}\right|_{x=x^*}
	& = -x_i^*+\left[\sum_{j\in[n]} W_{i j} x_j^*+b_i\right]_{+}
	=-x_i^*+\left[\sum_{j \in [n]\setminus\{k\}} W_{i j} x_j^*+b_i\right]_{+} \\
	& =-\widehat{x}_i^*+\left[\sum_{j \in [n]\setminus\{k\}} \widehat{W}_{i j} 	\widehat{x}_j^*+\widehat{b}_i\right]_{+} 
	= \left.\frac{d x_i}{d t}\right|_{x=\widehat{x}^*} = 0 \\
\end{align*}
where the second equality once again follows from the assumption that $x_k^* =0$, and the third equality from definition of  $\widehat{W}, \widehat{b},\widehat{x}$.
\end{proof}

We then easily see that neurons with negative input do not participate of any fixed point support:

\begin{corollary}\label{lemma:nonpositive-input}
Let $(W,b)$ be a competitive non-degenerate TLN. If $b_k\leq 0$ for some $k \in [n]$, then $\FP(W,b) = \FP(W_{[n]\setminus\{k\}},b|_{[n]\setminus\{k\}})$. In particular, $k\notin \sigma$ for all $\sigma \in \FP(W,b)$.
\end{corollary}

\begin{proof} 	Denote $\widehat{W} = W_{[n]\setminus\{k\}}$, $\widehat{b} = b|_{[n]\setminus\{k\}}$ and let $\widehat{x}^*:=x^*|_{[n]\setminus\{k\}}$. Observe that
\begin{align*}
	\sigma \in \FP(W,b) &\Leftrightarrow \text{ there exists a fixed point } x^* \text{  of }(W,b)\text{ such that } \sigma=\operatorname{supp}\left(x^*\right) \\
	&\Leftrightarrow x_k^*=0 \text{, and } \widehat{x}^* \text{ is a fixed point of } (\widehat{W}, \widehat{b}) \\
	&\Leftrightarrow k \notin \sigma \text{, and } \sigma \in \FP(\widehat{W}, \widehat{b}).
\end{align*}
Where the second equivalence follows from Lemma \ref{lemma:restricted-fixed-point}. Thus, $\FP(W,b) = \FP(\widehat{W},\widehat{b})$ and $k \notin \sigma$ for all $\sigma \in \FP(W, b)$, as desired.
\end{proof}

The above lemma explains the fusion attractors we saw in last chapter. 

\paragraph*{Examples from Chapter \ref{ch:sequential-control}.}
\begin{figure}[!h]
\begin{center}
	\includegraphics[width=0.6\textwidth]{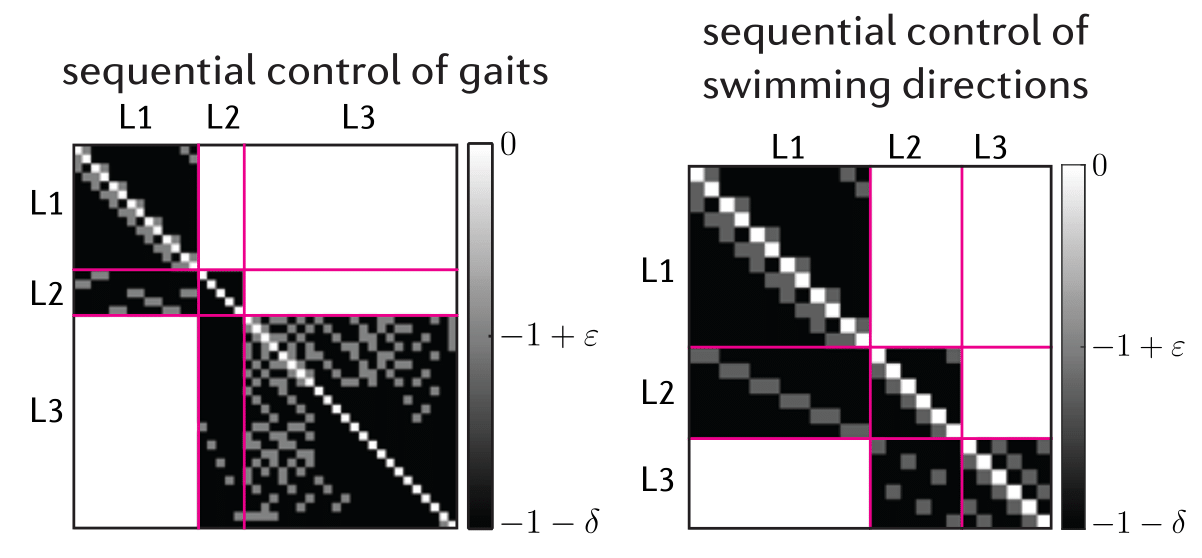}
\end{center}
\caption[Sequential control matrices]{Matrices from Chapter \ref{ch:sequential-control}.}
\label{fig:sequential_control_matrices}
\end{figure}

We begin by analyzing the network from Section \ref{sec:sequential-control-gaits}, the sequential control of gaits, whose connectivity matrix is reproduced in Figure \ref{fig:sequential_control_matrices}. We focus on deriving $\FP(W,\theta)$ outside of pulse times, which is precisely when the network settles in the attractors we observe. Since outside of pulses, $\theta_2 = 0$, by Lemma \ref{lemma:nonpositive-input}, $\FP(W,\theta) = \FP(W|_{L_1\cup L_3},b|_{L_1\cup L_3})$.
But \begin{equation}\label{eq:W-sequential-control}
	W|_{L_1\cup L_3} =  \left[\begin{array}{c|c}
		W_{L_1} & 0 \\
		\hline 
		0 & W_{L_3}
\end{array}\right],
\end{equation} and thus layers are decoupled and it is now easy to see that attractors from $L_1$ and $L_2$ will be preserved, since they do not really affect each other's dynamics.  

Analogously, the same mechanism is doing the work in the network of Section \ref{sec:sequential-control-mollusk}, as once again we have $\FP(W,\theta) = \FP(W|_{L_1\cup L_3},b|_{L_1\cup L_3})$, with $W$ as in Equation \ref{eq:W-sequential-control}. Layers $L_1$ and $L_3$ are transiently connected during pulse times through $L_2$, which allows $L_1$ to communicate its attractor transitions to $L_3$. Once the transition is communicated, and pulse ends, the network will settle in the chosen attractor. Neat.

The results in this section were inspired by a combination of the networks of Chapter \ref{ch:sequential-control}, and past results (in Secs. \ref{sec:technical-background} and \ref{sec:simply-embedded-CTLNs}). One of the things the proofs have in common is the following simple linear algebra lemma, which allow us to factor determinants whose top right block is rank 1:

\begin{lemma}\label{lemma:det-linear-algebra} Let 
\begin{figure}[H]
	\centering
	\includegraphics[width=0.15\textwidth]{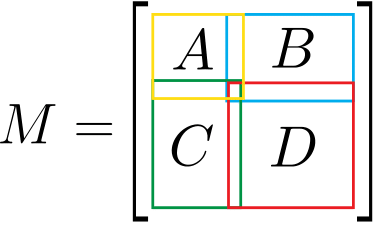}
\end{figure}
		
be an $(m+n-1)\times(m+n-1)$ matrix, consisting of the blocks $A$ ($m\times m$), $B$ ($m\times n$), $C$ ($n\times m$) and $D$ ($n\times n$),	where adjacent blocks overlap in one row/column. Note that $a_{m, m}=b_{m,_1}=c_{1 m}=d_{1,1}$ is the single entry where all four matrices overlap. If $B$ or $C$ is rank 1 and $a_{m,m} \neq 0$, then $$\det M=\frac{1}{a_{m,m}} \det A \det D.$$


\end{lemma}

\begin{proof}
Let \begin{align*}
		A&=\left[\begin{array}{ccc}
				a_{1,1} & \cdots & a_{1, m} \\
				\vdots & & \vdots \\
				a_{m, 1} & \cdots & a_{m, m}
			\end{array}\right], 
	 B=\left[\begin{array}{ccc}
				b_{1,1} & \cdots & b_{1, n} \\
				\vdots & & \vdots \\
				b_{m, 1} & \cdots & b_{m, n}
			\end{array}\right], \\
		C&=\left[\begin{array}{ccc}
				c_{1,1} & \cdots & c_{1, m} \\
				\vdots & \cdots & \vdots \\
				c_{n, 1} & \cdots & c_{n, m}
			\end{array}\right], 
	 D=\left[\begin{array}{ccc}
				d_{1,1} & \cdots & d_{m, n} \\
				\vdots & \cdots & \vdots \\
				d_{n, 1} & \cdots & d_{n, n}
			\end{array}\right].
\end{align*} 
Note that  
\begin{align*}
	\left[\begin{array}{c}
		a_{1, m} \\
		\vdots \\
		a_{m, m}
	\end{array}\right]=\left[\begin{array}{c}
		b_{1,1} \\
		\vdots \\
		b_{m, 1}
	\end{array}\right], 
	\left[\begin{array}{c}
		c_{1, m} \\
		\vdots \\
		c_{n, m}
	\end{array}\right]=\left[\begin{array}{c}
		d_{1,1} \\
		\vdots \\
		d_{n, 1}
	\end{array}\right],\left[\begin{array}{llll}
		b_{m, 1} & \cdots & b_{m, n}
	\end{array}\right]=\left[\begin{array}{llll}
		d_{1,1} & \cdots & d_{1, n}
	\end{array}\right]
\end{align*}  and $\left[\begin{array}{llll}
	c_{1,1} & \cdots & c_{1, m}
\end{array}\right] = \left[\begin{array}{llll}
	a_{m,1} & \cdots & a_{m,m}
\end{array}\right]$. 
Assume first that $B$ is rank 1, then we can take all columns of $B$ to be multiples of its first column. Let $\beta_k$ be such that 
$$
\left[\begin{array}{c}
	b_{1, k} \\
	\vdots \\
	b_{m, k}
\end{array}\right]=\beta_k\left[\begin{array}{c}
	b_{1,1} \\
	\vdots \\
	b_{m, 1}
\end{array}\right],
$$ with $\beta_1 = 1$. Subtracting $\beta_k$ times the $m$-th column of $M$ from its $k$-th column, we get
\begin{align*}
	\det M & =\left[\begin{array}{ccc|ccc}
		a_{1,1} & \cdots & a_{1, m} & 0 & \cdots & 0 \\
		\vdots & & \vdots & \vdots & & \vdots \\
		a_{m, 1} & \cdots & a_{m, m} & 0 & \cdots & 0 \\
		\hline c_{1,1} & \cdots & d_{2,1} & d_{2,2}-\beta_2 d_{2,1} & \cdots & d_{2, n}-\beta_n d_{2,1} \\
		\vdots & & \vdots & \vdots & & \vdots \\
		c_{n, 1} & \cdots & d_{n, 1} & d_{n, 2}-\beta_2 d_{n, 1} & \cdots & d_{n, n}-\beta_n d_{n, 1}
	\end{array}\right] 
	\\
	& =\det A \det\left[\begin{array}{cccc}
		d_{2,2}-\beta_2 d_{2,1} & \cdots & d_{2, n}-\beta_n d_{2,1} \\
		\vdots & & \vdots \\
		d_{n, 2}-\beta_2 d_{n, 1} & \cdots & d_{n, n}-\beta_n d_{n, 1}
	\end{array}\right] \\
	& = \det A \det\left[\begin{array}{c|ccc}
		1 & 0 & \cdots & 0 \\
		\hline
		&d_{2,2}-\beta_2 d_{2,1} & \cdots & d_{2, n}-\beta_n d_{2,1} \\
		*& \vdots & & \vdots \\
		& d_{n, 2}-\beta_2 d_{n, 1} & \cdots & d_{n, n}-\beta_n d_{n, 1}
	\end{array}\right] \\
	& =\det A \frac{1}{b_{m, 1}} \det\left[\begin{array}{c|ccc}
		b_{m, 1} & 0 & \cdots & 0 \\
		\hline d_{2,1} & d_{2,2}-\beta_2 d_{2,1} & \cdots & d_{2, n}-\beta_n d_{2,1} \\
		\vdots & \vdots & & \vdots \\
		d_{n, 1} & d_{n, 2}-\beta_2 d_{n,1} & \cdots & d_{n, n}-\beta_n d_{n, 1}
	\end{array}\right] \\
	& =\frac{1}{b_{m, 1}} \det A \det\left[\begin{array}{cccc}
		b_{m, 1} & \beta_2 b_{m, 1} & \cdots & \beta_n b_{m, 1} \\
		d_{2,1} & d_{2,2} & \cdots & d_{2, n} \\
		\vdots & \vdots & & \vdots \\
		d_{n, 1} & d_{n, 2} & \cdots & d_{n, n}
	\end{array}\right] \\
	& =\frac{1}{a_{m, m}} \det A \det\left[\begin{array}{cccc}
		d_{1,1} & d_{1,2} & \cdots & d_{1, n} \\
		d_{2,1} & d_{2,2} & \cdots & d_{2, n} \\
		\vdots & \vdots & & \vdots \\
		d_{n, 1} & d_{n, 2} & \cdots & d_{n, n}
	\end{array}\right] = \frac{1}{a_{m, m}} \det A \det D
\end{align*} because ${\left[\begin{array}{lll}
		b_{m, 1} & \cdots & b_{m, n}
	\end{array}\right]=\left[\begin{array}{llll}
		b_{m, 1} & \beta_2 b_{m, 1} & \cdots & \beta_n b_{m, 1}
	\end{array}\right] } 
=  {\left[\begin{array}{llll}
		d_{1,1} & \cdots & d_{1, n}
	\end{array}\right]}$ and $b_{m,1} = a_{m, m}$. If $C$ is rank 1, the proof is similar.
\end{proof}

This determinant lemma is a key technical tool for proofs in this section, as well as in \cite{fp-paper,Parmelee2022}, but also in Section \ref{sec:simply-embedded-CTLNs}. The fact that allows us to get a rank 1 block here, and in the past, is the simply-embeddedness of a given component network. This translates into said component receiving certain uniform inputs from other neurons. Below we make this notion precise by extending Definition \ref{def:simply-embedded} beyond CTLNs to apply more generally to TLNs:
\begin{definition}\label{def:simply-embedded}
	Let $(W,b)$ be a competitive non-degenerate TLN. We say that $\tau \subseteq [n]$ is simply-embedded in the network if for every $j \notin \tau$, $W_{ij} = W_{kj} = \gamma_j $ for all $i,k \in \tau$. That is, if $\tau = \{1,\dots,\ell\}$ and $[n]\setminus\tau = \{\ell+1,\dots,n\}$, then:
	\begin{align}
		W = \left[\begin{array}{c|ccc}
			& \gamma_{\ell+1} &  & \gamma_{n}\\
			W_{\tau} & \vdots & \cdots & \vdots \\
			& \gamma_{\ell+1} & & \gamma_{n} \\
			\hline
			\ast & \multicolumn{3}{c}{W_{[n]\setminus\tau}} 
		\end{array}\right]
	\end{align}
\end{definition}
	
With this definition, we can readily obtain the TLN version of Theorem \ref{thm:simply-added}, key in the proofs of \cite{fp-paper,Parmelee2022}, and in this section. The reasoning is as follows: if we can factor the $s_i$'s, we can inherit their signs to the component networks, if we can inherit their signs to the component networks then we can compare these signs to asses who is a fixed point support of the whole and parts via Theorem \ref{thm:sgn-conditions}. This is why the next lemma is so important.
\begin{theorem}\label{thm:simply-embedded-si-factorization}
	Let $\tau$ be simply-embedded in $(W,b)$ and $b_i = \theta$ for $i\in\tau$. Let $\omega = [n]\setminus\tau$. Then for any $\sigma \subseteq [n]$, we have
	$$s_i^\sigma = \frac{1}{\theta}s_i^{\sigma\cap\omega} s^{\sigma\cap\tau}_i = \alpha s^{\sigma\cap\tau}_i \quad\text{for each $i \in \tau$,}$$
	where $\alpha = \frac{1}{\theta}s_i^{\sigma\cap\omega}$ has the same value for every $i \in \tau$.
\end{theorem}
\begin{proof}
	Since $\tau$ is simply-embedded with uniform input, $W$ has the form
	$$W=\left[\begin{array}{c|c}
		W_\tau & \mathbbm{1}_\tau \gamma^\intercal \\
		\hline
		* & W_\omega
	\end{array}\right],$$ where $\mathbbm{1}_\tau$ is a column vector of all ones of size $|\tau|$, and $\gamma^\intercal = \left[\begin{array}{lll}
	\gamma_{1} & \cdots & \gamma_{m} 
	\end{array}\right]$ so that
 $$\mathbbm{1}_\tau \gamma^\intercal=\left[\begin{array}{lll}
		\gamma_1 &  & \gamma_m \\
		\vdots & \cdots & \vdots \\
		\gamma_1 &  & \gamma_m
		\end{array}\right].$$
	Let $i \in \tau$, then
	\begin{align*}
		s_{i}^{\sigma} & = \det\left(\left(I-W_{\sigma\cup\{i\}}\right)_{i};\theta\mathbbm{1}\right) 
		=\det\left[\begin{array}{c|c|ccc}
		 & \theta & \gamma_1 & & \gamma_m \\
			I-W_{\sigma\cap\tau} & \vdots & \vdots & \cdots & \vdots \\
		 & \theta & \gamma_1 &  & \gamma_m \\
		 \cline{1-2}
			\rotvert (I-W_{\sigma\cap\tau})_i \rotvert & \theta & \gamma_1 & \cdots & \gamma_m\\
			\hline
		 & \theta & & & \\
			* & \vdots& & I-W_{\sigma\cap\omega} & \\
		 & \theta & & & \\
		\end{array}\right]
	\end{align*}
	Since the upper right block,  $\mathbbm{1}_{\sigma \cap \tau} \left[\begin{array}{cccc}
		\theta & \gamma_{1} & \cdots & \gamma_m
	\end{array}\right]$, is rank 1 and $\theta \neq 0$, by Lemma \ref{lemma:det-linear-algebra}, 
	\begin{align*}
		s_{i}^{\sigma} & = \frac{1}{\theta} \det \left[\begin{array}{c|c}
			\multirow{ 2}{*}{$I-W_{(\sigma \cap \tau)\setminus\{i\}}$}  & \theta \\
			& \vdots \\
			\cline{1-1}\rotvert (I-W_{\sigma\cap\tau})_i \rotvert & \theta \\
		\end{array}\right] \det\left[\begin{array}{c|ccc}
			\theta & \gamma_1 & \cdots & \gamma_n\\
			\hline
			\theta & & & \\
			\vdots & & I-W_{\sigma\cap\omega}& \\
			\theta & & & \\
		\end{array}\right] \\
		& =\frac{1}{\theta}  \det((I-W_{\sigma\cap\tau})_i;\theta\mathbbm{1}) \det((I-W_{(\sigma\cap\omega) \cup \{i\}})_i;\theta\mathbbm{1})	\\
		& = \frac{1}{\theta}s_i^{\sigma\cap\tau} s^{\sigma\cap\omega}_i
	\end{align*}
	note that the last matrix, equal to $\alpha = \frac{s_i^{\sigma\cap\omega}}{\theta}$, does not depend on $i$.	
\end{proof}

With the factorization, we can now see how the signs of the $s_i$'s are inherited to simply-embedded components:

\begin{lemma}\label{lemma:simply-embedded-inheritance}
	Suppose $\tau_1,\cdots,\tau_N$ is a partition of the vertices of a TLN $(W,\theta)$, with $\theta$ constant input. For any $\sigma \subseteq [n]$, let $\sigma_\ell \od \sigma \cap \tau_\ell$.  If for every $\ell\in[N]$ we have $\tau_\ell$ simply-embedded in $(W,\theta)$, then for any $\sigma_i \neq \emptyset$,
	$$\sgn s_i^\sigma = \sgn s^\sigma_j \quad \Leftrightarrow \quad \sgn s_i^{\sigma_\ell} = \sgn s_j^{\sigma_\ell}, \quad\text{for all $i,j \in \tau_\ell$}.$$
\end{lemma}
\begin{proof}
	Since each $\tau_\ell$ is simply-embedded in $(W,\theta)$, and input is assumed to be uniform across all components, by Theorem \ref{thm:simply-embedded-si-factorization} we have $s_i^\sigma = \alpha s_i^{\sigma_\ell}$ for all $i \in \tau_\ell$, where $\alpha = \frac{1}{\theta}s_i^{\sigma\setminus\tau_\ell}$ has the same value for every $i \in \tau_\ell$.
	Hence, for all $i,j \in \tau_\ell$, we have that $\sgn s_i^\sigma = \sgn s_j^\sigma$ if and only if  $\sgn \alpha s_i^{\sigma_\ell} = \sgn \alpha s_j^{\sigma_\ell}$  if and only if $\sgn s_i^{\sigma_\ell} = \sgn s_j^{\sigma_\ell}$.	
\end{proof}

With the inheritance, we can now start to see how the supports of the parts relate to supports of the whole. The next theorem generalizes Theorem 1.4 in \cite{Parmelee2022} for TLNs with a simply-embedded structure, but constant input. Now we know that a fixed point of the whole network must have come from supports of the component parts.

\begin{theorem}[TLN version of Theorem 1.4 in \cite{Parmelee2022}]\label{thm:menu} Let $(W,\theta)$ be composed of layers $\tau_1,\dots,\tau_N$, such that each layer $\tau_\ell$ is simply-embedded in the network, with $\theta$ uniform input. For any $\sigma \subseteq[n]$, let $\sigma_\ell \od \sigma \cap \tau_\ell$. Then $$\sigma \in \FP(W,\theta) \quad \Rightarrow \quad \sigma_\ell \in \FP(W_{\tau_\ell},\theta)\cup \{\emptyset\}~~\text{ for all } \ell \in [N]. $$
	In other words, every fixed point support of $(W,\theta)$ is a union of component fixed point supports $\sigma_\ell$, at most one per component.
\end{theorem} 

\begin{proof}
	For $\sigma \in \FP(W,\theta)$, we have
	$$\sgn s_i^\sigma = \sgn s_j^\sigma = -\sgn s_k^\sigma$$
	for any $i,j \in \sigma_\ell$ and $k \in \tau_\ell\setminus\sigma_\ell$, by Theorem~\ref{thm:sgn-conditions} (sign conditions). Then by Lemma \ref{lemma:simply-embedded-inheritance}, we see that whenever $\sigma_\ell \neq \emptyset$,
	$$\sgn s_i^{\sigma_\ell} = \sgn s_j^{\sigma_\ell} = -\sgn_k^{\sigma_\ell},$$  and so $\sigma_\ell$ satisfies the sign conditions in $(W_{\tau_\ell},\theta)$. Thus $\sigma_\ell \in \FP(W_{\tau_\ell},\theta)$ for every nonempty $\sigma_\ell$.
\end{proof}

The next lemma generalizes Lemma \ref{lemma:full-factorization-CTLN} to TLNs, the structure below is the analog of a strongly simply-embedded partition. Recall that this lemma assumed several simply-\textit{embeddednesses} to obtain a full factorization of the $s_i$'s. Also recall the sets of surviving ($S_\sigma = \FP(W|_\sigma,b|_\sigma) \cap \FP(W,b)$) and dying fixed points supports ($D_\sigma = \FP(W|_\sigma,b|_\sigma)\setminus S_\sigma$).

\begin{lemma}\label{lemma:full-factorization}
	Let $(W,\theta)$ be a TLN on $n$ nodes with uniform input. And suppose that for $j\in [n]$, if $j \in \tau_\ell$, we have $W_{ij} = \gamma_j$ for all $i \notin \tau_\ell$. That is, if $\tau_j = \{t_{j-1}+1,\dots,t_j\}$, $W$ has the form
	\begin{align}
		W = \left[\begin{array}{ccc|ccc|c|ccc} 
			&  &  & \vert &  & \vert &  & \vert &  & \vert\\
			& W_{\tau_1} &  & \gamma_{{t_1}+1} & \cdots & \gamma_{t_2} & \cdots & \gamma_{{t_{N-1}}+1} & \cdots & \gamma_{t_N}\\
			&  &  & \vert &  & \vert &  & \vert &  & \vert\\
			\hline
			\vert &  & \vert &  &  &  &  & \vert &  & \vert\\
			\gamma_{1} & \cdots & \gamma_{t_1} &  & W_{\tau_2} &  & \cdots & \gamma_{{t_{N-1}}+1} & \cdots & \gamma_{t_N}\\
			\vert &  & \vert &  &  &  &  & \vert &  & \vert\\
			\hline
			& \vdots &  &  & \vdots &  & \ddots &  & \vdots & \\
			\hline
			\vert &  & \vert & \vert &  & \vert &  &  &  & \\
			\gamma_{1} & \cdots & \gamma_{t_1} & \gamma_{{t_1}+1} & \cdots & \gamma_{t_2} & \cdots &  & W_{\tau_N} & \\
			\vert &  & \vert & \vert &  & \vert &  &  &  & \\
		\end{array}\right]
	\end{align}
	For any $\sigma \subseteq [n]$, denote $\sigma_\ell \od \sigma \cap \tau_\ell$, and $\sigma_{\ell_1\dots \ell_k} \od \sigma_{\ell_1} \cup \dots \cup \sigma_{\ell_k}$ and let $I = \{\ell \in [N]~|~ \sigma_\ell \neq \emptyset\}$. Then for every $i\in [n]$, $$s_i^\sigma = \frac{1}{\theta^{|I|-1}}\prod_{\ell\in I} s_i^{\sigma_\ell},$$
	where $s_i^{\sigma_\ell}$ has the same value for every $i \in [n]\setminus \tau_\ell$. Moreover, for any $\sigma_\ell \in \FP(G|_{\tau_\ell})$ and $j \in \tau_\ell$:
	$$\sgn s_j^{\sigma_\ell} = \begin{cases}
		\phantom{-}\idx(\sigma_\ell) & \text{if } j \in \sigma_\ell \\
		-\idx(\sigma_\ell) & \text{if } j \in \tau_\ell\setminus \sigma_\ell
	\end{cases}$$
	while for any $k\notin \tau_\ell$,
	$$\sgn s_k^{\sigma_\ell} = \begin{cases}
		-\idx(\sigma_\ell) & \text{if } \sigma_\ell \in S_{\tau_\ell}\\
		\phantom{-}\idx(\sigma_\ell) & \text{if } \sigma_\ell \in D_{\tau_\ell}
	\end{cases}$$	
\end{lemma}

\begin{proof}
	Since $\tau_1$ is simply-embedded in $[n]\setminus\tau_1$, $$s_i^\sigma = \frac{1}{\theta} s_i^{\sigma_{2\dots N}}s_i^{\sigma_1} \text{ for all } i \in \tau_1$$
	by Theorem~\ref{thm:simply-embedded-si-factorization}.  On the other hand, since $[n]\setminus\tau_1$ is also simply-embedded, we also have
	$$s_i^\sigma = \frac{1}{\theta} s_i^{\sigma_1} s_i^{\sigma_{2\dots N}} \text{ for all } i \in [n]\setminus\tau_1.$$
	Therefore, the above factorization holds for all $i \in [n]$. Similarly, since both $\tau_2$ and $[n]\setminus\tau_2$ are simply-embedded, 
	$$s_i^{\sigma_{2\dots N}}= \frac{1}{\theta} s_i^{\sigma_2} s_i^{\sigma_{3\dots N}} \text{ for all } i \in [n]$$
	by Theorem~\ref{thm:simply-embedded-si-factorization}, and so $s_i^\sigma = \frac{1}{\theta^2} s_i^{\sigma_1} s_i^{\sigma_2} s_i^{\sigma_{3\dots N}}$.  Continuing in this fashion, we see that for any $i \in [n]$,
	$$s_i^{\sigma} =\frac{1}{\theta^{N-1}} s_i^{\sigma_1}\dots s_i^{\sigma_N}.$$
	Note that if $\sigma_\ell = \emptyset$, then $s_i^{\sigma_\ell} = s_i^{\emptyset} = s_i^{\{j\}} = \theta$, and thus for all $i \in [n]$,
	$$s_i^{\sigma} = \frac{\theta^{N-|I|}}{\theta^{N-1}}\prod_{i \in I}s_i^{\sigma_\ell} = \frac{1}{\theta^{|I|-1}}\prod_{i\in I} s_i^{\sigma_\ell}.$$
	The fact that $s_i^{\sigma_\ell}$ has the same value for every $i \in [n]\setminus \tau_\ell$ is a direct consequence of Theorem \ref{thm:simply-embedded-si-factorization}.
	
	Finally, to prove the last statements about the signs of $s_i^{\sigma_\ell}$, observe that for $i \in \tau_\ell$, the values of $\sgn s_i^{\sigma_\ell}$ are fully determined by Theorem \ref{thm:sgn-conditions} (sign conditions) since $\sigma_\ell \in \FP(G|_{\tau_\ell})$ by hypothesis.   In particular, if $\sigma_\ell \in S_{\tau_\ell}$, then $\sigma_\ell$ survives the addition of every $k \notin \tau_\ell$, and so $\sgn s_k^{\sigma_\ell} = -\idx(\sigma_\ell)$ by Theorem \ref{thm:sgn-conditions} (sign conditions).  On the other hand, if $\sigma_\ell \in D_{\tau_\ell}$ then $\sigma_\ell$ dies in $G$ and so there is some $k \notin \tau_\ell$ for which $\sgn s_k^{\sigma_\ell} = \idx(\sigma_\ell)$.  But by the first part of the theorem, all the $s_k^{\sigma_\ell}$ values are identical for $k \in [n] \setminus \tau_\ell$, and thus $\sgn s_k^{\sigma_\ell} = \idx(\sigma_\ell)$ for all such $k$.
\end{proof}

Finally, we arrive to the main theorem of last section, Theorem \ref{thm:bidir-sa-partition}, TLN version. This is the big result. It generalizes Theorem \ref{thm:bidir-sa-partition} from \cite{Parmelee2022}, and also Theorem \ref{thm:bidir-sa} from \cite{fp-paper}. The proof is a generalized version of what it was in both of those publications. 

\begin{theorem}\label{thm:strongly-simply-layered}
	Let $(W,b)$ be such that \begin{align}
		W = \left[\begin{array}{ccc|ccc|c|ccc} 
			&  &  & \vert &  & \vert &  & \vert &  & \vert\\
			& W_{\tau_1} &  & \gamma_{{t_1}+1} & \cdots & \gamma_{t_2} & \cdots & \gamma_{{t_{N-1}}+1} & \cdots & \gamma_{t_N}\\
			&  &  & \vert &  & \vert &  & \vert &  & \vert\\
			\hline
			\vert &  & \vert &  &  &  &  & \vert &  & \vert\\
			\gamma_{1} & \cdots & \gamma_{t_1} &  & W_{\tau_2} &  & \cdots & \gamma_{{t_{N-1}}+1} & \cdots & \gamma_{t_N}\\
			\vert &  & \vert &  &  &  &  & \vert &  & \vert\\
			\hline
			& \vdots &  &  & \vdots &  & \ddots &  & \vdots & \\
			\hline
			\vert &  & \vert & \vert &  & \vert &  &  &  & \\
			\gamma_{1} & \cdots & \gamma_{t_1} & \gamma_{{t_1}+1} & \cdots & \gamma_{t_2} & \cdots &  & W_{\tau_N} & \\
			\vert &  & \vert & \vert &  & \vert &  &  &  & \\
		\end{array}\right]
	\end{align} and $b = \theta\mathbbm{1}$. Let  $\sigma_i \od \sigma \cap \tau_i$ for any $\sigma \subseteq [n]$.  Then $\sigma \in \FP(W,\theta)$ if and only if $\sigma_i \in \FP(W_{\tau_i},\theta) \cup \{\emptyset\}$ for each $i \in [N]$, and either
	\begin{enumerate}
		\item[{\rm(a)}] every $\sigma_i$ is in $\FP(W,\theta) \cup \{ \emptyset\}$, or
		\item[{\rm(b)}] none of the $\sigma_i$ are in $\FP(W,\theta) \cup \{ \emptyset\}$.
	\end{enumerate}
	In other words, $\sigma \in \FP(W,\theta)$ if and only if $\sigma$ is either a union of surviving fixed points $\sigma_i$, at most one per component, or it is a union of dying fixed points, exactly one from every component.
\end{theorem}

\begin{proof}
	Note that $(W,\theta)$ satisfies the conditions for Lemma \ref{lemma:full-factorization} above, and thus, assuming without loss of generality that $\theta=1$, we have that for $I \od \{i ~|~ \sigma_i \neq \emptyset \}$ and $\sigma\subseteq[n]$, $$s_j^\sigma = \prod_{i \in I} s_j^{\sigma_i},$$ with $s_j^{\sigma_i}$ is constant across $j \in [n]\setminus\tau_i$ for each $i \in [N]$.
	
	($\Rightarrow$) Suppose $\sigma \in \FP(W,\theta).$ By Lemma \ref{lemma:full-factorization}, $\sgn s_i^{\sigma} = \prod_{i \in I} \sgn s_j^{\sigma_i}$.
	\begin{figure}[!h]
		\centering
		\includegraphics[width=0.85\textwidth]{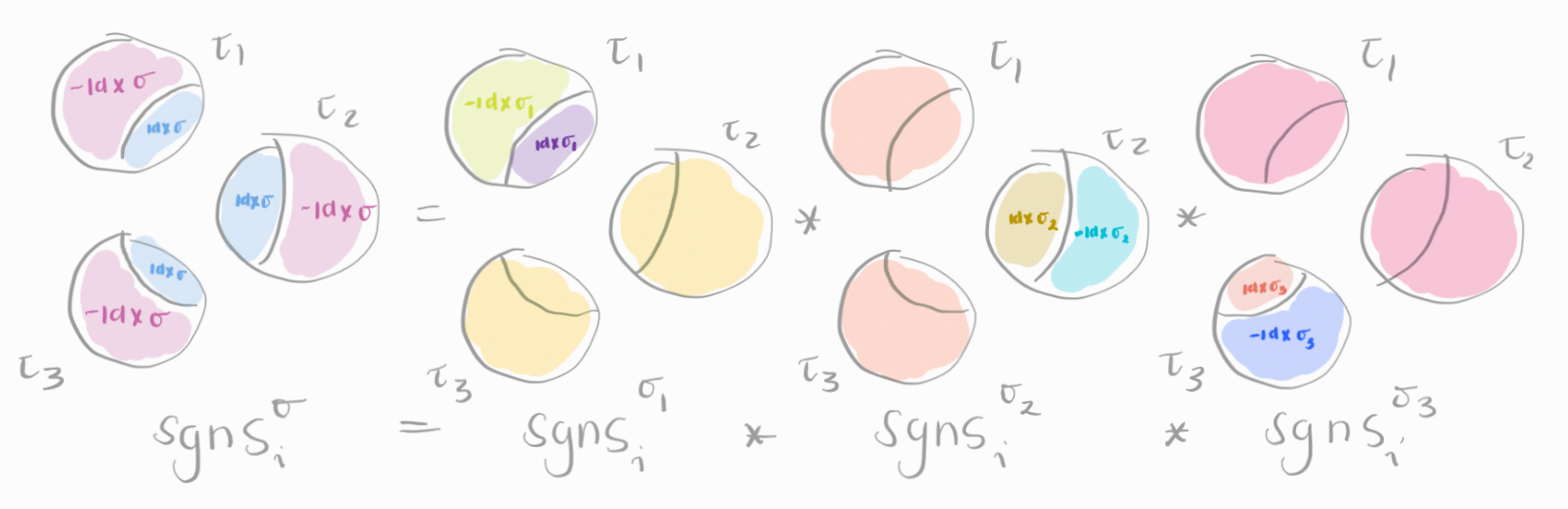}
	\end{figure}
	Next, by Theorem \ref{thm:menu}, $\sigma_i \in \FP(W_{\tau_i},\theta)$ for every $i \in I$, and by the $\sgn$ formula of Lemma \ref{lemma:full-factorization}. Denote by $\S \od \{a\in I~|~\sigma_a \in S_a\}$. For any $j \in \sigma$, there exists $i \in I$ such that $j \in \sigma_i$, then we have
	\begin{equation}\label{eq:sgn-s_j}
		\sgn s_j^\sigma = \idx(\sigma_i) \prod_{a \in \S\setminus\{i\}} -\idx(\sigma_a)  \prod_{b \in \S^c\setminus\{i\}} \idx(\sigma_b) = (-1)^{|\S \setminus \{i\}|}\prod_{\ell\in I} \idx(\sigma_\ell),
	\end{equation}
	
	Now, observe that if $\sigma$ contained a mix of $\sigma_a \in S_a$ and $\sigma_b \in D_b$, then there would be $i, j \in \sigma$ such that $i \in \sigma_a$ for some $a\in \S$, while $j \in \sigma_b$ for some $b \notin \S$ and thus $\sgn s_i^{\sigma} = (-1)^{|S|-1} \prod_{\ell \in I} \idx \sigma_\ell$ and  $\sgn s_j^{\sigma} = (-1)^{|S|} \prod_{\ell \in I} \idx \sigma_\ell = -\sgn s_i^{\sigma}$. But this contradicts the fact that $\sigma \in \FP(W,\theta)$ by Theorem \ref{thm:sgn-conditions} (since we assumed $i,j \in \sigma$). Thus, we must have either $\sigma_i \in S_{\tau_i}$ for all $i \in I$, as in (a), or $\sigma_i \in D_{\tau_i}$ for all $i \in I$ as in (b).
	
	Now to see that when $\sigma_i \in D_{\tau_i}$ for all $i \in I$, we must have $I = [N]$, suppose to the contrary that $I \subsetneq [N]$ so that there is some $m \in [N]$ such that $\tau_m \cap \sigma = \emptyset$. Then, for $k\in\tau_m$ (so $k \notin \sigma$), we have $\sgn s_k^{\sigma_\ell} = \idx(\sigma_\ell)$ for all $\ell \in I$, by Lemma~\ref{lemma:full-factorization}, since $\sigma_\ell \in D_{\tau_\ell}$.  Thus
	$\sgn s_k^\sigma = \prod_{\ell \in I} \idx(\sigma_\ell)$.
	
	On the other hand, if $j \in \sigma$, with $j \in \tau_i$, we have $$\sgn s_j^{\sigma} =(-1)^{|\S \setminus \{i\}|}\prod_{\ell\in I} \idx(\sigma_\ell) = \prod_{\ell\in I} \idx(\sigma_\ell).$$ Where the first equality follows from Equation \ref{eq:sgn-s_j}, and the second because $\S = \emptyset$ in this case.
	\begin{figure}[!h]
		\centering
		\includegraphics[width=0.85\textwidth]{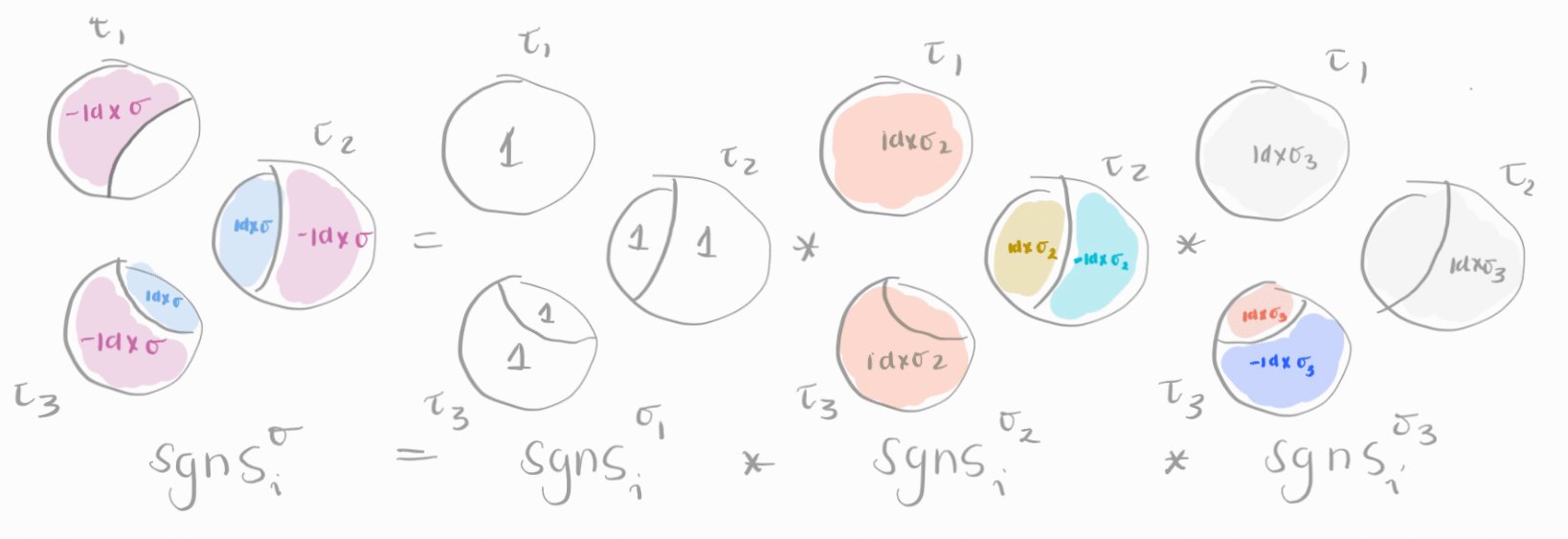}
	\end{figure}
	
	But then $\sgn_k^\sigma = \sgn_j^\sigma$ for $k \notin \sigma$ and $j \in \sigma$, which contradicts $\sigma\in\FP(W,\theta)$ by Theorem \ref{thm:sgn-conditions}. Thus $I=[N]$ when $\sigma_\ell \in D_\ell$ for all $\ell \in I$.
	
	($\Leftarrow$) Suppose (a) holds and so $\sigma_i \in S_{\tau_i}$ for all $i \in I$. Let us check the sign conditions for $\sigma \od \bigcup_{i \in I} \sigma_i$. 
	
	For any $j \in \sigma$, there exists $i \in I$ such that $j \in \tau_i$.  Then by Equation \ref{eq:sgn-s_j}, we have
	$$\sgn s_j^{\sigma} = (-1)^{|\S \setminus \{i\}|}\prod_{\ell\in I} \idx(\sigma_\ell) = (-1)^{|I|-1}\prod_{\ell\in I}\idx\sigma_\ell,$$
	since $\S = I$ in this case.
	\begin{figure}[!h]
		\centering
		\includegraphics[width=0.85\textwidth]{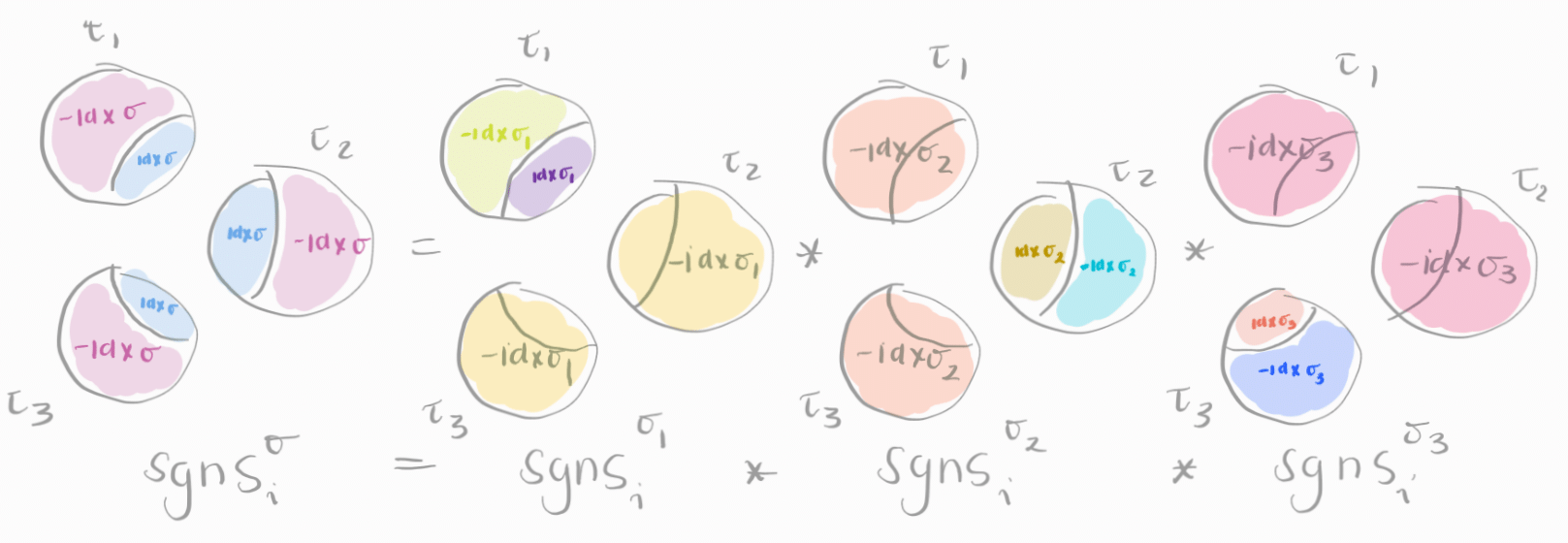}
	\end{figure}
	On the other hand, for $k \notin \sigma$, we have $\sgn s_k^{\sigma_\ell} = -\idx \sigma_\ell$ for all $\ell \in I$, by Lemma~\ref{lemma:full-factorization}, since $\sigma_\ell \in S_{\tau_\ell}$.  Thus
	$$\sgn s_k^{\sigma} = \prod_{\ell\in I}(-\idx\sigma_\ell) = (-1)^{|I|}\prod_{\ell\in I}\idx\sigma_\ell = -\sgn s_j^\sigma.$$
	Therefore $\sigma \in \FP(W,\theta)$ by Theorem \ref{thm:sgn-conditions} (sign conditions).
	
	Next, suppose (b) holds so $\sigma_\ell \in D_{\tau_\ell}$ for all $\ell \in [N]$ (so $I = [N]$). Then for any $j \in \sigma$, there is $i \in [N]$ such that $j \in \sigma_i$ and by Equation \ref{eq:sgn-s_j}, we have
	$$\sgn s_j^{\sigma}=(-1)^{|\S \setminus \{i\}|}\prod_{\ell\in [N]} \idx(\sigma_\ell) = \prod_{\ell\in [N]} \idx(\sigma_\ell),$$
	since $\S=\emptyset$.
	
	Now, let $k \notin \sigma$, with $k \in \tau_m$. Since $\sigma_m \in \FP(W_{\tau_m},\theta)$ with $\sigma_m \neq \emptyset$ (because $I = [N]$), we have $\sgn s_k^{\sigma_m} = -\idx(\sigma_m)$ and thus
	$$\sgn s_k^{\sigma}=\sgn s_k^{\sigma_m} \hspace{-.15in}\prod_{\ell\in [N]\setminus\{m\}}  \hspace{-.15in}\sgn s_k^{\sigma_\ell} = -\idx(\sigma_m)  \hspace{-.15in}\prod_{\ell\in [N]\setminus\{m\}}  \hspace{-.15in}\idx(\sigma_\ell) = -\prod_{\ell\in [N]} \idx(\sigma_\ell) = -\sgn s_j^\sigma.$$
	Thus sign conditions are satisfied, and so $\sigma \in \FP(W,\theta)$.
\end{proof}

Note that the structure of the network in Theorem \ref{thm:strongly-simply-layered} is such that the graph can be decomposed as shown in Figure \ref{fig:simplyembeddedlayers}B. In contrast, in Figure \ref{fig:simplyembeddedlayers}A we have the less restrictive structure of Theorem \ref{thm:menu}, where the edges to two different components might be different.

\begin{figure}[!h]
	\begin{center}
		\includegraphics[width=0.6\textwidth]{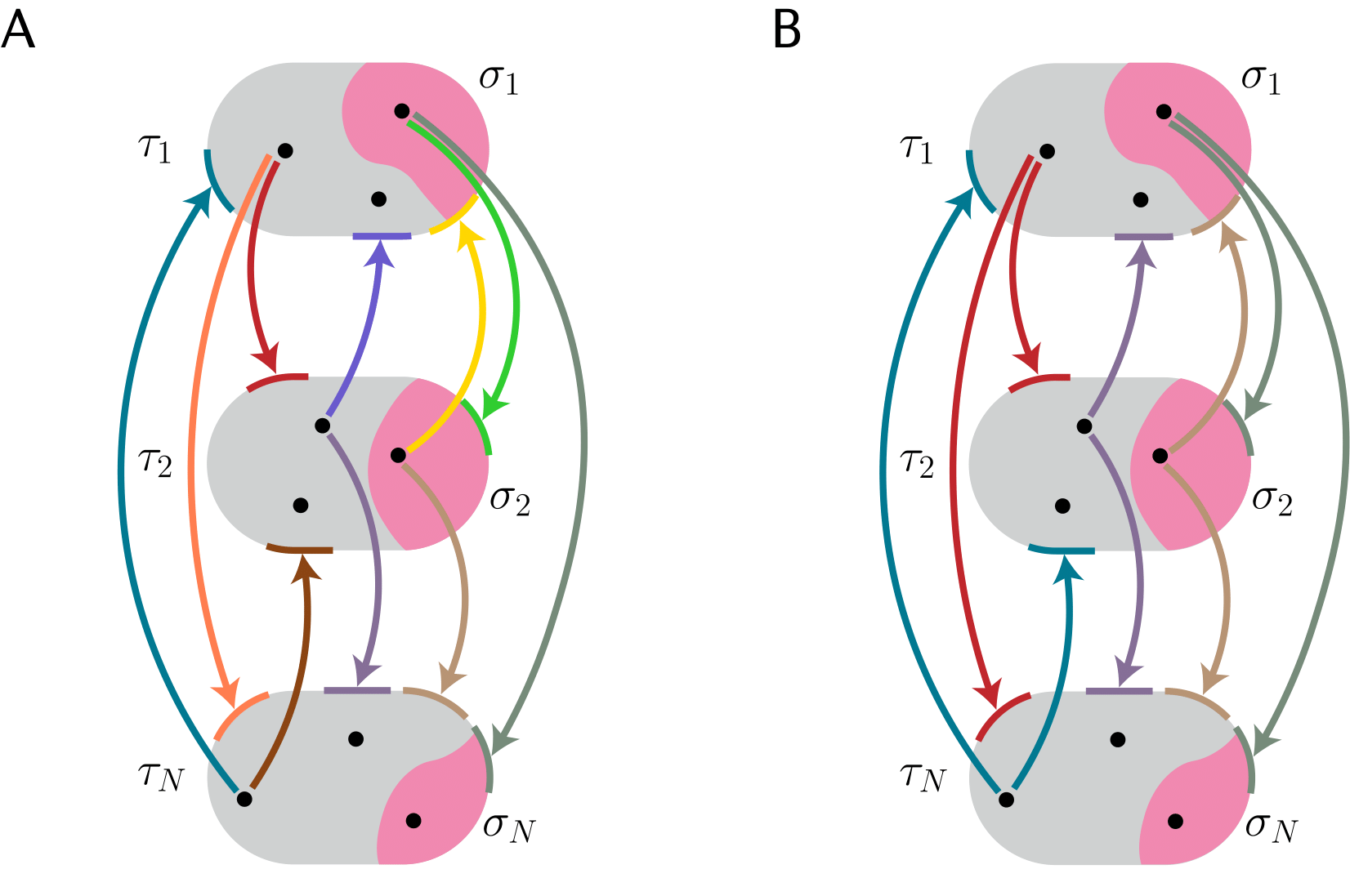}
	\end{center}
	\caption[Simply-embedded layers cartoon]{\textbf{Simply-embedded layers cartoon}. 
	(A) The structure of Theorem \ref{thm:menu}.
	(B) The structure of Theorem \ref{thm:strongly-simply-layered}.}
	\label{fig:simplyembeddedlayers}
\end{figure}

We can also consider a mid point between these two structures, where nested components are simply-embedded, to obtain:

\begin{theorem}\label{thm:simply-embedded-nested}
Let $(W,\theta)$ such that $\rho_k = \bigcup_{j=1}^{k} \tau_j$ is simply-embedded in $(W,\theta)$ for every $k = 1,\dots,N$. That is, if $\tau_j = \{t_{j-1}+1,\dots,t_j\}$ then $W$ has the form 
\begin{align}
	\left(\begin{array}{c|ccc|c|ccc}
		& \vert &  &  \vert & &\vert &  & \vert \\
		W_{\tau_1} & \gamma_{t_1+1} & \cdots & \gamma_{t_2} & \cdots & \gamma_{t_{N-1}+1} & \cdots & \gamma_{t_N} \\
		& \vert &  & \vert & &\vert &  & \vert \\
		\hline
		&  &  &  & &\vert &  & \vert \\
		* &  & W_{\tau_2} &  & \cdots &  \gamma_{t_{N-1}+1} & \cdots & \gamma_{t_N} \\
		& &  &  & &\vert &  & \vert \\
		\hline
		\vdots & & \vdots&  & \ddots &  &\vdots & \\
		\hline
		* & & * &  & * & &W_{\tau_N} &
	\end{array}\right).
\end{align} Then for any $\sigma \subseteq [n]$ and any $k \in [N]$ we have $$\sigma\in\FP(W,\theta) \Rightarrow \sigma \cap \rho_k \in \FP(W_{\rho_k},\theta) \cup \{\emptyset\}$$ where $\rho_k = \bigcup_{j=1}^{k} \tau_j$.
\end{theorem}
This means that the fixed point supports of the whole network must come from supports of ``ordered, nested'' parts.
\begin{proof}
Since $\sigma\in\FP(W,\theta)$, by Theorem \ref{thm:sgn-conditions} we have that $$\forall j,m \in \sigma \forall l \notin \sigma \quad \sgn s_j^\sigma = \sgn s_m^\sigma = -\sgn s_l^\sigma.$$ By Lemma \ref{lemma:simply-embedded-inheritance}, since for every $k$, $\rho_k$ is simply-embedded, we have $$\forall j,m \in \rho_k \forall l \notin \rho_k-\sigma \quad \sgn s_j^{\sigma\cap\rho_k} = \sgn s_m^{\sigma\cap\rho_k} = -\sgn s_l^{\sigma\cap\rho_k}.$$ This means that $\sigma\cap\rho_k$ satisfies the sign conditions in  $\rho_k$ and so $\sigma \cap \rho_k \in \FP(W_{\rho_k},\theta) \cup \{\emptyset\}$.
\end{proof}

\begin{corollary}
Let $\tau$ be simply-embedded in $(W,\theta)$. Then for any $\sigma \subseteq [n]$,
$$\sigma\in\FP(W,\theta) \Rightarrow \sigma \cap \tau \in \FP(W_{\tau},\theta) \cup \{\emptyset\}.$$ 
\end{corollary}

\subsection{Updated results on simply-embedded partitions}\label{sec:updated-SIADs-results}

As mentioned before, the results in this section generalized those published in \cite{Parmelee2022}, for which we omitted the proofs when we presented them in Section \ref{sec:simply-embedded-CTLNs}. Here we include references to which theorems from Section \ref{sec:layered-ctlns} will prove theorems of Section \ref{sec:simply-embedded-CTLNs}. 

First, note that Definition \ref{def:simply-embedded} is generalizing Definition \ref{def:sa-partition} now. That's how we straightforwardly get:

\begingroup
\def\thetheorem{\ref{lemma:simply-added}}
\begin{lemma}
Let $G$ have a simply-embedded partition $\{\tau_1|\cdots|\tau_N\}$, and consider $\sigma \subseteq [n]$. Let $\sigma_i \od \sigma \cap \tau_i$.  Then for any $\sigma_i \neq \emptyset$,
$$\sgn s_j^\sigma = \sgn s^\sigma_k \quad \Leftrightarrow \quad \sgn s_j^{\sigma_i} = \sgn s_k^{\sigma_i}, \quad\text{for all $j,k \in \tau_i$}.$$
\end{lemma}
\addtocounter{theorem}{-1}
\endgroup
\begin{proof}
Follows directly from Lemma \ref{lemma:simply-embedded-inheritance}.
\end{proof}

\begingroup
\def\thetheorem{\ref{thm:menu-CTLN}}
\begin{theorem}[$\FP(G)$ menu for simply-embedded partitions]
Let $G$ have a simply-embedded partition $\{\tau_1|\cdots|\tau_N\}$.  For any $\sigma \subseteq [n]$, let $\sigma_i \od \sigma \cap \tau_i$.  Then 
$$\sigma \in \FP(G) \quad \Rightarrow \quad \sigma_i \in \FP(G|_{\tau_i})\cup \{\emptyset\}~~\text{ for all } i \in [N]. $$ 
In other words, every fixed point support of $G$ is a union of component fixed point supports $\sigma_i$, at most one per component.
\end{theorem}
\addtocounter{theorem}{-1}
\endgroup
\begin{proof}
Follows directly from Theorem \ref{thm:menu}.
\end{proof}

\begingroup
\def\thetheorem{\ref{lemma:full-factorization-CTLN}}
\begin{lemma}
Let $G$ be a graph on $n$ nodes with a strongly simply-embedded partition $\{\tau_1|\dots|\tau_N\}$. For any $\sigma \subseteq [n]$, denote $\sigma_i \od \sigma \cap \tau_i$, and $\sigma_{i_1\dotsi_k} \od \sigma_{i_1} \cup \dots \cup \sigma_{i_k}$ and let $I = \{i \in [N]~|~ \sigma_i \neq \emptyset\}$. Then for every $j\in [n]$, $$s_j^\sigma = \frac{1}{\theta^{|I|-1}}\prod_{i\in I} s_j^{\sigma_i},$$ 
where $s_j^{\sigma_i}$ has the same value for every $j \in [n]\setminus \tau_i$.\\
Moreover, for any $\sigma_i \in \FP(G|_{\tau_i})$ and $j \in \tau_i$:
$$\sgn s_j^{\sigma_i} = \begin{cases} 
	\phantom{-}\idx(\sigma_i) & \text{if } j \in \sigma_i \\
	-\idx(\sigma_i) & \text{if } j \in \tau_i\setminus \sigma_i
\end{cases}$$ 
while for any $k\notin \tau_i$, 
$$\sgn s_k^{\sigma_i} = \begin{cases} 
	-\idx(\sigma_i) & \text{if } \sigma_i \in S_{\tau_i}\\
	\phantom{-}\idx(\sigma_i) & \text{if } \sigma_i \in D_{\tau_i}
\end{cases}$$
\end{lemma}
\addtocounter{theorem}{-1}
\endgroup
\begin{proof}
Follows directly from Lemma \ref{lemma:full-factorization}.
\end{proof}

\begingroup
\def\thetheorem{\ref{thm:bidir-sa-partition}}
\begin{theorem}
Suppose $G$ has a strongly simply-embedded partition $\{\tau_1|\dots|\tau_N\}$, and let $\sigma_i \od \sigma \cap \tau_i$ for any $\sigma \subseteq [n]$.  
Then $\sigma \in \FP(G)$ if and only if $\sigma_i \in \FP(G|_{\tau_i}) \cup \{\emptyset\}$ for each $i \in [N]$, and either
\begin{enumerate}
	\item[(a)] every $\sigma_i$ is in $\FP(G) \cup \{ \emptyset\}$, or 
	\item[(b)] none of the $\sigma_i$ are in $\FP(G) \cup \{ \emptyset\}$.
\end{enumerate}
In other words, $\sigma \in \FP(G)$ if and only if $\sigma$ is either a union of surviving fixed points $\sigma_i$, at most one per component, or it is a union of dying fixed points, exactly one from every component.
\end{theorem}
\addtocounter{theorem}{-1}
\endgroup
\begin{proof}
Follows directly from Theorem \ref{thm:strongly-simply-layered}.
\end{proof}



\chapter{Further theoretical results, and open questions}\label{ch:open-questions}

In this chapter, we explore two seemingly unrelated results originating from our work in Chapters \ref{ch:CPGs} and \ref{ch:new-theoretical-results}. The first part of this chapter emerges from our attempt to extend our five-gait quadruped network approach to model hexapod gaits. We wondered whether CLTNs could learn hexapod gaits using a combination of theoretical insights from CTLNs and computational machine learning techniques. While we did not ultimately answer this question, during our exploration, we made an interesting observation: many different networks could support the same attractor. This observation isn't entirely novel \cite{Goaillard2009, marder2021, Kamaleddin2022-rx}, which prompted us to study this phenomenon using CTLNs from a dynamical systems perspective. What conditions are sufficient for two different networks to support the same attractor? Our progress in addressing this question is discussed in Section \ref{sec:degeneracy}.

The second half of the chapter derived from Chapter \ref{ch:new-theoretical-results}. Recall that in that chapter, we repeatedly invoked the ability to fully factorize the $s_i$ values from Equation \ref{eq:s_i} when using a simply-embedded structure. This has been the case in 2016 \cite{fp-paper}, 2020 \cite{Parmelee2022}, and 2023 \ref{sec:layered-ctlns}. Now we extend this approach to even more general structures. As it turns out, these $s_i$ values are elements of a \emph{chirotope}. Section \ref{sec:chirotope} explores the concept of chirotopes, their relation to TLNs, and partially computes the chirotope of a TLN that has some simply-embedded subnetworks.

These projects are still in their early stages, which explains their inclusion, together, in the final chapter.

\section{Degeneracy}\label{sec:degeneracy}

One of the aims of neuroscience is to identify the relationship between function and the underlying structure responsible for that function. However, a wide variety of different structures can produce the same function, a concept known as \emph{degeneracy} \cite{Kamaleddin2022-rx}. Degeneracy manifests in various contexts in neuroscience, ranging from single cells with different properties generating the same behavior \cite{marder2021} to different neuronal circuits producing similar circuit performance \cite{Goaillard2009}. That is, there is not a unique, one-to-one correspondence between function and structure. This holds true for models as well. A myriad of different parameters in a single model can yield the same output \cite{Prinz2004,Gjorgjieva2016-hl}. What fraction of neurons need to be observed to reconstruct a network's behavior? 

In the context of TLNs, the degeneracy problem translates into: for a given pattern of firing $x(t)$, how many TLNs $(W,b)$ can generate that pattern? Furthermore, is there a precise notion under which two distinct networks $(W^{(0)},b^{(0)})$ and $(W^{(1)},b^{(1)})$ are considered equivalent? Can we derive a general principle that yields a family of "equivalent" networks? The degeneracy problem also raises the question: if several different networks can reproduce the same attractor, how do we identify conditions for accurate model fitting or parameter estimation? This would also require a precise notion under which two attractors are the same, and then conditions that facilitate this verification.

In the special case of TLNs, we have observed that even networks as small as 3 neurons can reproduce the same limit cycle, under different parameters. Figure \ref{fig:degenerateTLNs} shows an example. On top, there is a binary-synapse matrix $W^{(0)}$, arising as the CTLN connectivity matrix of a 3-cycle graph, and an all uniform input $b^{(0)}$ defining a network that supports a limit cycle, as shown by the rate curves. Below, a slightly different matrix $W^{(1)}$ and a non-uniform input $b^{(1)}$ is reproducing (apparently) the same limit cycle. 
\begin{figure}[h!]
    \centering
    \includegraphics[width=\textwidth]{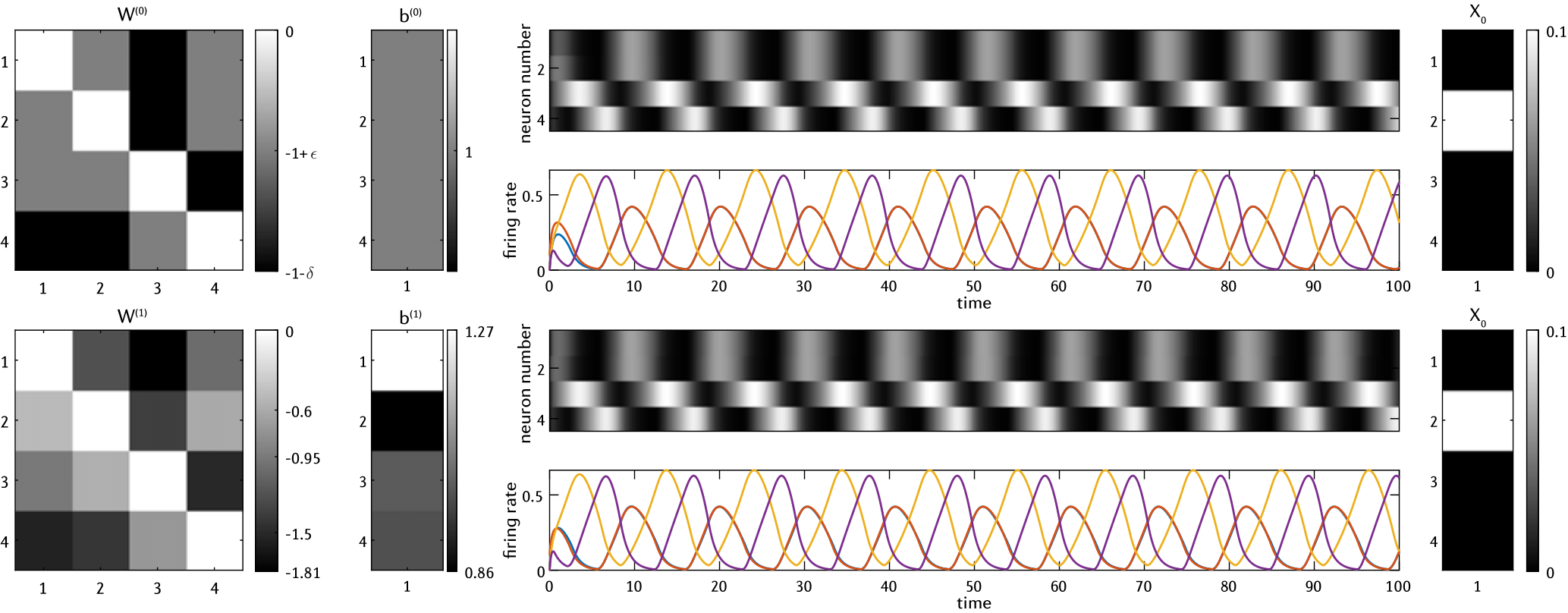}
    \caption[Two different TLNs reproducing the same limit cycle]{\textbf{Two different TLNs reproducing the same limit cycle}.} 
    \label{fig:degenerateTLNs}
\end{figure}

One partial result in this direction states that if the vector fields of two TLNs match at some point, then they match at that point for all TLNs that are convex combinations of the original two TLNs. More precisely: 

\begin{definition}
	Let $(W^{(0)}, b^{(0)})$ and $(W^{(1)}, b^{(1)})$ be two TLNs on $n$ nodes, with vector fields
	$$\frac{d x}{d t}=v^{(a)}(x) \stackrel{\text { def }}{=}-x+\left[W^{(a)} x+b^{(a)}\right]_{+}$$
	for $a=0,1$. For any $s \in[0,1]$, define the \textit{convex combination} $(W^{(s)}, b^{(s)})$ as the TLN with
	$$W^{(s)}=(1-s) W^{(0)}+s W^{(1)} \text { and } b^{(s)}=(1-s) b^{(0)}+s b^{(1)}.$$
\end{definition}

\begin{lemma}\label{lemma:vector-field-interpolation}
Let $v^{(s)}(x)=-x+\left[W^{(s)} x+b^{(s)}\right]_{+}$ be the vector field of a TLN convex combination $(W^{(s)}, b^{(s)})$, for $s \in[0,1]$. Suppose there exists a certain point $x_0 \in \mathbb{R}^n$ such that the vector fields match: $
v^{(1)}\left(x_0\right)=v^{(0)}\left(x_0\right)$. Then, for all $s \in[0,1], v^{(s)}\left(x_0\right)=v^{(0)}\left(x_0\right)$.
\end{lemma}

\begin{proof}
First, observe that if $\left[y_{1}\right]_{+}=\left[y_{2}\right]_{+}$, then for any $c_{1}, c_{2}>0$ we also have  $$\left[c_{1} y_{1}+c_{2} y_{2}\right]_{+}=c_{1}\left[y_{1}\right]_{+}+c_{2}\left[y_{2}\right]_{+}.$$ Moreover, this will hold for $y_{1}, y_{2} \in \mathbb{R}^{n}$, where some entries are positive and others are negative, and $y_{1} \neq y_{2}$ (as negative entries need not agree). Now suppose $v^{(1)}\left(x_{0}\right)=v^{(0)}\left(x_{0}\right)$. It follows that $$\left[W^{(1)} x_{0}+b^{(1)}\right]_{+}=\left[W^{(0)} x_{0}+b^{(0)}\right]_{+}.$$ and hence, for $s \in[0,1]$,
\begin{align*}
	{\left[W^{(s)} x_{0}+b^{(s)}\right]_{+} } & =\left[(1-s) W^{(0)} x_{0}+s W^{(1)} x_{0}+(1-s) b^{(0)}+s b^{(1)}\right]_{+}\\
	& =\left[(1-s)\left(W^{(0)} x_{0}+b^{(0)}\right)+s\left(W^{(1)} x_{0}+b^{(1)}\right)\right]_{+}\\
	& =(1-s)\left[W^{(0)} x_{0}+b^{(0)}\right]_{+}+s\left[W^{(1)} x_{0}+b^{(1)}\right]_{+}\\
	& =\left[W^{(0)} x_{0}+b^{(0)}\right]_{+}.
\end{align*}
This implies that $v^{(s)}\left(x_{0}\right)=-x_{0}+\left[W^{(s)} x_{0}+b^{(s)}\right]_{+}=-x_{0}+\left[W^{(0)} x_{0}+b^{(0)}\right]_{+}=v^{(0)}\left(x_{0}\right)$.
\end{proof}

This implies in particular that fixed points survive to interpolations in convex combinations:

\begin{corollary}\label{cor:fixed-point-interpolation}
If $x^{*}$ is a fixed point of both $(W^{(0)}, b^{(0)})$ and $(W^{(1)}, b^{(1)})$, then it is a fixed point  for all convex combinations $(W^{(s)}, b^{(s)})$.
\end{corollary}
\begin{proof}
If $v^{(1)}\left(x^{*}\right)=v^{(0)}\left(x^{*}\right)=0$, then $v^{(s)}\left(x^{*}\right)=0$ for all $s \in[0,1]$.
\end{proof}

Also, any attractor that appears identically in two distinct TLNs - that is, with  identical vector field along all point in the attractor - must also appear in all ``in between'' TLNs formed as convex combinations of the original two.

\begin{corollary}\label{cor:attractor-interpolation} If $x(t)$, for $t \in\left(t_{0}, t_{1}\right)$, is a trajectory of both $(W^{(0)}, b^{(0)})$ and $(W^{(1)}, b^{(1)})$, then  it is also a trajectory for all convex combinations $(W^{(s)}, b^{(s)})$.
\end{corollary}

Several questions arise from our exploration of attractor degeneracy in the context of TLNs. First, can any of these results be generalized to any convex hull, beyond line segments? If so, can we identify distinct "chunks" of TLN parameter space $(W,b)$ that yield the same attractor? If such chunks exist, can we systematically characterize them to obtain partitions of TLN space $(W,b)$ by attractors? We could begin tackling this from a computational perspective by developing algorithms to efficiently identify and characterize regions of parameter space that produce equivalent attractors.

Some answers could also help elucidate the relationship between structure and function. For example, are there specific structural properties or network configurations that lead to the emergence of these shared attractors? This is a challenging question, but computationally, we could at least attempt to answer how different network architectures impact the existence and distribution of shared attractors.

Further-reaching questions include: what implications do these findings have for our understanding of neural circuit function and plasticity? How might the presence of degenerate attractors influence information processing and computational capabilities within neural networks? Are there practical applications of understanding attractor degeneracy in the design of artificial neural networks? These questions go beyond the immediate scope of our current investigation but are worth exploring to gain deeper insights into neural network dynamics and their computational properties.

\section{Chirotope}\label{sec:chirotope}

\subsection{Background}

Given any set of vectors $E = \{z_1,\ldots,z_m\} \subseteq \RR^{d}$, the Grassmann-Plucker (G-P) relations give us a series of identities relating determinants formed from elements of $E$. Specifically, given any $x_1,\ldots,x_{d} \in E$ and $y_1,\ldots,y_{d} \in E$,
\begin{equation}\label{eq:G-P}
	\det(x_1,\ldots,x_{d})\det(y_1,\ldots,y_{d}) = 
	\sum_{\ell=1}^{d}\det(x_1,\ldots,x_{d-1},y_\ell)\det(y_1,\ldots,y_{\ell-1},x_{d},y_{\ell+1},\ldots,y_{d}).
\end{equation}

For different choices of $x_i$'s and $y_i$'s, Equation \ref{eq:G-P} yields all possible G-P relations on $E$. Note that if $x_d = y_j$ for some $j=1,\ldots,d$, then the G-P relation is trivial.

What do G-P relations tell us about TLNs? Recall that the rows of the $n \times (n+1)$ matrix $[-I+W \mid b]$ are the vectors
$$h_i = (W_{i1},W_{i2},\ldots,W_{i,i-1}, -1, W_{i,i+1},\ldots,W_{in},b_i).$$
For a TLN $(W,b)$ of dimension $n$, we are interested in the set of vectors
$$E = E(W,b) = \{h_1,\ldots,h_n,e_1,\ldots,e_{n+1}\} \subseteq \RR^{n+1},$$ 
corresponding to $d = n+1$. G-P relations for TLNs will relate determinants formed from these vectors, and this in turn will produce relationships among elements of the associated chirotope $\chi$:

\begin{definition}
	The {chirotope} $\chi = \chi_{(W,b)}$ of a TLN $(W,b)$ is the map $\chi:  E_{(W,b)}^{n+1} \to \{-1,0,1\}$, 
	where 
	$$\chi(v_1,\ldots,v_{n+1}) = \sgn \det(v_1,\dots,v_{n+1}),$$
	for any $v_1,\ldots,v_{n+1} \in E_{(W,b)}$. 
\end{definition}

The reason we are interested in the chirotope $\chi$ is that it encodes the full combinatorial geometry of a TLN. In particular, $\chi$ encodes all fixed point supports in $\FP(W,b)$. 

\begin{proposition}\label{prop:chirotope-FP}
	Let $(W,b)$ be a TLN on $n$ neurons, and let $\chi: E_{(W,b)}^{n+1} \to  \{-1,0,1\}$ be its associated chirotope. Then
	\begin{eqnarray*}
		\quad \quad \sigma \in \FP(W,b) &\;\Leftrightarrow \; & \chi(v_1^\sigma,\ldots,v_n^\sigma,e_{i}) 
		= \chi(v_1^\sigma,\ldots,v_n^\sigma,e_{n+1})  \text{ for all } i \in \sigma, 
		\text{ and}\\
		& & \chi(v_1^\sigma,\ldots,v_n^\sigma,h_{k}) 
		= - \chi(v_1^\sigma,\ldots,v_n^\sigma,e_{n+1})   \text{ for all }  k \in [n] \setminus \sigma,\nonumber
	\end{eqnarray*}
	where
	\begin{equation}\label{eq:vi^sigma}
		v_\ell^\sigma \od \left\{\begin{array}{cc} h_i & \text{ for } \ell \in \sigma,\\
			e_\ell & \text{ for } \ell \notin \sigma.\end{array}\right.
	\end{equation}
\end{proposition}

Recall that for $\sigma \subseteq [n]$, we use the notation $v_i^\sigma$
to indicate $h_i$ or $e_i$, with 
$v_i^\sigma = h_i$ for $i \in \sigma$ and $v_i^\sigma = e_i$ for $i \notin \sigma$ (see Eq. \ref{eq:vi^sigma}). We will also use the facts:
\begin{eqnarray} \label{eq:det-facts}
	0 &=& \det(v_1^\sigma,\ldots,v_n^\sigma,e_i), \text{ for } i \in [n]\setminus \sigma, \nonumber \\
	0 &=& \det(v_1^\sigma,\ldots,v_n^\sigma,h_i),  \text{ for } i \in \sigma, \nonumber\\
	W_{ij} &=& \det(e_1,\ldots,e_{j-1},h_i,e_{j+1},\ldots,e_{n+1}), \; \text{for } i \in [n], i \neq j, \nonumber\\
	-1 &=&  \det(e_1,\ldots,e_{i-1},h_i,e_{i+1},\ldots,e_{n+1}), \; \text{for } i \in [n], \nonumber\\
	b_i &=& \det(e_1,\ldots,e_{n},h_i),  \text{ and} \nonumber\\
	1 &=& \det(e_1,\ldots,e_n,e_{n+1}), \nonumber
\end{eqnarray}

Also, recall that:
$$\det A_j^{\{i,j\}} > 0\;\; \Leftrightarrow \;\; i \to j \quad \text{and} \quad 
\det A_i^{\{i,j\}} > 0\;\; \Leftrightarrow \;\; j \to i.$$

 Finally, we will use the notation:
\begin{equation}
	\boxed{\quad
		\begin{aligned}
			\Delta_\ell^{ij} &\od  \det(e_1,\ldots,e_{\ell-1}, h_i, e_{\ell+1}, \ldots, e_n,h_j) = - \Delta_\ell^{ji}\\
			&= W_{i\ell} \, b_j - W_{j\ell} \, b_i \quad \text{if } i,j \neq \ell.
		\end{aligned}\quad}
\end{equation}
Note that for $i = j$, we always have $\Delta_\ell^{ij} = 0$. Assuming $i \neq j$, then for $\ell = i$ we obtain
\begin{eqnarray}\label{eq:Delta-Aj}
	\Delta_i^{ij} &= & \det(e_1,\ldots,e_{i-1}, h_i, e_{i+1}, \ldots, e_n,h_j)\\
	&=& \det A_j^{\{i\}} = - \det A_j^{\{i,j\}} \nonumber \\
	&=& -b_i W_{ji} - b_j, \nonumber
\end{eqnarray}
and for $\ell = j$ we have
\begin{eqnarray}
	\Delta_j^{ij} &= & \det(e_1,\ldots,e_{j-1}, h_i, e_{j+1}, \ldots, e_n,h_j)\\
	&=& -\det A_i^{\{j\}} = \det A_i^{\{i,j\}} \nonumber \\
	&=& b_jW_{ij} + b_i. \nonumber
\end{eqnarray}
So altogether, we get:
$$\boxed{\;\;\Delta_\ell^{ij} = \left\{\begin{array}{cc} 
		b_jW_{i\ell} - b_iW_{j\ell}, & \text{if } i,j \neq \ell\\
		-b_j - b_i W_{ji}, & \text{if } i \neq j, \ell = i\\
		-b_j W_{ij} + b_i, & \text{if } i \neq j, \ell = j\\
		0, & \text{if } i = j.\end{array}\right.\;\;}$$


For any $\tau \subseteq [n]$, and $i,j,k \in [n]$, we can define:
$$\boxed{\;\Delta_k^{ij}(\tau) \od \det(v_1^{\tau\setminus\{i,j\}},v_2^{\tau\setminus\{i,j\}},\ldots,v_{k-1}^{\tau\setminus\{i,j\}},h_i,v_{k+1}^{\tau\setminus\{i,j\}},\ldots,v_n^{\tau\setminus\{i,j\}},h_j).\;}$$
Note that $v_i^{\tau\setminus\{i,j\}} = e_i$ and $v_j^{\tau\setminus\{i,j\}} = e_j$, so the above determinant does not repeat $h_i$ or $h_j$ terms, provided $i \neq j$, and is generically expected to be nonzero.
For $\tau = \{i,j\}$, this new determinant reduces to:
$$\Delta_k^{ij}(\{i,j\}) = \det(e_1,\ldots,e_{k-1},h_i,e_{k+1},\ldots,e_n,h_j) = \Delta_k^{ij}.$$

\subsection{New results}

If you follow the proofs of Chapter \ref{ch:new-theoretical-results}, you will notice that there is a recurrent strategy: assume a component is simply-embedded to obtain a factorization of the $s_i$'s, whose signs greatly control the set of fixed points supports of a network. In this section we aim to leverage again the simply-embedded structure to see if we can easily completely compute the chirotope for networks that have some simply-embedded structure in them. The first results in this direction find that there is plenty of elements that end up vanishing:

\begin{lemma}\label{lemma:generalizeddeltas}
	Consider a CTLN with graph $G$ on $n$ nodes. Let $\tau \subseteq [n]$ be simply-embedded in $G$. Then  
	$$\Delta_k^{ij}(\tau) = 0,$$
	for all $i,j \in \tau$, and $k \notin \tau$.
\end{lemma}

\begin{proof}
	If $i = j$, then $\Delta^{ij}_k = 0$ trivially ($h_i$ gets repeated). So suppose $i \neq j$.
	without loss of generality, let $\tau = \{1,\ldots,i,i+1\}$, let $j = i+1,$ and $k > i+1$. Then,
	\begin{eqnarray*}
		\Delta_k^{ij}(\tau) &=& \det(h_1,\ldots,h_{i-1},e_i,\ldots,e_{k-1},h_i,e_{k+1},\ldots,e_n,h_j)\\
	&=&\det \left(\begin{array}{cccccccccccc} 
		-1 & W_{12}  & \cdots & W_{1,i-1} & 0 & \cdots & 0 & W_{1,k} & 0 & \cdots & 0 & b_1\\
		W_{21} & -1 & \cdots & W_{2,i-1} & 0 & \cdots & 0 & W_{2,k} & 0 & \cdots & 0 & b_2\\
		\vdots &  &\ddots & \vdots & & \ddots &  & \vdots & & \ddots &  & \vdots\\
		W_{i-1,1} & W_{i-1,2} & \cdots & -1 & 0 & \cdots & 0 & W_{i-1,k} & 0 & \cdots & 0 & b_{i-1}\\
		0 & 0 & \cdots & 0 & 1 & \cdots & 0 & 0 & 0 & \cdots & 0 & 0\\
		\vdots & \vdots &  & \vdots & &  \ddots & & \vdots & \vdots &  & \vdots & \vdots\\
		0 & 0 & \cdots & 0 & 0 & \cdots & 1 & 0 & 0 & \cdots & 0 & 0\\
		W_{i,1} & W_{i,2} & \cdots & W_{i,i-1} & 0 & \cdots & 0 & W_{i,k} & 0 & \cdots & 0 & b_i\\
		0 & 0 & \cdots & 0 & 0 & \cdots & 0 & 0 & 1 & \cdots & 0 & 0\\
		\vdots & \vdots &  & \vdots & \vdots &   & \vdots & \vdots & & \ddots &  & \vdots\\
		0 & 0 & \cdots & 0 & 0 & \cdots & 0 & 0 & 0 & \cdots & 1 & 0\\
		W_{j,1} & W_{j,2} & \cdots & W_{j,i-1} & 0 & \cdots & 0 & W_{j,k} & 0 & \cdots & 0 & b_j\\
	\end{array}\right).
\end{eqnarray*}
Now, recall that for a CTLN we have $b_\ell = \theta$ for all $\ell \in [n]$, so the last column is of the form
$$\theta(1,1,\ldots,1,0,\ldots,0,1,0,\ldots,0,1)^T.$$
Furthermore, since $\tau$ is simply-embedded and $k \notin \tau$ all nonzero entries in the $k$-th column are identical and equal to $W_{i,k}$, so that this column has the form
$$W_{i,k}(1,1,\ldots,1,0,\ldots,0,1,0,\ldots,0,1)^T.$$
In other words, the $k$-th column is a scalar multiple of the last column, so $\Delta_k^{ij}(\tau) = 0$.
\end{proof}

In fact, we can prove that there are far more elements that vanish:
\begin{lemma}\label{lemma:vanishingdets}
Consider a CTLN with graph $G$. Let $\tau \subseteq [n]$ be simply-embedded in $G$. Any chirotope element whose only $h_i$ elements come from $\tau$ ($i\in\tau$), and does not contain $e_p,e_q$ for some $p,q\notin\tau$, will vanish.
\end{lemma}

\begin{proof}
Let $\sigma = \{1,...,k\} \subseteq \tau$. It suffices to prove:
\begin{equation}\label{eq:vanishingdet}
	\det(h_1,\dots,h_k,e_{s_{1}},\dots,\widehat{e_p},\dots,\widehat{e_q}\dots,e_{s_{n+1-k}}) = 0,
\end{equation} as other determinants satisfying the conditions in the theorem are either permutations or index relabelings of the LHS of Equation \ref{eq:vanishingdet}. Let $\rho = \tau \setminus \sigma = \{k+1,\dots,l\}$, and $[n]\setminus\tau = \{l+1,\dots,p,\dots,q,\dots,n\}$ then
\begin{align*}
	&\det(h_1,\dots,h_k,e_{s_{1}},\dots,\widehat{e_p},\dots,\widehat{e_q}\dots,e_{s_{n+1-k}}) \\
	= &\det\left(\begin{array}{c|c|ccccccc|c}
		& & W_{1,l+1} & & W_{1,p} & &W_{1,q}& &W_{1,n}&\theta\\
		\sigma \rightarrow \sigma & \rho \rightarrow \sigma &\vdots& \cdots & \vdots & \cdots &\vdots& \cdots &\vdots&\vdots\\
		& & W_{k,l+1} & & W_{k,p} & &W_{k,q}& &W_{k,n}&\theta\\
		\hline
		& &  & & 0 & &0& & &\\
		*&*&*& \cdots & \vdots & \cdots &\vdots& \cdots &*&*\\
		& &   & & 0 & &0& & &
	\end{array}\right)
\end{align*}
But since $\tau$ is simply-embedded, $W_{i,j} = W_{j}$ for all $i \in \tau$ and $j \notin\tau$ and therefore columns $p$ and $q$ are linearly dependent:
\begin{align*}
	&\det\left(\begin{array}{c|c|ccccccc|c}
		& & W_{l+1} & & W_{p} & &W_{q}& &W_{n}&\theta\\
		\sigma \rightarrow \sigma & \rho \rightarrow \sigma &\vdots& \cdots & \vdots & \cdots &\vdots& \cdots &\vdots&\vdots\\
		& & W_{l+1} & & W_{p} & &W_{q}& &W_{n}&\theta\\
		\hline
		& &  & & 0 & &0& & &\\
		*&*&*& \cdots & \vdots & \cdots &\vdots& \cdots &*&*\\
		& &   & & 0 & &0& & &
	\end{array}\right) = 0.
\end{align*}
Note however that the other columns may or may not be linearly dependent, as the * indicates the possible presence of a $1$, so it is indeed necessary to have $e_p$ and $e_q$ not appear in the determinant. Also, note that the proof still holds if $q=n+1$.
\end{proof}

Lemma \ref{lemma:generalizeddeltas} now follows from Lemma \ref{lemma:vanishingdets}:
\begin{align*}
\Delta_k^{ij}(\tau) = &\det(h_1,\ldots,h_{i-1},e_i,e_{i+1},e_{i+2}\ldots,e_{k-1},h_i,e_{k+1},\ldots,e_n,h_j)\\
= \pm &\det(h_1,\ldots,h_{i-1},h_i,h_{j},e_{i+2},\ldots,e_{k-1},e_i,e_{k+1},\ldots,e_n,e_{i+1})\\
= \pm &\det(h_1,\ldots,h_{j=i+1},e_i,e_{i+1},e_{i+2}\ldots,e_{k-1},e_{k+1},\ldots,e_n).
\end{align*}
It is clear from the last determinant that $e_k$ and $e_{n+1}$ are missing, with $k,n+1 \notin \tau = \{1,\ldots,i,i+1\}$. In this case, $\sigma = \tau$ in the proof above. Note that the indices above give a total of $n+1$ elements: $1,\dots,k-1,k+1,\dots,n$ with an extra $i,i+1$ for a total of $n-1+2 = n+1$ elements.

We can still push a little further:

\begin{lemma}\label{lemma:vanishingdetsv2} 
Let $\tau \subseteq [n]$ and suppose one of the following holds:
\begin{enumerate}
	\item There exist $p,q \in [n] \setminus \tau$, $(p \neq q)$ such that $\tau$ is simply-embedded in $\tau \cup \{p,q\}$. That is, $W_{i,p} = W_{j,p}$  and $W_{i,q} = W_{j,q}$ for all $i,j \in \tau$. (Assume no restrictions on $b$).
	\item There exists $p \in [n] \setminus \tau$ such that $\tau$ is simply-embedded in $\tau \cup \{p\}$, $q = n+1$, and $b_i = \theta$ for all $i \in \tau$ (it can be anything outside of $\tau$).
\end{enumerate}
Then, any chirotope determinant whose only $h_i$ elements come from $\tau$ (i.e. $i \in \tau$), and does not contain $e_p$ or $e_q$, will vanish.
\end{lemma}

\begin{proof}
Let $\sigma = \{1,...,k\} \subseteq \tau$. It suffices to prove:
\begin{equation}\label{eq:vanishingdetv2}
	\det(h_1,\dots,h_k,e_{s_{1}},\dots,\widehat{e_p},\dots,\widehat{e_q}\dots,e_{s_{n+1-k}}) = 0,
\end{equation} as other determinants satisfying the conditions in the theorem are either permutations or index relabelings of the LHS of Equation \ref{eq:vanishingdetv2}. 
First, suppose (i) holds. Let $\rho = \tau \setminus \sigma = \{k+1,\dots,l\}$, and $[n]\setminus\tau = \{l+1,\dots,p,\dots,q,\dots,n\}$, then
\begin{align*}
	&\det(h_1,\dots,h_k,e_{s_{1}},\dots,\widehat{e_p},\dots,\widehat{e_q}\dots,e_{s_{n+1-k}}) \\
	= &\det\left(\begin{array}{c|c|ccccccc|c}
		& & W_{1,l+1} & & W_{1,p} & &W_{1,q}& &W_{1,n}&b_1\\
		\sigma \rightarrow \sigma & \rho \rightarrow \sigma &\vdots& \cdots & \vdots & \cdots &\vdots& \cdots &\vdots&\vdots\\
		& & W_{k,l+1} & & W_{k,p} & &W_{k,q}& &W_{k,n}&b_n\\
		\hline
		& &  & & 0 & &0& & &\\
		*&*&*& \cdots & \vdots & \cdots &\vdots& \cdots &*&*\\
		& &   & & 0 & &0& & &
	\end{array}\right),
\end{align*} where the $*$ are blocks of mostly zeros and some ones, coming from the $e_{s_i}$ rows.
Now, since $\tau$ is simply-embedded in $\tau \cup \{p,q\}$, it follows that columns $p$ and $q$ are linearly dependent and thus the determinant vanishes.

For (ii), let $\rho = \tau \setminus \sigma = \{k+1,\dots,l\}$, and $[n]\setminus\tau = \{l+1,\dots,p,\dots,n\}$. Then the determinant looks like
\begin{align*}
	&\det(h_1,\dots,h_k,e_{s_{1}},\dots,\widehat{e_p},\dots,\widehat{e_{n+1}}\dots,e_{s_{n+1-k}}) \\
	= &\det\left(\begin{array}{c|c|ccccc|c}
		& & W_{1,l+1} & & W_{1,p} & &W_{1,n}&\theta\\
		\sigma \rightarrow \sigma & \rho \rightarrow \sigma &\vdots& \cdots & \vdots & \cdots &\vdots&\vdots\\
		& & W_{k,l+1} & & W_{k,p} & &W_{k,n}&\theta\\
		\hline
		& &  & & 0 & & &0\\
		*&*&*& \cdots & \vdots & \cdots &*&\vdots\\
		& &   & & 0 & & &0
	\end{array}\right)
\end{align*} and so in this case columns $p$ and $n+1$ are linearly dependent too, and the determinant vanishes.
\end{proof}

As mentioned above, our goal was to completely compute the chirotope for networks that have some simply added structure. More specifically, if a given CTLN has a strongly simply-embedded partition, can the chirotope be factorized, just like the $s_i$'s can? We still don't know.
%
\appendix
\titleformat{\chapter}[display]{\fontsize{30}{30}\selectfont\bfseries\sffamily}{Appendix \thechapter\textcolor{gray75}{\raisebox{3pt}{|}}}{0pt}{}{}
\Appendix{Supplementary materials}\label{ch:appendixA}

\section{Matrices and parameters of models}

This section contains the detailed data and parameters necessary for reproducing the basic simulations included in this thesis. MATLAB code for basic plots included in this dissertation can be found at \url{https://github.com/juliana-londono/phd-thesis-basic-plots}. In any case, below you can find the adjacency matrices $A$ of all the models, along with the $\varepsilon, \delta$ parameters needed to obtain the connectivity matrix $W$. Recall that $W$ is obtained from $A$ as 
\begin{equation*} 
W_{ij} = \left\{\begin{array}{ll} 
	\phantom{-}0 & \text{ if } i = j, \\ 
	-1 + \varepsilon & \text{ if } A_{ij} = 1,\\ 
	-1 -\delta & \text{ if } A_{ij} = 0. 
	\end{array}\right. 
\end{equation*} Alternatively, $W$ can be easily computed as $W = (-1+\varepsilon)A +(-1+\delta)(\mathbf{1}-I-A)$.

Finally, recall the dynamics of the network:
\begin{equation*}
	\dfrac{dx_i}{dt} = -x_i + \left[\sum_{j=1}^n W_{ij}x_j+b_i \right]_+, \quad i = 1,\ldots,n,
\end{equation*} 

\textbf{CTLN counter: }
\begin{footnotesize}
	\begin{equation}\label{eq:counter-matrix}
	A =\begin{bmatrix}
		0 & 1 & 0 & 0 & 0 & 0 & 0 & 0 & 0 & 0 & 1 & 0 \\
		1 & 0 & 0 & 0 & 0 & 0 & 0 & 0 & 0 & 0 & 0 & 1 \\
		1 & 0 & 0 & 1 & 0 & 0 & 0 & 0 & 0 & 0 & 0 & 0 \\
		0 & 1 & 1 & 0 & 0 & 0 & 0 & 0 & 0 & 0 & 0 & 0 \\
		0 & 0 & 1 & 0 & 0 & 1 & 0 & 0 & 0 & 0 & 0 & 0 \\
		0 & 0 & 0 & 1 & 1 & 0 & 0 & 0 & 0 & 0 & 0 & 0 \\
		0 & 0 & 0 & 0 & 1 & 0 & 0 & 1 & 0 & 0 & 0 & 0 \\
		0 & 0 & 0 & 0 & 0 & 1 & 1 & 0 & 0 & 0 & 0 & 0 \\
		0 & 0 & 0 & 0 & 0 & 0 & 1 & 0 & 0 & 1 & 0 & 0 \\
		0 & 0 & 0 & 0 & 0 & 0 & 0 & 1 & 1 & 0 & 0 & 0 \\
		0 & 0 & 0 & 0 & 0 & 0 & 0 & 0 & 1 & 0 & 0 & 1 \\
		0 & 0 & 0 & 0 & 0 & 0 & 0 & 0 & 0 & 1 & 1 & 0 
	\end{bmatrix}
		\end{equation}
\end{footnotesize}
 with $\varepsilon=0.25, \delta=0.5$.

\textbf{CTLN signed counter:} 
\begin{footnotesize}
	\begin{equation}\label{eq:signed-counter-matrix}
	A = \begin{bmatrix}
		0 & 1 & 0 & 0 & 0 & 0 & 0 & 0 & 0 & 0 & 1 & 0 \\
		1 & 0 & 0 & 1 & 0 & 0 & 0 & 0 & 0 & 0 & 0 & 0 \\
		1 & 0 & 0 & 1 & 0 & 0 & 0 & 0 & 0 & 0 & 0 & 0 \\
		0 & 0 & 1 & 0 & 0 & 1 & 0 & 0 & 0 & 0 & 0 & 0 \\
		0 & 0 & 1 & 0 & 0 & 1 & 0 & 0 & 0 & 0 & 0 & 0 \\
		0 & 0 & 0 & 0 & 1 & 0 & 0 & 1 & 0 & 0 & 0 & 0 \\
		0 & 0 & 0 & 0 & 1 & 0 & 0 & 1 & 0 & 0 & 0 & 0 \\
		0 & 0 & 0 & 0 & 0 & 0 & 1 & 0 & 0 & 1 & 0 & 0 \\
		0 & 0 & 0 & 0 & 0 & 0 & 1 & 0 & 0 & 1 & 0 & 0 \\
		0 & 0 & 0 & 0 & 0 & 0 & 0 & 0 & 1 & 0 & 0 & 1 \\
		0 & 0 & 0 & 0 & 0 & 0 & 0 & 0 & 1 & 0 & 0 & 1 \\
		0 & 1 & 0 & 0 & 0 & 0 & 0 & 0 & 0 & 0 & 1 & 0 \\
	\end{bmatrix}
	\end{equation}
\end{footnotesize}
 with $\varepsilon=0.25, \delta=0.5$.

\textbf{Dynamic attractor chain:} 
\begin{footnotesize}
	\begin{equation}\label{eq:dynamic-attractor-chain-matrix}
	\begin{bmatrix} A = 
		0 & 0 & 0 & 1 & 1 & 0 & 0 & 0 & 0 & 0 & 0 & 0 & 0 & 0 & 0 & 0 & 1 & 1 \\
		1 & 0 & 0 & 0 & 1 & 0 & 0 & 0 & 0 & 0 & 0 & 0 & 0 & 0 & 0 & 0 & 0 & 1 \\
		1 & 1 & 0 & 0 & 0 & 0 & 0 & 0 & 0 & 0 & 0 & 0 & 0 & 0 & 0 & 0 & 0 & 0 \\
		0 & 1 & 1 & 0 & 0 & 0 & 1 & 1 & 0 & 0 & 0 & 0 & 0 & 0 & 0 & 0 & 0 & 0 \\
		0 & 0 & 1 & 1 & 0 & 0 & 0 & 1 & 0 & 0 & 0 & 0 & 0 & 0 & 0 & 0 & 0 & 0 \\
		0 & 0 & 0 & 1 & 1 & 0 & 0 & 0 & 0 & 0 & 0 & 0 & 0 & 0 & 0 & 0 & 0 & 0 \\
		0 & 0 & 0 & 0 & 1 & 1 & 0 & 0 & 0 & 1 & 1 & 0 & 0 & 0 & 0 & 0 & 0 & 0 \\
		0 & 0 & 0 & 0 & 0 & 1 & 1 & 0 & 0 & 0 & 1 & 0 & 0 & 0 & 0 & 0 & 0 & 0 \\
		0 & 0 & 0 & 0 & 0 & 0 & 1 & 1 & 0 & 0 & 0 & 0 & 0 & 0 & 0 & 0 & 0 & 0 \\
		0 & 0 & 0 & 0 & 0 & 0 & 0 & 1 & 1 & 0 & 0 & 0 & 1 & 1 & 0 & 0 & 0 & 0 \\
		0 & 0 & 0 & 0 & 0 & 0 & 0 & 0 & 1 & 1 & 0 & 0 & 0 & 1 & 0 & 0 & 0 & 0 \\
		0 & 0 & 0 & 0 & 0 & 0 & 0 & 0 & 0 & 1 & 1 & 0 & 0 & 0 & 0 & 0 & 0 & 0 \\
		0 & 0 & 0 & 0 & 0 & 0 & 0 & 0 & 0 & 0 & 1 & 1 & 0 & 0 & 0 & 1 & 1 & 0 \\
		0 & 0 & 0 & 0 & 0 & 0 & 0 & 0 & 0 & 0 & 0 & 1 & 1 & 0 & 0 & 0 & 1 & 0 \\
		0 & 0 & 0 & 0 & 0 & 0 & 0 & 0 & 0 & 0 & 0 & 0 & 1 & 1 & 0 & 0 & 0 & 0 \\
		1 & 1 & 0 & 0 & 0 & 0 & 0 & 0 & 0 & 0 & 0 & 0 & 0 & 1 & 1 & 0 & 0 & 0 \\
		0 & 1 & 0 & 0 & 0 & 0 & 0 & 0 & 0 & 0 & 0 & 0 & 0 & 0 & 1 & 1 & 0 & 0 \\
		0 & 0 & 0 & 0 & 0 & 0 & 0 & 0 & 0 & 0 & 0 & 0 & 0 & 0 & 0 & 1 & 1 & 0 \\
	\end{bmatrix}
		\end{equation}
\end{footnotesize}
 with $\varepsilon=0.51, \delta=1.76$.

\textbf{Quadruped gaits isolated:} 
\begin{footnotesize}
	\begin{equation}\label{eq:bound-matrix}
		A_{\text{bound}} = \begin{bmatrix}
			0 & 0 & 0 & 1 & 1 & 0 & 0 & 0 & 1 & 0 \\
			0 & 0 & 1 & 0 & 0 & 1 & 0 & 0 & 0 & 1 \\
			0 & 1 & 0 & 0 & 0 & 0 & 1 & 0 & 0 & 1 \\
			1 & 0 & 0 & 0 & 0 & 0 & 0 & 1 & 1 & 0 \\
			1 & 0 & 0 & 0 & 0 & 0 & 0 & 1 & 1 & 0 \\
			0 & 1 & 0 & 0 & 0 & 0 & 1 & 0 & 0 & 1 \\
			0 & 0 & 1 & 0 & 0 & 1 & 0 & 0 & 0 & 1 \\
			0 & 0 & 0 & 1 & 1 & 0 & 0 & 0 & 1 & 0 \\
			0 & 1 & 1 & 0 & 0 & 1 & 1 & 0 & 0 & 0 \\
			1 & 0 & 0 & 1 & 1 & 0 & 0 & 1 & 0 & 0 \\
		\end{bmatrix}
	\end{equation}
\end{footnotesize}
\begin{footnotesize}
\begin{equation}\label{eq:pace-matrix}
	A_{\text{pace}} = \begin{bmatrix}
		0 & 1 & 0 & 0 & 1 & 0 & 0 & 0 & 0 & 1 \\
		1 & 0 & 0 & 0 & 0 & 1 & 0 & 0 & 0 & 1 \\
		0 & 0 & 0 & 1 & 0 & 0 & 1 & 0 & 1 & 0 \\
		0 & 0 & 1 & 0 & 0 & 0 & 0 & 1 & 1 & 0 \\
		1 & 0 & 0 & 0 & 0 & 1 & 0 & 0 & 0 & 1 \\
		0 & 1 & 0 & 0 & 1 & 0 & 0 & 0 & 0 & 1 \\
		0 & 0 & 1 & 0 & 0 & 0 & 0 & 1 & 1 & 0 \\
		0 & 0 & 0 & 1 & 0 & 0 & 1 & 0 & 1 & 0 \\
		1 & 1 & 0 & 0 & 1 & 1 & 0 & 0 & 0 & 0 \\
		0 & 0 & 1 & 1 & 0 & 0 & 1 & 1 & 0 & 0 \\
	\end{bmatrix}
\end{equation}
\end{footnotesize}
\begin{footnotesize}
\begin{equation}\label{eq:trot-matrix}
	A_{\text{trot}} = \begin{bmatrix}
		0 & 0 & 0 & 0 & 1 & 0 & 1 & 0 & 1 & 0 \\
		0 & 0 & 0 & 0 & 0 & 1 & 0 & 1 & 0 & 1 \\
		0 & 0 & 0 & 0 & 1 & 0 & 1 & 0 & 1 & 0 \\
		0 & 0 & 0 & 0 & 0 & 1 & 0 & 1 & 0 & 1 \\
		1 & 0 & 1 & 0 & 0 & 0 & 0 & 0 & 1 & 0 \\
		0 & 1 & 0 & 1 & 0 & 0 & 0 & 0 & 0 & 1 \\
		1 & 0 & 1 & 0 & 0 & 0 & 0 & 0 & 1 & 0 \\
		0 & 1 & 0 & 1 & 0 & 0 & 0 & 0 & 0 & 1 \\
		0 & 1 & 0 & 1 & 0 & 1 & 0 & 1 & 0 & 0 \\
		1 & 0 & 1 & 0 & 1 & 0 & 1 & 0 & 0 & 0 \\
	\end{bmatrix}
\end{equation}
\end{footnotesize}
\begin{footnotesize}
\begin{equation}\label{eq:walk-matrix}
	A_{\text{walk}} = \begin{bmatrix}
		0 & 0 & 0 & 0 & 1 & 0 & 0 & 0 & 1 & 0 & 0 & 0 & 0 & 0 & 0 & 1 \\
		0 & 0 & 0 & 0 & 0 & 1 & 0 & 0 & 0 & 1 & 0 & 0 & 1 & 0 & 0 & 0 \\
		0 & 0 & 0 & 0 & 0 & 0 & 1 & 0 & 0 & 0 & 1 & 0 & 0 & 0 & 1 & 0 \\
		0 & 0 & 0 & 0 & 0 & 0 & 0 & 1 & 0 & 0 & 0 & 1 & 0 & 1 & 0 & 0 \\
		1 & 0 & 0 & 0 & 0 & 0 & 0 & 0 & 1 & 0 & 0 & 0 & 0 & 0 & 0 & 1 \\
		0 & 1 & 0 & 0 & 0 & 0 & 0 & 0 & 0 & 1 & 0 & 0 & 1 & 0 & 0 & 0 \\
		0 & 0 & 1 & 0 & 0 & 0 & 0 & 0 & 0 & 0 & 1 & 0 & 0 & 0 & 1 & 0 \\
		0 & 0 & 0 & 1 & 0 & 0 & 0 & 0 & 0 & 0 & 0 & 1 & 0 & 1 & 0 & 0 \\
		1 & 0 & 0 & 0 & 1 & 0 & 0 & 0 & 0 & 0 & 0 & 0 & 0 & 0 & 0 & 1 \\
		0 & 1 & 0 & 0 & 0 & 1 & 0 & 0 & 0 & 0 & 0 & 0 & 1 & 0 & 0 & 0 \\
		0 & 0 & 1 & 0 & 0 & 0 & 1 & 0 & 0 & 0 & 0 & 0 & 0 & 0 & 1 & 0 \\
		0 & 0 & 0 & 1 & 0 & 0 & 0 & 1 & 0 & 0 & 0 & 0 & 0 & 1 & 0 & 0 \\
		1 & 0 & 0 & 0 & 1 & 0 & 0 & 0 & 1 & 0 & 0 & 0 & 0 & 0 & 0 & 0 \\
		0 & 1 & 0 & 0 & 0 & 1 & 0 & 0 & 0 & 1 & 0 & 0 & 0 & 0 & 0 & 0 \\
		0 & 0 & 0 & 1 & 0 & 0 & 0 & 1 & 0 & 0 & 0 & 1 & 0 & 0 & 0 & 0 \\
		0 & 0 & 1 & 0 & 0 & 0 & 1 & 0 & 0 & 0 & 0 & 1 & 0 & 0 & 0 & 0 \\
	\end{bmatrix}
\end{equation}
\end{footnotesize}
\begin{footnotesize}
\begin{equation}\label{eq:pronk-matrix}
	A_{\text{pronk}} = \begin{bmatrix}
		0 & 1 & 0 & 1 & 0 & 1 \\
		1 & 0 & 1 & 0 & 0 & 1 \\
		0 & 1 & 0 & 1 & 0 & 1 \\
		1 & 0 & 1 & 0 & 0 & 1 \\
		1 & 1 & 1 & 1 & 0 & 0 \\
		0 & 0 & 0 & 0 & 1 & 0 \\
	\end{bmatrix}.
\end{equation}
\end{footnotesize}
All of them with $\varepsilon=0.25, \delta=0.5$.

\textbf{Five-gait network:} 
\begin{footnotesize}
\begin{equation}\label{eq:five-gait-matrix}
	A =\begin{bmatrix}
		0 & 1 & 0 & 1 & 1 & 0 & 1 & 0 & 1 & 0 & 0 & 0 & 1 & 0 & 0 & 1 & 1 & 0 & 0 & 0 & 0 & 1 & 0 & 1 \\
		1 & 0 & 1 & 0 & 0 & 1 & 0 & 1 & 0 & 1 & 0 & 0 & 0 & 1 & 0 & 1 & 0 & 1 & 1 & 0 & 0 & 0 & 0 & 1 \\
		0 & 1 & 0 & 1 & 1 & 0 & 1 & 0 & 0 & 0 & 1 & 0 & 0 & 1 & 1 & 0 & 1 & 0 & 0 & 0 & 1 & 0 & 0 & 1 \\
		1 & 0 & 1 & 0 & 0 & 1 & 0 & 1 & 0 & 0 & 0 & 1 & 1 & 0 & 1 & 0 & 0 & 1 & 0 & 1 & 0 & 0 & 0 & 1 \\
		1 & 0 & 1 & 0 & 0 & 1 & 0 & 1 & 1 & 0 & 0 & 0 & 1 & 0 & 0 & 1 & 1 & 0 & 0 & 0 & 0 & 1 & 0 & 0 \\
		0 & 1 & 0 & 1 & 1 & 0 & 1 & 0 & 0 & 1 & 0 & 0 & 0 & 1 & 0 & 1 & 0 & 1 & 1 & 0 & 0 & 0 & 0 & 0 \\
		1 & 0 & 1 & 0 & 0 & 1 & 0 & 1 & 0 & 0 & 1 & 0 & 0 & 1 & 1 & 0 & 1 & 0 & 0 & 0 & 1 & 0 & 0 & 0 \\
		0 & 1 & 0 & 1 & 1 & 0 & 1 & 0 & 0 & 0 & 0 & 1 & 1 & 0 & 1 & 0 & 0 & 1 & 0 & 1 & 0 & 0 & 0 & 0 \\
		1 & 0 & 0 & 0 & 1 & 0 & 0 & 0 & 0 & 0 & 0 & 0 & 0 & 0 & 0 & 0 & 0 & 0 & 0 & 0 & 1 & 0 & 0 & 0 \\
		0 & 1 & 0 & 0 & 0 & 1 & 0 & 0 & 0 & 0 & 0 & 0 & 0 & 0 & 0 & 0 & 0 & 0 & 1 & 0 & 0 & 0 & 0 & 0 \\
		0 & 0 & 1 & 0 & 0 & 0 & 1 & 0 & 0 & 0 & 0 & 0 & 0 & 0 & 0 & 0 & 0 & 0 & 0 & 0 & 1 & 0 & 0 & 0 \\
		0 & 0 & 0 & 1 & 0 & 0 & 0 & 1 & 0 & 0 & 0 & 0 & 0 & 0 & 0 & 0 & 0 & 0 & 0 & 1 & 0 & 0 & 0 & 0 \\
		0 & 1 & 1 & 0 & 0 & 1 & 1 & 0 & 0 & 0 & 0 & 0 & 0 & 0 & 0 & 0 & 0 & 0 & 0 & 0 & 0 & 0 & 0 & 0 \\
		1 & 0 & 0 & 1 & 1 & 0 & 0 & 1 & 0 & 0 & 0 & 0 & 0 & 0 & 0 & 0 & 0 & 0 & 0 & 0 & 0 & 0 & 0 & 0 \\
		1 & 1 & 0 & 0 & 1 & 1 & 0 & 0 & 0 & 0 & 0 & 0 & 0 & 0 & 0 & 0 & 0 & 0 & 0 & 0 & 0 & 0 & 0 & 0 \\
		0 & 0 & 1 & 1 & 0 & 0 & 1 & 1 & 0 & 0 & 0 & 0 & 0 & 0 & 0 & 0 & 0 & 0 & 0 & 0 & 0 & 0 & 0 & 0 \\
		0 & 1 & 0 & 1 & 0 & 1 & 0 & 1 & 0 & 0 & 0 & 0 & 0 & 0 & 0 & 0 & 0 & 0 & 0 & 0 & 0 & 0 & 0 & 0 \\
		1 & 0 & 1 & 0 & 1 & 0 & 1 & 0 & 0 & 0 & 0 & 0 & 0 & 0 & 0 & 0 & 0 & 0 & 0 & 0 & 0 & 0 & 0 & 0 \\
		1 & 0 & 0 & 0 & 1 & 0 & 0 & 0 & 1 & 0 & 0 & 0 & 0 & 0 & 0 & 0 & 0 & 0 & 0 & 0 & 0 & 0 & 0 & 0 \\
		0 & 1 & 0 & 0 & 0 & 1 & 0 & 0 & 0 & 1 & 0 & 0 & 0 & 0 & 0 & 0 & 0 & 0 & 0 & 0 & 0 & 0 & 0 & 0 \\
		0 & 0 & 0 & 1 & 0 & 0 & 0 & 1 & 0 & 0 & 0 & 1 & 0 & 0 & 0 & 0 & 0 & 0 & 0 & 0 & 0 & 0 & 0 & 0 \\
		0 & 0 & 1 & 0 & 0 & 0 & 1 & 0 & 0 & 0 & 1 & 0 & 0 & 0 & 0 & 0 & 0 & 0 & 0 & 0 & 0 & 0 & 0 & 0 \\
		1 & 1 & 1 & 1 & 0 & 0 & 0 & 0 & 0 & 0 & 0 & 0 & 0 & 0 & 0 & 0 & 0 & 0 & 0 & 0 & 0 & 0 & 0 & 0 \\
		0 & 0 & 0 & 0 & 0 & 0 & 0 & 0 & 0 & 0 & 0 & 0 & 0 & 0 & 0 & 0 & 0 & 0 & 0 & 0 & 0 & 0 & 1 & 0 \\
	\end{bmatrix}
\end{equation}
\end{footnotesize}
 with $\varepsilon=0.05, \delta=0.15$. 

\textbf{Induced gaits:} For the induced gaits we used the matrix for the five-gait network restricted to the following subsets of neurons:
\begin{align*}
	\text{bound} &: \{1,2,3,4,5,6,7,8,13,14\} \\
	\text{pace} &: \{1,2,3,4,5,6,7,8,15,16\} \\
	\text{trot} &: \{1,2,3,4,5,6,7,8,17,18\} \\
	\text{walk} &: \{1,2,3,4,5,6,7,8,9,10,11,12,19,20,21,22\} \\
	\text{pronk} &: \{1,2,3,4,23,24\}.
\end{align*} 
All of them with $\varepsilon=0.25, \delta=0.5$.

\textbf{Mollusk octahedral network:} 
\begin{footnotesize}
\begin{equation}\label{eq:mollusk-matrix}
	A = \begin{bmatrix}
		0 & 0 & 1 & 0 & 0 & 1 \\
		1 & 0 & 0 & 1 & 0 & 0 \\
		0 & 1 & 0 & 0 & 1 & 0 \\
		0 & 0 & 1 & 0 & 0 & 1 \\
		1 & 0 & 0 & 1 & 0 & 0 \\
		0 & 1 & 0 & 0 & 1 & 0 \\
	\end{bmatrix}
\end{equation}
\end{footnotesize} with $\varepsilon=0.07, \delta=0.3$.

\textbf{Sequential control of gaits and swimming directions: } These are both TLNs whose diagonal blocks are CTLNs. The connectivity matrix $W$ (not $A$!) for each is given by
\begin{figure}[!h]
\begin{center}
	\includegraphics[width=0.7\textwidth]{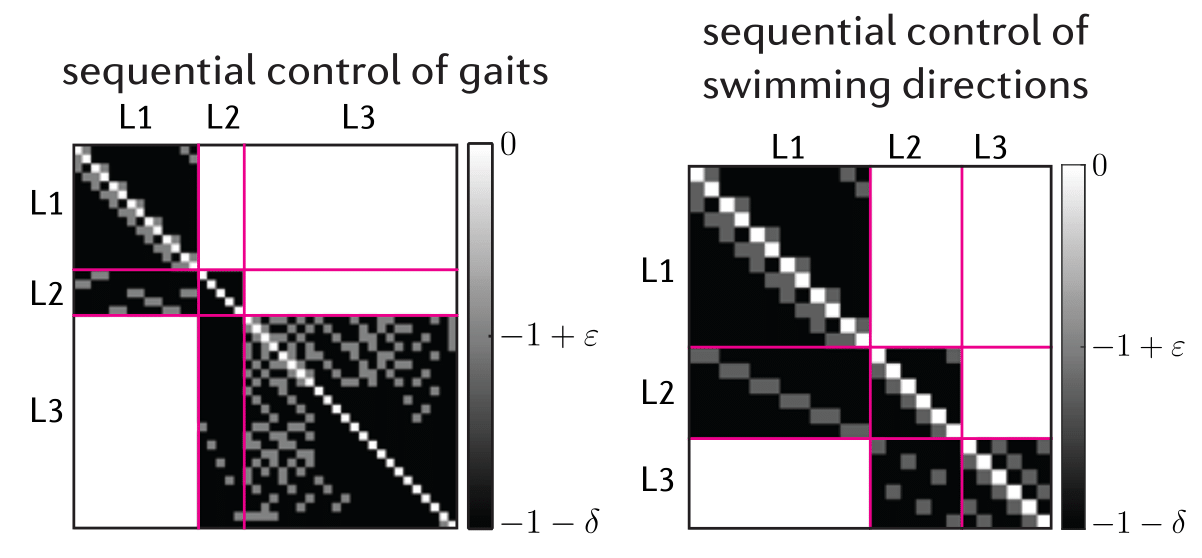}
\end{center}
\end{figure} 

\noindent where $\varepsilon=0.25, \delta=0.5$ for the sequential control of gaits and $\varepsilon=0.07, \delta=0.3$. for the sequential control of swimming directions. More precisely, these are TLNs whose diagonal blocks are CTLNs. The connectivity matrix for both is given by 
\begin{equation}\label{eq:sequential-control-matrices}
	W = \begin{bmatrix}
		W_1 & 0 & 0 \\
		W_{2,1} & W_2 & 0 \\
		0 & W_{3,2} & W_3 \\
	\end{bmatrix}
\end{equation} where $W_1$ is the CTLN connectivity matrix of the unsigned counter using seven and six 2-cliques respectively, $W_2$ is the CTLN connectivity matrix of an independent set in five and six nodes respectively, and $W_3$ is the CTLN connectivity matrix of the respective CPG (either quadruped gaits obtained from the adjacency matrix in Eq. \ref{eq:five-gait-matrix} or the mollusk swimming directions obtained from the adjacency matrix in Eq. \ref{eq:mollusk-matrix}).

\section{$\FP(G)$ calculations for quadruped gaits}

\begin{table}[ht]
	\caption[Minimal sets in $\FP(G)$ for each \emph{induced} isolated gait]{Minimal sets in $\FP(G)$ for each \emph{induced} isolated gait. Nodes are uniquely colored by gait.}
	\centering
	\resizebox{\columnwidth}{!}{%
		\begin{tabular}{c|c|c|c|c}
			\includegraphics[scale=0.25]{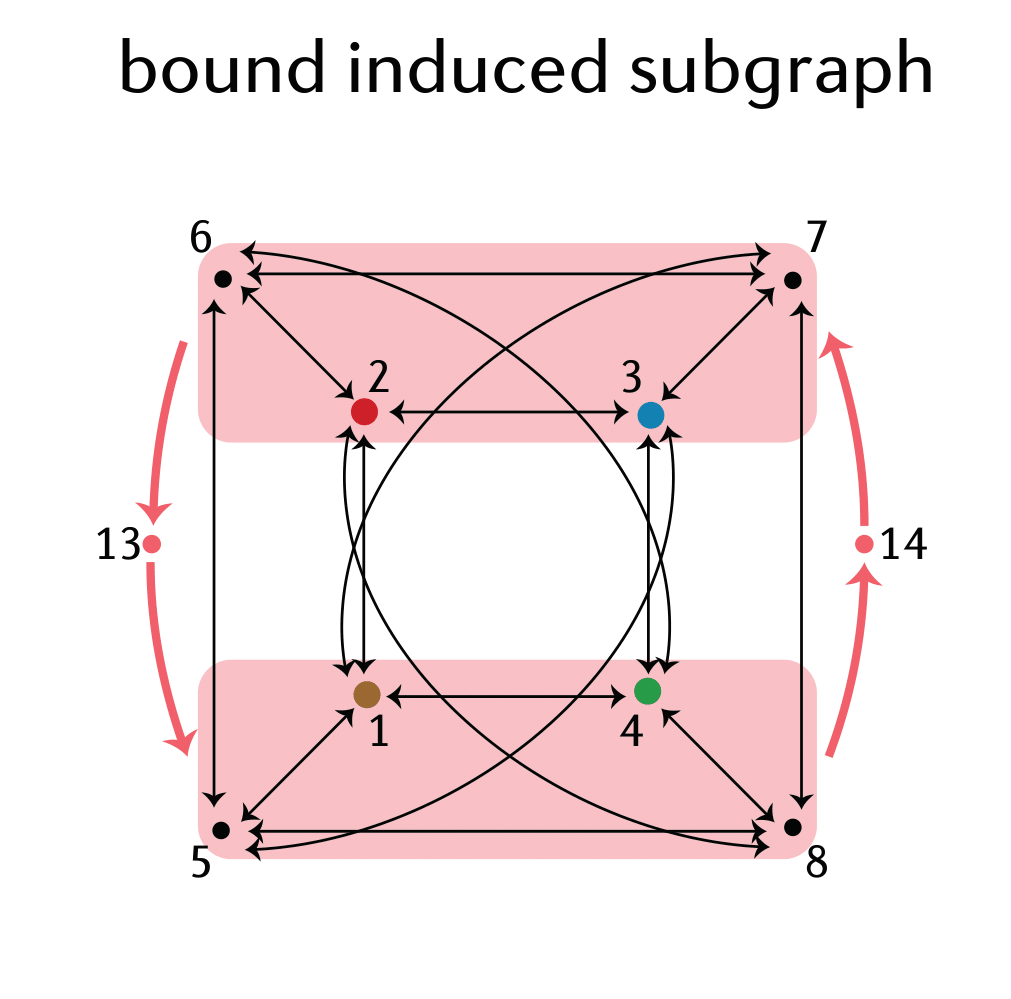}&\includegraphics[scale=0.25]{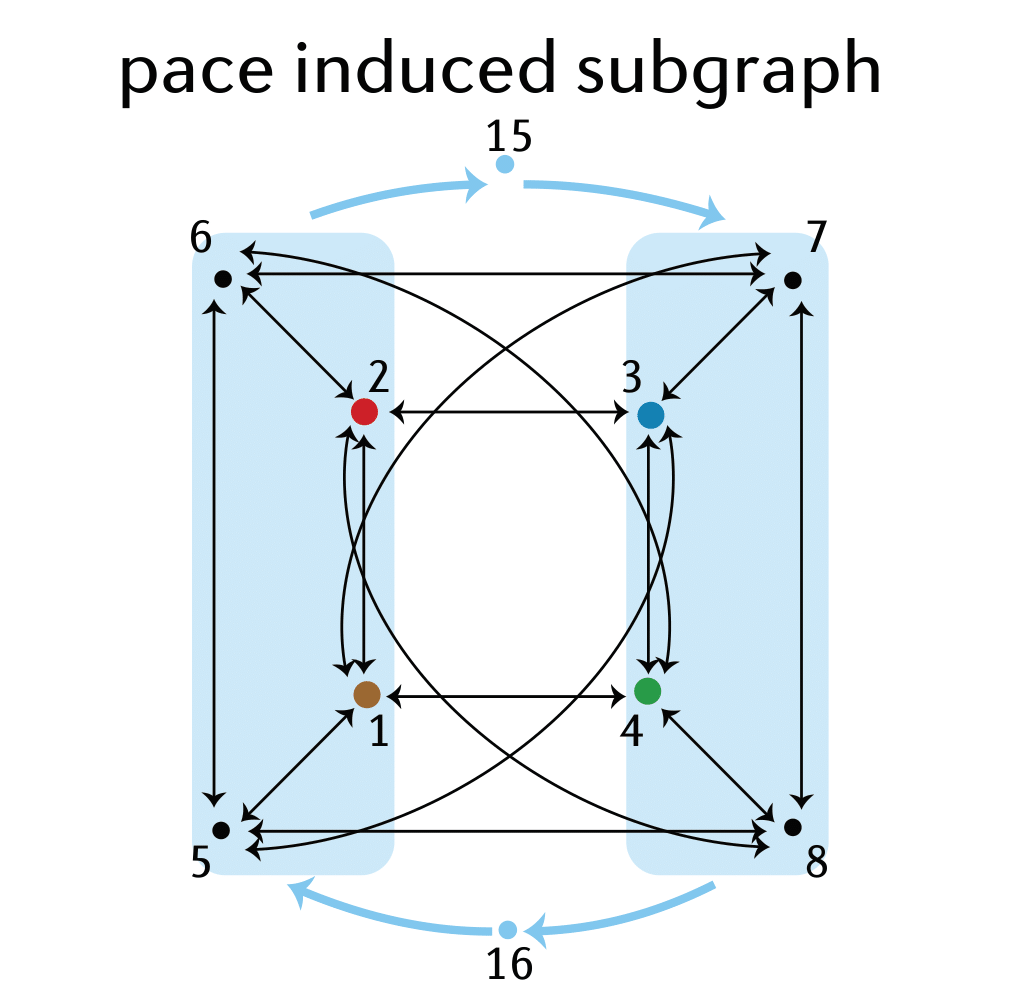}&\includegraphics[scale=0.25]{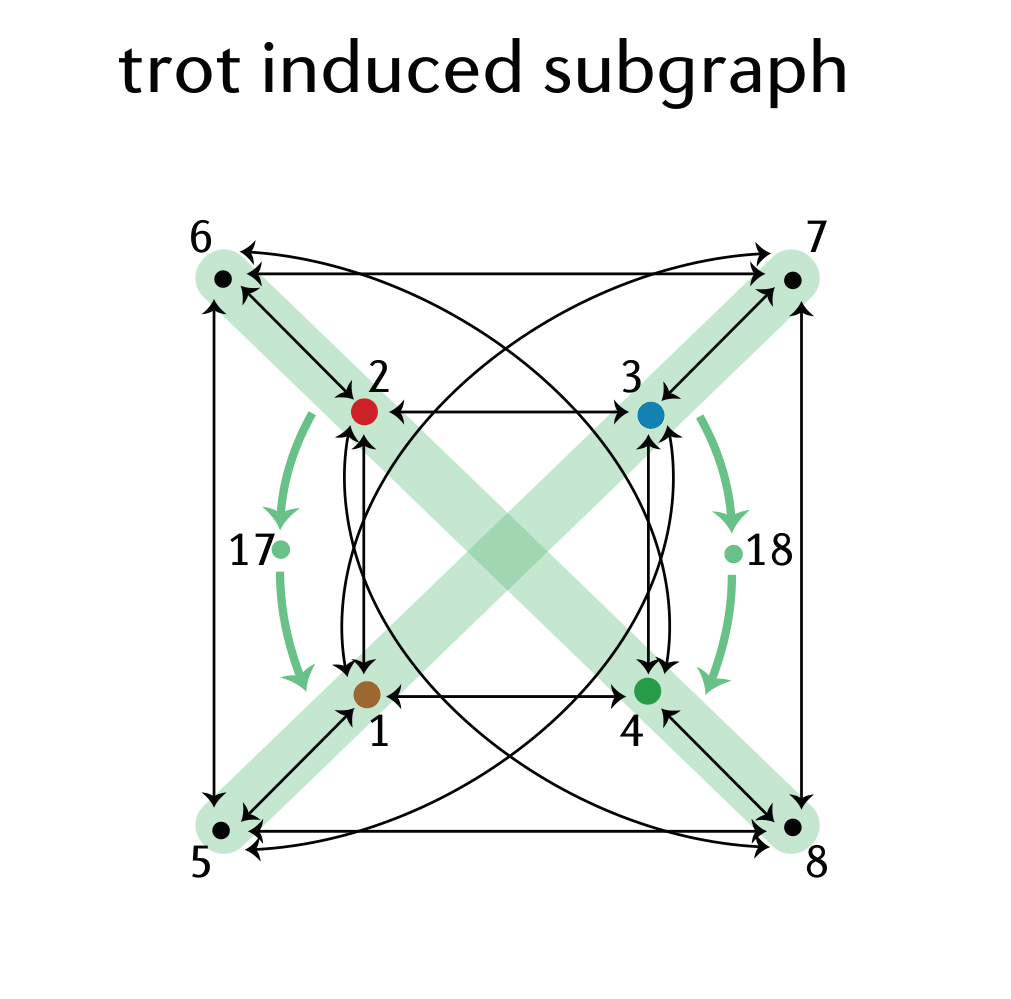}&\includegraphics[scale=0.25]{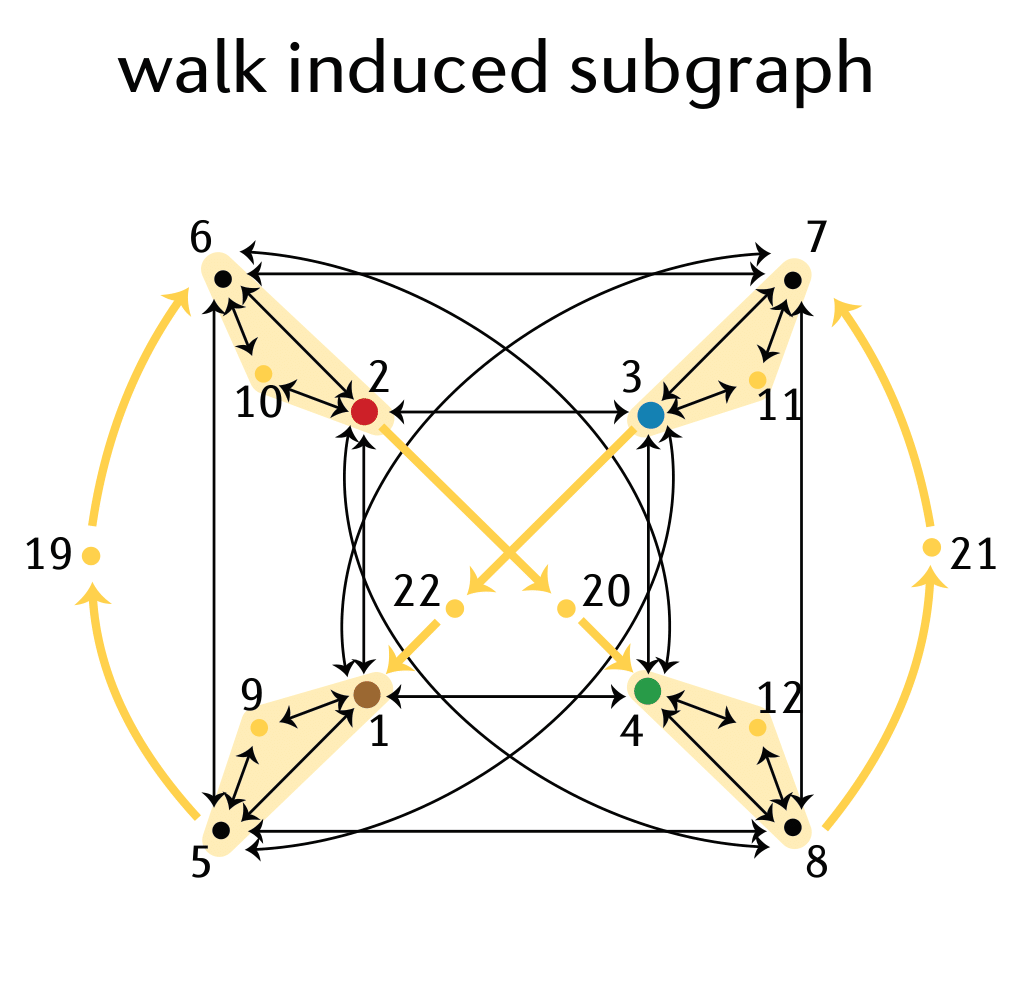}&\includegraphics[scale=0.25]{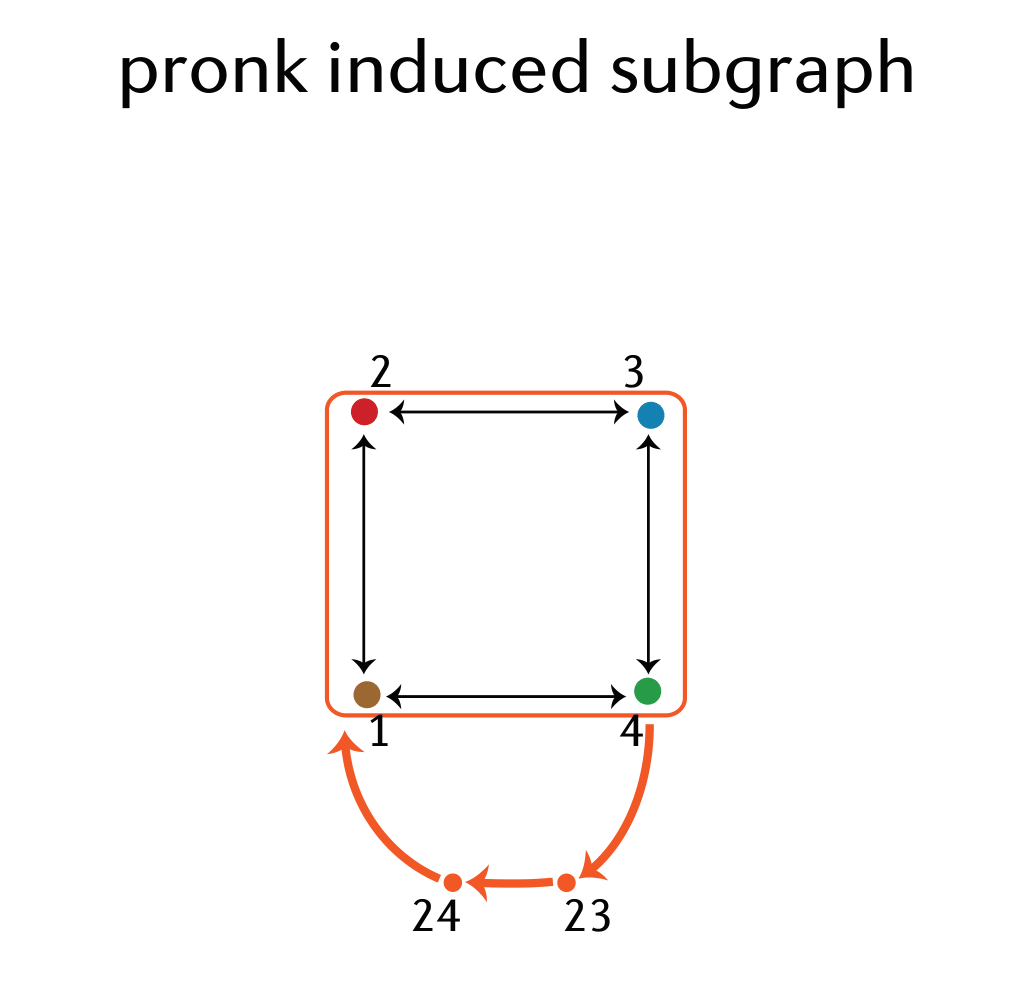}\\
			\hline
			\{\legone{1}, \legtwo{2}\}&\{\legone{1}, \legfour{4}\}&\{\legone{1}, \legtwo{2}\}&\{\legone{1}, \legtwo{2}\}&\{\legone{1}, \legtwo{2}, 5, 6\}\\
			\{\legone{1}, 7\}&\{\legone{1}, 7\}&\{\legone{1}, \legfour{4}\}&\{\legone{1}, \legfour{4}\}&\{\legone{1}, \legfour{4}, 5, 6\}\\
			\{\legtwo{2}, 8\}&\{\legtwo{2}, \legthree{3}\}&\{\legtwo{2}, \legthree{3}\}&\{\legone{1}, 7\}&\{\legtwo{2}, \legthree{3}, 5, 6\}\\
			\{\legthree{3}, \legfour{4}\}&\{\legtwo{2}, 8\}&\{\legthree{3}, \legfour{4}\}&\{\legtwo{2}, \legthree{3}\}&\{\legthree{3}, \legfour{4}, 5, 6\}\\
			\{\legthree{3}, 5\}&\{\legthree{3}, 5\}&\{5, 6\}&\{\legtwo{2}, 8\}& \\
			\{\legfour{4}, 6\}&\{\legfour{4}, 6\}&\{5, 8\}&\{\legthree{3}, \legfour{4}\}& \\
			\{5, 6\}&\{5, 8\}&\{6, 7\}&\{\legthree{3}, 5\}& \\
			\{7, 8\}&\{6, 7\}&\{7, 8\}&\{\legfour{4}, 6\}& \\
			& & &\{5, 6\}& \\
			& & &\{5, 8\}& \\
			& & &\{6, 7\}& \\
			& & &\{7, 8\}& \\
			\hline
		\end{tabular}
	}
	\label{tab:gt}
\end{table}

\begin{table}[h!]
	\caption[Gaits' $\FP_\text{core}(G)$]{Gaits' $\FP_\text{core}(G)$, built individually as explained in Chapter \ref{ch:CPGs}, and the resulting network with its $\FP_\text{core}(G)$.}
	\centering
		\resizebox{\columnwidth}{!}{%
	\begin{tabular}{c|c|c|c}
		Network $G$ & $|\FP(G)|$ & $|\FP_\text{core}(G)|$ & $\FP_\text{core}(G)$ \\
		\hline\hline
		\raisebox{-.5\totalheight}{\includegraphics[scale=0.25]{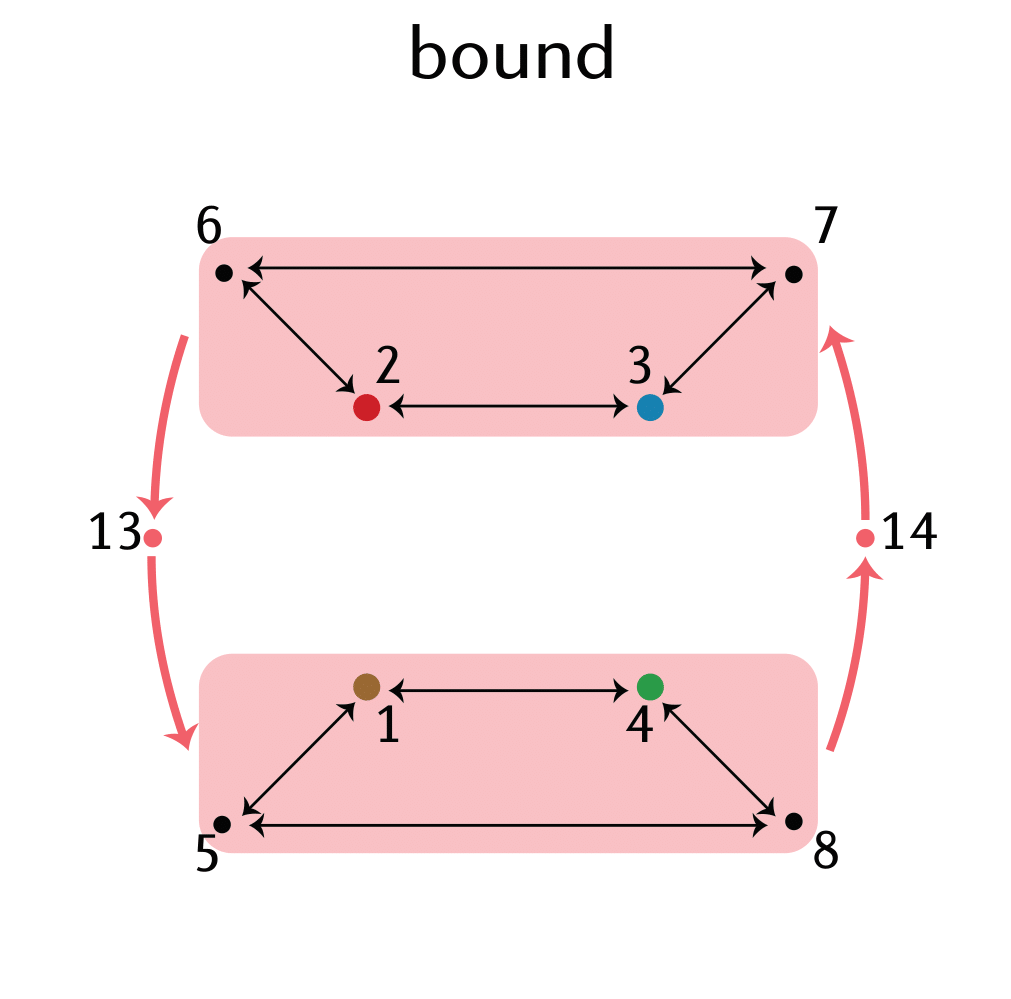}} & $81$ & $16$ & \makecell{\small $\sigma_1 \cup \{13\} \cup \sigma_2 \cup \{14\}$, \\
			where $\sigma_1 \in \{\{2,3\},\{2,6\},\{6,7\},\{3,7\}\}$ \\ 
			and $\sigma_2 \in \{\{1,4\},\{1,5\},\{5,8\},\{4,8\}\}$} \\
		\hline
		\raisebox{-.5\totalheight}{\includegraphics[scale=0.25]{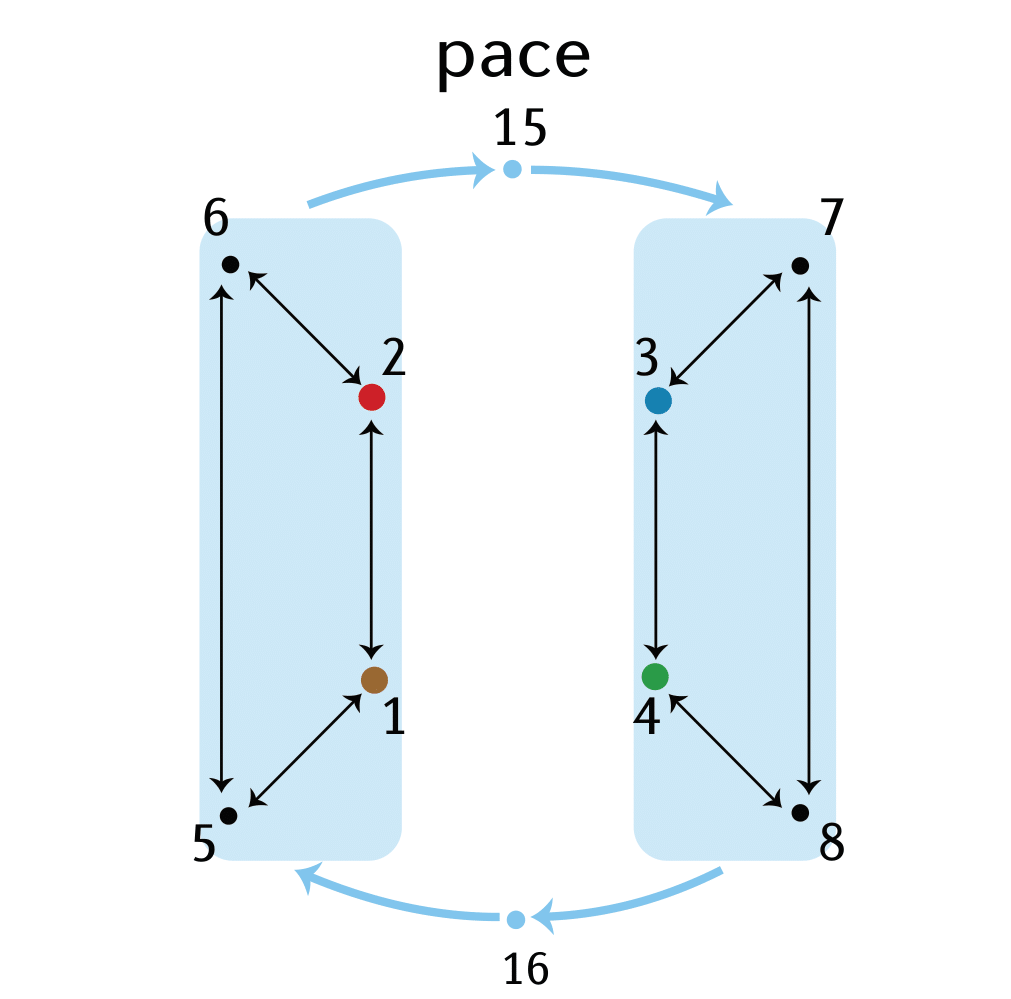}} & $81$ & $16$ & \makecell{\small $\sigma_1 \cup \{15\} \cup \sigma_2 \cup \{16\}$, \\
			where $\sigma_1 \in \{\{1,2\},\{2,6\},\{5,6\},\{5,1\}\}$ \\ 
			and $\sigma_2 \in \{\{3,4\},\{3,7\},\{7,8\},\{4,8\}\}$} \\
		\hline
		\raisebox{-.5\totalheight}{\includegraphics[scale=0.25]{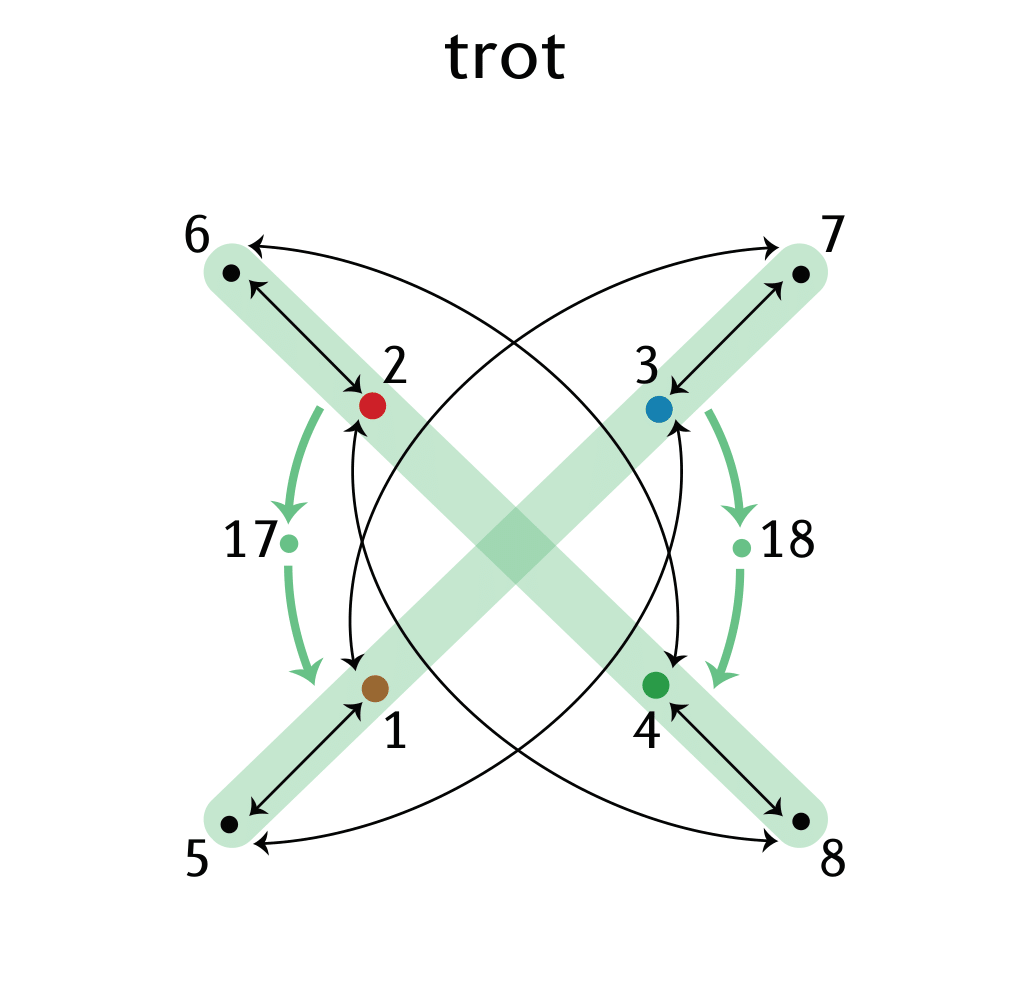}} & $81$ & $16$ & \makecell{\small $\sigma_1 \cup \{17\} \cup \sigma_2 \cup \{18\}$, \\
			where $\sigma_1 \in \{\{2,6\},\{2,8\},\{8,4\},\{4,6\}\}$ \\ 
			and $\sigma_2 \in \{\{1,5\},\{5,3\},\{3,7\},\{1,7\}\}$} \\
		\hline
		\raisebox{-.5\totalheight}{\includegraphics[scale=0.25]{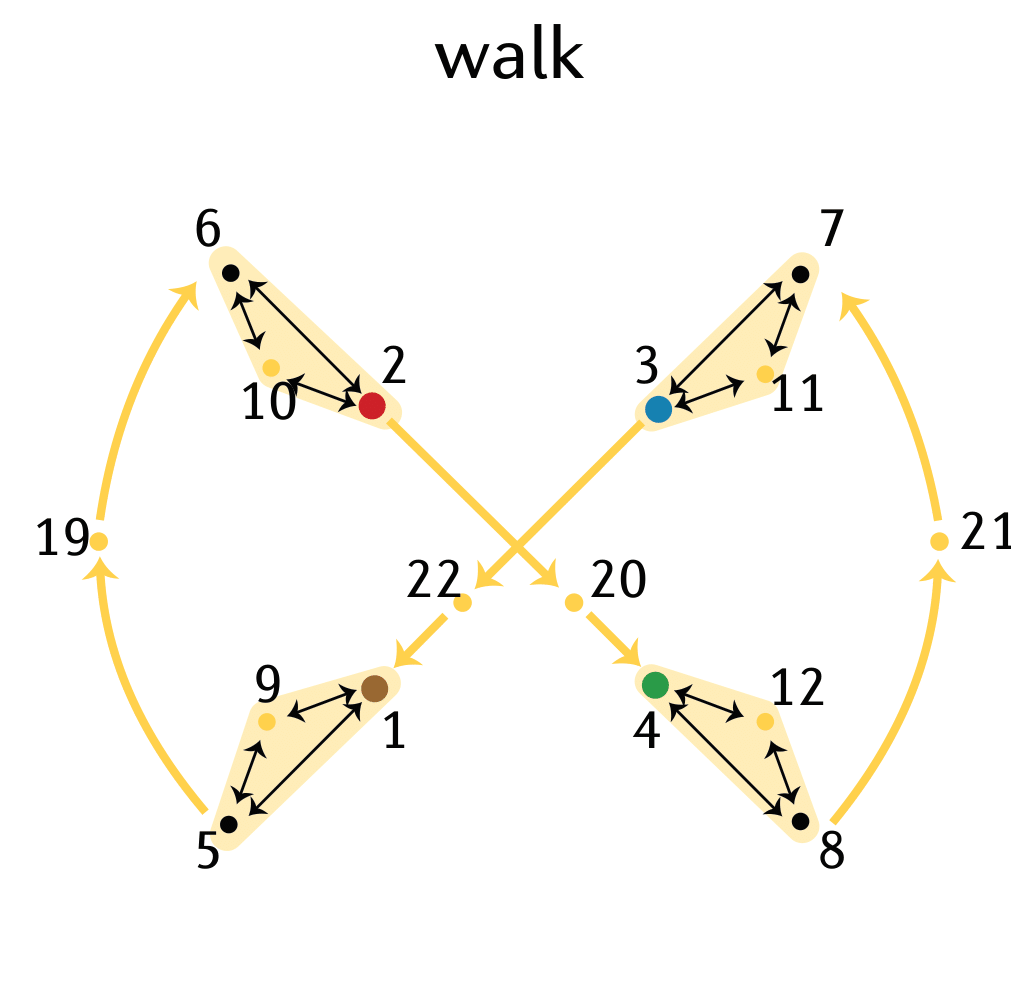}} & $1$ & $1$ & \makecell{\small $\{1,5,9\}\cup \{19\}\cup \{2,6,10\}\cup \{20\}\cup $ \\
			$\{4,8,12\}\cup \{21\}\cup \{3,7,11\}\cup \{22\}$} \\
		\hline
		\raisebox{-.5\totalheight}{\includegraphics[scale=0.25]{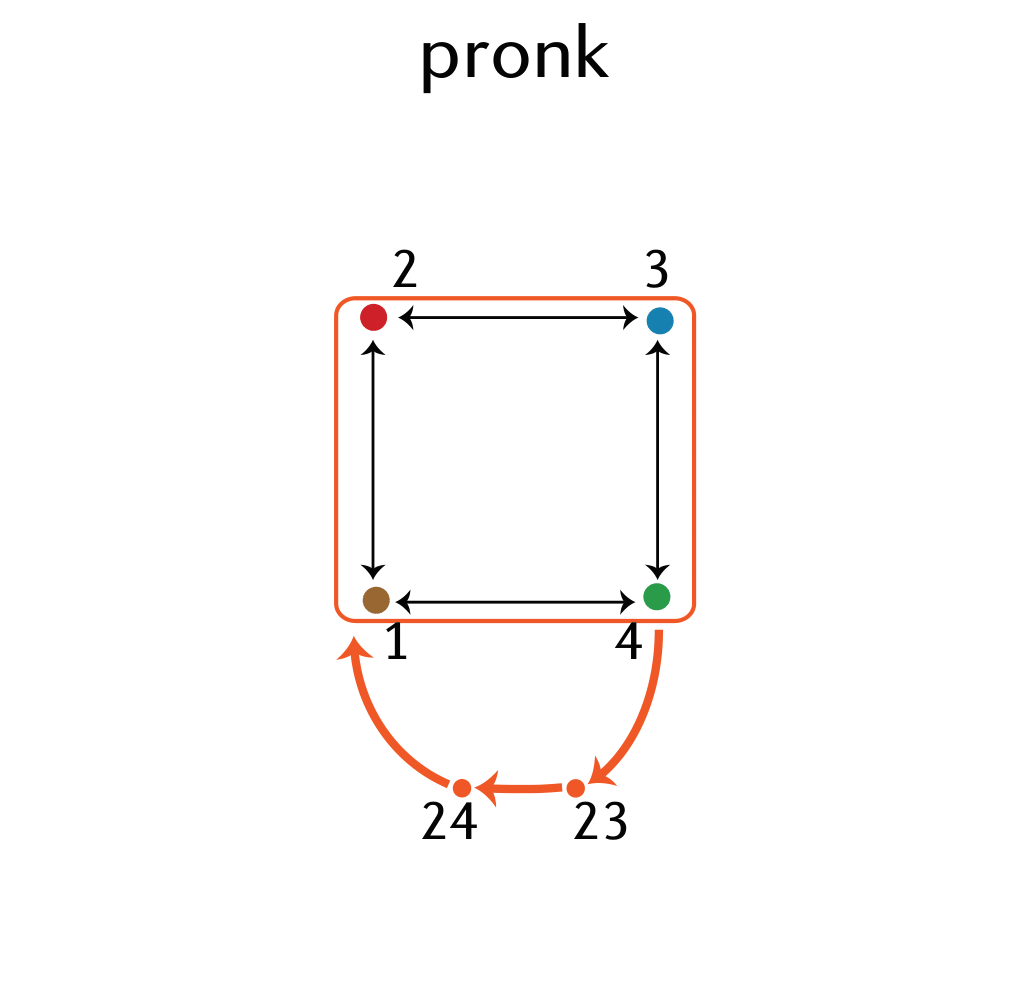}} & $9$ & $4$ & \makecell{\small $\sigma_1 \cup \{23\} \cup \{24\}$, \\
			where $\sigma_1 \in \{\{1,2\},\{2,3\},\{3,4\},\{4,1\}\}$} \\ 
		\hline
		\raisebox{-.5\totalheight}{\includegraphics[scale=0.25]{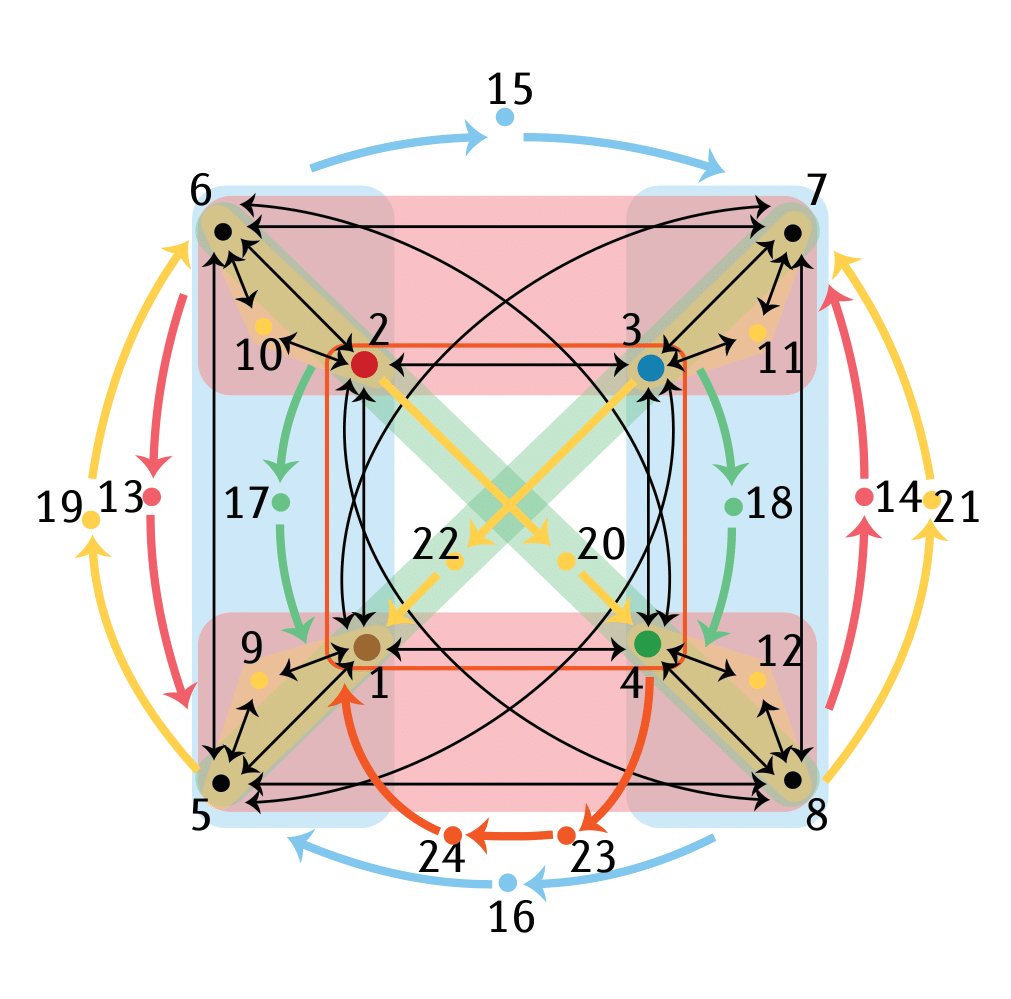}} & $875$ & $8$ & \makecell{\small $\{1, 2, 23, 24\}, \{1, 4, 23, 24\}, \{2, 3, 23, 24\}, \{3, 4, 23, 24\},$ \\
			\small $\{1, 6, 7, 8, 13, 16, 18\}, \{2, 5, 7, 8, 14, 16, 17\},$ \\
			\small $\{3, 5, 6, 8, 14, 15, 18\}, \{4, 5, 6, 7, 13, 15, 17\}$} 
	\end{tabular}
}
	\label{tab:quadrupedgaitsFP}
\end{table}

\begin{table}[!h]
\caption[Minimal sets in $\FP(G)$ for each isolated gait and merged gaits network]{Minimal sets in $\FP(G)$ for each isolated gait and merged gaits network. Nodes are colored according to Figure \ref{fig:five-gait-network-and-transitions}A. Bolded nodes (1 - 4) are shared by all gaits. Leg nodes in bold.}
\centering
\resizebox{\columnwidth}{!}{%
\begin{tabular}{c|c|c|c|c}
\includegraphics[scale=0.25]{Figures-png/bound_graph-1.png}&\includegraphics[scale=0.25]{Figures-png/pace_graph-1.png}&\includegraphics[scale=0.25]{Figures-png/trot_graph-1.png}&\includegraphics[scale=0.25]{Figures-png/walk_graph-1.png}&\includegraphics[scale=0.25]{Figures-png/pronk_graph-1.png}\\
\hline 
\hline 
\multicolumn{5}{c}{Minimal sets in $\FP(G)$ for isolated gaits} \\
\hline
\{\legtwo{2},\legthree{3},\legone{1},\legfour{4},\boundaux{13},\boundaux{14}\}&\{\legone{1},\legtwo{2},\legthree{3},\legfour{4},\paceaux{15},\paceaux{16}\}&\{\legtwo{2},6,\legone{1},5,\trotaux{17},\trotaux{18}\}&\{\legone{1},5,\walkaux{9},\walkaux{19},\legtwo{2},6,\walkaux{10},\walkaux{20},&\{\legone{1},\legtwo{2},\pronkaux{23},\pronkaux{24}\}\\
\{\legtwo{2},\legthree{3},\legone{1},5,\boundaux{13},\boundaux{14}\}&\{\legone{1},\legtwo{2},\legthree{3},7,\paceaux{15},\paceaux{16}\}&\{\legtwo{2},8,\legone{1},5,\trotaux{17},\trotaux{18}\}& \legfour{4},8,\walkaux{12},\walkaux{21},\legthree{3},7,\walkaux{11},\walkaux{22}\}&\{\legone{1},\legfour{4},\pronkaux{23},\pronkaux{24}\}\\
\{\legtwo{2},6,\legone{1},\legfour{4},\boundaux{13},\boundaux{14}\}&\{\legone{1},\legtwo{2},\legfour{4},8,\paceaux{15},\paceaux{16}\}&\{\legtwo{2},6,\legone{1},7,\trotaux{17},\trotaux{18}\}& &\{\legtwo{2},\legthree{3},\pronkaux{23},\pronkaux{24}\}\\
\{\legtwo{2},6,\legone{1},5,\boundaux{13},\boundaux{14}\}&\{\legone{1},\legtwo{2},7,8,\paceaux{15},\paceaux{16}\}&\{\legtwo{2},8,\legone{1},7,\trotaux{17},\trotaux{18}\}& &\{\legthree{3},\legfour{4},\pronkaux{23},\pronkaux{24}\}\\
\{\legthree{3},7,\legone{1},\legfour{4},\boundaux{13},\boundaux{14}\}&\{\legone{1},5,\legthree{3},\legfour{4},\paceaux{15},\paceaux{16}\}&\{\legfour{4},6,\legone{1},5,\trotaux{17},\trotaux{18}\}& & \\
\{\legthree{3},7,\legone{1},5,\boundaux{13},\boundaux{14}\}&\{\legone{1},5,\legthree{3},7,\paceaux{15},\paceaux{16}\}&\{\legfour{4},8,\legone{1},5,\trotaux{17},\trotaux{18}\}& & \\
\{6,7,\legone{1},\legfour{4},\boundaux{13},\boundaux{14}\}&\{\legone{1},5,\legfour{4},8,\paceaux{15},\paceaux{16}\}&\{\legfour{4},6,\legone{1},7,\trotaux{17},\trotaux{18}\}& & \\
\{6,7,\legone{1},5,\boundaux{13},\boundaux{14}\}&\{\legone{1},5,7,8,\paceaux{15},\paceaux{16}\}&\{\legfour{4},8,\legone{1},7,\trotaux{17},\trotaux{18}\}& & \\
\{\legtwo{2},\legthree{3},\legfour{4},8,\boundaux{13},\boundaux{14}\}&\{\legtwo{2},6,\legthree{3},\legfour{4},\paceaux{15},\paceaux{16}\}&\{\legtwo{2},6,\legthree{3},5,\trotaux{17},\trotaux{18}\}& & \\
\{\legtwo{2},\legthree{3},5,8,\boundaux{13},\boundaux{14}\}&\{\legtwo{2},6,\legthree{3},7,\paceaux{15},\paceaux{16}\}&\{\legtwo{2},8,\legthree{3},5,\trotaux{17},\trotaux{18}\}& & \\
\{\legtwo{2},6,\legfour{4},8,\boundaux{13},\boundaux{14}\}&\{\legtwo{2},6,\legfour{4},8,\paceaux{15},\paceaux{16}\}&\{\legtwo{2},6,\legthree{3},7,\trotaux{17},\trotaux{18}\}& & \\
\{\legtwo{2},6,5,8,\boundaux{13},\boundaux{14}\}&\{\legtwo{2},6,7,8,\paceaux{15},\paceaux{16}\}&\{\legtwo{2},8,\legthree{3},7,\trotaux{17},\trotaux{18}\}& & \\
\{\legthree{3},7,\legfour{4},8,\boundaux{13},\boundaux{14}\}&\{\legthree{3},\legfour{4},5,6,\paceaux{15},\paceaux{16}\}&\{\legfour{4},6,\legthree{3},5,\trotaux{17},\trotaux{18}\}& & \\
\{\legthree{3},7,5,8,\boundaux{13},\boundaux{14}\}&\{5,6,\legthree{3},7,\paceaux{15},\paceaux{16}\}&\{\legfour{4},8,\legthree{3},5,\trotaux{17},\trotaux{18}\}& & \\
\{6,7,\legfour{4},8,\boundaux{13},\boundaux{14}\}&\{5,6,\legfour{4},8,\paceaux{15},\paceaux{16}\}&\{\legfour{4},6,\legthree{3},7,\trotaux{17},\trotaux{18}\}& & \\
\{6,7,5,8,\boundaux{13},\boundaux{14}\}&\{5,6,7,8,\paceaux{15},\paceaux{16}\}&\{\legthree{3},7,\legfour{4},8,\trotaux{17},\trotaux{18}\}& & \\
\hline 
\hline 
\multicolumn{5}{c}{Minimal sets in $\FP(G)$ for supernetwork (sorted according to which gait-specific node participates in it)} \\
\hline
\{\legone{1},5,6,7,\boundaux{13}\}&\{\legone{1},5,7,8,\paceaux{16}\}&\{\legtwo{2},8,\legone{1},5,\trotaux{17}\}& & \{\legone{1},\legtwo{2},\pronkaux{23},\pronkaux{24}\}\\
\{\legtwo{2},6,5,8,\boundaux{14}\}&\{\legtwo{2},6,7,8,\paceaux{16}\}&\{\legtwo{2},6,\legone{1},7,\trotaux{18}\}& & \{\legone{1},\legfour{4},\pronkaux{23},\pronkaux{24}\}\\
\{\legthree{3},7,5,8,\boundaux{14}\}&\{5,6,\legthree{3},7,\paceaux{15}\}&\{\legfour{4},6,\legone{1},5,\trotaux{17}\}& & \{\legtwo{2},\legthree{3},\pronkaux{23},\pronkaux{24}\}\\
\{6,7,\legfour{4},8,\boundaux{13}\}&\{5,6,\legfour{4},8,\paceaux{15}\}&\{\legfour{4},8,\legone{1},7,\trotaux{18}\}& & \{\legthree{3},\legfour{4},\pronkaux{23},\pronkaux{24}\}\\
& &\{\legtwo{2},6,\legthree{3},5,\trotaux{18}\}& & \\
& &\{\legtwo{2},8,\legthree{3},7,\trotaux{17}\}& & \\
& &\{\legthree{3},5,\legfour{4},8,\trotaux{18}\}& & \\
& &\{\legfour{4},6,\legthree{3},7,\trotaux{17}\}& & \\
\hline 
\multicolumn{5}{c}{supports sharing multiple gait-specific auxiliary nodes} \\
\hline 
\multicolumn{5}{c}{\{\legone{1},6,7,8,\boundaux{13},\boundaux{14},\trotaux{18}\}} \\
\multicolumn{5}{c}{\{\legtwo{2},5,7,8,\boundaux{14},\paceaux{16},\trotaux{17}\}} \\
\multicolumn{5}{c}{\{\legthree{3},5,6,8,\boundaux{14},\paceaux{15},\trotaux{18}\}} \\
\multicolumn{5}{c}{\{\legfour{4},5,6,7,\boundaux{13},\paceaux{15},\trotaux{17}\}} \\
\hline 
\multicolumn{5}{c}{supports not involving any gait-specific auxiliary nodes} \\
\hline 
\multicolumn{5}{c}{\{\legone{1},\legtwo{2},7,8\}} \\
\multicolumn{5}{c}{\{\legone{1},\legfour{4},6,7\}} \\
\multicolumn{5}{c}{\{\legtwo{2},\legthree{3},5,8\}} \\
\multicolumn{5}{c}{\{\legthree{3},\legfour{4},5,6\}} \\
\multicolumn{5}{c}{\{5,6,7,8\}} \\
\hline
\end{tabular}
}
\label{tab:isolated-gaits-fp}
\end{table}

\begin{figure}[h]
  \begin{center}
\includegraphics[width=\textwidth]{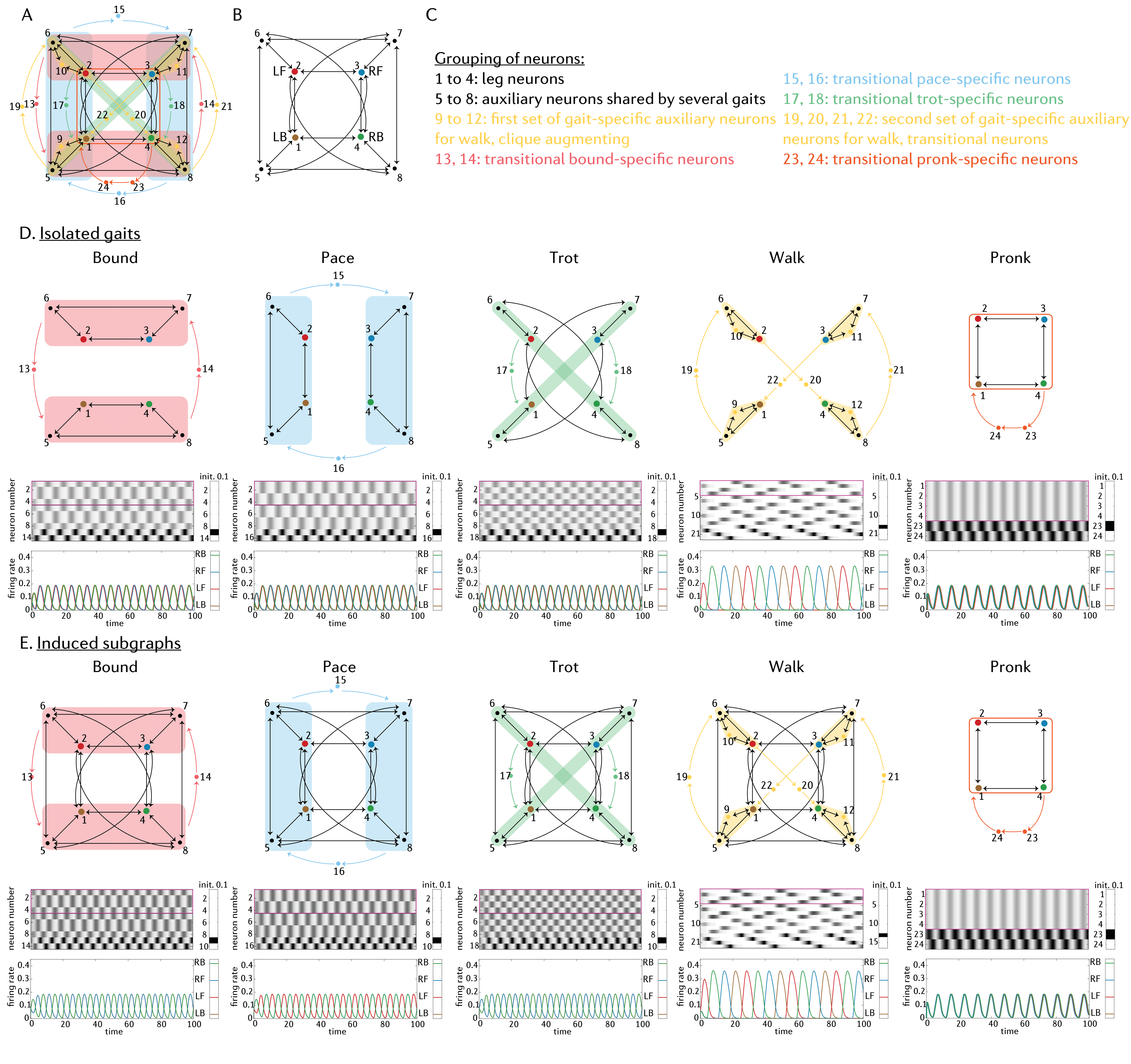}
  \end{center}
  \caption[Simulation of five gaits isolated and induced]{Simulation of five gaits isolated (as opposed to embedded into super network), and as \emph{induced} subgraphs of 5-gait network.
  (A) 5-gait network. 
  (B) Induced subgraph of the 5-gait network nodes 1 - 8. These nodes are shared by two or more gaits.
  (C) Grouping of neurons by type. Types are assigned on the basis of function.
  (D) Graphs, greyscales and rate curves of all isolated gaits. These are not induced subgraphs. 
  (E) Graphs, greyscales and rate curves of induced subgraphs for all gaits. The attractors corresponding to each gait are still perfectly preserved.}
\label{fig:all-gaits-isolated}
\end{figure}
} 

	\begin{singlespace}
	\bibliographystyle{plain}
	\addcontentsline{toc}{chapter}{Bibliography}
	\bibliography{Biblio-Database}

\begin{thebibliography}{10}

\bibitem{purves2001}
{\em Neuroscience}.
\newblock Sinauer Associates, 2 edition, 2001.

\bibitem{Alexander1984}
R.~McN. Alexander.
\newblock The gaits of bipedal and quadrupedal animals.
\newblock {\em The International Journal of Robotics Research}, 3(2):49--59,
  1984.

\bibitem{Amit-ANNs}
Daniel~J. Amit.
\newblock {\em Modeling brain function: {T}he world of attractor neural
  networks}.
\newblock Cambridge University Press, 1989.

\bibitem{Inhibition-hippocampus}
M.~Arriaga and E.~B. Han.
\newblock Dedicated hippocampal inhibitory networks for locomotion and
  immobility.
\newblock {\em J. Neurosci.}, 37:9222--9238, 2017.

\bibitem{Ashwin2016}
P.~Ashwin, S.~Coombes, and R.~Nicks.
\newblock {{M}athematical {F}rameworks for {O}scillatory {N}etwork {D}ynamics
  in {N}euroscience}.
\newblock {\em J Math Neurosci}, 6(1):2, Dec 2016.

\bibitem{Horacio-paper}
A.~Bel, R.~Cobiaga, W.~Reartes, and H.~G. Rotstein.
\newblock Periodic solutions in threshold-linear networks and their
  entrainment.
\newblock {\em SIAM J. Appl. Dyn. Syst.}, 20(3):1177--1208, 2021.

\bibitem{Biswas2022}
Tirthabir Biswas and James~E. Fitzgerald.
\newblock Geometric framework to predict structure from function in neural
  networks.
\newblock {\em Phys. Rev. Res.}, 4:023255, Jun 2022.

\bibitem{Buono2001}
P.~L. Buono and M.~Golubitsky.
\newblock {{M}odels of central pattern generators for quadruped locomotion.
  {I}. {P}rimary gaits}.
\newblock {\em J Math Biol}, 42(4):291--326, Apr 2001.

\bibitem{BUTTNER20061}
U.~B\"{u}ttner and J.A. B\"{u}ttner-Ennever.
\newblock Present concepts of oculomotor organization.
\newblock In J.A. B\"{u}ttner-Ennever, editor, {\em Neuroanatomy of the
  Oculomotor System}, volume 151 of {\em Progress in Brain Research}, pages
  1--42. Elsevier, 2006.

\bibitem{Cain2012}
Nicholas Cain and Eric Shea-Brown.
\newblock Computational models of decision making: integration, stability, and
  noise.
\newblock {\em Current Opinion in Neurobiology}, 22(6):1047--1053, 2012.
\newblock Decision making.

\bibitem{tesis-joaquin}
Carlos~Joaquín Castañeda~Castro.
\newblock Ctlns definidos por torneos, 2023.

\bibitem{Collins1993}
J.~J. Collins and I.~N. Stewart.
\newblock Coupled nonlinear oscillators and the symmetries of animal gaits.
\newblock {\em Journal of Nonlinear Science}, 3(1):349--392, Dec 1993.

\bibitem{net-encoding}
C.~Curto, A.~Degeratu, and V.~Itskov.
\newblock Encoding binary neural codes in networks of threshold-linear neurons.
\newblock {\em Neural Comput.}, 25:2858--2903, 2013.

\bibitem{stable-fp-paper}
C.~Curto, J.~Geneson, and K.~Morrison.
\newblock Stable fixed points of combinatorial threshold-linear networks.
\newblock Available at \verb!https://arxiv.org/abs/1909.02947!

\bibitem{fp-paper}
C.~Curto, J.~Geneson, and K.~Morrison.
\newblock Fixed points of competitive threshold-linear networks.
\newblock {\em Neural Comput.}, 31(1):94--155, 2019.

\bibitem{Curto2016}
Carina Curto and Katherine Morrison.
\newblock Pattern completion in symmetric threshold-linear networks.
\newblock {\em Neural Computation}, 28:2825--2852, 12 2016.

\bibitem{Notices}
Carina Curto and Katherine Morrison.
\newblock Graph rules for recurrent neural network dynamics.
\newblock {\em Notices of the American Mathematical Society}, 70(04):536–551,
  2023.

\bibitem{extended-notices}
Carina Curto and Katherine Morrison.
\newblock Graph rules for recurrent neural network dynamics: extended version,
  2023.

\bibitem{Dayan2001}
1965 Dayan, Peter and L.~F. Abbott.
\newblock {\em Theoretical neuroscience : computational and mathematical
  modeling of neural systems}.
\newblock Computational neuroscience. Massachusetts Institute of Technology
  Press, Cambridge, Mass., 2001.

\bibitem{Durstewitz2023}
Daniel Durstewitz, Georgia Koppe, and Max~Ingo Thurm.
\newblock Reconstructing computational system dynamics from neural data with
  recurrent neural networks.
\newblock {\em Nature Reviews Neuroscience}, 24(11):693--710, Nov 2023.

\bibitem{Dutta2019}
Sourav Dutta, Abhinav Parihar, Abhishek Khanna, Jorge Gomez, Wriddhi
  Chakraborty, Matthew Jerry, Benjamin Grisafe, Arijit Raychowdhury, and Suman
  Datta.
\newblock Programmable coupled oscillators for synchronized locomotion.
\newblock {\em Nature Communications}, 10(1):3299, Jul 2019.

\bibitem{Dzeladini2018}
Florin Dzeladini, Nadine Ait-Bouziad, and Auke Ijspeert.
\newblock {\em CPG-Based Control of Humanoid Robot Locomotion}, pages 1--35.
\newblock Springer Netherlands, Dordrecht, 2018.

\bibitem{Ermentrout2010}
G.~Bard Ermentrout and David~H. Terman.
\newblock {\em Firing Rate Models}, pages 331--367.
\newblock Springer New York, New York, NY, 2010.

\bibitem{Feudel2018}
Ulrike Feudel, Alexander~N. Pisarchik, and Kenneth Showalter.
\newblock Multistability and tipping: From mathematics and physics to climate
  and brain—minireview and preface to the focus issue.
\newblock {\em Chaos: An Interdisciplinary Journal of Nonlinear Science},
  28(3):033501, 2018.

\bibitem{Gambaryan1974}
P.~P. Gambaryan.
\newblock {\em How mammals run: anatomical adaptations}.
\newblock Wiley, New York, 1974.

\bibitem{Gjorgjieva2016-hl}
Julijana Gjorgjieva, Guillaume Drion, and Eve Marder.
\newblock Computational implications of biophysical diversity and multiple
  timescales in neurons and synapses for circuit performance.
\newblock {\em Curr Opin Neurobiol}, 37:44--52, 4 2016.

\bibitem{marder2021}
Jean Goaillard and Eve Marder.
\newblock Ion channel degeneracy, variability, and covariation in neuron and
  circuit resilience.
\newblock {\em Annu Rev Neurosci}, 44:335--357, 7 2021.

\bibitem{Goaillard2009}
Jean Goaillard, Adam Taylor, David Schulz, and Eve Marder.
\newblock Functional consequences of animal-to-animal variation in circuit
  parameters.
\newblock {\em Nature Neuroscience}, 12:1424--1430, 11 2009.

\bibitem{goldman_robust_2003}
M.~S. Goldman.
\newblock Robust {Persistent} {Neural} {Activity} in a {Model} {Integrator}
  with {Multiple} {Hysteretic} {Dendrites} per {Neuron}.
\newblock {\em Cerebral Cortex}, 13(11):1185--1195, November 2003.

\bibitem{goldman_neural_2010}
Mark~S Goldman, A.~Compte, and X.~J. Wang.
\newblock {\em Neural {Integrator} {Models}}, pages 165--178.
\newblock Elsevier Ltd, 2010.

\bibitem{Golubitsky-nature}
M.~Golubitsky, I.~Stewart, P-L. Buono, and J.J. Collins.
\newblock Symmetry in locomotor central pattern generators and animal gaits.
\newblock {\em Nature}, 401:693--695, 1999.

\bibitem{Golubitsky1998}
Martin Golubitsky, Ian Stewart, Pietro-Luciano Buono, and J.J. Collins.
\newblock A modular network for legged locomotion.
\newblock {\em Physica D: Nonlinear Phenomena}, 115(1):56--72, 1998.

\bibitem{Graybiel1998}
Ann~M. Graybiel.
\newblock The basal ganglia and chunking of action repertoires.
\newblock {\em Neurobiology of Learning and Memory}, 70(1):119--136, 1998.

\bibitem{CPG-review}
S.~Grillner and P.~Wall\'{e}n.
\newblock Cellular bases of a vertebrate locomotor system -- steering,
  intersegmental and segmental co-ordination and sensory control.
\newblock {\em Brain Res. Rev.}, 40:92--106, 2002.

\bibitem{Seung-Nature}
R.~H. Hahnloser, R.~Sarpeshkar, M.A. Mahowald, R.J. Douglas, and H.S. Seung.
\newblock Digital selection and analogue amplification coexist in a
  cortex-inspired silicon circuit.
\newblock {\em Nature}, 405:947--951, 2000.

\bibitem{HahnSeungSlotine}
R.~H. Hahnloser, H.S. Seung, and J.J. Slotine.
\newblock Permitted and forbidden sets in symmetric threshold-linear networks.
\newblock {\em Neural Comput.}, 15(3):621--638, 2003.

\bibitem{Hahnloser2003-ma}
Richard Hahnloser, H.~Sebastian Seung, and Jean-Jacques Slotine.
\newblock Permitted and forbidden sets in symmetric threshold-linear networks.
\newblock {\em Neural Comput}, 15:621--638, 3 2003.

\bibitem{Hartline1958}
H.~K. HARTLINE and F.~RATLIFF.
\newblock {{S}patial summation of inhibitory influences in the eye of
  {L}imulus, and the mutual interaction of receptor units}.
\newblock {\em J Gen Physiol}, 41(5):1049--1066, May 1958.

\bibitem{Hopfield1}
J.J. Hopfield.
\newblock Neural networks and physical systems with emergent collective
  computational abilities.
\newblock {\em Proc. Natl. Acad. Sci.}, 79(8):2554--2558, 1982.

\bibitem{Hopfield2}
J.J. Hopfield.
\newblock Neurons with graded response have collective computational properties
  like those of two-sate neurons.
\newblock {\em Proc. Natl. Acad. Sci.}, 81:3088--3092, 1984.

\bibitem{Ijspeert2008}
Auke~Jan Ijspeert.
\newblock Central pattern generators for locomotion control in animals and
  robots: A review.
\newblock {\em Neural Networks}, 21(4):642--653, 2008.
\newblock Robotics and Neuroscience.

\bibitem{Kamaleddin2022-rx}
Mahdi Kamaleddin.
\newblock Degeneracy in the nervous system: from neuronal excitability to
  neural coding.
\newblock {\em Bioessays}, 44:e2100148, 1 2022.

\bibitem{Khona2022}
Mikail Khona and Ila~R. Fiete.
\newblock Attractor and integrator networks in the brain.
\newblock {\em Nature Reviews Neuroscience}, 23(12):744--766, Dec 2022.

\bibitem{Kornysheva2014}
Katja Kornysheva and Jörn Diedrichsen.
\newblock Human premotor areas parse sequences into their spatial and temporal
  features.
\newblock {\em eLife}, 3:e03043, aug 2014.

\bibitem{Koulakov2002}
Alexei~A. Koulakov, Sridhar Raghavachari, Adam Kepecs, and John~E. Lisman.
\newblock Model for a robust neural integrator.
\newblock {\em Nature Neuroscience}, 5(8):775--782, Aug 2002.

\bibitem{Lappalainen2023}
Janne~K. Lappalainen, Fabian~D. Tschopp, Sridhama Prakhya, Mason McGill,
  Aljoscha Nern, Kazunori Shinomiya, Shin ya~Takemura, Eyal Gruntman, Jakob~H.
  Macke, and Srinivas~C. Turaga.
\newblock Connectome-constrained deep mechanistic networks predict neural
  responses across the fly visual system at single-neuron resolution.
\newblock {\em bioRxiv}, 2023.

\bibitem{Lashley1951}
Karl~S. Lashley.
\newblock The problem of serial order in behavior.
\newblock 1951.

\bibitem{Latorre2019}
Roberto Latorre, Pablo Varona, and Mikhail~I. Rabinovich.
\newblock Rhythmic control of oscillatory sequential dynamics in heteroclinic
  motifs.
\newblock {\em Neurocomputing}, 331:108--120, 2019.

\bibitem{Logiaco2021}
Laureline Logiaco, L.F. Abbott, and Sean Escola.
\newblock Thalamic control of cortical dynamics in a model of flexible motor
  sequencing.
\newblock {\em Cell Reports}, 35(9):109090, 2021.

\bibitem{Long2010}
Michael~A. Long, Dezhe~Z. Jin, and Michale~S. Fee.
\newblock Support for a synaptic chain model of neuronal sequence generation.
\newblock {\em Nature}, 468(7322):394--399, Nov 2010.

\bibitem{Marder-CPG}
E.~Marder and D.~Bucher.
\newblock Central pattern generators and the control of rhythmic movements.
\newblock {\em Curr. Bio.}, 11(23):R986--996, 2001.

\bibitem{Mazurek2003}
Mark~E. Mazurek, Jamie~D. Roitman, Jochen Ditterich, and Michael~N. Shadlen.
\newblock {A Role for Neural Integrators in Perceptual Decision Making}.
\newblock {\em Cerebral Cortex}, 13(11):1257--1269, 11 2003.

\bibitem{book-chapter}
K.~Morrison and C.~Curto.
\newblock {\em Predicting neural network dynamics via graphical analysis}.
\newblock Book chapter in Algebraic and Combinatorial Computational Biology,
  edited by R. Robeva and M. Macaulay. Elsevier, 2018.

\bibitem{CTLN-preprint}
K.~Morrison, A.~Degeratu, V.~Itskov, and C.~Curto.
\newblock Diversity of emergent dynamics in competitive threshold-linear
  networks.
\newblock Available at \verb!https://arxiv.org/abs/1605.04463!

\bibitem{Muybridge1891}
Eadweard. Muybridge.
\newblock {\em Attitudes of Animals in Motion, Illustrated with the
  Zoopraxiscope}.
\newblock University of Pennsylvania, 1891.

\bibitem{Nikitchenko2008}
Maxim Nikitchenko and Alexei Koulakov.
\newblock Neural integrator: a sandpile model.
\newblock {\em Neural computation}, 20(10):2379--2417, Oct 2008.
\newblock 18533820[pmid].

\bibitem{OliveiraSantos2023}
Sara Oliveira~Santos, Nils Tack, Yunxing Su, Francisco Cuenca-Jim{\'e}nez,
  Oscar Morales-Lopez, P.~Antonio Gomez-Valdez, and Monica~M. Wilhelmus.
\newblock Pleobot: a modular robotic solution for metachronal swimming.
\newblock {\em Scientific Reports}, 13(1):9574, Jun 2023.

\bibitem{Panchin1995neuronalmechanisms}
Y.~V. Panchin, Y.~I. Arshavsky, T.~G. Deliagina, L.~B. Popova, and G.~N.
  Orlovsky.
\newblock Control of locomotion in marine mollusk clione limacina. ix. neuronal
  mechanisms of spatial orientation.
\newblock {\em Journal of Neurophysiology}, 73(5):1924–1937, 1995.

\bibitem{Parmelee2022}
Caitlyn Parmelee, Juliana~Londono Alvarez, Carina Curto, and Katherine
  Morrison.
\newblock Sequential attractors in combinatorial threshold-linear networks.
\newblock {\em SIAM Journal on Applied Dynamical Systems}, 21(2):1597--1630,
  2022.

\bibitem{new-cores-paper}
Caitlyn Parmelee, Joaquin Castañeda, Katherine Morrison, and Carina Curto.
\newblock New core motifs paper.

\bibitem{rule-of-thumb}
Caitlyn Parmelee, Samantha Moore, Katherine Morrison, and Carina Curto.
\newblock Core motifs predict dynamic attractors in combinatorial
  threshold-linear networks.
\newblock {\em PLOS ONE}, 17(3):1--21, 03 2022.

\bibitem{Penn2016}
Yaron Penn, Menahem Segal, and Elisha Moses.
\newblock Network synchronization in hippocampal neurons.
\newblock {\em Proceedings of the National Academy of Sciences},
  113(12):3341--3346, 2016.

\bibitem{Pisarchik2014}
Alexander~N. Pisarchik and Ulrike Feudel.
\newblock Control of multistability.
\newblock {\em Physics Reports}, 540(4):167--218, 2014.
\newblock Control of multistability.

\bibitem{Prinz2004}
Astrid~A. Prinz, Dirk Bucher, and Eve Marder.
\newblock Similar network activity from disparate circuit parameters.
\newblock {\em Nature Neuroscience}, 7:1345--1352, 12 2004.

\bibitem{Mikhail2018}
Mikhail~I. Rabinovich and Pablo Varona.
\newblock Discrete sequential information coding: Heteroclinic cognitive
  dynamics.
\newblock {\em Frontiers in Computational Neuroscience}, 12, 2018.

\bibitem{Sakai2003}
Katsuyuki Sakai, Katsuya Kitaguchi, and Okihide Hikosaka.
\newblock Chunking during human visuomotor sequence learning.
\newblock {\em Exp. Brain Res.}, 152(2):229--242, September 2003.

\bibitem{AppendixE}
H.S. Seung and R.~Yuste.
\newblock {\em Principles of Neural Science}, chapter Appendix {E}: Neural
  networks, pages 1581--1600.
\newblock McGraw-Hill Education/Medical, 5th edition, 2012.

\bibitem{SEUNG2000259}
H.Sebastian Seung, Daniel~D. Lee, Ben~Y. Reis, and David~W. Tank.
\newblock Stability of the memory of eye position in a recurrent network of
  conductance-based model neurons.
\newblock {\em Neuron}, 26(1):259--271, 2000.

\bibitem{Cui2020}
Cui Su and Jun Pang.
\newblock Sequential temporary and permanent control of boolean networks.
\newblock In {\em Computational Methods in Systems Biology: 18th International
  Conference, CMSB 2020, Konstanz, Germany, September 23–25, 2020,
  Proceedings}, page 234–251, Berlin, Heidelberg, 2020. Springer-Verlag.

\bibitem{Tsodyks1997}
M.~V. Tsodyks, W.~E. Skaggs, T.~J. Sejnowski, and B.~L. McNaughton.
\newblock {{P}aradoxical effects of external modulation of inhibitory
  interneurons}.
\newblock {\em J Neurosci}, 17(11):4382--4388, Jun 1997.

\bibitem{Varona2002}
Pablo Varona, Mikhail~I. Rabinovich, Allen~I. Selverston, and Yuri~I.
  Arshavsky.
\newblock Winnerless competition between sensory neurons generates chaos: A
  possible mechanism for molluscan hunting behavior.
\newblock {\em Chaos: An Interdisciplinary Journal of Nonlinear Science},
  12(3):672--677, 2002.

\bibitem{Suresh2010}
Suresh Vasa, Tao Ma, Kiran~V. Byadarhaly, Mithun Perdoor, and Ali~A. Minai.
\newblock A spiking neural model for the spatial coding of cognitive response
  sequences.
\newblock In {\em 2010 IEEE 9th International Conference on Development and
  Learning}, pages 140--146, 2010.

\bibitem{Kopell1}
J.A. White, C.C. Chow, J.~Ritt, C.~Soto-Trevi{\~n}o, and N.~Kopell.
\newblock Synchronization and oscillatory dynamics in heterogeneous, mutually
  inhibited neurons.
\newblock {\em J. Comput. Neurosci.}, 5(1):5--16, 1998.

\bibitem{Kopell2}
M.A. Whittington, R.D. Traub, N.~Kopell, B.~Ermentrout, and E.H. Buhl.
\newblock Inhibition-based rhythms: experimental and mathematical observations
  on network dynamics.
\newblock {\em Int. J. Psychophysiol.}, 38(3):315--336, 2000.

\bibitem{Williams2010}
Thelma~L Williams.
\newblock A new model for force generation by skeletal muscle, incorporating
  work-dependent deactivation.
\newblock {\em J. Exp. Biol.}, 213(4):643--650, February 2010.

\bibitem{Wong2019}
Aaron~L Wong and John~W Krakauer.
\newblock Why are sequence representations in primary motor cortex so elusive?
\newblock {\em Neuron}, 103(6):956--958, September 2019.

\bibitem{XieHahnSeung}
X.~Xie, R.~H. Hahnloser, and H.S. Seung.
\newblock Selectively grouping neurons in recurrent networks of lateral
  inhibition.
\newblock {\em Neural Comput.}, 14:2627--2646, 2002.

\bibitem{Yokoi2019}
Atsushi Yokoi and Jörn Diedrichsen.
\newblock Neural organization of hierarchical motor sequence representations in
  the human neocortex.
\newblock {\em Neuron}, 103(6):1178--1190.e7, 2019.

\bibitem{Yuste-CPG}
R.~Yuste, J.N. MacLean, J.~Smith, and A.~Lansner.
\newblock The cortex as a central pattern generator.
\newblock {\em Nat. Rev. Neurosci.}, 6:477--483, 2005.

\end{thebibliography}
	\end{singlespace}





\backmatter


\begin{center}
	\textbf{Education}
\end{center}
Ph.D. in Mathematics (expected), The Pennsylvania State University \hfill 2024\\
B.Sc. in Mathematics, Universidad Nacional de Colombia \hfill 2018\\
\begin{center}
	\textbf{Awards and honors}
\end{center}
AMS Sectional Meeting Graduate Student travel award \hfill
Fall 2023\\
Pritchard Dissertation Fellowship \hfill
Fall 2023\\
COSYNE new attendee travel grant\hfill
Spring 2023\\
Departmental teaching award \hfill Fall 2022\\
\begin{center}
	\textbf{Publications}
\end{center}
\begin{itemize}
	\item \textbf{J. Londono Alvarez}, K. Morrison, C. Curto. Sequential control of coexistent attractors using combinatorial threshold-linear networks. \emph{In preparation}. 
	\item C. Lienkaemper, \textbf{J. Londono Alvarez}, H. Santa Cruz, C. Curto. A novel notion of rank for neural data analysis. \emph{In preparation}.
	\item C. Parmelee, \textbf{J. Londono Alvarez}, C. Curto, K. Morrison. Sequence generation in inhibition-dominated neural networks (summary of \emph{Sequential attractors in combinatorial threshold-linear networks}). \emph{\url{https://dsweb.siam.org/The-Magazine/Article/sequence-generation-in-inhibition-dominated-neural-networks} {\textbf{DSWeb Dynamical Systems Magazine}}}
	\item C. Parmelee, \textbf{J. Londono Alvarez}, C. Curto, K. Morrison. Sequential attractors in combinatorial threshold-linear networks. \emph{\url{https://doi.org/10.1137/21M1445120}{\textbf{SIAM Journal on Applied Dynamical Systems}, Vol. 21, Issue 2, pp. 735-1661.}}
\end{itemize}

\begin{center}
	\textbf{Teaching}
\end{center}
Instructor - MATH220 - Matrices \hfill Fall 2021,2023 \\
Grader - MATH535 - Linear Algebra and Its Applications \hfill Fall 2022

\begin{center}
	\textbf{Talks}
\end{center}
Special Session of AMS Sectional Meeting, Creighton University \hfill October 2023
\\ 
CCN Junior Theoretical Neuroscientist’s Workshop, Flatiron \hfill June 2023
\\ 
CoNNExINS, NYU \hfill June 2023
\\
Diversity in Math Bio, SMB, online \hfill June 2023
\\ 
Theoretical biology seminar, Penn State \hfill October 2022
\\ 
Dynamical Principles of Biological and Artificial Neural  \hfill January 2022
\\
Networks Workshop, BIRS

\end{document}